# Eléments de mécanique quasi-statique des milieux granulaires mouillés ou secs

*Pierre EVESQUE*

**Notation:**

*Mécanique:*

| | |
|---|---|
| $\varphi$ | frottement solide |
| $\sigma$ | contrainte |
| F | force |
| v | vitesse |
| p | pression |
| $\varepsilon$ | déformation |

*Géométrique et écoulement*

| | |
|---|---|
| $\rho$ | densité |
| $\rho_s$ | densité du solide |
| $\kappa$ | perméabilité |
| $\Phi$ | porosité |
| e | indice des vides |
| $\nu$ | volume spécifique |

avec $\nu=1+e$ ; $e=\Phi/(1-\Phi)$

# 1. Introduction

Les milieux granulaires composent notre environnement immédiat. L’existence de toute espèce passe plus ou moins directement par leur utilisation et leur compréhension: à l’image du poisson qui sait nager ou du castor qui sait bâtir des barrages, le fourmilion sait construire des tas à la limite de glissement pour chasser. La compréhension de cet environnement paraît tellement intrinsèque à la survie de l’homme que peu de civilisations ont déifié le concept de milieu granulaire.

Ainsi, depuis longtemps, de nombreux procédés industriels reposent sur l’exploitation des poudres et des grains. Ces procédés, pourtant simples et faciles à appréhender (stockages, transports...), ont généralement nécessité, pour leur mise au point, de nombreuses années d'amélioration patiente, la germination d'idées astucieuses et l'utilisation d'associations d'idées et de résultats expérimentaux plus ou moins simples. L’activité économique autour du milieu granulaire ne se limite cependant pas à ces techniques industrielles anciennes. On note actuellement l’émergence de nombreuses applications nouvelles, d’obtention de brevets (céramique, terre armée, pneu-sol, ancre pour du sable, matelas auto-serrant sous vide pour le transport de blessés,...) accompagnés d’une connaissance théorique accrue du comportement rhéologique des milieux granulaires (stabilité des pentes, liquéfaction pendant les tremblements de terre...).

Le champ d’utilisation des milieux granulaires présente ainsi une grande diversité. Cependant, la compréhension des mécanismes mis en jeu dans ces applications reste simple, voire intuitive. Dans une phase de modélisation, il semble naturel d’associer à cette grande diversité de comportements un nombre suffisant de paramètres de contrôle. Or la multiplication de ces paramètres s’oppose à la simplicité recherchée du modèle. Dans cet ouvrage, nous proposons une vue cohérente et unificatrice de la mécanique d'un milieu granulaire, basée sur un nombre très restreint de paramètres de contrôle (contrainte intergranulaire, déformation du milieu, histoire des contacts, pression du liquide interstitiel). Le champ d’évolution de ces paramètres traduit la diversité de comportements recherchée. Nous verrons par ailleurs que i) certains états de la matière, aux grandes déformations, sont des attracteurs et que ii) si l'on ramène la description du comportement dans des axes précis, sa description devient beaucoup plus simple. Le modèle proposé retraduit tout à la fois la multitude de réponses différentes observables, la simplicité de la compréhension du comportement du milieu dans le cas de manutentions simples et répétitives, et les “aberrations” de comportements que l'on observe parfois quand la réponse réelle fournie diffère trop de la réponse prévue.

Ce livre se limite à la description des propriétés mécaniques des milieux granulaires saturés ou non en liquide ou fluide, qui subissent des déformations lentes (régime quasi statique). Le chapitre 2 est introductif; il est consacré à un certain nombre de rappel de mécanique (tenseurs des contraintes et des déformations, équation d'Euler et de Navier-Stokes, loi de Darcy, perméabilité et porosité des empilement, viscosité des suspensions,…). Ces sujets ne seront pas utilisés par la suite, mais ils permettent de recadrer ce livre dans un contexte plus général et de permettre au lecteur d'accéder à ces autres domaines plus facilement.

Dans le troisième chapitre on aborde le domaine propre à ce livre c'est à dire la mécanique quasi-statique des milieux granulaires. On le fera par un exposé didactique des propriétés mécaniques expérimentales; cet exposé est très influencé par l'approche de Roscoe et de l'Ecole Anglaise: après un rappel des principales caractéristiques mécaniques de ces milieux granulaires (frottement solide, cohésion, dilatance, présentation de Mohr-Coulomb), puis le rappel du rôle primordial du fluide interstitiel (gaz ou liquide), nous introduirons l'approximation de Terzaghi qui décompose le milieu en deux milieux effectifs couplés, l'un comprenant l'ossature granulaire, l'autre décrivant l'état du fluide (pression hydrostatique combinée à la loi de Darcy pour décrire l'écoulement). Ceci nous permettra de rappeler un certain nombre d'effets très simples que la présence d'eau peut provoquer (liquéfaction, sable mouvant, stabilité de pente, tic-tac du sablier). Une fois que l'intérêt de la notion de tenseur intergranulaire sera acquise, nous pourront alors décrire la mécanique quasi-statique des milieux granulaires. Un premier jalon sera donné dans la dernière partie du chapitre 3 où nous utiliserons l'approche de Roscoe et al.(1958) et de l'Ecole anglaise (Rowe 1962, Atkinson & Bransby 1977) comme fil directeur: nous nous intéresserons tout d'abord à définir la notion de "sol normalement consolidé" en montrant qu'il correspond à l'état de densité la plus faible possible sous un état de contrainte donné; puis nous décrirons les caractéristiques des différents types d'essais qui servent à étudier la mécanique de ces milieux, en prenant soin de donner leurs conditions aux limites précises; nous nous limiterons à des chemins de compression simple. Nous aborderons ensuite le comportement mécanique précis dans les différents cas précédents en mettant l'accent sur les qualités de contractance ou de dilatance du matériau. Bien qu'incomplète et succincte, cette description présente l'avantage de montrer que le caractère dilatant-contractant du matériau dépend non seulement des conditions initiales du milieu mais aussi du chemin contrainte-déformation qu'il subit. Nous séparerons ainsi les types d'essais en deux grandes classes (essai normalement consolidé, essai surconsolidé), puis nous distinguerons pour chacune d'elles les essais drainés des essais à volume imposé constant (non drainés). Cela nous permettra d'introduire les notions de surface de Hvorslev et de Roscoe (Atkinson & Bransby 1977). Nous conclurons en décrivant comment les mécaniciens des sols récapitulent ces résultats puis en discutant certaines limites du modèle proposé.

Bien que nous espérons que cet exposé du chapitre 3 soit aussi didactique que possible, il utilise d'amblé le langage de la mécanique des sols. Il risque donc de décourager le lecteur non spécialiste. Ce dernier sera alors peut-être totalement découragé s'il entreprend tout de suite la lecture des chapitres 4 et 6, car ceux-ci détaillent encore plus ces résultats expérimentaux; de plus, il les résume à l'aide des conclusions auxquels les mécaniciens des sols sont arrivés. Un tel lecteur aura probablement intérêt à sauter ces deux chapitres et à commencer par la lecture des chapitres 5 et 7 qui sont consacrés à une description théorique simple des propriétés mécaniques des milieux granulaires, quitte à revenir par la suite sur les chapitres 4 et 6. En effet, il n'est pas nécessaire d'avoir assimiler les concepts des chapitres 4 et 6 pour entreprendre la lecture de ces chapitre 5 & 7.

Dans le chapitre 4, nous poursuivrons la description des propriétés mécaniques des milieux granulaires sous chargement simple, telles que les expérimentateurs les décrivent. Nous utiliserons pour cela les essais triaxiaux; nous donnerons les différentes lois de variation auxquelles leurs essais les ont conduits. Cela nous permettra d'introduire précisément les notions de frottement solide, d'état critique, de dilatance. Nous verrons de plus que le volume spécifique critique $v_c$ dépend de la contrainte moyenne mais pas l'angle de frottement. Nous introduirons la notion de stabilité pour montrer quels essais sont stables et quels autres sont potentiellement instables. Ensuite, nous donnerons la loi de dilatance de Rowe (1962) qui relie la dilatance à l'état de contrainte; nous préciserons la notion de surface de Hvorslev en donnant la relation qui lie le rapport q/p du déviateur q de contraintes divisé par la pression intergranulaire moyenne p au pic de contrainte et à l'indice des vides initial.

Nous donnerons ensuite, au chapitre 5, un modèle simple (Evesque & Stéfani 1991) basée sur l'existence d'une fonction de dissipation qui ne dépend que de l'état de contrainte et du mode de déformation (dilatance) pour décrire les lois expérimentales observées au triaxial; ceci permettra de préciser la notion d'état caractéristique et de déduire un certain nombre de traits caractéristiques qui relient les courbes {contrainte-déformation axiale} et {déformation volumique-déformation axiale}. Nous finirons en utilisant un argument issue de la théorie des systèmes dynamiques pour montrer que les surfaces de Hvorslev et de Roscoe sont deux parties de la même surface (Evesque 1999). Nous montrerons que notre prédiction est en accord avec les données expérimentales

Le chapitre 6 abordera le problème des essais cycliques du point de vue expérimental et donnera une clé de leur interprétation. On distinguera les comportements mécaniques obtenus à volume constant (essais non drainés) de ceux obtenus à volume non constant (essais drainés). On montrera dans ce dernier cas l'intérêt de la notion d'état caractéristique, qui permettra de différencier les comportements dilatants (ou sur-caractéristiques) des comportements contractants (ou sous-caractéristiques) (Luong 1978, Habib & Luong 1978). On verra aussi que la différence entre ces deux comportements a lieu pour un rapport de contrainte précis. Enfin on montrera le rôle joué par une permutation ou une rotation de la direction des contraintes majeure et mineure; ceci prendra tout son sens dans le problème de la liquéfaction des sables et des sols. Ce chapitre sera bâti sur l'approche proposé par Luong et Habib.

Le chapitre 7 proposera une modélisation simple des essais triaxiaux sous compression simple; elle est spécialement adaptée à une approche par éléments finis. Ce modèle postulera l'existence d'une loi incrémentale linéaire par zone entre les incréments de contrainte $\delta\sigma$ et les incréments de déformation $\delta\varepsilon$. Pour simplifier, on supposera dans la majeure partie de ce chapitre que la réponse du milieu est isotrope. On montrera qu'une loi aussi simple permet de décrire la majeure partie des résultats des chapitres 4 & 6, et permet même la description d'autres essais (Evesque 1997 & 1999). La seule hypothèse nécessaire pour obtenir ce modèle simple est de considérer un pseudo-coefficient de Poisson $\nu$ qui dépend de l'état de contrainte appliqué au matériau, ou plus exactement du rapport des contraintes principales majeures et mineures $\sigma_1/\sigma_3$. On déterminera les limites de validité du modèle. On montrera par exemple que la réponse devient anisotrope lorsque les déformations deviennent suffisamment grandes. Ceci sera aussi l'occasion de préciser et de généraliser la notion d'état caractéristique. On montrera aussi que la réponse incrémentale $\delta\varepsilon=f(\delta\sigma)$ ne peut être purement linéaire pour permettre de décrire les phénomènes d'hystérésis et les effets de mémoire. On verra enfin que ce modèle est en accord avec la description en terme de système dynamique des surfaces de Hvorslev et de Roscoe, approche que nous avions proposée au chapitre 5. Une des caractéristiques de ce modèle est le faible nombre de paramètre ajustable, puisque le comportement du matériau granulaire y est connu dès qu'on en connaît la réponse contrainte-déformation au triaxial, et que ce comportement est lui-même bien caractéristique. Il est étonnant qu'un modèle aussi simple puisse décrire sans grand calcul les essais oedométriques (à rayon constant) les essais à volume constant, et qu'il permette de rendre compte de la constante de Jaky des terres au repos. Aucun modèle précédent n'a ces capacités et sa simplicité. On cherchera ensuite à étendre ce modèle pour qu'il décrive aussi les essais cycliques. Pour cela, il faudra introduire une variable d'écrouissage; nous chercherons à le faire le plus simplement du point de vue physique. C'est pourquoi, nous nous inspirerons d'une approche proposée initialement par Boutreux et de Gennes pour décrire la densification au tap-tap; mais nous la modifierons pour qu'elle colle plus à la réalité tout en diminuant le nombre de paramètre ajustable.

Le chapitre 8 proposera une approche de la mécanique des contacts à l'échelle microscopique; on y développera une modélisation simple de la distribution des forces de contact basée sur la physique statistique et sur une hypothèse d'entropie maximum (Evesque 1999). Bien entendu ce modèle ne peut décrire un grand nombre de phénomène: à ce stade, il ne permet pas encore de construire la physique statistique des mouvements des grains lors d'une déformation du matériau.

Nous ne parlerons pas de l'écoulement du sablier bien que ce soit un problème très important. En effet les travaux développés indépendamment par Savage et Nedderman se suffisent à eux-mêmes. De plus le livre de Nedderman (1992) traite amplement ce sujet.

Par contre, nous incluons en appendice quelques notions sur la théorie des systèmes dynamiques de manière à ce que la deuxième partie du chapitre 5 soit accessible à un lecteur non spécialiste.

Enfin, nous proposerons dans la conclusion un nouveau schéma de présentation de la mécanique quasi statique des milieux granulaires. Ce schéma de présentation sera basé sur les résultats récents expliqués dans ce livre. Le modèle théorique proposé permet de décrire simplement la majeur partie des résultats expérimentaux et des caractéristiques mécaniques des milieux granulaires. Nous pensons que ce schéma permet de gagner en simplicité et en efficacité et nous espérons qu'il sera la base de l'enseignement futur de mécanique des sols.

## Bibliographie:

# 2. Rappels introductifs

Ce chapitre compile un certain nombre de résultats que l'on trouve exposer de façon éparse dans la littérature. Il rappelle les notions de base de la mécanique des milieux homogènes ; puis il passe à des rappels de mécanique des fluides, au rappel de la loi de Darcy et à l'expression de la perméabilité d'un milieu granulaire.

## 2.1 contraintes, déformations

### *2.1.1. Tenseur des déformations:*

Un matériau homogène se déforme sous l'effet de forces extérieures. Pour caractériser cette déformation, on peut repérer chacun de ses points par son vecteur position **r** dans un repère fixe lorsqu'aucune force n'est appliquée au matériau, puis repérer la nouvelle position **r'** de chacun de ses points lorsqu'un champ de force lui est appliqué. On définit le vecteur déplacement **u**=**r'**-**r** et le tenseur de déformation

(2.1) $$\varepsilon_{ij} = ( \partial u_i/\partial x_j+\partial u_j/\partial x_i)/2$$

De par cette définition $\underline{\underline{\varepsilon}}$ est un tenseur symétrique d'ordre 2 (cf. Landau & Lifchitz 1967).

Lorsque le matériau n'est pas homogène, le champ de déplacement n'est plus uniforme ; il faut donc définir le tenseur de déformation à l'aide de moyenne.

### *2.1.2. Tenseur des contraintes*

Soit un corps en équilibre et considérons un élément de volume de ce milieu. La résultante des forces qui agissent sur ce volume peut être considérée comme l'intégrale sur ce volume des forces **F** qui agissent de l'extérieur de l'élément sur lui car les forces internes s'annulent du fait même du principe de l'action et de la réaction (cf. Landau & Lifchitz 1967) :

(2.2) $$\iiint \underline{\mathbf{F}}dv$$

Si ces forces agissent par l'intermédiaire de la surface de l'élément, ce qui est le cas pour les forces de contact ou à courte portée, cette intégrale triple se transforme en une intégrale de surface. Comme on l'apprend en analyse vectorielle, l'intégrale d'un scalaire étendue à un volume peut être transformée en intégrale de surface si ce scalaire est la divergence d'un certain vecteur. Ici F est un vecteur, l'intégrale de ce vecteur sur un volume peut être transformée en intégrale de surface si F est la divergence d'un tenseur d'ordre 2 noté $\underline{\underline{\sigma}}$ et appelé tenseur des contraintes:

(2.3) $$F_i=\Sigma_j\partial\sigma_{ij}/\partial x_j$$

On pourra ainsi écrire:

(2.4) $$\iiint F_i dv= \iiint \Sigma_j\partial\sigma_{ij}/\partial x_j dv = \iint \Sigma_j \sigma_{ij}\, ds_j$$

où **ds** est un vecteur orienté suivant la normale intérieure à la surface, de norme égale à l'aire de cet élément de surface. Ainsi, la force exercée sur cet élément de surface est:

(2.5) $$\underline{\mathbf{F}} = \underline{\underline{\sigma}}\ \underline{\mathbf{ds}}$$

La force exercée par ce volume sur l'extérieur est égale et opposée à celle-ci:

$$\underline{\mathbf{F}} = -\ \underline{\underline{\sigma}}\ \underline{\mathbf{ds}}$$

On montre aussi que l'équilibre des moments impose que le tenseur $\underline{\underline{\sigma}}$ est symétrique:

$$\sigma_{ij}=\sigma_{ji}$$

## 2.2. Equation d'Euler de la dynamique:

Lorsque le corps est un milieu hétérogène, les forces locales peuvent fluctuer d'un point à un autre ainsi que les mouvements des particules. Cependant si l'on se focalise sur un milieu granulaire, on peut considérer que des moyennes représentatives peuvent être définies lorsqu'on considère des éléments de volume suffisamment grands. Considérons donc un tel élément de volume et appliquons lui l'équation de la dynamique; cet élément est soumis à des forces de surface définies par $\underline{\underline{\sigma}}$, à des forces de volumes tels que son poids ρg dv (nous négligerons les autres forces à distance).
En se rappelant que la différentielle totale d/dt s'écrit aussi $(\partial/\partial t + \Sigma_j\ v_j\ \partial/\partial x_j)$, l'équation de la dynamique impose:

(2.6) $$-\Sigma_j \partial\sigma_{ij}/\partial x_j + \rho g_i = \rho(\partial v_i /\partial t + \Sigma_j\ v_j\ \partial\ v_i /\partial x_j)$$

Si l'élément de volume est en équilibre mécanique, cette équation devient :

(2.7) $$-\Sigma_j \partial\sigma_{ij}/\partial x_j + \rho g_i = 0$$

Un autre problème s'ajoute à celui-ci lorsque par exemple le milieu granulaire est saturé en eau et que cette eau s'écoule dans le poreux; dans ces cas il existe deux phases, l'une liquide, l'autre granulaire. Chacune obéit à une équation d'Euler particulière qui contient en plus un terme de couplage entre les deux phases car le mouvement de chacune de ces deux phases est freiné par l'autre.

## 2.3. Mécanique des fluides: équation de Navier-Stokes

Lorsque le milieu est un fluide incompressible visqueux, le tenseur de contrainte au repos est isotrope et égal à p ; et la contrainte visqueuse est proportionnelle au gradient de la vitesse (viscosité cinématique μ, viscosité dynamique η=μ/ρ). Dans ces conditions, en notant $\Delta = \Sigma_j\ \partial^2/\partial x_i^2$ , l'équation de la mécanique devient l'équation de Navier Stokes:

(2.8) $$\rho\partial v_i \partial t + \Sigma_j\ v_j\ \partial\ v_i /\partial x_j = -\partial p/\partial x_j + \mu\Delta v_i + \rho g_i$$

ou encore

(2.9) $\partial \underline{\mathbf{v}} \partial t + (\underline{\mathbf{v}} \bullet \nabla)\, \underline{\mathbf{v}} = -\nabla p/\rho + \eta \Delta \underline{\mathbf{v}} + \mathbf{g}$

## 2.4. Quelques résultats, définitions et applications

### *2.4.1. nombre de Reynolds*

Les équations 2.8 et 2.9 sont non linéaires par rapport à la vitesse d'écoulement. Ceci est lié à la présence du terme inertiel $(\underline{\mathbf{v}} \bullet \nabla)\, \underline{\mathbf{v}}$. Cependant, ce terme peut être négligé par rapport au terme de frottement visqueux $\eta \Delta \mathbf{v}$ aux faibles vitesses. Considérons un objet de taille l plongé dans un écoulement de vitesse relative v, cet objet perturbe l'écoulement sur une distance de l'ordre de sa taille; on peut donc évaluer $(\underline{\mathbf{v}} \bullet \nabla)\, \underline{\mathbf{v}}$ à $v^2/l$ et $\eta \Delta \mathbf{v}$ à $\eta v/l^2$. Le rapport de ces deux quantités s'appellent le nombre de Reynolds $R_e = vl/\eta = \rho v l/\mu$ . Lorsque $R_e$ est grand, l'écoulement est dominé par les forces inertielles, son comportement est non linéaire et turbulent; Lorsque $R_e$ est petit (<1), l'écoulement est dominé par les forces visqueuses, son comportement est linéaire.

### *2.4.2. traînée d'une sphère et sédimentation d'un grain:*

C'est à partir de ces équations qu'on peut calculer la force de traînée F des objets. Par exemple pour une sphère de rayon R se déplaçant à une vitesse v dans un liquide, on trouve:

(2.10) $F = 6\pi\eta R v (1 + 3Rv/8\eta)$

Dans le cas où v est faible, le terme en $v^2$ peut être négligé et l'on obtient l'équation de Stokes.

Lors de la sédimentation cette force agit sur la particule en sens contraire de la force de pesanteur $(4\pi/3)(\rho_s - \rho_l) g R^3$ . Il existe une vitesse $v_s$ pour laquelle ces deux forces sont égales. C'est la vitesse de sédimentation $v_s$:

(2.11) $v_s = 2(\rho_s - \rho_l) g R^2/(9\eta)$

Cette sphère est par ailleurs soumise à une fluctuation de force liée aux fluctuations de pressions du fluide l'entourant; elle se trouve donc animée d'un mouvement brownien de vitesse caractéristique $v_{th}$. La thermodynamique indique que $v_{th}$ est fonction de la température T et de la masse m de la particule: $v_{th} = (2kT/m)^{1/2} = (3kT/[2\pi\rho R^3])^{1/2}$ . Ceci a plusieurs implications: tout d'abord cela indique qu'en plus du mouvement de sédimentation la particule est animée d'un mouvement brownien d'autant plus important que la masse de la particule est petite. Pour calculer le coefficient de diffusion on peut calculer le temps nécessaire pour que le frottement visqueux stoppe le mouvement brownien et la longueur parcourue pendant ce temps; ces deux quantités caractériserons le pas élémentaire de la marche au hasard. On peut aussi utiliser la relation d'Einstein qui relie le coefficient de diffusion D à la mobilité de la particule b : D=kTb, où T est la température. b est justement le coefficient qui relie la vitesse d'une particule à la force qui est exercée sur cette particule ; soit dans notre cas $b=1/(6\pi\eta R)$ d'après l'Eq. (2.10). Ainsi le coefficient de diffusion d'une particule de rayon R est : $D=kT/(6\pi\eta R)$. A cause de ce mouvement brownien les particules peuvent se rencontrer, puis s'agréger si la force interparticulaire est suffisante; ceci donnera lieu à des agrégats plus ou moins ténus suivant le processus réel d'agrégation (agrégation limitée par la diffusion, agrégation par diffusion d'amas, agrégation balistique,…). Ces agrégats peuvent être décrits en général par une dimension fractale qui dépend du processus d'agrégation et de la dimension de l'espace.

Par ailleurs, lorsque les particules sont petites, la suspension ne sédimente pas complètement, car le mouvement brownien est responsable du mélange des particules lorsque la suspension est à l'équilibre thermodynamique. En effet, à l'équilibre thermodynamique le profil de concentration c doit être tel que le nombre de particules qui traversent une strate horizontale à cause de la sédimentation doit être égal au nombre de particules qui remontent vers le haut du fait de l'agitation thermique et de la différence de concentration entre deux couches horizontales successives. Si on appelle D le coefficient de diffusion, $v_s$ la vitesse de sédimentation et qu'on oriente z vers le bas, ceci conduit à l'équation :

$$v_s\, c(z)=D\partial c/\partial z$$

Soit à un profil de concentration $c=c_o \exp(v_s z/D)= c_o \exp[(4\pi(\rho_s-\rho_l)gR^3)z/( 3T)]$ . Pour que la suspension ne sédimente pas trop et reste fluide, il faut que le terme exponentiel ne varie pas trop vite sur une hauteur égale à la taille d'une particule, c'est-à-dire qu'il faut que $4\pi(\rho_s-\rho_l)gR^4/3=\delta mgR$ reste inférieur ou très inférieur à kT.

En fait, lorsque la solution est en écoulement, d'autres phénomènes de mélange peuvent avoir lieu ; elle peut par exemple être le lieu de turbulences qui mélangent les particules comme le ferait un mouvement brownien et on peut définir un coefficient de diffusion de turbulence $D_{turbulence}$ ; cette turbulence permet de “ solvater ” des particules plus grosses et d'en assurer le transport ; c'est en partie ce qui se passe dans les rivières et les torrents (cf. P.Y. Julien, 1992).

### *2.4.3. Viscosité d'une suspension*

Les suspensions très concentrées ont des rhéologies très compliquées pouvant imposer un ordre à plus ou moins courte distance. Nous nous limiterons à citer le résultat d'Einstein pour la viscosité d'une suspension contenant une fraction volumique c de particules sphériques de diamètre R ($c=4\pi R^3 n/3$), où n est le nombre de particules par unité de volume. Dans la limite des faibles concentrations, on obtient:

(2.12a) $$\eta= \eta_o[1+c(\eta_o +5\eta_1 /2)/(\eta_o + \eta_1)]$$

où $\eta_o$ est la viscosité du liquide pur et $\eta_1$ la viscosité des gouttes (cf. G.K. Batchelor). On retrouve la formule d'Einstein lorsque les gouttes sont des sphères rigides ($\eta_1=\infty$):

(2.12b) $$\eta= \eta_o (1+5c/2)$$

Lorsqu'on applique cette formule à des bulles de faible viscosité cinématique ($\eta_1 << \eta_o$), on trouve:

(2.12c) $$\eta = \eta_o (1+c)$$

***2.4.4. débit d'une conduite:***

Le débit Q d'une conduite circulaire de diamètre D, de longueur l soumis à une différence de pression $\delta p$ est un problème classique de mécanique des fluides. On trouve (cf. L. Landau et E. Lifchitz, G.K.Batchelor) que la vitesse v du liquide à une distance r du centre de la conduite est:

(2.13a) $$v = \delta p(R^2-r^2) /(4\eta l)$$

et que le débit volumique vaut:

(2.13b) $$Q=\pi \delta p R^4/(8\eta l)$$

***2.4.5. liquides dans les poreux:***

Un milieu granulaire comporte un ensemble de pores qui forment un réseau continu connecté à l'extérieur et à travers lequel peut s'écouler un fluide. C'est ce qu'on appelle la porosité ouverte. Il ne faut pas confondre cette quantité avec la porosité fermée, ni avec la porosité totale. En effet, certains matériaux contiennent des pores bouchés et donc non reliés à l'extérieur, d'autres des grains contenant des vides comme dans le cas de la pierre ponce, soit parce que certains pores sont fermés par des dépots imperméable,…. On parle dans ce cas de porosité fermée. La porosité totale est évidemment la somme de ces deux porosités.

Nous ne considérerons ici que la porosité ouverte qui peut donc être complètement saturée en eau et qui peut être le lieu d'un écoulement de liquide. Nous la noterons $\Phi$. Supposons de plus que les grains sont fixes, liés les uns aux autres de manière à former une structure rigide et considérons le cas où l'écoulement fluide est permanent et de vitesse faible, c'est-à-dire telle que est telle que le nombre de Reynolds $R_e = v\, d_p /\eta$ est inférieur à 1 si la taille typique des pores est $d_p$. Dans ce cas, l'Eq. (2.9) devient:

(2.14) $$\nabla p = \mu \Delta \underline{v} + \rho_f\ \mathbf{g}$$

où $\mu$ et $\rho_f$ sont la viscosité dynamique et la densité du fluide. On peut alors considérer que l'écoulement à travers le poreux est représenté par sa vitesse typique $\mathbf{v}$, que cette vitesse typique est fonction du gradient de pression moyen, de la section typique des pores, de la géométrie de leur connection et de leur statistique. On posera ainsi :

(2.15) $$\nabla \mathbf{p} = -(\mu/\kappa)\ \mathbf{v} + \rho_f\ \mathbf{g}$$

où $\mathbf{v}$ et $\mathbf{p}$ sont des valeurs moyennes de la vitesse et de la pression du liquide et où $\kappa$ est un coefficient de proportionalité appelée perméabilité. $\underline{\underline{\kappa}}$ est a priori un tenseur d'ordre 2, mais peut être un scalaire si le poreux est isotrope . L'Eq. (2.15) est appelée loi de Darcy. Pour préciser la définition de v, il est bon d'écrire le débit Q de fluide à travers une surface S de poreux orientée perpendiculairement à l'écoulement. Avec cette défintion on a: Q= S $\mathbf{v}$
De plus, un liquide étant incompressible et les pores étant saturés, la conservation de la matière impose par ailleurs:

(2.16) $$\nabla.\underline{v} = 0$$

Dans le cas où les conditions aux limites sont telles que les pressions sont imposées au bord, la combinaison des Eqs. (2.15) et (2.16) donne donc:

(2.17) $$\Delta\, \mathbf{p} = 0$$

Ayant résolu l'Eq. (2.17) pour les conditions aux limites de pressions données, on déduit le champ de vitesse de la loi de Darcy (Eq. 2.15) et les caractéristiques de l'écoulement.

***•2.4.6. Perméabilité d'un empilement de grains***
Certains modèles permettent de calculer la perméabilité d'un ensemble de grains. La théorie de Carman (1937) & Kozeny (1927) conduit à la perméabilité:

(2.18) $$\kappa = (D_{p2})^2\, \Phi^3 / [180(1-\Phi)^2]$$

avec $D_{p2} = \int (D_p)^3\, h(D_p)\, dD_p \;/ \int (D_p)^2\, h(D_p)\, dD_p$

où $h(D_p)$ est la distribution de taille de billes de diamètre $D_p$. Cette expression est d'autant meilleure que la distribution est resserrée et que les grains sont sphériques. (*cf.* D.A. Nield & A. Bejan, p. 6)

***2.4.7. Extension de la loi de Darcy aux mouvements rapides ou accélérés***
Dans un poreux, la turbulence hydrodynamique ne peut apparaître qu'à l'échelle inférieure au pore.
L'extension de la loi de Darcy aux mouvements accélérés ne peut donc pas se faire de façon analogue à celle aboutissant à l'équation d'Euler, car le terme $\underline{v}\bullet\underline{\nabla}(\underline{v})$ n'a aucun sens physique du fait de l'existence des parois. De même le terme $\partial v/\partial t$ ne doit exprimer qu'un terme de relaxation des régimes transitoires dus à la viscosité du liquide et au mouvement du fluide dans les pores. On trouve ainsi que l'équation dynamique doit prendre la forme:

(2.19) $$\rho_f\,.c_a\,.\partial\, \mathbf{v} /\partial t = -\,\underline{\nabla}\, \mathbf{p} - (\mu/\kappa)\mathbf{v}$$

avec $$c_a = \gamma^2\, (D_{p2})^2/(\kappa\lambda_1)$$

où $\rho_f$ est la densité du fluide, $\gamma$ une constante et où $\lambda_1$ =2.405 est la plus petite racine positive de la fonction de Bessel $J_o$ de première espèce. (cf. D.A. Nield & A. Bejan, p. 8).
Enfin, lorsque la vitesse typique du fluide est grande (10>Re>1), il est nécessaire d'introduire un terme quadratique en fonction de la vitesse pour tenir compte des effets inertiels dans l'équation de Darcy:

(2.20) $\underline{\nabla}\ \mathbf{p} = -(\mu/\kappa)\ \mathbf{v} - c_f\ \kappa^{1/2} \rho_f\ \mathbf{v}^2$

où $c_f$ est un coefficient. L'équation qui est considérée comme présentant la meilleure corrélation entre résultats expérimentaux est due à Ergund (1952):

(2.21) $\underline{\nabla}\ \mathbf{p} = -E_1\ \mathbf{v} - E_2\ \mathbf{v}^2$

avec $E_1 = (15\mu/18\kappa)$ et $E_2 = 1{,}75\ \rho_f\,(1-\Phi)/(D_{p2}\,\Phi^3)$

qui se réécrit en fonction du nombre de Reynolds $R_e = \rho_f\ \mathbf{v}\ D_{p2}/[\mu(1-\Phi)]$:

(2.22) $\underline{\nabla}\ \mathbf{p} = -E_1\ \mathbf{v}\ (1 - 0{,}0117\ R_e)$

### *2.4.8. Fluidisation d'un milieu granulaire*

Dans certains procédés industriels, le milieu granulaire est traversé par un courant d'air ou de liquide dirigé vers le haut. Si le courant est suffisant, il peut y avoir équilibtre entre les forces de pesanteurs et la trainée visqueuse. Si on appelle $\rho_s$ et $\rho_f$ les densités du solide et du liquide, ceci a lieu lorsque:

(2.23) $(1-\Phi)(\rho_s - \rho_f)g = \underline{\nabla}\ \mathbf{p} = (\mu/\kappa)\ \mathbf{v}$

ce qui fixe la vitesse $\mathbb{Y}$ du fluide et donc le débit $Q=S\mathbb{Y}$ par l'intermédiaire des équations 2.15 & 2.18. Comme la perméabilité $\kappa$ (i.e. $\kappa = (D_{p2})^2\ \Phi^3/[180(1-\Phi)^2]$) dépend de la porosité du milieu de façon non linéaire, le système pourra être stable ou instable au seuil de fluidisation suivant qu'on travaille à $\nabla p$ imposé ou à débit constant: à $\nabla p$ imposé une diminution de la porosité augmente les forces de pesanteurs et supprime la fluidisation; à débit imposé une diminution de la porosité augmente les forces de pesanteurs, mais augmente encore plus la traînée visqueuse, ce qui fait passer le système au delà du seuil de fluidisation.

## Bibliographie:

# 3. Introduction à la mécanique quasi-statique des milieux granulaires:

Le but de ce chapitre est d’introduire les principes de base de la mécanique quasi-statique des milieux granulaires secs ou mouillés. Nous rappellerons donc tour à tour ce qu’est le frottement solide, la cohésion et la dilatance. Nous montrerons ensuite comment tenir compte des effets de l’eau. Cela nous permettra d'introduire la notion de contrainte effective, de discerner les notions de contrainte totale, notée $\sigma$,de contrainte réellellement supportée par le milieu granulaire, notée $\sigma'$, et de pression du liquide saturant les pores, noté $u_w$.
Puis nous décrirons quelques méthodes pour mesurer le frottement, la cohésion et la dilatance et donnerons la définition de quelques chemins de contraintes simples. Pour cela nous commencerons par définir un sable normalement consolidé et son comportement. C’est une notion qui est peu utilisée dans la pratique, car il est très difficile de fabriquer un sable suffisamment lâche pour réaliser facilement des expériences sur cet état. Cependant, c’est une notion très utile, qui permet d’élargir les résultats du sable aux argiles. Ensuite nous étudierons le comportement d’un sable surconsolidé. Nous utiliserons quelques uns de ces chemins pour chercher à quantifier l’évolution des propriétés mécaniques au cours de ces chemins; nous représenterons cette évolution dans un diagramme volume spécifique v, contrainte de cisaillement $q=\sigma_1-\sigma_3=\sigma'_1-\sigma'_3$, et pression $p'=(\sigma'_1+\sigma'_2+\sigma'_3)/3$. L’intérêt d’utiliser cet espace à trois dimensions est de montrer qu’il est suffisant pour décrire la mécanique d’un milieu granulaire dans la majorité des cas classiques. Cependant, on ne peut pas considérer que l’espace (v,q,p) est l’espace des phases complet puisqu’il ne peut pas tenir compte de l’évolution de l’anisotropie des distributions de contacts, paramètre qui joue un rôle important dans certtains mécanismes.
Nous rappelons que l’indice des vides, noté e, correspond au rapport du volume des vides et du volume du solide; il s’exprime en fonction de la porosité $\Phi$ comme $e=\Phi/(1-\Phi)$ . Par définition, le volume spécifique est relié linéairement à l’indice des vides v=1+e. Il est facile de présenter les résultats dans l’espace (q,p,e) à partir de la représentation (q,p,v) et l’on utilisera l’un ou l’autre de ces schémas alternativement.

## 3.1. Le milieu granulaire sec:

### *3.1.1. Frottement*

On sait depuis longtemps que la mécanique d’un milieu granulaire est régie par trois grandeurs essentielles: le frottement entre grains, la cohésion intergranulaire et la dilatance. On sait par exemple qu’un talus peut présenter une surface libre inclinée. Coulomb (1773) introduisit un coefficient de frottement solide pour rendre compte de ce phénomène; il montra en particulier qu’une couche parallèle à la surface libre est en équilibre tant que la composante T de son poids parallèle à cette surface est inférieure à une fraction k de sa composante N normale à cette surface ($T < k N$). k est appelé le coefficient de frottement solide (Fig. 3.1); selon ce modèle, k est relié à l'angle maximal $\theta_{max}$ du talus par $k=\tan\varphi=\tan\theta_{max}$. Un point remarquable est que l'on ne sait toujours pas calculer φ à partir des propriétés mécaniques de chaque grain et de leur forme.

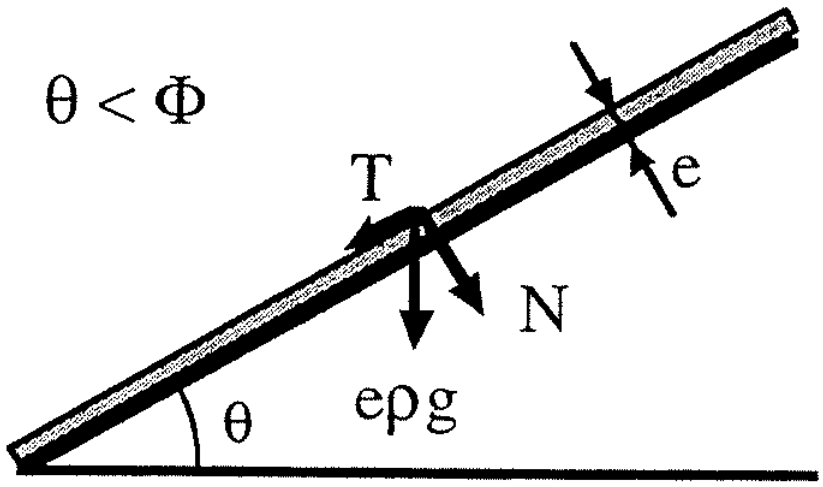

***Figure 3.1*** *- Equilibre d'un talus naturel: une couche de terrain, d'épaisseur e et de masse volumique ρ, inclinée d'un angle θ par rapport à l'horizontale reste stable tant que les composantes du poids de la couche, parallèle* (T) *et perpendiculaire* (N) *à la pente, sont telles que* $T < kN$ *où* k *est le coefficient de frottement caractéristique de l'interface étudiée. Il existe donc un angle maximal de talus* $\theta_{max}$ *tel que* $k = \tan\varphi$.

### *3.1.2. Frottement et cohésion*

En fait, Coulomb a aussi montré que ce résultat pouvait être modifié par la présence d'une certaine cohésion entre les grains et qu'il ne s'appliquait en toute rigueur que pour les couches très profondes, là où la cohésion intergranulaire devenait négligeable devant les forces de frottement. C'est pourquoi ce résultat doit s'appliquer dans le cas d'un talus présentant une pente très longue, où seule la pente locale peut être plus importante, voire verticale par endroits. Ainsi, on ne peut bâtir des châteaux de sable avec des murs verticaux que de hauteur limitée ou créer des voûtes et des surplombs que de taille finie, (si l'on exclut les effets de dilatance que nous verrons par la suite).

Coulomb a proposé de tenir compte de ces deux phénomènes (frottement solide et cohésion) en ajoutant leur contribution indépendamment l'une de l'autre. Il supposa de plus que la cohésion est isotrope. Dans ces conditions, un moyen simple de poser le problème mathématique est de raisonner en termes de contraintes. Soit $\sigma_1$, $\sigma_2$ et $\sigma_3$ les contraintes principales au sein du matériau, qui sont a priori non isotropes. On appellera $\sigma_1$ la contrainte principale majeure et $\sigma_3$ la contrainte mineure. L'équilibre mécanique sera assuré si pour n'importe quelle orientation θ du plan choisi, la contrainte de cisaillement $\tau_p$ à laquelle est soumis ce plan, diminuée de la cohésion c (car elle renforce la structure), est inférieure au frottement solide engendré par la contrainte normale $\sigma_p$:

$$(3.1) \qquad |\tau_p| - c < k\,|\sigma_p|$$

Dans le cas contraire, l'équilibre n'est pas préservé, ce qui se traduit par l'apparition d'effondrements ou de bandes de glissement. Il faut donc chercher l'orientation θ du plan pour lequel $[|\tau_p| - c] / |\sigma_p|$ est le plus grand. $\tau_p$ et $\sigma_p$ s'expriment en fonction de l'orientation θ du plan considéré par rapport à la direction de la contrainte principale majeure $\sigma_1$, de $\sigma_1$ et de la contrainte principale mineure $\sigma_3$ (équations traditionnelles proposées par exemple dans Schofield & Wroth (1968), Cordary (1994)). En effet si **n** et **t** sont les vecteurs normal et tangentiel à ce plan, on a $\sigma_p = \mathbf{n\sigma n} = \sigma_1 \cos^2\theta + \sigma_3 \sin^2\theta$ et $\tau_p = \mathbf{t\sigma n} = (\sigma_1-\sigma_3)\sin\theta\cos\theta$ , soit:

$$\tau_p = (\sigma_1-\sigma_3)\,[\sin(2\theta)]/2 \qquad (3.2)$$

$$\sigma_p = (\sigma_1+\sigma_3)/2 + (\sigma_1-\sigma_3)[\cos(2\theta)]/2$$

### *3.1.3. Plan de Mohr-Coulomb, cercle de Mohr*

Le lieu des points à considérer dans le plan $(\tau_p, \sigma_p)$ est donc un cercle, appelé cercle de Mohr, de rayon $R = (\sigma_1-\sigma_3)/2$, centré sur l'axe des σ à l'abscisse: $(\sigma_1+\sigma_3)/2$.

Le lieu où l'inégalité (3.1) d'équilibre est vérifiée est donné par la section du plan délimitée par les deux demi-droites d'équation $\tau_p - c = k\,\sigma_p$ et $\tau_p + c = -k\,\sigma_p$. Si le cercle de Mohr est totalement à l'intérieur de ce secteur, le système est stable; s'il coupe l'une ou l'autre des demi-droites, il est instable pour certaines orientations θ des plans. L'équilibre limite du système est réalisé quand le cercle de Mohr est tangent à ces droites soit quand (Fig. 3.2):

$$(\sigma_1-\sigma_3)/2 + c = k\,[(\sigma_1+\sigma_3)/2 + (\sigma_1-\sigma_3)[\cos(2\theta)]/2] \qquad (3.3)$$

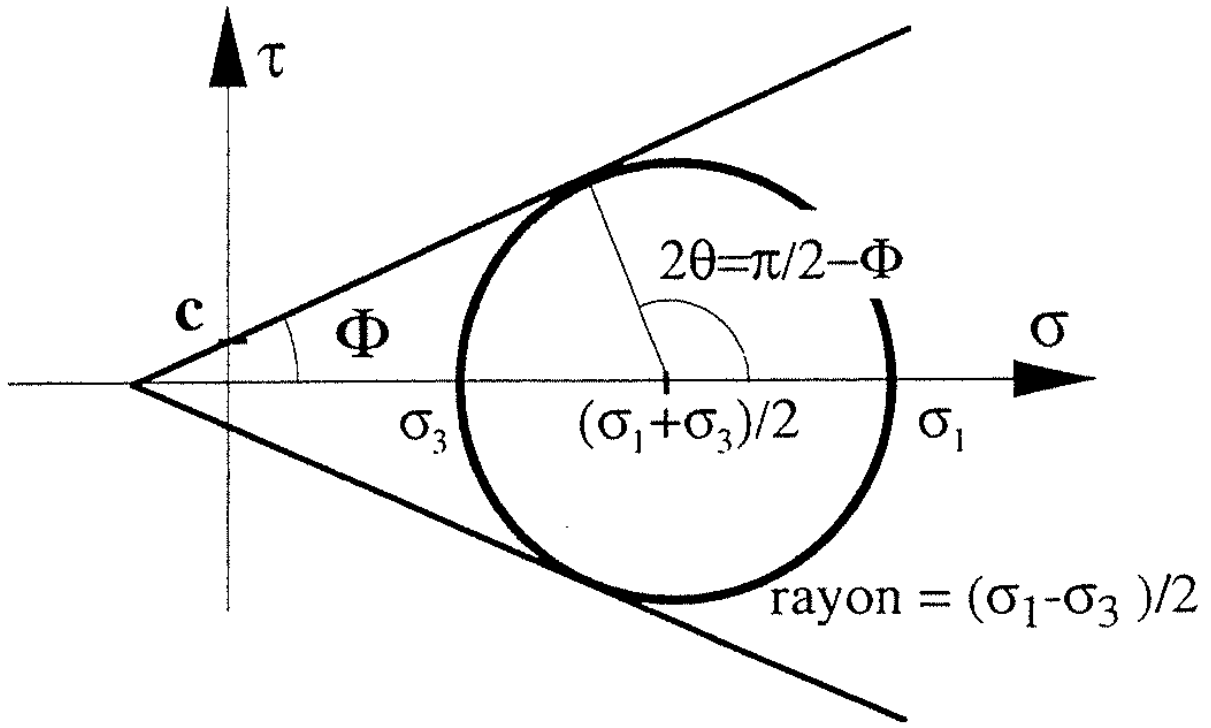

***Figure 3.2*** *- Cercle de Mohr, Plan de Mohr-Coulomb: l'Eq.(3.2) fournit, en fonction des contraintes principales majeure et mineure, l'expression des contraintes parallèle et normale à une facette d'orientation* θ *par rapport à la direction de la contrainte principale majeure. Le lieu de ces points est un cercle (cercle de Mohr). L'équilibre mécanique, donné par l'Eq. (3.1), est assuré tant que le cercle de Mohr précédent reste à l'intérieur du secteur délimité par deux demi-droites paramétrées par* c *et* k = tan φ. *Lorsque le cercle de Mohr est tangent à ces frontières, le glissement s'oriente dans une direction formant un angle* π/4-φ/2 *avec la direction de la contrainte majeure.*

Si le système est à l'équilibre limite, cet équilibre limite n'est réalisé que pour un plan d'orientation θ bien défini: $\varphi+(\pi-2\theta) = \pi/2$ . Le glissement s'il a lieu se fera parallèlement à ce plan; soit dans une direction:

(3.4) $$\theta = \pi/4 - \varphi/2$$

par rapport à la direction de la contrainte principale majeure. **Attention:** θ est l'angle que fait la normale au plan de glissement avec cette contrainte principale majeur

### *3.1.4. Dilatance*

Reynolds (1885) fit une série de constatations remarquables qui lui permit de conclure qu'un milieu granulaire ne pouvait se déformer notablement qu'en se dilatant préalablement: en effet, il faut que certains grains puissent se glisser dans les vides laissés entre les autres pour que les grains puissent bouger les uns par rapport aux autres; ceci nécessite que les pores soient de grande taille et donc que le milieu soit suffisamment dilaté lors de la déformation. Ainsi le milieu devra en général se dilater avant que la déformation n'ait pu avoir lieu. Reynolds appela ce phénomène l'effet de dilatance. Il expliqua ainsi l'assèchement que l'on observe sur le pourtour du pied lorsqu'on pose le pied sur une plage humide; cette sensation d'assèchement est créée par l'augmentation du volume des pores, sans augmentation du volume de l'eau contenue dans les pores.

Bien entendu, un matériau granulaire peut être fabriqué à différentes densités suivant la méthode de tassement utilisé; ceci veut dire que l'effet de dilatance observé sera d'autant plus fort que la densité initiale du tas sera grande; de même, nous verrons que la déformation peut engendrer une diminution du volume total du tas, c'est-à-dire une contractance, lorsque le tas est trop lâche.

En fait, le phénomène de dilatance est encore un peu plus complexe: nous verrons (i) qu'un milieu granulaire sous contrainte isotrope doit d'abord se contracter aux très faibles déformations, (autrement son état mécanique serait instable), (ii) qu'il se dilate ensuite (ou au contraire poursuit sa contraction s'il est trop peu dense) et (iii) que la dilatation globale du milieu est fonction non seulement de la densité initiale, mais aussi de la contrainte moyenne qui lui est appliquée: plus cette contrainte sera faible, plus le phénomène de dilatance sera important à densité initiale donnée. Ces derniers points n'ont pas été observés par Reynolds, mais ont été étudiés plus récemment grâce à l'utilisation de nouveaux moyens de mesures tels que le test triaxial.

## 3.2 Influence de l'eau

Il semble, de prime abord, que le comportement macroscopique d'un matériau granulaire doive être très différent lorsque ce dernier est sec ou au contraire totalement saturé de liquide. C'est ce premier a priori qu'il nous semble nécessaire de lever le plus rapidement possible. En effet, toutes les études expérimentales prouvent que l'on peut traiter l'eau et le milieu granulaire comme deux phases séparées tant que les grains restent en contact les uns avec les autres et que les forces de contact sont non nulles. Le tenseur de contrainte totale est égal dans ce cas à la somme des tenseurs de contrainte correspondant à chacune des phases.

C'est Terzaghi (1925) qui fit le premier cette approximation. Deux cas peuvent alors se présenter suivant qu'il existe ou non un courant de liquide à travers les pores du milieu granulaire.

### *3.2.1. Absence de courant*

Si aucun courant n'a lieu, le tenseur de contrainte totale $\{\sigma_{tot1}, \sigma_{tot2}, \sigma_{tot3}\}$ appliquée sur l'échantillon se décompose en une pression $u_w$ exercée sur l'eau interstitielle et un tenseur $\{\sigma_1', \sigma_2', \sigma_3'\}$ qui décrit les contraintes supportées par le squelette granulaire de telle sorte que la contrainte totale s'écrit :

$$(3.5) \qquad \{\sigma_{tot1}, \sigma_{tot2}, \sigma_{tot3}\} = \{\sigma_1' + u_w, \sigma_2' + u_w, \sigma_3' + u_w\}$$

Dans ces conditions, les lois d'équilibre du tenseur $\sigma'$ restent les mêmes que dans le paragraphe précédent, c'est-à-dire qu'elles sont données en première approximation par les lois de Mohr-Coulomb, auxquelles on peut rajouter les effets de dilatance.

### *3.2.2. Effet d'un courant*

Si un courant $\underline{\mathbf{j}}$ traverse le milieu granulaire dans la direction r, il est lié à l'existence d'un gradient de la pression de liquide $d(u_w)/dr$. Ce courant massique $\underline{\mathbf{j}}$ est donné par la loi de Darcy:

$$(3.6) \qquad \underline{\mathbf{j}} = \eta_l \rho_l \underline{\underline{k}}\, \mathbf{grad}\, (u_w)$$

où $\eta_l$ et $\rho_l$ sont la viscosité cinématique du liquide et sa masse volumique et où $\underline{\underline{k}}$ est la perméabilité du milieu granulaire. $\underline{\underline{k}}$ dépend non seulement de la dispersion granulométrique du milieu granulaire, mais aussi de la structure réelle de l'empilement et donc de la densité réelle locale et de l'anisotropie du milieu; c'est cette dernière dépendance qui impose sa nature tensorielle à la perméabilité.

Bien entendu, on sait que l'écoulement d'un liquide à travers un milieu poreux peut être plus complexe, suivant la complexité réelle de la topologie locale des " conduites ", et par exemple de la nature fractale du matériau; dans ce cas la perméabilité du milieu n'est plus une grandeur homogène et peut dépendre de l'échelle de longueur de l'expérience. Cependant, les mêmes remarques pourraient être faite aussi en ce qui concerne la cohésion et le frottement solide. Ainsi les lois de cohésion et de frottement solide que nous avions utilisées au paragraphe précédent imposent que le système puisse être considéré comme homogène à l'échelle de la mesure; de la même manière, on doit garder la loi de Darcy sous la forme de l'Eq. (3.6) pour rester cohérent avec ce degré d'approximation. La même remarque tient aussi pour l'Eq. (3.5).

La perte de charge du liquide exprimée par la loi de Darcy est bien entendu engendrée par le frottement visqueux de l'eau. En conséquence, elle correspond à un accroissement de la contrainte appliquée sur les grains, tout au moins si l'on néglige les termes inertiels.

### *3.2.3. Equation de continuité*

Lorsque le milieu granulaire est complètement saturé d'un liquide, une condition supplémentaire existe, donnée par l'équation de conservation de la matière. Si on appelle $\rho_o$ la masse volumique d'un grain, $\rho_s$ celle du milieu granulaire sec, ($\rho_s$ dépend de la porosité du matériau), $\rho_l$ celle du liquide, et $\alpha$ le rapport de la densité d'un grain solide à celle du liquide, on doit avoir dans le référentiel lié à la structure granulaire:

(3.7) $\alpha \operatorname{div} \mathbf{j} + d(\rho_s)/dt = 0$

Dans le cas d'un gaz de masse volumique $\rho_g$, le milieu est compressible et l'Eq. (3.7) n'est plus valable et doit être remplacée par:

(3.8) $\rho_o \{\operatorname{div} \mathbf{j} + d(\rho_g)/dt\} + \rho g\ d(\rho_s)/dt = 0$

à laquelle doit être associée la loi de variation du volume du gaz avec la pression, loi qui dépend elle-même de la nature de la transformation thermodynamique du gaz (adiabatique, isotherme,...).

### *3.2.4. Conditions drainées et non drainées*

On peut comprendre pourquoi le même échantillon peut avoir des comportements différents suivants les conditions opératoires. Comparons en effet ce qui se passe lorsque l'on utilise un sable sec ou un sable saturé d'eau.

Un sable sec peut se dilater ou se contracter facilement. Par contre, pour qu'un sable mouillé puisse se contracter ou se dilater, il faut qu'il évacue de l'eau de ses pores ou qu'il s'en imbibe. Ceci peut être rendu plus ou moins facile suivant la vitesse à laquelle cette contraction ou dilatation doit être réalisée et selon le volume total concerné: En effet, la loi de Darcy montre que le courant est proportionnel au gradient de pression d'eau. Supposons par exemple que le centre d'un massif subissant une contrainte intergranulaire $\sigma$ doive se dilater brutalement. Pour que cela puisse se faire, il faut créer un débit d'eau j et donc un gradient de pression d'eau. Il en résulte une dépression $\delta u_w$ de la pression de l'eau interstitielle au sein du massif qui ne peut être réalisée que par le rééquilibrage entre contraintes intergranulaires et pression hydraulique ($\sigma_{tot} = \sigma' + u_w + \delta u_w = c^{ste}$). Il en résulte en particulier une diminution de la contrainte isotrope supportée par le milieu granulaire et une diminution de la limite de résistance du milieu au cisaillement. L'importance de cette diminution dépendra des conditions d'évacuation et d'imbibition de l'eau.

Les mécaniciens des sols se limitent en général à l'étude expérimentale des deux cas extrêmes drainé et non drainé, puis se servent des résultats pour traiter un problème quelconque en utilisant des calculs par éléments finis.

### *3.2.5. Exemples*

- *Premier exemple: stabilité d'une pente sèche et mouillée*

La pente limite est $\theta_{max}=\varphi$ si l'on néglige les phénomènes de dilatance. Si le milieu est saturé d'eau et que l'eau ne s'écoule pas, la contrainte de cisaillement que doit supporter une section plane parallèle à la surface (pente $\theta$) est celle produite par l'eau et par le milieu granulaire: $(m_w+m_g)g \sin\theta$ ; la contrainte intergranulaire normale à la surface reste la même: $m_g g\cos\theta$; la pente maximum est donnée par la valeur de la contrainte de cisaillement maximale admissible: $\tau = m_g\ g\ \cos\theta\ tg\varphi$ . Ainsi la pente maximum admissible est donnée par $tg\theta = tg\varphi\{m_g/(m_w+m_g)\}$ qui dépend de la porosité du milieu et de la densité relative de l'eau et des grains. C'est pourquoi les pentes en fond de vallée qui sont souvent saturées d'eau sont plus faibles que sur les côtés. La situation se complique s'il y a écoulement d'eau parallèlement à la surface, ce qui rajoute un cisaillement supplémentaire et diminue la stabilité de la pente. C'est pourquoi les glissements naturels de terrain ont lieu après les pluies.

- *Autres exemples: liquéfaction, sable mouvant, tic-tac du sablier,...*

D'autres phénomènes bien connus sont liés à la présence de l'eau: par exemple, on sait qu'un séisme peut liquéfier un sol saturé d'eau. Le mécanisme mis en jeu dans ce phénomène est le suivant: on verra au chapitre 6 que le cisaillement horizontal cyclique qui correspond à la secousse tellurique tend à densifier le sable, mais ceci ne peut arriver à cause de la présence de l'eau qui sature les pores; il en résulte une augmentation de la pression de l'eau interstitielle et une diminution de la contrainte intergranulaire, qui s'annule même par intermittence lorsque le séisme est suffisamment important. Ceci peut déstabiliser les immeubles et provoquer une éruption de l'eau et la formation d'un cratère.

Le cas des sables mouvants fait intervenir un mécanisme probablement différent; ces sables sont le lieu de courants verticaux ascendants d'eau, sous l'effet d'une mise en charge locale, qui mettent le sable en suspension; ce dernier perd ainsi toute capacité portante.

Le “ tic-tac ” du sablier est provoqué par l'existence d'une surpression périodique de l'air à l'embouchure qui stabilise pendant un instant une voûte de sable et provoque l'arrêt de l'écoulement du sable. Puis l'air filtre lentement à travers les pores, faisant diminuer la différence de pression d'air entre chambre haute et chambre basse, ce qui déstabilise la voûte précédente et permet l'amorçage d'un nouvel écoulement de sable; celui-ci provoque à son tour une augmentation de la pression de l'air dans la chambre du bas et la restabilisation de la voûte. Et ainsi de suite.

Les effets inverses sont possibles: Reynolds pour démontrer le phénomène de dilatance a utilisé une outre remplie de sable et juste saturée en eau. Il a montré i) l'augmentation très forte de la résistance mécanique d'un milieu granulaire soumis à une compression uniaxiale si on l'oblige à garder son volume constant, et sinon ii) la sensation d'assèchement que l'observateur ressent visuellement au moment où il pose son pied sur une plage de sable humide.

Enfin, la présence d'eau dans un massif peut permettre de stabiliser temporairement les parois d'une excavation (tranchée par exemple). En effet, le mouvement du terrain cherche à créer une dilatation du milieu, engendrant ainsi une diminution de la pression interstitielle, ce qui rend le milieu “cohésif”.

## 3.3. Les différents types d’essais

Comme nous l’avons montré à l’aide de quelques exemples, la rhéologie d’un milieu granulaire dépend des contraintes qu’il subit. Le caractère polyphasique de ce milieu granulaire (grains et fluides interstitiels) permet d’envisager de multiples conditions expérimentales:

♣ Différentes conditions de drainage:

i) Le drainage du fluide interstitiel et la conservation du volume du milieu granulaire sont liés:
   - en condition **non drainée**, le sable est astreint à garder son volume constant. La pression de l’eau interstitielle peut rester constante ou non suivant les cas pratiques étudiés;
   - en condition **drainée**, l’eau interstitielle est libre de circuler afin d’équilibrer sa pression. Si les déformations sont très lentes, et la taille des pores grande, on parle alors de sable **parfaitement drainé**.

♣ Autres différents types d’essais:

ii) On peut imposer l’absence de dilatation latérale en exploitant la symétrie; l’essai **oedométrique**, durant lequel on confine l’échantillon de sable dans un moule cylindrique rigide, en est un exemple.

iii) On impose à la pression moyenne $p=(\sigma_1+\sigma_2+\sigma_3)/3$ ou bien seulement à la contrainte latérale $\sigma_2=\sigma_3$, dans le cas où une symétrie axiale existe, de rester constante. On parlera ainsi d’essai à pression moyenne p constante, à $\sigma_3$ constant,... De telles conditions sont envisageables sur un appareil **triaxial**.

De nombreux appareillages sont utilisés pour caractériser le comportement des matériaux granulaires (boîte de cisaillement ou de Casagrande, scissomètre, oedomètre, pressiomètre, pénétromètre, triaxial...). Nous nous limiterons à la description de deux d’entre eux.

### *3.3.1. L’oedomètre*

L’essai oedométrique consiste à placer un échantillon dans une cellule cylindrique, considérée rigide, et à le solliciter en compression. Cet essai est réalisé en conditions drainées, des pierres poreuses circulaires étant interposées entre l’échantillon et les pistons. La hauteur de l’échantillon évolue sous le chargement axial choisi et selon la cinétique de drainage (perméabilité du milieu),. Ceci permet l’identification du tassement de l’échantillon en traçant point par point la courbe de tassement au fur et à mesure des incréments de chargement. Cette évolution peut être représentée dans le repère $(\sigma',v)$ où $\sigma'$ est la contrainte axiale effective appliquée et v le volume spécifique atteint (v = (volume total) / (volume des grains)). La Fig. 3.3 correspond à la réponse typique obtenue pour un sol soumis à un cycle de chargement.

L’échantillon est soumis à une charge croissante de A à C, puis décroissante de C à D et enfin croissante de D à E. On remarque que le comportement de l’échantillon n’est pas purement élastique. L’évolution du volume spécifique v se décompose en une partie quasiment réversible (courbes AB ou CD et DC) et une partie irréversible (courbes BC et CE). La courbe BCE représente l'état limite. L'échantillon ne peut présenter un état d'équilibre correspondant à (σ',v) au-delà de cette frontière.

- Un échantillon présentant un état (σ',v) en deçà de cette courbe de consolidation est dit **surconsolidé**.
- Un échantillon présentant un état (σ',v) sur la courbe de consolidation est dit **normalement consolidé**.

La transcription des résultats obtenus dans un repère semi logarithmique (log(σ'),v) en simplifie l'exploitation. Dans ce repère, la courbe de consolidation et la courbe correspondant à l'évolution réversible du volume spécifique sont assimilées à des droites. Le point d'intersection de ces deux droites fournit la **contrainte de préconsolidation**.

*Constante de Jaky :* On constate aussi une évolution du rapport des contraintes $\sigma'_3/\sigma'_1$; cette évolution ainsi que la valeur asymptotique finale de $\sigma'_3/\sigma'_1$ sont différentes lorsque la contrainte croît ou décroît : $\sigma'_3/\sigma'_1$ tend vers $k_J=1-\sin\varphi$ approximativement quand σ' croît ; $\sigma'_3/\sigma'_1$ tend vers la valeur de rupture, *i.e.* $tg^2(\pi/4-\varphi/2)$, approximativement quand σ' décroît. $k_J$ est appelé la constante de Jaky (1944); nous donnerons une explication de ces valeurs dans le chapitre 7.

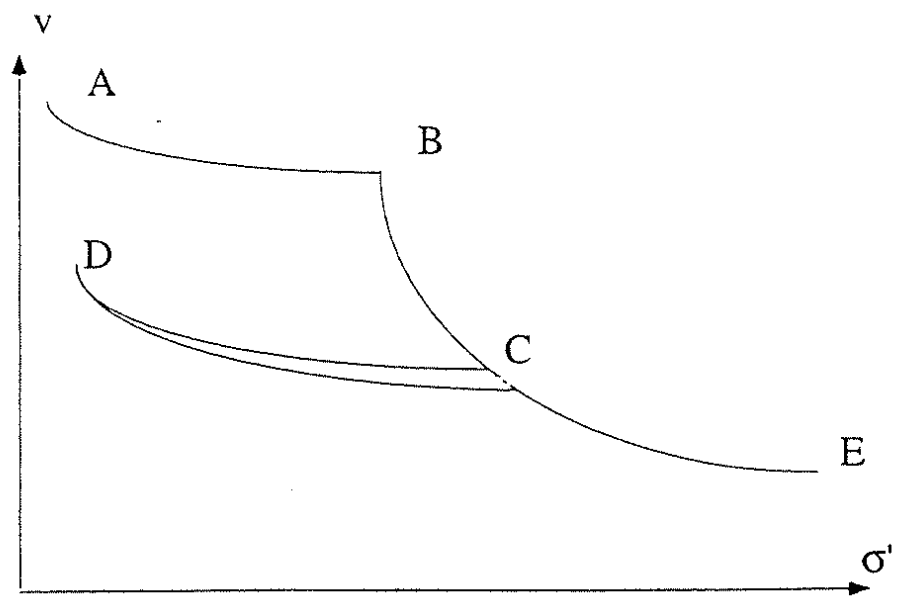

**Figure 3.3** - *Courbe de tassement: L'évolution du volume spécifique* v *d'un échantillon cylindrique sollicité en compression isotrope en fonction de la contrainte effective intergranulaire* σ' *se décompose en deux parties. L'une est quasiment réversible (courbe* AB *ou* CD *et* DC*). L'autre est irréversible (courbe* BC *et* CE*). La courbe* BCE *dite courbe de consolidation représente l'état limite. Nous verrons dans le chapitre 4 que ces courbes deviennent des droites lorsque l'on utilise* log(σ').

*Oedomètre : Des résultats similaires sont obtenus avec l'oedomètre . Cependant, dans ce cas, le rapport des contraintes principales* $\sigma'_3/\sigma'_1$ *évolue au cours de l'augmentation de pression pour tendre rapidement vers une constante lorsqu'on augmente constamment* p' *; cette valeur est la constante de Jaky et vaut 1-sinφ à peu près. A la décompression le rapport des contraintes tend vers la valeur de rupture plastique :* $\sigma'_3/\sigma'_1=tg^2(\pi/4+\varphi/2)$.

### *3.3.2. Compression isotrope*

S'il est intéressant de connaître le comportement d'un sol soumis à une compression uniaxiale pour évaluer l'ampleur des tassements du milieu sollicité sous une fondation, il peut s'avérer intéressant de s'affranchir de la direction de sollicitation. Pour cela, il est préférable d'exploiter les données d'un essai de compression isotrope obtenu par le confinement de l'échantillon; il fournit l'évolution de v en fonction de la pression effective $p' = p - u_W$ , *cf.* Fig.3.3. Les résultats obtenus avec un tel chargement sont semblables à ceux de l'oedomètre. La courbe d'états limites ainsi obtenue est nommée **courbe de consolidation vierge** et la courbe correspondant à l'évolution réversible du volume spécifique est dite courbe de "gonflement". L'intersection de ces deux courbes est nommée **pression de préconsolidation**. La courbe de consolidation vierge est généralement modélisée par une équation du type:

(3.9) $$v = \Gamma - \lambda \ln p'$$

La contrainte de préconsolidation et la pression de préconsolidation sont souvent considérées comme des indicateurs de l'état de contrainte effective extrême vécu par le matériau durant son histoire. Des états de contrainte au-delà de ces valeurs impliquent des modifications importantes et irréversibles de l'arrangement granulaire du milieu.

### *3.3.3. L'appareil triaxial*

L'appareil triaxial est un des appareillages le plus couramment utilisé en laboratoire. Il est schématisé sur la Fig. 3.4: l'échantillon cylindrique de sable à étudier est entouré d'une poche plastique (membrane), fermée en ses deux extrémités par deux pistons ce qui permet d'appliquer une surcharge verticale q variable sur l'échantillon.

$p' = (\sigma_1 + 2.\sigma_3)/3 - u_w$
$p = (\sigma_1 + 2.\sigma_3)/3$
$q = \sigma_1 - \sigma_3 = \sigma'_1 - \sigma'_3$

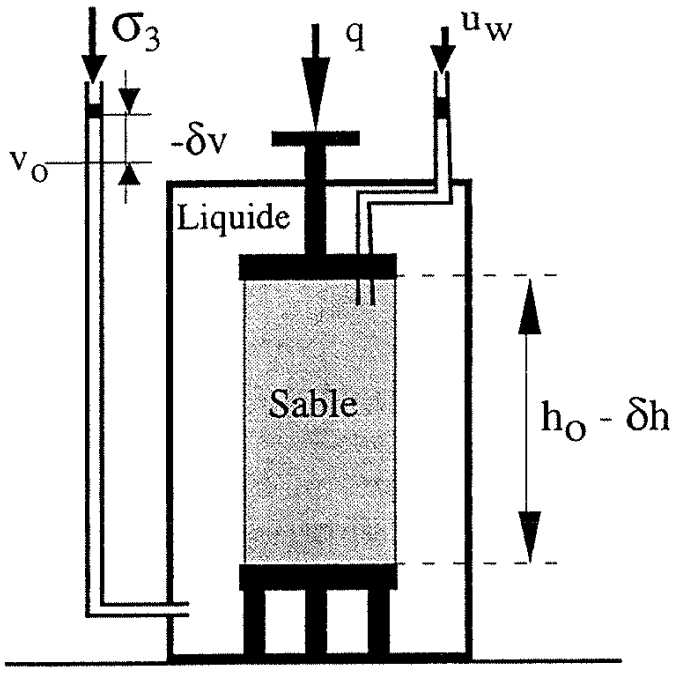

***Figure 3.4*** - *Appareil triaxial: un échantillon cylindrique entouré d'une membrane est placé dans une cellule remplie d'eau. La mise en pression de l'eau de la cellule fournit une pression isotrope* $\sigma_3$ *sur l'échantillon. Une surcharge q est appliquée progressivement à l'échantillon à l'aide du piston. La contrainte axiale de l'échantillon est* $\sigma_1 = \sigma_3 + q$. *Les variations du volume de l'échantillon sont indiquées par l'évolution du volume d'eau de cellule. Le déplacement du piston fournit l'évolution de la hauteur de l'échantillon. Un conduit est relié directement à l'échantillon. Il permet la mesure des variations du volume d'eau interstitielle ou de sa pression* $u_w$.

On connaît à tout moment l'écrasement de l'échantillon par la mesure de la distance entre ces deux pistons. La poche est immergée dans une cellule contenant de l'eau et alimentée depuis l'extérieur de telle sorte que l'on puisse en maîtriser la pression et le volume durant l'essai. On connaît ainsi le tenseur de contrainte appliqué à l'échantillon ($\sigma_1$=$\sigma_3$+ q, $\sigma_2$ = $\sigma_3$), et les variations de hauteur δh et de volume δv de ce dernier. Un ou deux autres conduits permettent d'alimenter l'intérieur de la poche en eau à une pression donnée $u_w$, voire d'imposer un courant. Le tenseur des contraintes effectives intergranulaires est donc $\underline{\underline{\sigma}}' = \underline{\underline{\sigma}} - u_w \underline{\underline{I}}$ , où $\underline{\underline{I}}$ est la matrice unité. De même le tenseur des déformations est caractérisé par $\varepsilon_v$= -δv/$v_o$ et $\varepsilon_1$= -δh/$h_o$, où nous avons choisi de garder la convention classique de mécanique des sols (toute diminution de volume et de hauteur est comptée comme une déformation positive).

En assimilant l'échantillon testé à un volume élémentaire nous construisons à chaque instant le tenseur des contraintes correspondant à l'état du milieu. Pour représenter l'évolution des contraintes on préfère au plan de Mohr une représentation dans le plan (p,q) ou (p',q). Dans ce plan le chemin de contrainte forme une courbe. Des exemples de réponses d'essais triaxiaux obtenues avec diverses conditions expérimentales sont cités dans le paragraphe suivant. Ils servent de support à la construction d'un modèle général de comportement d'un milieu granulaire.

Un résultat important obtenu grâce à cet appareillage est la validation de l'approximation de Terzaghi qui découple le tenseur des contraintes en la somme d'un tenseur de contraintes intergranulaires et d'un tenseur hydraulique. Ceci reste vrai lorsque les forces "d'adhésion" de l'eau sont mises en jeu, c'est-à-dire lorsque l'eau est en dépression ($u_w$ devient négatif).

L'appareil triaxial est également utilisable pour identifier les paramètres liés à la consolidation. Il suffit de réaliser un essai isotrope en faisant évoluer la pression de confinement $\sigma_3$ et d'enregistrer les variations de volume spécifique tout en permettant bien sur le drainage dans l'échantillon, *i.e.* $u_w$=$c^{ste}$.

### 3.4. Comportement mécanique d'un milieu granulaire

Le comportement mécanique d'un matériau granulaire ne peut être correctement modélisé si l'on se limite à l'analyse des variations du tenseur de contrainte. Le critère de Coulomb n'est pas suffisant. Il faut relier les variations du tenseur de contrainte σ aux variations du tenseur des déformations, ou du moins à la variation de volume. On identifie expérimentalement trois paramètres de contrôle tels que (p',q,e). Nous nous limiterons, durant cette approche, à une modélisation du comportement sous faibles taux de déformation.

L'état initial du milieu granulaire, en particulier sa densité initiale, joue un rôle prépondérant dans l'évolution de son comportement. Le caractère non répétitif de la compacité du milieu, obtenu à la mise en place des échantillons, ne facilite pas la compréhension du comportement des milieux granulaires. Nous proposons donc de suivre l'approche de Roscoe *et al.* (1958) et de construire un modèle de comportement de ces milieux en synthétisant dans le repère (p', q, e) l'ensemble des informations obtenues lors d'essais oedométriques ou triaxiaux. Le comportement ainsi proposé nous permettra de comprendre, par l'illustration, les mécanismes de la dilatance ou de la mise en dépression des fluides interstitiels.

### *3.4.1. Essais sur sol normalement consolidé*

#### *3.4.1.1. Essais non drainés sur sol normalement consolidé*

Les essais non drainés sont réalisés sur un appareil triaxial en conservant un volume constant d'eau interstitielle dans l'échantillon. L'échantillon est initialement normalement consolidé, ce qui signifie que son état initial est représenté par un point de la courbe de consolidation vierge dans le plan (p',e). Cet état est obtenu par consolidation de l'échantillon sous une contrainte de confinement $\sigma'_1 = \sigma'_3$ choisie de la cellule triaxiale. Le cisaillement est induit par l'augmentation de $\sigma_1$, $\sigma_2=\sigma_3$ restant constant. p croît donc linéairement, avec $\Delta q/\Delta p = 3$ , *cf.* Fig. 3.5. Le volume restant constant, on enregistre simultanément à l'augmentation de $\sigma_1$ une mise en pression $u_w$ de l'eau interstitielle. On a représenté sur la Fig. 3.5 le chemin des contraintes effectives suivi, dans les différents plans $(q,\varepsilon_1)$, $(q/p',\varepsilon_1)$, (q,p), (q,p') et $(u_w,\varepsilon_1)$ ; on sait aussi que v=constante. q/p' et $u_w$ croissent et atteignent leurs valeurs maximales correspondant à la rupture (q/p'=M', point indiqué sur la Fig. 3.5). La rupture correspond à l'état critique pour lequel le critère de plasticité (Coulomb) est satisfait. On constate aussi que la pression p' reste d'abord constante, puis commence à décroître. Corrélativement, q commence par croître, passe par un maximum puis décroît jusqu'à ce qu'il atteigne sa valeur à la rupture q/p'=M'. L'évolution du système peut devenir instable après le maximum de q ; dans ce cas, le milieu ne reste pas homogène et l'on observe une localisation des déformations .

La répétition de cet essai pour des valeurs de v différentes fournit des chemins de contraintes effectives présentant la même allure générale. Ils sont souvent considérés comme affines entre eux.

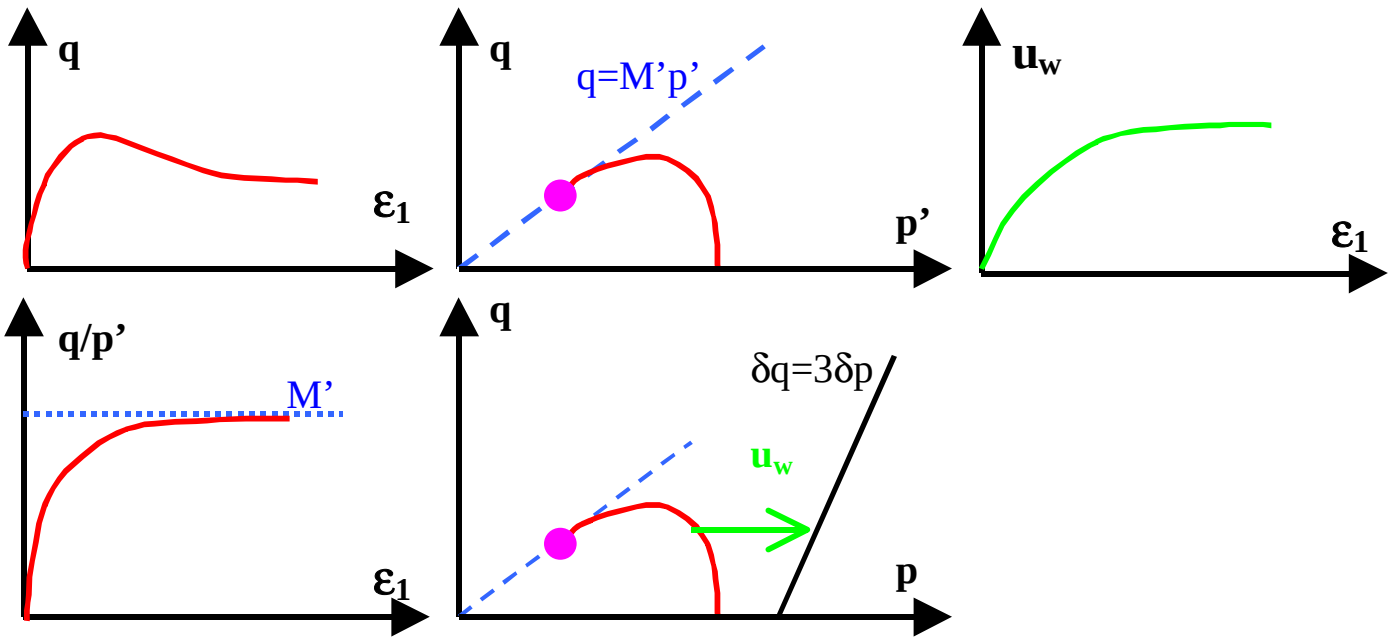

***Figure 3.5 :*** *Essais non drainés sur sol normalement consolidé: cet essai est réalisé à $\sigma_3$ constant et en maintenant constant le volume d'eau interstitielle dans l'échantillon. Cette figure donne l'allure typique des évolutions de* q *et* $u_w$ *en fonction de la déformation axiale* $\varepsilon_1$ *, à partir desquels on trace l'évolution des contraintes dans les plan* (q,p') *et* (q,p). *Sur un sol normalement consolidé, un tel essai se solde par la mise en pression du liquide interstitiel. Une modélisation simple de ces résultats sera donné au chapitre 7.*

#### *3.4.1.2. Essais drainés sur sol normalement consolidé*

Les essais drainés sont réalisés sur un appareil triaxial en laissant varier le volume d'eau interstitielle dans l'échantillon. De la même façon que précédemment, l'échantillon est initialement normalement consolidé, ce qui signifie que son état initial est représenté par un point de la courbe de consolidation vierge dans le plan (p',v). Le cisaillement est induit par l'augmentation de $\sigma_1$ , $\sigma_3$ et $\sigma'_3$ restant constants. p et p' croissent donc linéairement, avec $\Delta q/\Delta p = \Delta q/\Delta p'=3$ . Le drainage permet le maintient de la pression interstitielle $u_w$ à $u_{wo}$, donc $p = p'+u_{wo}$.
Le matériau voit p' croître. Il va réagir en se consolidant, c'est à dire en diminuant son volume spécifique (Fig. 3.6). Cependant, q croissant également, le chemin de contrainte quitte la courbe de consolidation vierge. Le volume spécifique décroît jusqu'à ce que q atteigne sa valeur maximale $q_{max}$, point qui correspond à la rupture. La rupture correspond à l'état critique pour lequel le critère de plasticité (Coulomb) est satisfait $q_{max} = M'p'$.
Ce comportement peut se résumer de la façon suivante: la "consolidation" d'un milieu granulaire apparaît plus vite (p' inférieur) pour un chargement anisotrope que pour un chargement isotrope. Autrement dit, pour densifier un tel milieu il est préférable de le cisailler en le confinant que de le confiner simplement.

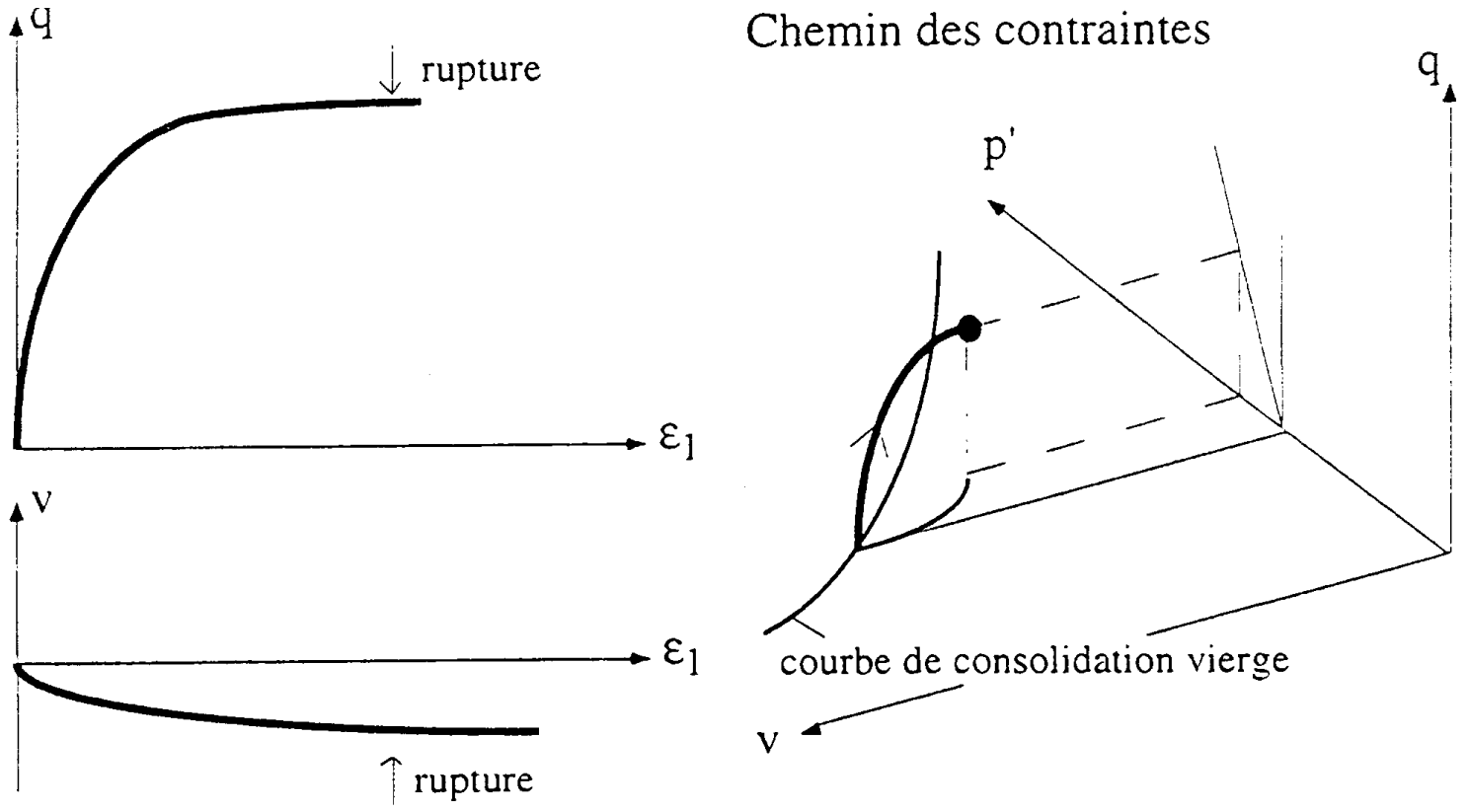

***Figure 3.6*** *- Essais drainés sur sol normalement consolidé: cet essai est réalisé à* $\sigma_3$ *constant et en permettant les variations du volume d'eau interstitielle de l'échantillon. Cette figure fournit l'allure traditionnelle des évolutions de* q *et* e *en fonction de la déformation axiale* $\varepsilon_1$. *Ces enregistrements permettent le tracé du chemin des contraintes dans l'espace* (p',q,e). *Sur un sol normalement consolidé, un tel essai se solde par une diminution de l'indice des vides* e *et du volume spécifique* v *de l'échantillon.*

### *3.4.1.3. Surface des états limites et courbe des états critiques*

Un milieu normalement consolidé peut être obtenu par un chargement isotrope en utilisant un appareil triaxial. L'ensemble des essais réalisés sur un tel milieu montre que les chemins de contraintes induits par une croissance de $\sigma_1$ quittent un point de la courbe de consolidation vierge, image sous chargement isotrope de la courbe BCE de la Fig. 3.3, qui est un point d'état limite, et rejoignent un point de rupture qui est appelé «état critique». Ou état de plasticité parfaite. L'ensemble des points «état critique » constitue la **courbe des états critiques** dans l'espace (p',q,v). Cette courbe est la représentation, dans l'espace choisi, du critère de plasticité (Schofield & Wroth 1968). A chaque point de cette courbe correspond un état de contrainte dit "de rupture" (ou de plasticité parfaite), c'est-à-dire produisant des grandes déformations à volume constant (variation de volume nulle). Pour un milieu granulaire dont le critère de plasticité est le critère de Coulomb (sans cohésion), la projection de cette courbe des états critiques sur le plan (p',q) est une droite passant par l'origine dont la pente M' est liée à l'angle de frottement interne φ:

(3.10) $$q = M' p' = 6\sin\varphi/(3-\sin\varphi)\ p'$$

ou $$q = M\sigma'_3 = 2\ \sigma'_3 \sin\varphi/(1-\sin\varphi)$$

Les relations entre M, M' & φ se déduisent des relations entre q, p', $\sigma'_1$ & $\sigma'_3$.
La réalisation d'essais non drainés sur des échantillons légèrement surconsolidés mais présentant le même volume spécifique v révèle que les chemins de contraintes effectives suivis convergent rapidement (p' quasi constant) vers le chemin obtenu sur un échantillon normalement consolidé. Par ailleurs, la rupture se produit pour la même valeur du couple ($p'$,$q_{max}$).
C'est pourquoi l'ensemble des chemins de contrainte obtenus pour des essais drainés ou non drainés, sur des échantillons normalement consolidés, constitue une **surface, appelée surface des états limites**. Cette surface est aussi nommée **surface de Roscoe** (Fig. 3.7). L'évolution des essais sur des échantillons légèrement consolidés converge assez vite vers cette surface de Roscoe. D'où l'intérêt de la notion de surface de Roscoe.

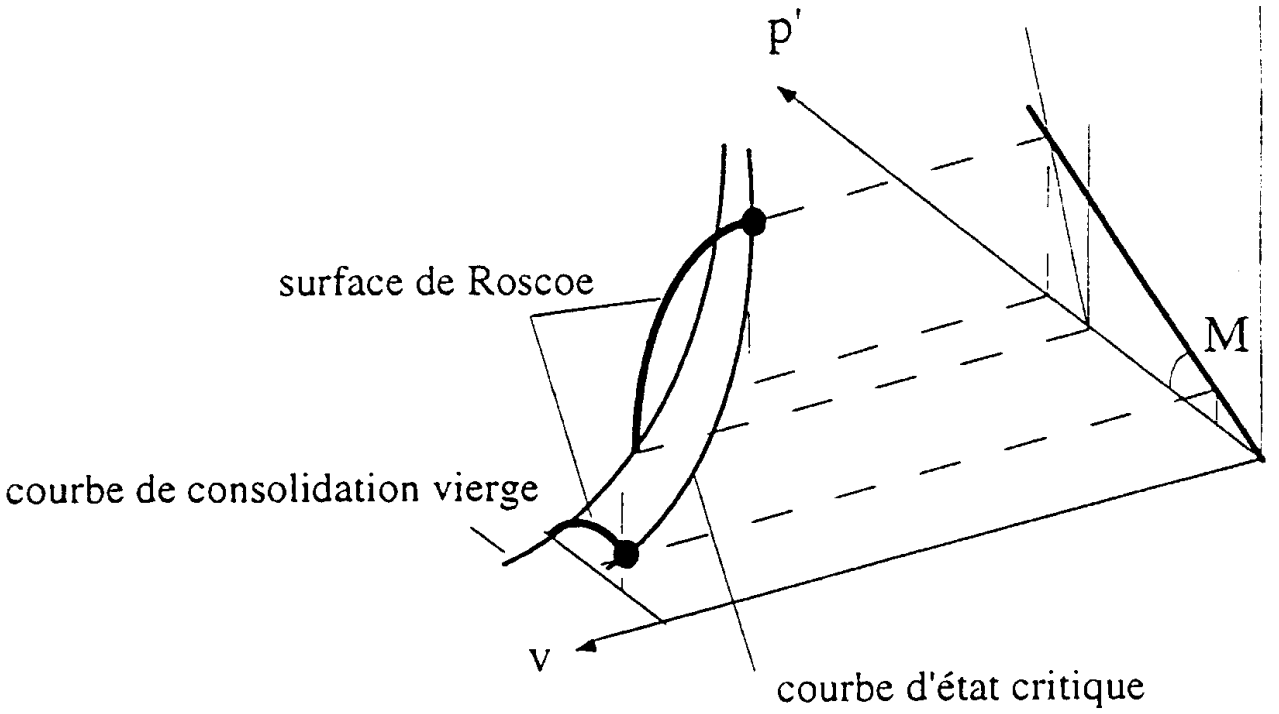

***Figure 3.7*** *- Surface des états limites, courbe des états critiques: dans l'espace* (p',q,v)*, l'ensemble des chemins de contrainte obtenus pour des essais drainés ou non drainés, sur des échantillons normalement consolidés, constitue une surface appelée surface des états limites ou surface de Roscoe. L'ensemble des états critiques constitue la courbe des états critiques.*

Enfin, tout état d'équilibre d'un matériau normalement consolidé ou faiblement surconsolidé est représenté par un point dans l'espace (p',q,v) à l'intérieur de la surface des états limites.

Un état d'équilibre limite correspondant à un point de la surface de Roscoe est caractérisé par une variation du volume spécifique négative (v tend à décroître).
L'affinité du chemin de contraintes effectives obtenu lors d'un essai non drainé implique une allure similaire de la projection de la courbe des états critiques à celle de la ligne de consolidation vierge, dans le plan (p',v) (linéaire en repère semi logarithmique).
La courbe des états critiques d'un sable est donc modélisée par les deux équations:

(3.11) $v = v_{co} - \lambda.\ln p'$ et $q = M' p'$ avec $\sin\varphi = 3M'/(6+M')$

ou $q = M\sigma'_3$ avec $\sin\varphi = 2M/(2+M)$

### 3.4.2. Essais sur sol surconsolidé

#### *3.4.2.1. Essais non drainés sur sol surconsolidé*

L'échantillon utilisé est initialement confiné par une contrainte $\sigma'_1 = \sigma'_3$ inférieure à la pression de préconsolidation correspondant à sa valeur initiale de volume spécifique v. L'essai est réalisé à drainage fermé (volume d'eau interstitielle dans l'échantillon maintenu constant). Le cisaillement est induit par l'augmentation de $\sigma'_1$ ; $\sigma_3$ reste constant mais $\sigma'_3$ peut varier. p croît donc linéairement, avec $\Delta q/\Delta p = 3$ . Le volume restant constant, on enregistre simultanément l'augmentation de $\sigma'_1$ et les évolution de p' et de la pression $u_w$ de l'eau interstitielle. Le chemin des contraintes effectives est représenté Fig. 3.8 dans le plan $v=c^{ste}$. q croît et atteint la droite q/p'=M' puis q suit cette droite et atteint sa valeur maximale correspondant à la rupture (point d'état critique). P' reste constant au départ, puis finalement croît jusqu'au point de rupture. La pression interstitielle $u_w$ croît (légère mise en pression) puis devient fortement décroissante. Cette mise en dépression "compense" le désir du milieu de se dilater.

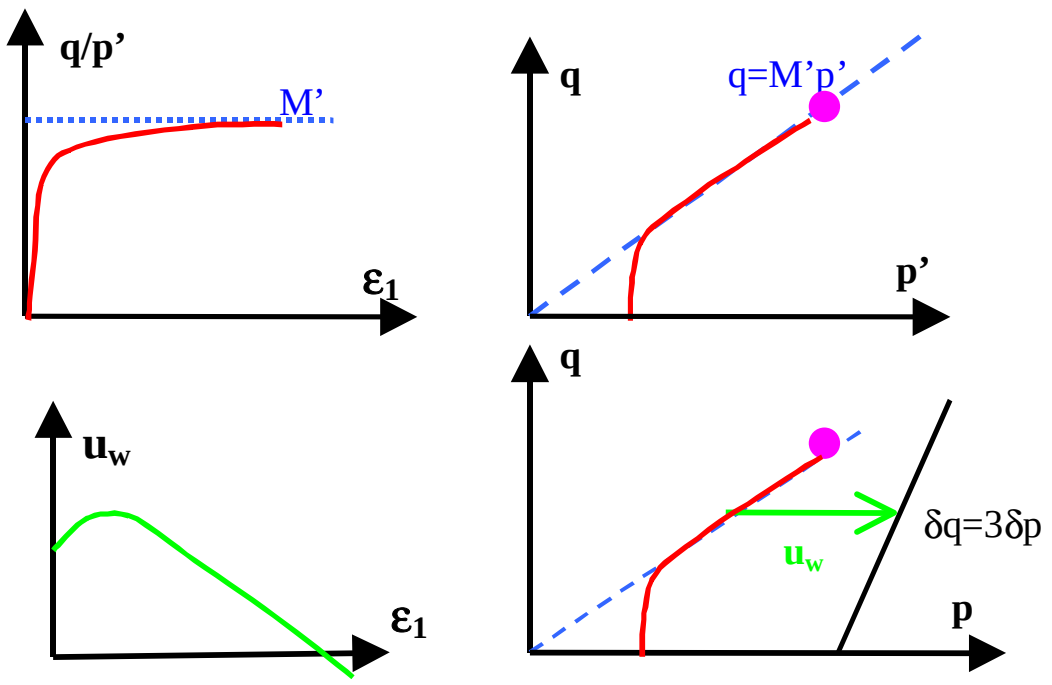

***Figure 3.8*** *- Essais non drainés sur sol surconsolidé: cet essai est réalisé à $\sigma_3$ constant et en maintenant constant le volume d'eau interstitielle de l'échantillon. Cette figure fournit l'allure générale des évolutions de* q *et* $u_w$ *en fonction de la déformation axiale* $\varepsilon_1$*. Ces enregistrements permettent le tracé du chemin des contraintes dans les plans* (p',q)*. Sur un sol surconsolidé, un tel essai se solde par une mise en dépression de l'eau interstitielle. Une modélisation simple de ces résultats sera donnée au chapitre 7.*

La répétition de ces essais sur des échantillons de même volume spécifique mais testés avec des pressions initiales de confinement différentes montre que les chemins de contraintes convergent rapidement (p' quasi constant) vers une courbe enveloppe relativement linéaire. La rupture se fait à la même valeur de $q_{max}$ pour tous les échantillons de même volume spécifique v.

La répétition de cet essai pour des valeurs de v différentes fournit des courbes enveloppes des chemins de contraintes présentant les mêmes allures générales. Ces courbes sont souvent considérées comme affines entre elles. La fin de ces trajectoires sont des droites, *i.e.* q=M'p', v=$c^{ste}$. L'ensemble de ces fins de trajectoire (des droites) constitue une seconde surface (réglée) d'états limites dans l'espace (p',q,v). Cette surface se nomme **surface de Hvorslev** (Fig. 3.9).

Certains mécaniciens considèrent que la fin de ces trajectoires est une droite qui ne passe pas par l'origine, mais par le point d'ordonnée $\alpha$ dans le plan (q,p') et que la pente de la droite est $\beta_o$ et non M'. Faut-il considérer $\alpha$ comme de la cohésioon et bo comme un seconde valeur du coefficient de frottement?

D'autres auteurs considèrent que la trajectoire réelle peut passer au dessus de cette droite, pour revenir dessus en suite. Nous ne rentrerons pas dans ces détails.

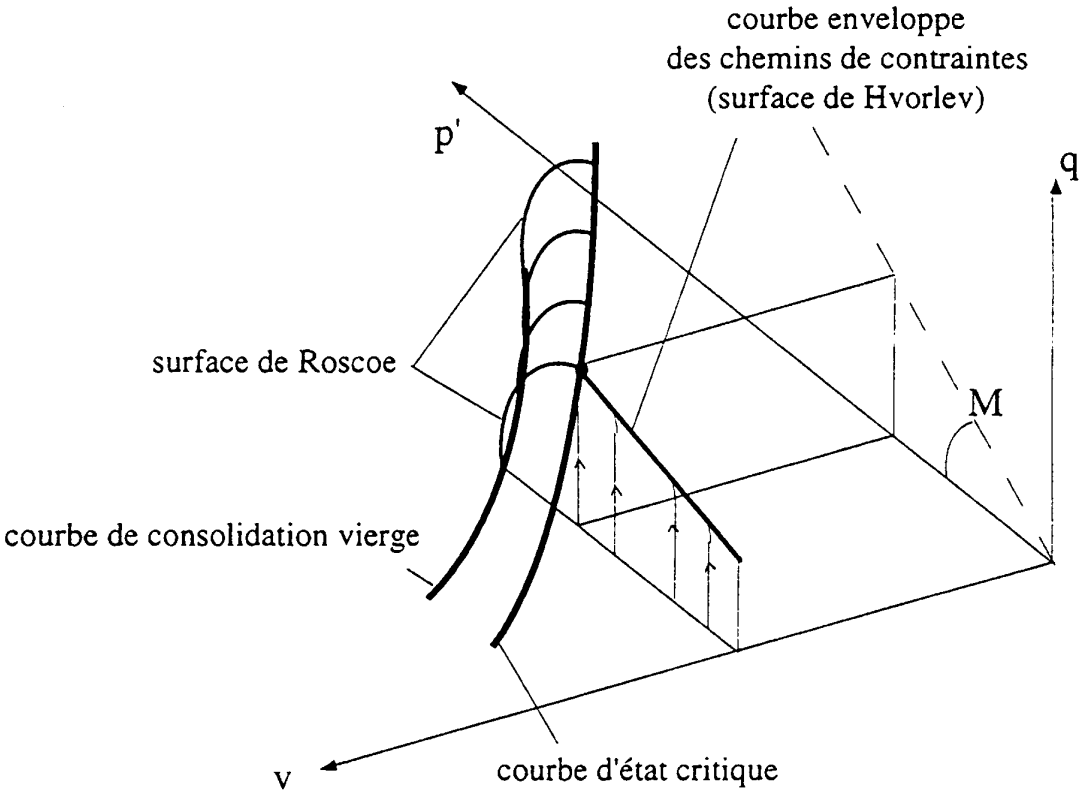

***Figure 3.9** - Surfaces limites: dans l'espace* (p',q,v), *les chemins de contraintes obtenus lors des essais non drainés sur des échantillons surconsolidés convergent vers une surface d'états limites. Cette surface est nommée surface de Hvorslev.*

### *3.4.2.2. Essais drainés à* $\sigma'_2=\sigma'_3=c^{ste}$ *sur sol surconsolidé*

L'état de départ est surconsolidé; cela veut dire que nous exploitons un échantillon qui a subi une pression de préconsolidation supérieure à la pression initiale de la cellule triaxiale. On parle alors facilement de sable "dense". Il convient cependant de rappeler que la surconsolidation est une notion liée aux contraintes et non au volume.

L'échantillon utilisé est initialement confiné à une contrainte $p'_i=\sigma'_1=\sigma'_3$ inférieure à la pression maximale $p'_{max}$ qu'il a subi au cours de son histoire; cette pression maximale est dite pression de préconsolidation $p'_{max}$ . Par la suite on supposera qu'elle correspond aussi à la valeur initiale de son volume spécifique v.

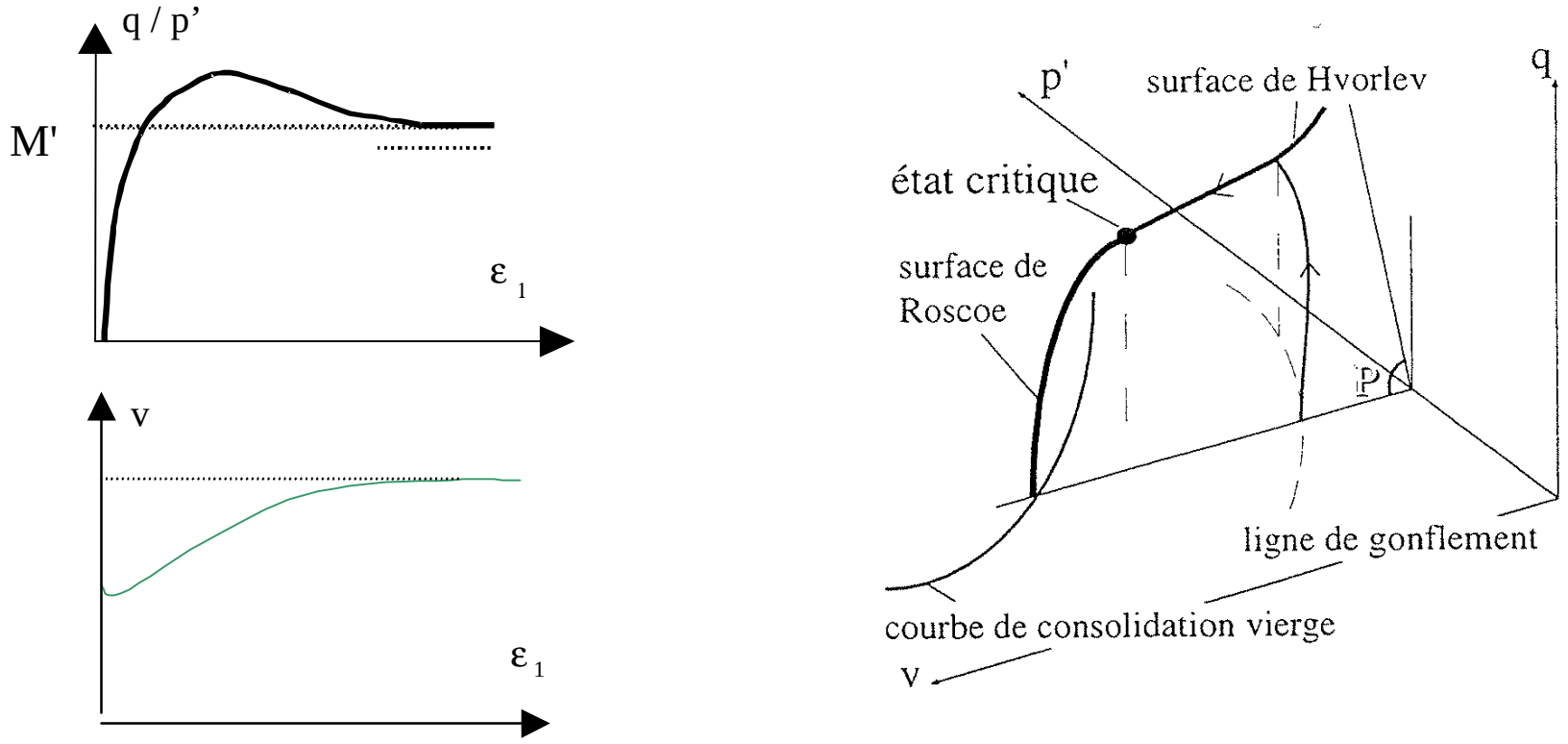

*Figure 3.10 - Essais drainés sur sol surconsolidé* ($\sigma'_2=\sigma'_3=c^{ste}$)*: cet essai est réalisé à* $\sigma_3$ *constant et en permettant les variations du volume d'eau interstitielle de l'échantillon. Cette figure fournit l'allure traditionnelle des évolutions de* q *et* v *en fonction de la déformation axiale* $\varepsilon_1$*. Ces enregistrements permettent le tracé du chemin de contrainte dans les plans* (p',q)*. Sur un sol surconsolidé, un tel essai se solde en début d'essai par une légère diminution de v puis une forte augmentation de* v *(dilatance).*

L'essai est réalisé à drainage ouvert. Le cisaillement est induit par l'augmentation de $\sigma_1$ , $\sigma_3$ restant constant. p' croît donc linéairement, avec $\Delta q/\Delta p' = 3$, la pression $u_w$ de l'eau interstitielle restant nulle (ou constante). q croît et passe par une valeur maximale. Au-delà de ce maximum on constate parfois la stabilisation de q, mais il décroît le plus souvent. En parallèle le volume spécifique évolue. Il décroît légèrement au début du chargement puis peut croître de façon importante. Ce phénomène est nommé dilatance. On le relie au désenchevêtrement des grains d'un milieu dense soumis à un cisaillement.

L'analyse du chemin des contraintes obtenu dans un tel essai (Fig. 3.10) permet de comprendre les mécanismes liés à la dilatance. De la même façon que pour les essais drainés déjà analysés, le chemin de contrainte de tels essais est porté dans le plan de chargement P incliné de telle façon que $\Delta q/\Delta p' = 3$.

Le matériau voyant p' et q croître va réagir en modifiant son volume spécifique, *cf.* courbe DC de la Fig. 3.3. Initialement, le matériau a tendance à suivre une ligne de "gonflement". Le volume spécifique décroît. A l'image de l'essai non drainé (Fig. 3.9), le chemin de contrainte va converger vers la surface limite (surface de Hvorslev) dès que q atteint $q_{max}$.

La Fig. 3.10 représente le chemin type suivi par le matériau. Il évolue dans le plan P défini plus haut; il commence à (q=0, v, $p'_i$). q et p' commencent à croître et v à décroître dans un premier temps; puis pendant une portion de trajet q, p' et v croissent, jusqu'à ce que q et p' passent par un maximum . Ceci arrive approximativement à l'intersection entre le plan (P) et la surface limite dite de "Hvorslev"; arrivée à ce stade, la trajectoire suit alors la courbe définie par l'intersection entre le plan (P) et la surface de Hvorslev, vers les v croissant; et elle se termine lorsqu'elle a rejoint l'intersection entre la courbe des états critiques et le plan (P). L'évolution du volume spécifique associée à ce modèle de chemin de contrainte révèle la dilatance.

*Modèlisation simplifiée:* Une modélisation simplifiée possible consiste à simplifier le comportement réel précédent en associant la partie contractante à l'augmentation de q, de 0 jusqu'à $q_{max}$, et la partie dilatante à la diminution de q, depuis $q_{max}$ jusqu'à $q_c$. Dans ce cas, la première partie de la trajectoire peut être associée à la réponse quasi-élastique observée dans le comportement de "gonflement" (*cf.* Fig. 3.3 et § 3.3.2): cette réponse élastique prévoit bien que le volume décroît lorsque la contrainte p' croît. Après qmax, la réponse du matériau change complètement et suit un comportement typique irréversible. Ce comportement typique est caractérisé par la surface de Hvorslev dans l'espace (q,p',v). La trajectoire réelle observée sera donc l'intersection de la surface de Hvorslev avec le plan (P) défini plus haut par $\delta p=\delta q/3$ et passant par le point ($q=0, p'=p'_i, v=v_i$). Dans ce modèle, la transition entre contractance et dilatance apparaît lorsque q atteind $q_{max}$ **et** lorsque la trajectoire élastique coupe la surface de Hvorslev.

L'intérêt de ce modèle simplifié réside (i) en ce qu'il révèle l'importance de la dilatance, (ii) en ce qu'il indique que l'amplitude de la croissance de v (la dilatance) dépend de la valeur de p initiale. Plus le sol sera surconsolidé, plus cette amplitude sera forte.

Cependant, ce modèle simplifié a plusieurs défauts: (i) il impose la simultanéité du maximum en q et du minimum en v. Ce point est contredit par les résultats expérimentaux, comme nous l'avons vu il y a deux paragraphes. (ii) il suppose que la contractance initial est lié à un comportement quasi-réversible. Ceci est faux, car dès le départ le comporterment n'est pas réversible. Pour rendre compte de ces faits, différentes approches sont possibles (Darve 1987). Citons aussi en exemple l'approche énergétique proposée par Evesque et Stefani (1991), qui est à la base du modèle proposé au chapitre 5.

Par ailleurs, l'apparition de la dilatance n'impose pas le passage obligé de q par un maximum. Il est possible de trouver des chemins de contraintes effectives différents de ceux évoqués précédemment ($\sigma'_2=\sigma'_3=c^{ste}$) pour lesquels le matériau demeure dilatant (v croissant) et q constamment croissant; pour cela il faut choisir la loi de variation liant $\sigma'_2=\sigma'_3$ à $\sigma'_1$: $\sigma'_2=\sigma'_3=f(\sigma'_1)$. Le modèle qui est proposé au chapitre 7 permet de trouver facilement ce chemin.

Il est admis qu'un état d'équilibre limite correspondant à un point de la surface de Hvorslev (ou du moins voisin de cette surface) est caractérisé par une variation du volume spécifique positive (v tend à croître). Nous verrons cependant dans les chapitres 5 &7 que ce point n'a pas énormément de sens, car les surfaces de Roscoe et de Hvorslev sont 2 parties de la même surface; en un point donné, on peut passer d'un comportement dilatant à un comportement contractant en changeant la direction de l'incrément de contrainte. Ceci peut se déduire simplemement du modèle proposé aux chapitres 5 et 7.

En fait un traitement plus complet nécessite de retraduire l'évolution de v en fonction de $\varepsilon_1$ aussi; les trajectoires peuvent être vues comme l'évolution du système dans l'espace (q,p',v) et $\varepsilon_1$ joue le rôle du temps. On peut alors appliquer la théorie des systèmes dynamiques pour déduire certaines propriétés des trajectoires , *cf.* chaps. 5.2, 7.5 et appendice.

Ceci sera abordé par la suite au chapitre 5. Pour conclure cette section, l'analyse de la Fig. 3.10 conduit aux conclusions suivantes:

- la valeur maximum en q ne correspond pas à un état critique (plastique);
- la valeur stabilisée de q (après dilatance) correspond à l'état critique;
- l'existence d'un maximum en cisaillement dépend essentiellement du mode de chargement choisi (relation entre p' et q).

### 3.5. Description plus détaillée du comportement au triaxial:

Nous ne détaillerons pas le comportement rhéologique précis du matériau granulaire dans ce chapitre. Ceci sera fait dans le chapitre suivant. Nous donnons simplement dans la Fig. (3.11) la manière générale qui est utilisée par les mécaniciens des sols pour récapituler leurs résultats. Ceci se fait à l'aide de quatres projections. Pour être complet, il faudrait aussi discuter le rôle de la contrainte intermédiaire si $\sigma_1>\sigma_2>\sigma_3$, le rôle de l'anisotropie; mais cela déborde le cadre de ce chapitre et les descriptions aux quelles ces rétudes donnent lieu ne sont pas admises par toute la communauté….

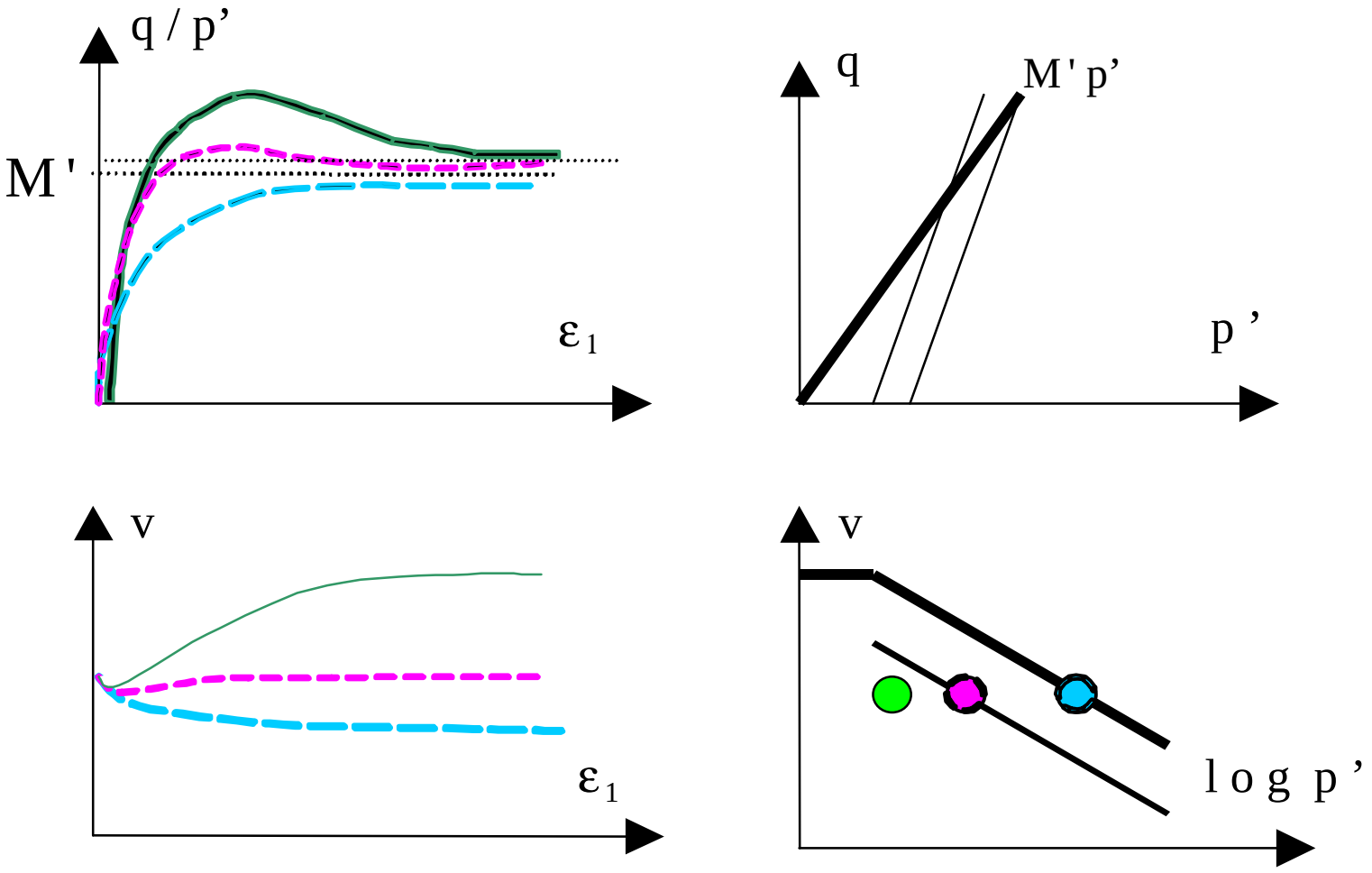

***Figure 3.11:*** *Les résulats se récapitulent en étudiant les trajectoires du système; pour cela on utilise les quatres projections* (q, $\varepsilon_1$), (q,p'), (v, $\varepsilon_1$) *ou* ($\varepsilon_v$,$\varepsilon_1$) *et* (v, lnp') *[ou* (v, p')*]. On sait en effet grâce à la droite des états critiques si le matériau doit se contracter ou se dilater pour sa densité initiale.*

### 3.6. Discussion sur le comportement spécifique des sables et conclusion

Les essais sur les sables sont généralement réalisés en condition drainée (Figs. 3.6 et 3.10), sauf quand on veut étudier leur réponse à des cycles de contraintes et leur liquéfaction. Les sables présentent généralement des comportements relativement proches du modèle simplifié construit précédemment. On peut cependant noter les éléments suivants:

- Les sables sont relativement peu compressibles, ce qui signifie que la courbe de consolidation vierge est peu pentue.
- Les sables présentent des courbes de "gonflement" (réponse pseudoélastique) également très plates.
- Dans le plan (p’,v) la projection de la courbe d’état critique est très proche de la courbe de consolidation vierge.

Les conséquences de ces trois caractéristiques majeures du comportement d’un sable sont multiples:

- L’identification d’une pression de préconsolidation est rendue difficile du fait d’une différence peu marquée entre la pente d’une ligne de "gonflement" et la pente de la ligne de consolidation vierge.
- Pour un sable lâche (v important) le domaine possible de pression effective (plan (p’,v)) correspond majoritairement à des états initiaux peu surconsolidés du milieu. Dans ce cas un chargement conduit à un état limite sur la surface de Roscoe et donc à une diminution de v si le drainage est permis, ou une mise en pression du fluide interstitiel si le drainage est fermé. (Attention, *cf.* ce qui est dit aux §3.4 et §7.5 pour une discussion de la notion de surface de Roscoe et de Hvorslev).
- Pour un sable dense (v faible) le domaine possible de pression effective (plan (p’,v)) correspond majoritairement à des états initiaux surconsolidés du milieu. Dans ce cas un chargement conduit à un état limite sur la surface de Hvorslev et donc, à terme, à une augmentation de v si le drainage est permis (dilatance), ou une mise en dépression du fluide interstitiel si le drainage est fermé. (Attention, *cf.* ce qui est dit aux §3.4 et §7.5 pour une discussion de la notion de surface de Roscoe et de Hvorslev).
- Un essai drainé sur un sable surconsolidé (en conséquence dense) débute par une faible diminution de v du fait de l’allure des lignes de "gonflement".
- L’influence de la granulométrie d’un sable sur son comportement peut être analysée de la façon suivante. Comparons deux sables, le premier constitué de grains de même diamètre, le second de grains de diamètres différents. Pour une même pression de préconsolidation, le volume spécifique du premier sable sera supérieur à celui du second. En conséquence, une granulométrie étroite du sable induira un domaine expérimental possible de pression effective (plan (p’,v)) qui correspond majoritairement à des états initiaux surconsolidés du milieu. Ce sable sera plutôt dilatant. Par contre, une granulométrie étalée du sable induira un domaine expérimental possible de pression effective (plan (p’,v)) qui correspond majoritairement à des états initiaux faiblement surconsolidés du milieu. Ce sable sera plus faiblement dilatant, voire contractant pour des pressions suffisantes.

Les lois de comportement exposées pour illustrer le comportement mécanique d'un sable sont assez simple. Elles permettent de comprendre les influences des conditions expérimentales (conditions de chargement mais surtout conditions de drainage) sur la réponse d'un massif de sable soumis à des chargements. Nous avons envisagé des cas extrêmes de drainage ou de non drainage. Cependant, durant l'étude du comportement réel d'un massif, ces conditions de drainage peuvent évoluer. Citons par exemple le cas du tic tac d'un sablier. Ceci se traduit par un chemin de contraintes qui peut être très compliqué, inhomogène, autodéterminé et donc difficilement intuité.

Par ailleurs, on constate, sur les échantillons de sable cisaillés, que le phénomène de dilatance conduit à une localisation des déformations dans une zone voisine des plans de cisaillement (épaisseur de cette zone estimée entre 10 à 18 fois la taille moyenne des grains du milieu). Cette localisation des déformations crée des plans de "glissement" qui sont prédits par le modèle mécanique retenu (Fig. 3.2). Ce phénomène valide donc l'approche initiale de Coulomb. Les caractéristiques de la bande de cisaillement restent liées aux comportements mécaniques locaux des grains et ne peuvent être prises en compte par l'approche macroscopique présentée. Une telle perte d'homogénéité dans l'échantillon peut être à l'origine d'une non stabilisation de q à sa valeur correspondant à l'état critique en fin d'essai, car la direction de glissement peut être lié à $q_{max}$ et non à $q_{\text{état critique}}$.

Biarez et Hicher (1994) fournissent une présentation expérimentale plus complète et largement illustrée des lois de comportement que nous venons de proposer. Ce dernier s'adapte globalement correctement aux réponses d'un milieu granulaire en écoulement plastique quasi statique. Bien que n'intégrant pas la notion d'endommagement (Mazard 1984), l'adaptation de ces lois aux milieux cohésifs constitue une première approche acceptable du comportement des roches ou du béton durci. Ces lois peuvent être transposées à l'étude d'écoulement de milieux granulaires (béton frais, pâtes céramiques...). Dans ce cas la vitesse relative entre grains, induite par le cisaillement du milieu, est en concurrence avec la vitesse relative du fluide interstitiel (soumis à un gradient de pression) par rapport aux grains (Davis 1995). La vitesse d'écoulement et la perméabilité du milieu conditionnent ainsi la conservation de l'homogénéité du milieu en écoulement. On peut trouver quelques éléments relatifs aux écoulements de milieux granulaires sont rassemblés dans Thornton (1993).

## Bibliographie:

# 4. Compléments de mécanique quasi statique des milieux granulaires sous chargements simples

Nous décrirons dans ce chapitre un certain nombre de résultats expérimentaux obtenus sur des matériaux granulaires. Nous montrerons dans les chapitres 5 & 7 que ces observations expérimentales peuvent avoir une explication théorique simple.

## 4.1. Cas des états très lâches sous contraintes isotropes ou anisotropes :

Nous avons vu que l'espace des phases qui permettait de décrire la mécanique d'un milieu granulaire était l'espace (v, q, p'). De plus le volume spécifique v=1+e (où e est l'indice des vides) d'un empilement stable est toujours inférieur à une certaine valeur qui dépend de manière logarithmique de la pression moyenne $p'=(\sigma'_1+\sigma'_2+\sigma'_3)/3$. Pour un milieu granulaire non cohésif, on trouve de plus que c'est lorsque la contrainte est isotrope que la densité minimale réalisable est la plus faible. Cette limite est appelée sable normalement consolidé sous chargement isotrope. Le lieu de ces points $v_{max}(p')$ est une droite de pente $-\lambda$ dans le plan {v, ln(p')}. Cependant, comme la densité d'un tas formé de grains en contact ne peut tendre vers 0 aux très faibles pressions, ceci impose que le volume spécifique $v_{max}(p')$ tend vers une constante à faible pression.

### *4.1.1. Etats normalement consolidés sous chargement anisotrope :*

On peut de même tracer le lieu des volumes spécifiques maximum réalisables sous chargement anisotrope. On trouve de la même façon une droite parallèle à la précédente, placée au dessous d'elle à une distance qui croit avec l'amplitude relative $\eta=q/p'$. Une de ces droites, celle pour laquelle $\eta=M'$, est le lieu des points critiques. On trouve que ces droites ont pour équation (Roscoe et al., 1958):

$$v_{max}(\eta)= v_{nco}(p_o)- \lambda \ln(p'/p'_o) - \lambda_d \ln(1+\eta^2/M'^2) \qquad (4.1a)$$

$\lambda$ varie d'un sable à un autre et dépend de la granulométrie. $v_{nco}$ et $p'_o$ sont deux constantes, caractéristiques du sable considéré. Cette pente est caractérisée par la valeur des volumes spécifiques minimum et maximum. Au contraire, $\lambda_d$ est toujours de l'ordre de 0,14 quel que soit le sable.

### *4.1.2. Etats critiques :*

Pour $\eta=q/p'=M'$, on a la droite des états critiques d'équation :

$$\eta=q/p'=M' \qquad (4.1b) \qquad \&$$

$$v_c= v_{nco}(p_o)- \lambda \ln(p'/p'_o) - \lambda_d \ln(2) = v_c(p'_o)- \lambda \ln(p'/p'_o) \qquad (4.1c)$$

C'est la même équation que l'Eq. (3.11)

### *4.1.3. Courbe de déchargement :*

Lorsque l'empilement est situé sur la droite des états normalement consolidés isotrope, il correspond donc bien à l'empilement le plus lâche réalisable. si on part d'un de ces états et qu'on diminue la pression p', le milieu granulaire se dilate légèrement ; si maintenant on comprime une autre fois le matériau, le volume spécifique diminue et suit pratiquement le trajet précédent en sens inverse jusqu'à ce qu'il atteigne le volume spécifique de l'état normalement consolidé. Cette déformation peut donc être considérée comme élastique ou réversible, puisque le matériau récupère son volume spécifique initial. Ce trajet fait intervenir l'élasticité de la structure granulaire dont le module d'Young est très élevé et le trajet considéré apparaît donc comme un segment de droite presque horizontale dans le diagramme précédent {v, ln(p')}.

### *4.1.4. Réponse élastique ou pseudo élastique:*

On constate que tout essai triaxial sur un sable isotrope part de façon tangente à cette courbe de déchargement dans le plan {v, ln(p')} ou (v, p'), quelque soit le chemin de chargement.

***Interprétation :*** Supposons le matériau isotrope et modélisons son comportement par sa réponse incrémentale liant $\delta\sigma$ et $\delta\varepsilon$. Si on appelle E et $\nu$ le pseudo module d'Young et le pseudo coefficient de Poisson de la structure granulaire, la modélisation incrémentale est alors analogue à celle de l'élasticité isotrope, ce qui donne (Landau et Lifchitz, 1990):

$$(4.2) \qquad \varepsilon_v = \varepsilon_{11} + \varepsilon_{22} + \varepsilon_{33} = [(1-2\nu)/E](\sigma_{11} + \sigma_{22} + \sigma_{33}) = (1-2\nu)p'/E$$

L'Eq. (4.2) montre que la variation de volume est proportionnelle à la pression moyenne, et seulement à cette pression moyenne, avec un coefficient de proportionnalité $(1-2\nu)/E$ . Ceci implique donc que tout essai triaxial partant d'un empilement isotrope, mais suivant n'importe quel chemin de chargement, doit débuter par un segment de courbe tangent à la courbe de déchargement isotrope.

Dans la pratique, le coefficient de Poisson $\nu$ est une constante qui vaut entre 0,2 et 0,3 , indépendante de la pression. On trouve que le module d'Young E dépend de la pression de confinement exercée sur le sable.

$$(4.3.a) \qquad E \approx E_o(p'/p'_o)^{1/2} \quad \text{et } \nu = 0.25$$

On considère en général que cette dépendance «anormale» du module d'Young en fonction de la pression est liée à la forme tridimensionnel des grains qui forment le milieu. Dans ce cas en effet, la surface de contact entre les grains varie en fonction de la force de contact F et/ou de l'écrasement h. Si l'on fait l'hypothèse de contacts élastiques, on trouve comme Hertz que F varie comme $h^{3/2}$ (Landau & Lifschitz 1990, pp.45-50). Après moyennage sur l'ensemble des grains, on trouve que la pression moyenne p' doit varier comme $\varepsilon^{3/2}$. Ceci conduit à une dépendance du module d'Young avec p' :

(4.3.b) $E/E_o=(p'/p'_o)^{1/3}$

On peut expliquer de différentes manière la différence entre le modèle élastique et la loi expérimentale. Tout d'abord, on peut considérer que la densité ρ du nombre de contacts dépend elle aussi de façon non linéaire de la pression p' appliquée à l'ossature granulaire. Ceci modifie la relation entre F et p', et donc celle entre (F,h) et (E,p'). D'autre part, si l'on avait considéré des grains de forme cylindrique, la loi de puissance trouvée aurait été: $E/E_o=(p'/p'_o)^{1/2}$ . Ainsi, on peut expliquer la différence en considérant que les surfaces réelles en contact ont une forme très allongée ; on pourrait aussi compliquer en considérant que l'ellipticité de ces contacts dépend de la pression p'. Par ailleurs, on pourrait supposer que chaque grain est recouvert d'une pellicule élastique ; ceci modifierait aussi la loi F vs. h. Enfin, il n'est pas sur que le calcul correcte du module d'Young consiste en une moyenne simple, comme nous l'avons fait. On sait par exemple que cette moyenne donne de bons résultats pour un polycristal pour lequel chaque cristallite a une réponse relativement isotrope, mais que ce calcul est faux dans des cas plus anisotrope (Landau & &Lifschitz 1990, pp.58-59). Nous cesserons là cette discussion, sachant qu'il existe probablement bien d'autres modèles théoriques qui pourraient expliquer la dépendance de l'Eq (4.3.a).

***Remarque importante :*** L'intégration de l'Eq. (4.2) en tenant compte de l'Eq. (4.3) donne $\ln(v/v_o)=2(1-2\nu)(p'_o)^{1/2}\,(p'_o-p')^{1/2}$ . Cette variation n'est donc pas linéaire ; elle ne donne pas le comportement expérimental dans le plan {v, ln(p')}. Cependant, l'amplitude des variations de v est très faible lors de la décompression, beaucoup plus faible que la plage de densité réalisable. Ainsi, on peut considérer que la trajectoire des points est presque une droite horizontale dans cette gamme de volume. En fait, le comportement linéaire de v en fonction de ln(p') à la décompression n'a été vraiment étudié que dans le cas des argiles, il y est beaucoup plus visible, sa pente $\lambda_s$ est comparable à la pente λ des états critiques et à celle des états normalement consolidés isotropes. Et c'est par référence à ce comportement que l'on garde la même présentation en mécanique des sols.

La Fig. 4.1 récapitule les différents comportements du milieu granulaire à déviateurs relatifs constants. La Fig. 4.1.a décrit la réponse élastique. Nous allons maintenant expliciter la position relative de ces états dans le plan {v,ln(p')} grâce à différents exemples de compression et à la Fig.4.1.b.

### *4.1.5. Cisaillement drainé:*

Considérons un état normalement consolidé isotrope ; le lieu de ces états est donné par l'Eq. (4.1a) avec η=0 ; c'est la droite NC sur la Fig. 4.1.b d'équation : $v_{nc}= v_{nco}(p_o)- \lambda\ \ln(p'/p'_o)$. Appliquons à un de ces états normalement consolidés isotropes une compression drainée à $\sigma'_2=\sigma'_3=c^{ste}$ . L'état initial est caractérisé par la pression $p'_i=\sigma'_3$ de compression initiale. Il subit donc une surcharge définie par le déviateur de contrainte $q=\sigma_1-\sigma_3=\sigma'_1-\sigma'_3$. Dans ces conditions, la densité du milieu évolue conformément à l'Eq . (4.1.a) ; elle finit par se stabiliser lorsque le milieu atteint l'état critique, c'est-à-dire lorsque η=M', ou ce qui est équivalent $q=M\sigma'_3=Mp'_i$. La pression finale $p'_f$ est alors donnée par $p'_f=q/3+\sigma'_3= q/3+p'_i = p'_f=(1+M/3)p'_i$; la densité finale est donnée par l'Eq. (4.1.a) avec $\eta=M'$ et $p'=p'_f$ ; soit : $v_f= v_{nco}(p_o)- \lambda \ln([M/3+1]p'_i /p'_o) - \lambda_d \ln(2)$.

Cet état final est un point du lieu des états critiques ; dans le plan {v, ln(p')}, ce lieu est une droite C parallèle à la droite (NC) et en dessous d'elle.

Si l'on part d'un état normalement consolidé isotrope NC défini par son volume spécifique $v_{nc}$ et sa pression $p'_i$, qu'on travaille en drainé et qu'on augmente le déviateur q, le volume spécifique v décroît et suit une trajectoire dans le plan (v, ln(p')) décrite par l'équation: $v_\eta=v_{nc} - \lambda_d \ln[1+(\eta/M)^2]$ avec $\eta=q/p'$ et $p'=p'_i+q/3$ ; c'est la formule de Roscoe.

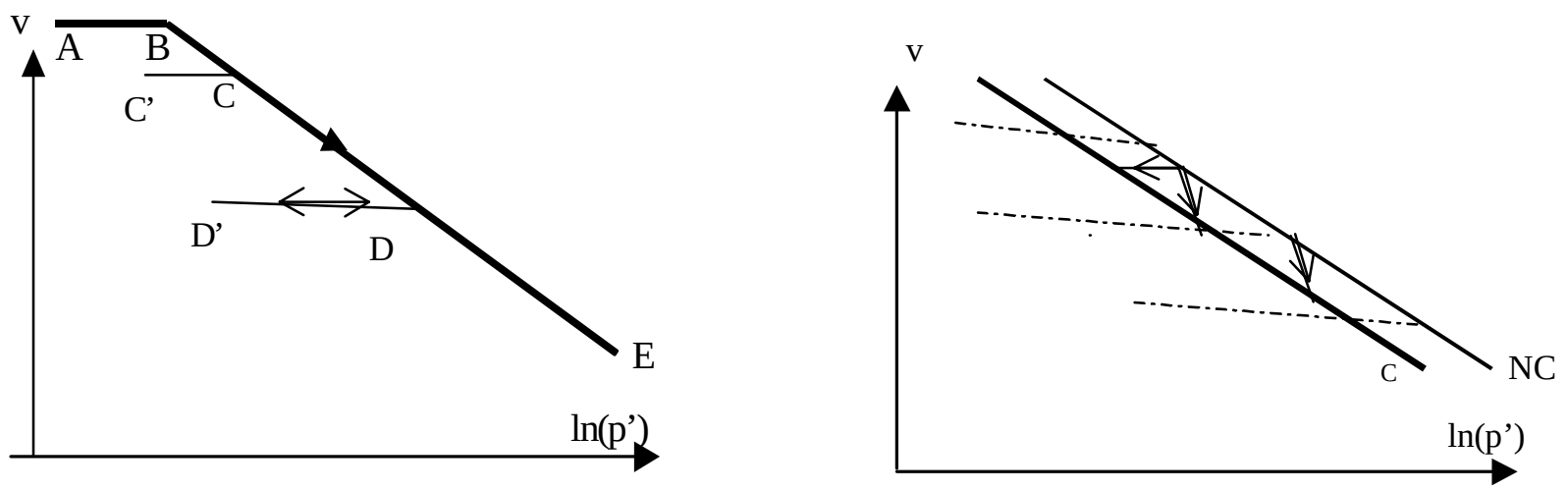

***Figure 4.1: 1.a: état normalement consolidé:*** *Par référence au comportement de l'argile, la densité minimale des empilements isotropes de sable dépend de la pression de confinement; c'est la droite* BCDE *dans le plan* v-ln(p'). *Elle est décrite par l'équation :* $v_{nc} = v_{nco} + \lambda \ln(p'/p'_o)$. *En fait le volume spécifique d'un sable de grains donnés ne peut pas excédé une certaine valeur vmax, ce qui impose l'existence de la droite horizontale* AB *pour finir le diagramme des états les plus lâches. De plus, le volume spécifique ne peut être inférieur à une certaine valeur* $v_{min}$; $v_{min}$ *est supérieur à 1 par définition, car c'est le volume spécifique du solide seul par définition; il ne peut . Ainsi, aucun point au dessus de ABCDE et au dessous de* $v= v_{min}$ *correspond à un état physique réalisable. On a reporté la droite* BCDE *sur* la Fig. 4.1.b où elle est marquée NC *(normalement consolidé).*

*Quoique l'expérience soit difficile à réaliser, en principe le parcours de la droite* BDE *ne peut avoir lieu que de gauche à droite, c'est-à-dire en augmentant la pression. C'est un parcours irréversible. Se pose alors le problème de la décompression.*

***Décompression:*** *Lorsqu'on part d'un état normalement consolidé isotrope* ($v_{nc1}$, *pression* $p'_1$), *(situé donc sur la droite* BDE, *par exemple en* D *ou en B) et qu'on diminue la pression p'* (sans cisailler), le volume spécifique augmente faiblement dans le plan (v,ln(p'*)) et l'on peut décrire son évolution par l'équation* $v = v_{nc1} - \lambda_1 \ln(p'/p'_1)$ *avec* $\lambda_1$ *très petit; c'est la trajectoire* DC *ou* BA. *Cette trajectoire est presque réversible quand on comprime à nouveau.*

***Figure 4.1.b :*** *Sous chargement anisotrope $\eta=q/p=cste$, on constate que les empilements les plus lâches sont plus denses que les précédents, ceci indique que les états de densité minimum précédents ne sont plus stables. A la place, on obtient un ensemble de droites parallèles à la précédente placé en dessous d'elle et qu'on peut paramétrer en fonction du déviateur relatif de contrainte appliqué* $\eta=q/p'$ *de la façon suivante :* $v= v_{nco}(p_o)- \lambda \ln(p/p_o) - \lambda_d \ln(1+\eta^2/M^2)$. *Seule la droite pour* $\eta=M$ *est représentée sur la figure 1.b ; elle est notée* C *et est appelée droite des points critiques. Les autres droites pour* $0<\eta<M$, *sont parallèles à ces deux droites et situées entre elles. Pour mémoire, on reporte en pointillés sur cette figure les courbes de décompression qui sont presque des droites horizontales.*

Cette équation implique que le lieu des points atteint en imposant un déviateur q/p' donné en partant d'un état normalement consolidé est une droite parallèle à la droite normalement consolidée et à la droite des points critiques. De plus ces droites sont aussi les lieux des états normalement consolidés sous un cisaillement $\eta$=q/p' donné.
Cette équation implique aussi que la trajectoire démarre de façon tangente à la droite des états normalement consolidés.

### *4.1.6. Cisaillement non drainé:*

Dans ce cas on impose que le volume spécifique reste constant v. La trajectoire représentant l'évolution est donc une droite horizontale dans le plan (v, p') mais la pression p' doit s'ajuster tout au long de la compression. La trajectoire est donc une droite horizontale qui rejoint la droite des états critiques. D'après le diagramme (Fig. 4.1), ceci ne peut se faire que si p' décroît ; en d'autres termes, si on travaille à $p=p'+u_w=c^{ste}$ , cela veut dire que la pression $u_w$ de l'eau augmente. C'est ce que nous avions mentionné dans le paragraphe 3.4.1.

### *4.1.7. Compression à déviateur relatif constant :*

Compte tenu de l'ensemble de ces résultats, on peut prévoir le comportement d'un milieu granulaire de densité initiale donnée $v_i$ et soumis à une compression sous déviateur relatif q/p' constant . Dans le plan (q,p'), la trajectoire est une droite de pente constante. Pour caractériser l'évolution du système, il suffit donc de représenter sa trajectoire dans le plan {v, ln(p')}. C'est ce que nous allons faire.
Supposons le milieu granulaire dense ; sa réponse doit donc être quasi élastique dans une première phase et obéira à l'Eq. (4.2). Il subira ainsi tout d'abord une compression quasi élastique lui permettant de rejoindre la droite représentant l'état normalement consolidé sous déviateur donné $\eta$. Cette première partie de trajectoire est donc une courbe pratiquement horizontale, *i .e.* de pente très faible, dans le plan {v, ln(p')} ; cette courbe est parallèle à la courbe C'C. Arrivé au point tel que: $v_i \approx v_{nco}(p_o) - \lambda \ln(p'/p'_o) - \lambda_d \ln(1+\eta^2/M'^2)$, l'état du sable a rejoint un état normalement consolidé anisotrope . A ce stade, si on continue à le comprimer, sa trajectoire obliquera et se poursuivra en glissant vers le bas le long de la droite des états normalement consolidés sous déviateur donné d'équation :

$$v = v_{nco}(p'_o) - \lambda \ln(p'/p'_o) - \lambda_d \ln(1+\eta^2/M'^2). \qquad (4.4)$$

A ce stade final, on peut vouloir le décomprimer. Dans ce cas son évolution suivra une décompression quasi élastique représentée par une courbe parallèle à DD' sur la Fig. 4.1.a .

## 4.2. Essai triaxial sur trajectoire déviatoire simple

Ces bases étant acquises, on peut chercher à comprendre et à représenter le comportement d'un milieu granulaire soumis à un essai triaxial drainé. En fait, il existe deux types d'essais triaxiaux simples, ceux qui travaillent à contrainte déviatoire imposée croissante et ceux qui travaillent à déformations imposées. Dans le premier cas, on mesure la variation de volume $\delta v$ et de hauteur $\delta h$ pour une compression $\delta q$ donnée et on reporte les résultats dans les plans (q, $\varepsilon_1$) et (v, $\varepsilon_1$) ou ($q/p'$, $\varepsilon_1$) et ($v$, $\varepsilon_1$). Dans le second cas, on impose des variations incrémentales $\delta h$ et on mesure les variations de q et de v nécessaires pour réaliser cette déformation. De la même façon, on reporte les résultats dans les plans ($q$, $\varepsilon_1$) et ($v$, $\varepsilon_1$) ou ($q/p'$, $\varepsilon_1$) et ($v$, $\varepsilon_1$).

Pour caractériser complètement la nature de l'essai, il faut préciser dans les deux cas précédents quelle est la variation de pression $\sigma'_2$ et $\sigma'_3$. En général on impose $\sigma'_2 = \sigma'_3 = c^{ste}$ tout le long de l'essai, mais on pourrait imposer $p' = c^{ste}$.

Nous donnons des résultats typiques obtenus sous déformation $\varepsilon_1$ croissante dans la Fig. 4.2 . La Fig. 4.2.a correspond à trois résultats différents, obtenus sur le même sable compacté à différentes densités soumis à la même pression p'. La Fig. 4.2.b correspond à trois résultats différents, obtenus sur le même sable compacté à la même densité initiale, mais soumis à la même pression initiale $p_i'$. Les six essais ont été réalisés en imposant $\sigma'_2 = \sigma'_3 = c^{ste}$ (et $p' = \sigma'_3 + q/3$).

### *4.2. 1 Aux grandes déformations ($\varepsilon_1$>30%) : état critique.*

On constate que le volume spécifique de chacun de ces essais tend vers une constante qui ne dépend que de la pression de confinement $\sigma'_2 = \sigma'_3 = c^{ste}$. Ce volume spécifique est donc celui de l'état critique $v_c$ (*cf.* Eq. 4.1.a) que nous avons déjà défini. De même, on constate que le rapport q/p' tend vers une constante M' aux grandes déformations. Cette valeur est indépendante des conditions initiales et de la pression moyenne. Elle caractérise la dissipation. C'est donc la mesure du frottement solide $k = tg\varphi$. En utilisant le cercle de Mohr, on sait que $\sin\varphi = (\sigma'_1 - \sigma'_3)/(\sigma'_1 + \sigma'_3)$. Ceci impose que $M' = 3(\sigma'_1 - \sigma'_3)/(\sigma'_1 + 2\sigma'_3)$ peut s'écrire :

(4.5.a) $$M' = (q/p')_{ec} = 6\sin\varphi/(3 - \sin\varphi) \quad \text{ou} \quad \sin\varphi = 3M'/(6 + M')$$

De la même façon, si on appelle M le rapport $q/\sigma'_3$ à l'état critique, la relation entre M et $\sin\varphi$ est:

(4.5.b) $$M = (q/\sigma'_3)_{ec} = 2\sin\varphi/(1 - \sin\varphi) \quad \text{ou} \quad \sin\varphi = M/(2 + M)$$

### *4.2. 2. Régime des petites déformations ($\varepsilon_1$<1%) :*

On constate i) que le déviateur croît très vite en fonction de $\varepsilon_1$ au début de la déformation $\varepsilon_1 < 0{,}5\%$ et ii) que la variation du volume spécifique est toujours négative. Nous verrons que cette deuxième caractéristique est nécessaire pour assurer la stabilité mécanique de l'empilement.

### *4.2. 3 Régimes des déformations intermédiaires* (1%<$\varepsilon_1$<30%) *:*

Dans cette gamme de déformation l'évolution du système dépend de la densité initiale $\rho$ et de la pression p' de confinement : si le milieu est s'il est relativement stable lâche pour la pression p' considéré ($\rho$ petit et p' grande), c'est-à-dire s'il est proche d'un état normalement consolidé, on constate que le volume spécifique décroît constamment jusqu'à tendre vers sa valeur finale $v_c$ et de même le rapport q/p' croît constamment pour tendre vers M' ; en revanche, si $\rho$ est plus grand et/ou p' plus petit,

i) le volume spécifique passe par un minimum (pour $\varepsilon_1 \approx 1$à2%) puis croît pour tendre asymptotiquement vers la constante $v_c$(p'), ii) le déviateur q/p' dépasse la valeur M', atteint un maximum $q_{max}$/p' correspondant à un "pseudo frottement de pic " $\varphi_{pic}$ , puis décroît pour atteindre assymptotiquement la valeur M'.

### *4.2.4. Description complète dans l'espace* (v,q,p')

Cependant il s'avère que pour bien comprendre l'évolution de l'échantillon les deux projections précédentes ne sont pas suffisantes. Il faut représenter l'évolution dans l'espace (q, p') et le plan {v, ln(p')}. Ceci est fait dans la Fig. 4.2. On voit alors:

- La trajectoire du système dans le plan (q, p') est une droite de pente 3 car $p'=(q+3\sigma'_3)/3$, par définition ; ceci donne $q=3p'-3\sigma'_3$. Cette droite peut couper la droite des états critiques q=M'p'.
- Dans le plan (q/p', $\varepsilon_1$) on remarque la valeur asymptotique M' du rapport q/p' qui ne dépend ni de p', ni de v initial. Dans le plan (v, $\varepsilon_1$) on remarque la valeur assymptotique $v_c$ de v qui dépend de v initial, mais pas de p'. Ces deux états limites définissent l'état critique ou de plasticité parfaite. On remarque aussi que v est contractant au début de la déformation puis il peut se dilater ; quand il se dilate le maximum de q/p' est plus grand que M'.

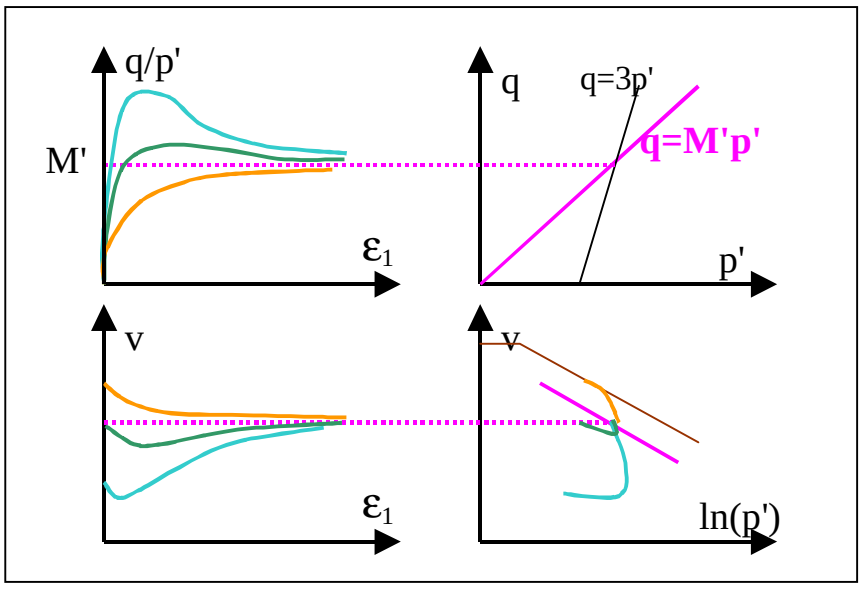

***Figure 4.2.a***

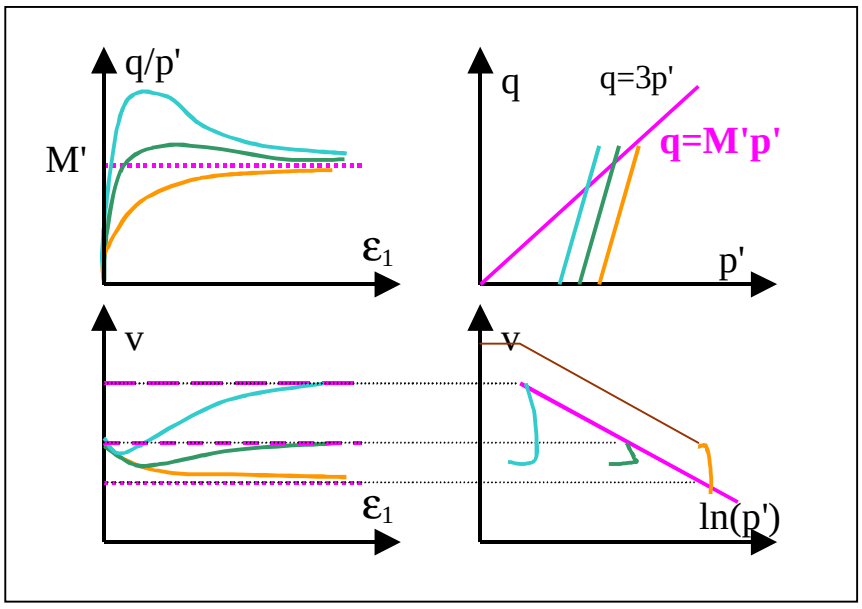

***Figure 4.2.b***

***Figure 4.2:*** *Résultats typiques d'essais triaxiaux à déformation $\varepsilon_1$ imposée et $\sigma'_2=\sigma'_3=c^{ste}$. Fig. 4.2.a: trois essais sur le même sable compacté à trois densités différentes et soumis à la même contrainte $\sigma'_3$ . Fig. 4.2.b: trois essais réalisés sur le même sable compacté à la même densité initiale, mais soumis à trois valeurs différentes de la contrainte $\sigma'_3$ .*

• Dans le plan (q,p’) la trajectoire de chaque échantillon est une droite de pente 3. Lorsque l’empilement est suffisamment dense, la trajectoire coupe la droite M’p’ atteint un maximum et redescend pour s’arrêter sur la droite M’p’. Si la densité initiale est suffisamment faible, la trajectoire ne coupe pas cette droite, mais l’atteint juste (i.e. elle s’arrête dessus).

Dans les plans (q, p’) et {v,ln(p’)}, toutes les trajectoires commencent à la même pression p’ mais à des v différents pour la Fig. 4.2.a, tandis qu’elles commencent à des v identiques, mais à des pressions différentes pour la Fig. 4.2.b. Il s’ensuit que dans le plan (q, p’), les trois trajectoires sont partiellement superposées pour la Fig. 4.2.a, mais totalement distinctes et parallèles dans la Fig. 4.2.b.

Dans le plan {v, ln(p’)}, compte tenu de ce qui a été dit pour la Fig. 4.1, chaque trajectoire doit débuter en partant à droite presque horizontalement, mais en descendant légèrement. Si $p'_i$ est la pression initiale d’un essai, compte tenu de la trajectoire dans le plan (q, p’), on trouve que le pression finale $p'_f$ de la trajectoire doit être $p'_f = 3p'_i/(3-M')$ , ce qui se traduit dans le plan {v, ln(p’)} par une distance constante sur l’axe horizontale : $\ln(p'_f)-\ln(p'_i)=\ln 3-\ln(3-M')$. La fin de chaque trajectoire doit se trouver sur la droite des états critiques. Ces deux conditions fixes le point final de la trajectoire. Si l’état initial est situé à gauche de la droite des états critiques et s’il est suffisamment éloigné d’elle, le point critique qui doit être atteint à la fin de l’évolution doit présenter un volume spécifique plus grand. Ainsi, le système sera obligé de se dilater pour rejoindre cette dernière droite. Par contre, si l’état initial est situé à droite ou très légèrement à gauche de la droite critique, le système diminuera toujours de volume.

Dans cette figure, il n’a pas été possible de conserver l’échelle des contractances : le volume spécifique final est très souvent plus grand que le volume initial dans la réalité.

## 4.3. Analyse de stabilité :

### *4.3.1. Instabilité de la zone située après le pic de contrainte :*

Même s’il n’y a pas de démonstration générale du principe que nous allons énoncer, on sait que pour qu’un état mécanique soit stable il faut que le travail d’ordre 2, c’est-à-dire $\delta^2W= \underline{\underline{\delta\sigma}}\ \underline{\underline{\delta\varepsilon}}$ , soit positif.

(4.6) Stabilité : $\delta^2W= \underline{\underline{\delta\sigma'}}\ \underline{\underline{\delta\varepsilon}} , > 0$

On peut appliquer ce principe à l’analyse de la stabilité des états successifs obtenus au triaxial pour montrer que l’état correspondant au maximum du rapport de q/p' est instable: Comme $\sigma'_2=\sigma'_3=c^{ste}$ dans le trajet de chargement choisi, on a $\delta^2W=\delta\sigma'_1\ \delta\varepsilon_1$. La loi de chargement impose de plus que $\delta\varepsilon_1$ est positif; ainsi, la stabilité n’est obtenue que si $\delta\sigma'_1$ croît. Ainsi, dès que le système passe par un maximum de $\sigma'_1$ , il devient instable.

C’est pourquoi il y a une grande différence entre un essai à chargement imposé et un essai à déformation imposée : dans le premier cas, il est impossible de poursuivre l'essai au delà du maximum de contrainte sans entraîner une déformation macroscopique incontrôlée; l'évolution du matériau devient incontrôlable. Dans le second cas, il est possible d'explorer la courbe de chargement complète si le matériau reste spontanément homogène; s'il ne reste pas homogène, son évolution globale reste contrôlée, mais pas son évolution locale et son inhomogénéité grandit.

#### *4.3.2. la stabilité du point de chargement isotrope impose que le système est contractant en ce point :*

Nous verrons au §4.2.1. (1er résultat) une autre application de ce principe de stabilité. Nous montrerons en effet qu'il implique qu'un échantillon isotrope de sable soumis à une charge isotrope et à un incrément $\delta\sigma'_1$ positif doit être contractant pour être stable.

### 4.4. valeur de $q_{max}$ ; Surface de Hvorslev :

Ainsi, les états pour lesquels le déviateur q est maximum sont instables. La valeur de $q_{max}$ dépend des conditions initiales, c'est à dire de $v_i$ et de $p'_i$ et des conditions de chargement. Ainsi $q_{max}$ est une fonction de $v_i$ et $p'_i$: $q_{max}(v_i , p'_i)$ pour un type de chargement. Cependant, la connaissance de la loi de chargement (par exemple $\sigma'_3=c^{ste}$) et la connaissance de $p'_i$ permettent de connaître la pression finale et/ou le volume finale $v_c$. On en conclut que connaître $p'_i$ et $v_i$ est équivalent à connaître $v_i$ et $v_c$ , pour une loi de chargement donnée. Ceci implique donc qu'il doit exister une relation entre $q_{max}$, $v_i$ et $v_c$: $q_{max}(v_i, v_c)$.
L'analyse des résultats expérimentaux à $\sigma'_3=c^{ste}$ a permis à Biarez et Hicher (1994) de trouver une relation intéressante entre la valeur de $q_{max}/p'$ et l'indice des vides initial e (ou le volume spécifique v, car v=1+e) :

$$(4.7.a) \qquad (v_{initial}-1)\ tg\varphi_{pic}= (v_c-1)\ tg\varphi$$

où $\varphi_{pic}$ est défini de façon identique au frottement solide parfait $\varphi$ que l'on trouve dans les équations (4.5) :

$$(4.7.b) \qquad \sin\varphi=3M'/(6+M') \qquad \text{et} \qquad \sin[\varphi_{pic}] = 3q_{max}/(6p'+q_{max})$$

Nous avons vu au chapitre précédent (§ 3.4.2 ) que le système évolue spontanément vers un état final, appelé état critique, et que l'ensemble formé par la fin des trajectoires possibles décrivent la surface de Hvorslev. Nous supposerons que la trajectoire décrivant l'échantillon rejoint cette surface de Hvorslev très vite, *i.e.* dès que $q_{max}$, a été dépassé. Nous donnerons plus loin une interprétation topologique (§4.2.2) de cette surface de Hvorslev en termes très généraux, avec la théorie des systèmes dynamiques. Et nous démontrerons que c'est bien une surface de dimension 2.

### 4. 5. Relation de Rowe :

Si l'on fait subir à du sable des essais triaxiaux sur chemin monotone tel que $\varepsilon_1$ est toujours croissant, on trouve expérimentalement que la variation de volume est reliée à la valeur du déviateur de contrainte par :

(4.8) $\sigma'_1/\sigma'_3=\tan^2(\pi/4+\varphi/2)(1-\partial\varepsilon_v/\partial\varepsilon_1)$

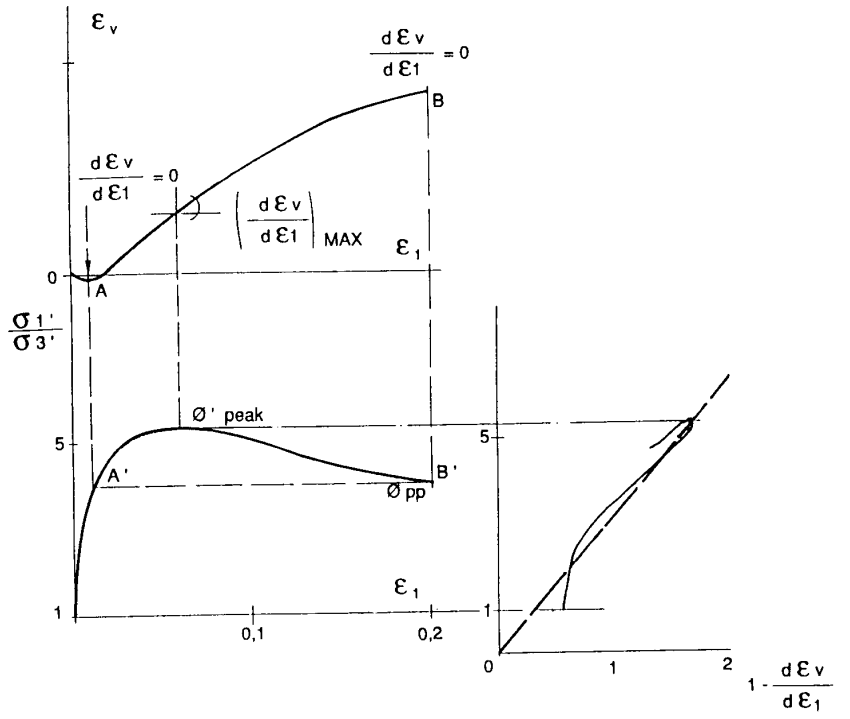

***Figure 4.3**: vérification expérimentale de la loi de Rowe*

Rowe a été le premier a trouvé cette relation en partant de considération théorique sur les rôles relatifs du frottement solide et de la dilatation lorsque le matériau granulaire subit une déformation. Son calcul est basé sur l'analyse du mouvement local des grains pour des empilements 2-d et réguliers de grains; son approche prédit l'existence d'une direction unique de glissement pour un état de contrainte donnée. C'est donc une théorie de plasticité à un seul mécanisme. Ensuite, il a pu reformuler son équation de mouvement, qui lie contraintes et déformation, de telle sorte qu'elle ne dépende plus du réseau choisi. C'est l'Eq. (4.8). Enfin, il a étendu ce résultat par un raisonnement de champ moyen à un réseau quelconque 2-d, puis 3-d; et il a vérifié expérimentalement l'efficacité de son modèle sur des essais triaxiaux à $\sigma'_3$=cste. Cette relation est relativement bien vérifiée (Frossard 1978), cf. Fig. 4.3.
Nous discuterons la validité de cette approche dans les chapitres 5 et 7. En effet, elle suppose en particulier que la déformation est gérée par un seul mécanisme de plasticité, ce qui ne nous semble pas acceptable dans le cas général, et ce qui n'est pas vérifié expérimentalement. Ceci dit, l'Eq. (4.8) de Rowe est vérifiée expérimentalement, et nous nous en servirons comme base pour construire notre modèle théorique du comportement du sable et des milieux granulaires, cf. Chapitres 5 et 7 .

### 4.6 Fracture des grains :

Lorsque la pression de confinement est trop importante, on observe le broyage et la fracture des grains au cours de la déformation. Pour du sable par exemple, cela arrive dès 10 MPa. Comme la contrainte dans un sable est distribuée de façon aléatoire à l'échelle locale, *cf.* chap. 8, les contraintes locales supportées par les contact peuvent être dix fois supérieures aux contraintes moyennes, cela veut dire que la contrainte seuil de rupture de ces grains est 100 MPa, approximativement. Ce résultat est assez bien corroboré aux résultats d'expériences de fracturation sur des grains isolés. Du point de vue expérimentale, ce mécanisme de fracturation a pour conséquence d’augmenter l’inclinaison des droites lieux des états normalement consolidés et critiques au dessus de 10 MPa dans un diagramme (v, ln(p’)); bien entendu ceci se fait progressivement, sans rupture de pente brusque (cf. Fig. 6.4 de Biarez-&-Hicher-1994).

## 4.7. Autres essais :

### *4.7.1. essai à volume constant :*

Avec l’appareil triaxial on peut faire bien d’autres essais. Par exemple, on pourrait chercher à travailler à volume constant, même si le milieu n’est pas saturé d’eau ; cela forcerait à réduire la pression latérale de confinement au fur et à mesure de l’augmentation du déviateur de contrainte ; dans ce cas, l’évolution de l’état du matériau serait semblable à celle obtenue en saturant les pores d’eau et nous reportons donc le lecteur intéressé à la section concernée du chapitre précédent (§ 3.4.1.1 & § 3.4.2.1).

### *4.7.2. essai à pression moyenne p’ constante :*

Si l’on travaille à pression $p'=(\sigma'_1+\sigma'_2+\sigma'_3)/3$ imposée, les résultats s'interprètent de la même manière que précédemment. Pour un matériau donné, à une pression de confinement donnée p’ donnée et à un volume spécifique initiale v donné, toutes les variations observées n’auront lieu que dans le plan (v,q). Cependant, pour comparer les résultats d’essais entre eux, il sera bon d’utiliser la troisième dimension p’. lorsqu’on applique un déviateur de contrainte, le matériau doit rejoindre la courbe de plasticité parfaite (états critiques) ; s’il est trop dense, il devra donc se dilater, et s’il n’est pas assez dense, il devra se contracter.
Certains exemples sont donnés dans Biarez & Hicher 1994, où les Fig. 4.10 et 4.11 concernent des argiles. On trouvera d’autres exemples dans Saïm (1997) (Fig. 3.10). Nous proposerons une théorie simple pour comprendre l'évolution des échantillons dans le chapitre 7.

### *4.7.3. essai oedométrique :*

On appelle chemin oedométrique un chemin de contrainte tel que le rayon de l’échantillon est constant. Ce chemin a une grande importance pratique pour le dimensionnement des fondations lors de la construction des maisons, …, puisqu’il permet l’étude et la simulation de l’enfoncement et de la compression des terres sous une fondation. C’est aussi un chemin intéressant les géologues, puisqu’il permet de caractériser l’évolution de l’état de contrainte d’un sol recouvert après de nombreuses sédimentations.
*En compression,* si on part d’un sable ou d’une argile relativement lâche, on trouve expérimentalement qu’après une compression suffisante le rapport des contraintes $\sigma'_1/\ \sigma'_3$ tend vers une constante notée $K_o$ (et appelée de temps en temps $K_o$, de temps en temps constante de Jaky). Si l’on poursuit la compression le volume spécifique du sable évolue en suivant l’Eq. (4.1a) avec $\eta=3(K_o-1)/(K_o+2)$

(4.9) $v_{max}(\eta)= v_{nco}(p_o)- \lambda \ln(p'/p'_o) - \lambda_d \ln(1+9(K_o-1)^2/((K_o+2)^2M'^2))$

Les mécaniciens des sols considèrent que $K_o$ n'est pas relié de façon évidente à l'angle de frottement de plasticité parfaite (ou critique φ), car ce n'est ni la valeur du rapport des contraintes à la poussée $\tan^2(\pi/4+\varphi/2)$ ni celui à la butée $\tan^2(\pi/4-\varphi/2)$. On trouve cependant que les valeurs expérimentales de $K_o$ sont corrélées à celles du frottement de plasticité parfaite φ par la relation (dite formule de Jaky):

(4.10) $K_o=1-\sin\varphi$

On appelle souvent $K_o$ la constante des terres au repos, car ce rapport $K_o$ est le rapport des contraintes dans une terre au repos ou de la matière déversée par strates horizontales dans un silo. C'est aussi le rapport des contraintes dans un silo que l'on vient de remplir, si ces murs sont indéformables. On la nomme aussi constante de Jaky, du nom du mécanicien qui a proposé la relation (4.10) grâce à une explication fausse. Nous donnerons une explication simple de cette relation à l'aide du formalisme incrémental dans le chapitre 7, ce qui permettra de relier ce résultat aux autres essais triaxiaux à $p'=c^{ste}$, à $\sigma'_3=c^{ste}$ et aux essais à volume constant .

*A la décompression*, le volume spécifique varie peu, comme nous l'avons expliqué dans la Fig. 4.1, jusqu'à ce que le rapport des contraintes atteigne la valeur de $\sigma'_3/\sigma'_1=\tan^2(\pi/4+\varphi/2)$; c'est-à-dire la valeur du rapport des contraintes lorsque le milieu a atteint le domaine de la plasticité parfaite. Pour que les expériences soient significatives, il faut que le moule n'aie pas un comportement élastique.

## Bibliographie :

# 5. Essai d'interprétation du comportement expérimental des milieux granulaires sous chargements quasi statiques simples :

Dans ce chapitre, nous chercherons à développer une approche théorique globale basée sur des règles énergétiques globales et sur le nombre de paramètres indépendants qui contrôlent l'évolution du milieu granulaire, *i.e.* espace des phases, pour obtenir des caractéristiques essentielles du comportement des milieux granulaires, pour les relier entre eux et pour relier certains concepts entre eux, tels ceux d'états critiques et d'états caractéristiques, ou surface de Hvorslev et surface de Roscoe. Dans le chapitre 7, nous proposerons une modélisation quantitative simple, basée sur une approche incrémentale $\delta\varepsilon=g(\delta\sigma)$, qui conservera les caractéristiques que nous aurons définies dans ce chapitre.

Nous avons vu que l'espace permettant de représenter l'évolution du milieu granulaire est l'espace (v, q, p'). Cependant, cet espace n'est pas complet ; en effet, il ne permet pas de déterminer par exemple l'énergie à fournir au milieu granulaire lors de son évolution car il manque l'information fondamentale qui est la variation de hauteur de l'échantillon (ou $\varepsilon_1$) comme l'indique l'équation :

(5.1) $$\delta W= \underline{\underline{\sigma}}\, \underline{\underline{\varepsilon}}= p'\delta v + q\delta h$$

Dans un premier temps, nous allons chercher à utiliser cette équation et des raisonnements énergétiques pour comprendre les caractéristiques des courbes contrainte-déformation verticale ($\varepsilon_1$) et volume (v)-déformation verticale ($\varepsilon_1$) des essais triaxiaux. Pour cela nous ferons abstraction de l'évolution du matériau dans les plans (q, p') et {v, ln(p')}. Nous verrons qu'en postulant une loi simple de dissipation énergétique, on retrouve l'essentielle des caractéristiques des courbes (q, $\varepsilon_1$ ) et (v, $\varepsilon_1$). Certaines des caractéristiques que nous déduirons sont connues depuis une trentaine d'années; c'est-à-dire qu'elles ont déjà été décrites et prédites par certaines théories et font maintenant office de référence comme critère de réussite d'un essai triaxial; cependant, les théories précédentes font appel à des hypothèses moins générales que celles que nous utiliserons. Par ailleurs, nous aboutirons aussi à d'autres caractéristiques, telle la constance de la dilatance à l'origine, qui ne font pas partie de ces critères établis; de plus, nous constaterons que les essais triaxiaux qui sont sélectionnés par les mécaniciens comme étant des essais réussis, vérifient ces nouveaux critères aussi ; ceci démontre tot à la fois la cohérence des résultats expérimentaux et la validité de l'approche que nous proposons.

Dans un second temps, nous chercherons à utiliser la théorie de l'évolution des systèmes dynamiques complexes pour décrire certaines caractéristiques des trajectoires observées dans le plan {v, ln(p')}.

## 5.1. Un modèle simple de dissipation d'énergie :

Si R et h sont les rayons et la hauteur de l'échantillon, on peut réécrire cette équation : $\sigma'_3$

(5.2) $$\delta W= (q–K\, \sigma'_3)\, \pi R^2 \delta h$$

si on définit la dilatance K comme :

(5.3) $$K= - (h/v)\partial v/\partial h= -\partial\varepsilon_v/\partial\varepsilon_1$$

Nous choisirons la convention de mécanique des sols de considérer positivement les diminutions de volume et de hauteur. $\delta W$ est l'énergie mécanique fournit au système par le travail des forces extérieures. Elle doit correspondre à de l'énergie interne stockée sous forme d'énergie potentielle et à de l'énergie dissipée par déformation plastique irréversible. Nous noterons $\delta W_f$ la somme de ces deux dernières quantités. Nous considérerons par ailleurs les grains comme presque rigides ; ainsi, $\delta W_f$ doit être positif puisqu'il correspond essentiellement à de l'énergie dissipée ; en conséquence $\delta W$ doit être positif pour que le système soit stable. Nous noterons aussi

(5.4) $$\delta W_f= D_{plast}\, \pi R^2 \delta h$$

Comme $\delta W_f$ doit être égal à $\delta W$, l'Eq. (5.4) permet d'écrire l'équilibre sous la forme :

(5.5) $$D_{plast}= (q–K\, \sigma'_3)$$

### *5.1.1 1^er^ résultat : le système est contractant (K<0) à q=0*

$D_{plast}$ est une dissipation; $D_{plast}$ doit être positif ; faisant q=0 dans l'Eq . (5.5) et sachant que $\sigma'_3$ est positif, on en conclut que la dilatance K est négative à l'origine.

### *5.1.2. 2^ème^ résultat : détermination expérimentale de la dilatance et de la dissipation*

A partir de la courbe expérimentale de variation du volume expérimentale en fonction de la déformation verticale, on peut déterminer la dilatance K pour chaque couple (q, $\sigma'_3$) et donc déterminer la valeur de $D_{plast}$ expérimentale par l'Eq. (5.5).

(5.6) $$K= - v_o h/(h_o v)\, \partial\varepsilon_v/\partial\varepsilon_1$$

### *5.1.3. 3^ème^ résultat : l'état caractéristique*

Nous allons faire maintenant une hypothèse fondamentale sur les propriétés de $D_{plast}$. Comme nous l'avons dit, $D_{plast}$ peut être considéré comme la somme d'une énergie de dissipation plastique et d'une énergie potentielle. En ce sens, on s'attend à ce que $D_{plast}$ dépende de q et de $\sigma'_3$ pour un sable donné. Mais on se doute aussi que $D_{plast}$ doit dépendre de la nature de la déformation, et donc de la dilatance K. Nous allons supposer que $D_{plast}$ ne dépend que de ces trois paramètres :

(5.7) $D_{plast}=f(q, \sigma'_3, K)$.

On pourrait supposer que $D_{plast}$ dépend aussi du volume spécifique, mais nous n'allons pas le faire pour simplifier ; on verra de plus que cette hypothèse donnerait des résultats incompatibles avec les faits expérimentaux.

***Remarque:*** Une justification de l'Eq. (5.7) pourrait être donnée en utilisant le calcul de Rowe pour déduire sa relation (Eq. 4.8). Nous verrons cependant que le modèle de Rowe n'est pas correcte, même s'il conduit à une relation (4.8) vérifiée expérimentalement. En effet, il conduit aussi à considérer que la déformation du milieu granulaire est toujours régi par un seul mécanisme de plasticité; ceci n'est pas acceptable surtout lorsque le chargement est presque isotrope; L'hypothèse de plasticité de Rowe est d'ailleurs contredite expérimentalement par les lois contrainte-déformation près du chargement isotrope, qui dépend bien du chemin de contrainte. Mais ceci est une autre histoire que nous discuterons plus loin, puis plus complètement au chap. 7.

Revenons donc à notre problème et partons des Eqs. (5.7) et (5.5); combinons les :

(5.8) $q-K\,\sigma'_3=f(q, \sigma'_3, K)$.

Considérons le cas K=0 à $\sigma'_3$ constant ; dans ce cas, l'Eq. 5.8 permet de calculer q à $\sigma'_3$ fixée, ceci impose en particulier que la valeur $q_o/\sigma'_3$ mesurée dans un essai triaxial lorsque la variation de volume est nulle est fixée. Tous les essais triaxiaux de la Fig. 4.2 tendent vers un volume spécifique constant à la fin de l'essai $v_\infty$; mais certains d'entre eux sont contractants puis dilatants, c'est-à-dire passent par un minimum de volume puis voit le volume augmenter. Le point de tangente horizontal (minimum de volume) est appelé l'état caractéristique. L'Eq. (5.8) permet donc de conclure que pour chacun de ces derniers essais, le rapport $q/\sigma'_3$ obtenu à l'état caractéristique doit être le même qu'à la fin de l'essai, c'est-à-dire à l'état critique : $v_\infty= v_o$.

Ceci est effectivement observé expérimentalement.

### *5.1.4. 4ème résultat : l'état critique*

Considérons l'Eq. (5.8) pour K=0, elle s'écrit : $q_o=f(q_o,\sigma'_3,0)$. Ceci impose que $q_o$ est indépendant de l'état initial, et en particulier de sa densité initiale. $q_o$ ne dépend donc que de $\sigma'_3$, et cette dépendance doit être linéaire pour que l'équation soit homogène ; ceci revient à choisir le rapport $q/\sigma'_3$ comme la quantité adéquate et revient à démontrer que $q_o/\sigma'_3$ est constante, indépendante des conditions initiales. Ce résultat doit être valable pour l'état critique et l'état caractéristique comme nous l'avons montré au paragraphe précédent. On en conclut donc que le rapport $q_\infty/\sigma'_3$ de l'état critique doit être constant, indépendant des conditions initiales, c'est-à-dire $q_\infty/\sigma'_3$ =M.

Ceci est effectivement observé expérimentalement ; c'est même la définition de l'état critique (Schofield et Wroth 1968) ou de plasticité parfaite (Biarez) et cela montre la compatibilité de ce modèle avec les résultats de mécanique des sols.

### *5.1.5. 5ème résultat : unicité de la dilatance à l'origine*

Considérons le cas q=0 et notons $K_1$ la dilatance correspondant à cet état de contrainte. L'Eq. (5.8) s'écrit $K_1\sigma'_3=-f(0,K_1,\sigma'_3)$, où la fonction f ne dépend que de $\sigma'_3$ et de $K_1$. La fonction f mesure la variation d'énergie dissipée et d'énergie potentielle ; Supposons que ces deux termes soient proportionnels à la pression $\sigma'_3$, la fonction f sera proportionnelle à $\sigma'_3$. Dans ce cas on peut réécrire $K_1$ sous la forme $K_1=-f(0,K_1,\sigma'_3)/\ \sigma'_3$ = constante. On en conclut que la pente à l'origine des courbes $(v,\varepsilon_1)$, doit être constante.
Cette conclusion $K_1=c^{ste}$ n'est pas un trait caractéristique reconnu du comportement des milieux granulaires au triaxial. Cependant, ce résultat semble être effectivement observé expérimentalement par les essais reconnus comme étant de qualité. De plus, cette pente à l'origine doit être la même quelle que soit le type d'essais , qu'il soit mené à pression moyenne $\sigma'_3$ constante ou à $\sigma'_2=\sigma'_3=c^{ste}$. C'est effectivement ce que l'on observe (*cf.* Figs. 3.9 et 5.27 de Saïm 1997) .
Nous verrons cependant, que la constance de la dilatance à q=0 n'est pas universelle, c'est-à-dire qu'elle dépend de la nature de l'essai, i.e. à volume constant, à rayon constant (oedométrique), à $\sigma'_3=c^{ste}$, à $\sigma'_3=\sigma'_2=c^{ste}$… Nous essayerons par la suite de décrire cette relation à l'aide d'une loi incrémentale reliant les incréments de contrainte aux incréments de déformation.
Si on applique cette théorie sur un milieu isotrope, on obtient qu'une déformation à pression moyenne constante donne une variation de volume nul. Ceci semble contredit par certains essais, (*cf.* Figs. 3.9 et 5.27 de Saïm 1997) pour lesquels la déformation volumique lors d'une déformation à $\sigma'_3$ constant est non nulle. Ceci indique simplement soit que le matériau est anisotrope, soit que l'expérience a été mal contrôlée, par exemple le milieu granulaire a été imparfaitement saturé par de l'eau de telle sorte qu'il reste de l'air compressible dans les pores.

### *5.1.6. 6ème résultat: instabilité*

L'instabilité du comportement mécanique apparaît lorsque le travail mécanique incrémental $\delta W$ qui correspond à une modification minime $\delta h$ de la hauteur h de l'empilement engendre un essai d'énergie mécanique qui ne peut être complètement dissipée par dissipation plastique $\delta W_f$. En termes mathématiques, ceci se traduit par l'équation :

$$\text{(5.9)} \qquad \text{instabilité si} \qquad d/dh\{q-Kp-f(q,\sigma'_3,K)\}>0 \qquad \text{à } q,\ \sigma'_3 \text{ fixés}$$

De façon générale, l'Eq. (5.9) impose que la limite de stabilité est obtenue pour dK/dh=0, c'est à dire pour un extremum de K, et que l'échantillon devient instable lors d'un changement de signe de dK/dh, c'est-à-dire pour un changement de signe de la courbure de v en fonction de h. Compte tenu de la forme générale des courbes $(v,\varepsilon_1)$ (*cf.* Fig. 4.2), cette extremum doit être un maximum et correspond au point d'inflexion de v en fonction de $\varepsilon_1$. Par ailleurs, nous avons vu que l'instabilité mécanique apparaissait au maximum de la courbe de $q/\sigma'_3$ en fonction de $\varepsilon_1$, ces deux points doivent coïncider et le maximum de $q/\sigma'_3$ et le point d'inflexion de v en fonction de $\varepsilon_1$ doivent être concomitants.
Ceci semble être effectivement observé expérimentalement . Nous avons représenter les caractéristiques de ces courbes sur la Fig. 5.1 .

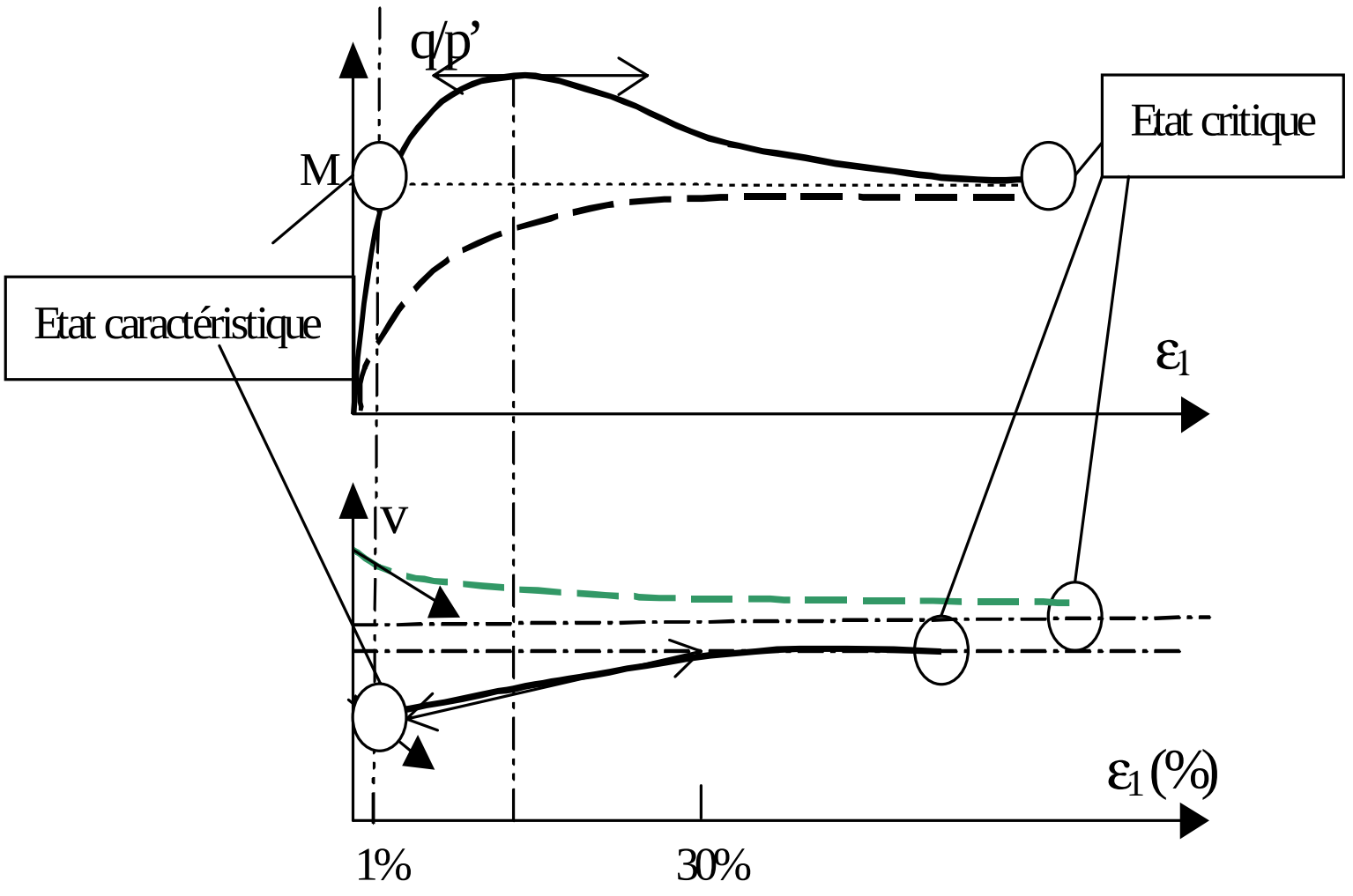

**Figure 5.1 :** *Caractéristiques typiques des courbes obtenues au triaxial : i) La pente à l'origine de la variation de volume en fonction de* $\varepsilon_1$ *doit être constante, indépendante du type d'essai* ($\sigma'_3$=c$^{ste}$ *ou* $\sigma'_2=\sigma'_3$=c$^{ste}$) *et de la densité initiale du sable. ii) l'état caractéristique correspond au minimum de volume ; il doit correspondre à un rapport* $q/\sigma'_3$ *égal au frottement solide de l'état critique (ou de plasticité parfaite)* M. *iii) La déformation* $\varepsilon_1$ *pour laquelle le rapport* $q/\sigma'_3$ *est maximum doit correspondre aussi au maximum de dilatance, c'est-à-dire au point d'inflexion de la courbe donnant* v *en fonction de* $\varepsilon_1$.

### *5.1.7. Lien avec la loi de Rowe :*

L'Eq. (5.8) fixe une relation entre les grandeurs q, $\sigma'_3$ et K. Nous avons vu par ailleurs qu'il existe déjà une relation proposée par Rowe (Eq. 4.8) entre ces trois grandeurs. Ces deux relations doivent donc être identiques.

$$\sigma'_1/\sigma'_3=\tan^2(\pi/4+\varphi/2)(1-\partial\varepsilon_v/\partial\varepsilon_1)=q/\sigma'_3+1=(1+K)(1+\sin\varphi)/(1-\sin\varphi) \quad (4.8)$$

on trouve ainsi :

$$f(q,\sigma'_3,K)=2\,(1+K)\sigma'_3\sin\varphi/(1-\sin\varphi) \quad (5.10)$$

### *5.1.7. Discussion :*

Cette modélisation explique la plupart de traits caractéristiques des courbes contrainte-déformation. Il reste cependant un certain nombre de problèmes que nous approfondirons par la suite (*cf.* chapitre 7 sur la modélisation incrémentale).

Nous noterons les quelques points suivants :

♣ L'équation de Rowe prédit bien que la pente de la variation de volume en fonction de $\varepsilon_1$ est bien constant. Mais il semble ce point n'a pas été utilisé comme l'un des tests pour sélectionner les résultats expérimentaux correctes.

♣ Le modèle énergétique de l'Ecole de Cambridge (Roscoe *et al.* 1958, Schofield & Wroth 1968) peut aboutir à des relations entre les courbes contrainte-déformation similaires à ce que nous avons trouvé, *i .e* . état caractéristique,… , *cf.* Atkinson & Bransby (1977). Cependant il utilise une loi énergétique particulière, qui ne vérifie pas la loi de Rowe . Cette loi de dissipation est un ingrédient fondamental de l'approche faite par cette Ecole ; ainsi, ce modèle n'est pas compatible avec les résultats expérimentaux et avec l'approche de Rowe.

♣ De plus l'approche de l'Ecole de Cambridge est basée sur la théorie de la plasticité parfaite à un mécanisme. Ce qui veut dire, en théorie, que l'état de contrainte défini le mode de déformation, et donc la valeur de la dilatance K. Ainsi, la valeur de K n'est pas ajustable dans ce modèle mais est fonction de q et p'. La formulation est ainsi très différente de celle que nous avons proposée, puisque la dilatance K est un paramètre de contrôle de la dissipation plastique dans notre approche. En d'autres termes, pour nous, la dilatance K contrôle le mode de déformation $\delta\varepsilon$, et/ou le mode d'incrémentation $\delta\sigma$; et l'énergie dissipée dépend de q, de p' , mais aussi de la valeur de K.

♣ On voit ainsi que notre approche se tourne vers l'utilisation d'une modélisation incrémentale $\delta\varepsilon$=g($\delta\sigma$) plutôt que vers une approche de plasticité parfaite. En ce sens, elle rompt donc aussi avec l'approche « plasticité à un mécanisme » de Rowe.

♣ Nous verrons au chapitre 7 que l'approche incrémentale est justifiée, c'est-à-dire que nous montrerons sur des exemples expérimentaux que le mode de déformation du matériau granulaire dépend réellement de la direction de l'incrément de contrainte. Nous interpréterons cette loi incrémentale comme étant liée à l'existence de plusieurs mécanismes de plasticité qui sont mobilisables conjointement.

♣ Un des résultats importants de notre modélisation est l'existence des états caractéristiques, tels que K=0 lorsque $q/\sigma'_3$=M. Peut-on en déduire quelque(s) propriété(s) particulières de l'évolution du matériau ?

♣ Enfin, d'après notre modèle, lorsque le matériau est soumis à une loi de chargement simple par une loi défini par q(t) et p'(q(t)), la connaissance de q ou de $\sigma'_3$ suffit à connaître $\partial v/\partial\varepsilon_1$ ; cependant, cela ne suffit pas à déterminer v. Pourquoi? Pourquoi la dilatance ne dépend-elle que du rapport des contraintes et ne dépend-elle pas de la densité du matériau?

Nous chercherons à donner une réponse à ces questions au chapitre 7.

## 5.2. Théorie des systèmes dynamiques :

Nous avons vu que l'évolution du système tendait vers un état appelé état critique lorsqu'on faisait subir à un échantillon granulaire une augmentation de cisaillement pour obtenir les grandes déformations. Cet état est caractérisé par un volume spécifique constant qui ne dépend que de la pression moyenne de confinement p'. A ce stade de déformation, si l'on garde le même taux relatif de cisaillement, mais que l'on change la pression moyenne de confinement, on passe continûment d'un état critique à un autre. Ces états forment une courbe dans l'espace (q,v,p') qui est pratiquement une droite dans le système de coordonnées {q,v, ln(p')}. Puisque ces états critiques forment une courbe dans l'espace (q,v,p'), un état critique est simplement défini par la donnée d'une seule de ces coordonnées, nous choisirons de le définir par son volume spécifique $v_c$ ou par la valeur de $p'_c$ sachant que les deux autres coordonnées se déduisent de celle là par l'Eq. (4.1.b) :

(4.1.b&c) $q_c/p'_c=M'$ et $v_c = v_{co} - \lambda \ln(p'_c/p'_{co})$.

Nous avons vu par ailleurs que chacun de ces états critiques s'atteignait directement à partir d'un état quelconque initial $(0,v_i,p'_i)$ en augmentant le déviateur de contrainte et en imposant soit p' égal constante, soit $\sigma'_2=\sigma'_3$ égale constante, soit une autre relation entre les contraintes q et p'….. Ainsi, on peut traduire ces faits dans le langage des systèmes dynamiques en disant que chaque état critique est un point attracteur de la dynamique, et qu'en conséquence la ligne d'états critiques est une ligne attractrice de l'évolution du système.
Nous allons maintenant utiliser ce langage des systèmes dynamiques pour compléter notre compréhension de l'évolution des systèmes granulaires et caractériser les surfaces de Hvorslev et de Roscoe. Nous rappelons en appendice les quelques notions de base nécessaires pour comprendre la suite de ce chapitre.

### *5.2.1. Rappel de la définition des surfaces de Hvorslev et de Roscoe :*

Les Eq. (4.1) définissent les états critiques par $\{q=M'p' , v=v_c=v_{co}(p'_o)- \lambda \ln(p'/p'_o)\}$, et les états normalement consolidés sous pression isotrope p' : $\{q=0 , v=v_c=v_{co}(p'_o)- \lambda \ln(p'/p'_o) +\lambda_d \ln(2)\}$.
Considérons des états très fortement consolidés ou au contraire normalement ou faiblement consolidés. Ils n'ont pas le même comportement au triaxial : les uns doivent se dilater pour arriver au point critique, les autres non. C'est pourquoi on considère qu'ils forment deux classes différentes, avec des comportements différents appelés comportement normalement consolidé et comportement surconsolidé. On peut caractériser ces comportements en utilisant l'espace {q,p',v} grâce à l'arrivée au point critique sous différentes conditions expérimentales ; c'est ce qu'ont fait Roscoe et Hvorslev les premiers.
Si on considère un état critique donné et l'ensemble des essais triaxiaux de compression uniforme, l'ensemble des trajectoires qui arrivent à cet état critique et qui partent des états normalement consolidés forment une surface de l'espace {q, p', v} appelée surface de Roscoe ; de même, l'arrivée au point critique se fait par cette même surface si l'ensemble de départ est très légèrement consolidé ; cette surface de Roscoe est donc une surface attractrice que les trajectoires rejoignent très vite pour converger vers l'attracteur qu'est le point critique.

De la même façon, Hvorslev a montré que l'ensemble des trajectoires qui partent de tous les points surconsolidés et qui arrivent à un même point critique convergent vers ce point en décrivant une surface de l'espace {q,p',v}. En d'autres termes, si l'on considère deux états surconsolidés différents soumis à un test triaxial de tel sorte qu'ils aboutissent au même état critique, la trajectoire de l'espace {q,p',v} correspondant à chacun de ces deux tests seront différentes dans un premier temps, puis se rejoindront pour converger vers l'état critique. Nous avons vu, *cf.* chap.4, qu'on considère que la partie commune de l'évolution commence juste après le maximum de q/p'.

Ces deux surfaces de Hvorslev et de Roscoe sont représentées en général comme étant distinctes, c'est-à-dire comme faisant un angle entre elles (*cf.* Figs. 11.13-11.16 & 13.4 de Atkinson & Bransby (1977*)* . Nous voulons discuter ici la validité de cette affirmation.

### *5.2.2. Arguments topologiques simples utilisant la théorie des systèmes dynamiques* :

Replaçons le problème dans l'espace des phases {ln(q), ln(p'), v}. Cet espace est donc de dimension $d_e=3$ ; l'ensemble des états finaux après un essai triaxial, est l'ensemble des états critiques ; il se représente par une droite dans l'espace {ln(q), ln(p'), v} ; cet ensemble est l'ensemble attracteur de l'évolution; il est de dimension $d_a=1$; l'évolution du système est une trajectoire de dimension 1 qui se termine en un point critique $(q_c,p'_c,v_c)$ ; chaque trajectoire évolue en fonction de la déformation axiale ; celle-ci joue donc un rôle analogue à celui du temps pour un système dynamique normal.

L'argument topologique est alors le suivant : d'un point de vue expérimental, l'ensemble des systèmes évoluent vers un ensemble de points appelés états critiques ; cet ensemble final forme une courbe de dimension $d_a=1$ appelée ligne des états critiques ; c'est donc un ligne attractrice de l'évolution sous l'ensemble des conditions expérimentales choisies. Une telle ligne ne peut pas attirer les points proches d'elle parallèlement à elle-même, autrement tout l'environnement convergerait vers un seul point et l'attracteur ne serait pas une ligne, mais un point. Ceci impose que les directions d'attraction définissent un sous espace de dimension $d_b$ inférieur ou égal à $d_e$-$d_a$ . De plus on doit avoir $d_b=d_e-d_a$ si l'évolution de tout point de l'entourage de la ligne critique est attirée par elle. Ceci est le cas ; on a donc $d_b=d_e-d_a=2$. Ainsi, l'ensemble des trajectoires qui aboutissent à un point critique forment donc un espace à 2 dimensions ; c'est donc une surface. Cette surface coupe la ligne attractrice au point critique considéré.

Compte tenu de ce que nous avons dit, deux surfaces arrivant à deux points critiques proches ne peuvent pas se couper, sinon les points communs à ces deux surfaces auraient deux évolutions distinctes possibles pour les mêmes conditions expérimentales ; ceci impliquerait que l'évolution du système n'est plus déterministe ce qui n'est pas acceptable. En conséquence, la direction des normales à ces surfaces doit évoluer continûment en fonction du point critique et deux surfaces voisines doivent être parallèle lorsqu'elles coupent la ligne des points critiques. Ces surfaces doivent être identifiées aux surfaces de Roscoe lorsque les états de départ sont peu denses et aux surfaces de Hvorslev pour les états surconsolidés.

♣ *Quelques remarques :*

i) Cela semble un résultat naturel : l'espace des phases est 3-d, les surfaces de Hvorslev et de Roscoe passant par le même point critique ont un point commun ; elles doivent donc voir une ligne commune de telle sorte qu'elles définissent une seule surface. Cette surface pourrait avoir quelques caractéristiques topologiques spécifiques :

ii) Par exemple, les surfaces pourraient être le lieu de discontinuités angulaires le long d'un certain nombre de lignes passant par le point critique ; ceci traduirait l'existence de sauts d'orientation de la normale à la surface.

iii) Le point critique pourrait être un accident topologique particulier, tel que la forme de la surface définie par l'ensemble des trajectoires pourrait être un cône dont le sommet est le point critique ; ainsi aucune arrivée au point critique ne pourrait se faire dans le sens inverse d'une autre, et les tangentes aux trajectoires arrivant au même point critique feraient toujours un angle différent de 0 ou de $\pi/2$.

iv) de tels défauts topologiques existent souvent lorsque l'espace des phases complet est de dimensionalité plus grande que l'espace de représentation et que celui-ci est obtenu par la projection de l'espace des phases réel sur un sous-espace de dimension inférieur; cette projection engendre alors des repliements,… qui font que les trajectoires peuvent ne pas être régulières. C'est pourquoi il est important d'étudier plus en détail la topologie des trajectoires et leur arrivée au point critique.

En fait l'existence d'un cône comme nous le supposons au point (iii) montrerait que les vecteurs propres du sous-espace choisi {v, ln(q), ln(p')} ne sont pas adaptés à notre étude: un changement adéquate de variable permettrait en effet de passer continûment d'un cône à un plan.

Si notre espace est complet et bien choisi, nous nous attendons donc à ce que les chemins surconsolidés et normalement consolidés arrivent en sens inverse au point critique considéré. C'est une hypothèse facile à tester expérimentalement et c'est ce que nous allons faire maintenant, en partant d'une collection de résultats triaxiaux obtenus par un certain nombre d'auteurs.

### *5.2.3. Comparaison avec les résultats expérimentaux :*

La question à laquelle nous cherchons à répondre est donc de savoir si la surface constituée par l'ensemble des trajectoires arrivant à un point critique est régulière ou non et si le point critique est un point régulier de cette surface.

Pour cela, il suffit de démontrer que les trajectoires arrivant par des côtés opposés ont la même direction mais sont dirigées en sens inverse, et ceci pour deux couples de trajectoires indépendantes car l'ensemble des trajectoires forment une surface à 2 dimensions, ou en d'autres termes, car les surfaces de Hvorslev et de Roscoe sont à 2 dimensions, (et qu'elles sont donc définies par deux vecteurs indépendants, et deux seulement, si elles forment une surface régulière). Les trajectoires que nous avons décidées d'étudier correspondent i) à des essais triaxiaux à volume constant (essais non drainés) et ii) à des essais de compression à $\sigma'_2=\sigma'_3=c^{ste}$.

♣ *Essais triaxiaux (à $\sigma'_2=\sigma'_3=c^{ste}$)*

Ce test impose $p'=q/3+p'_o$ ; la projection dans le plan (p',q) est donc une droite de pente 1/3, imposée par les conditions initiales. Dans ce plan, les trajectoires arrivent bien au point critique en sens inverse, puisque les états denses arrivent à q=M'p' par valeurs supérieures tandis que les états lâches y arrivent par valeurs inférieures et que la direction d'arrivée est $\delta p'=\delta q/3$ dans les deux cas.

Ainsi, il reste juste à démontrer que les trajectoires arrivent bien en sens opposé dans le plan (p',v) aussi. C'est ce que nous constatons dans la Fig. 4.4 . Beaucoup d'autres exemples sont disponibles dans la littérature ; voir par exemple, les Figs. 5.19a et b, 5,25 du livre de Biarez et Hicher pour du sable, la Fig. 5.24b de ce livre pour des agrégats ; voir aussi dans la thèse de Saïm les Figs. 5.5, 5.6, 5.7, 5.16, 5.25, 5.26, 5.39, 5.43 concernant du sable et les Figs. 3.12 à 3.14, 4.9, 5.2, 5.4 pour des argiles. Ces résultats sont meilleurs lorsqu'ils sont tracés dans le plan (v,ln(p')), ce qui est normal car la ligne des états critiques est alors une droite.

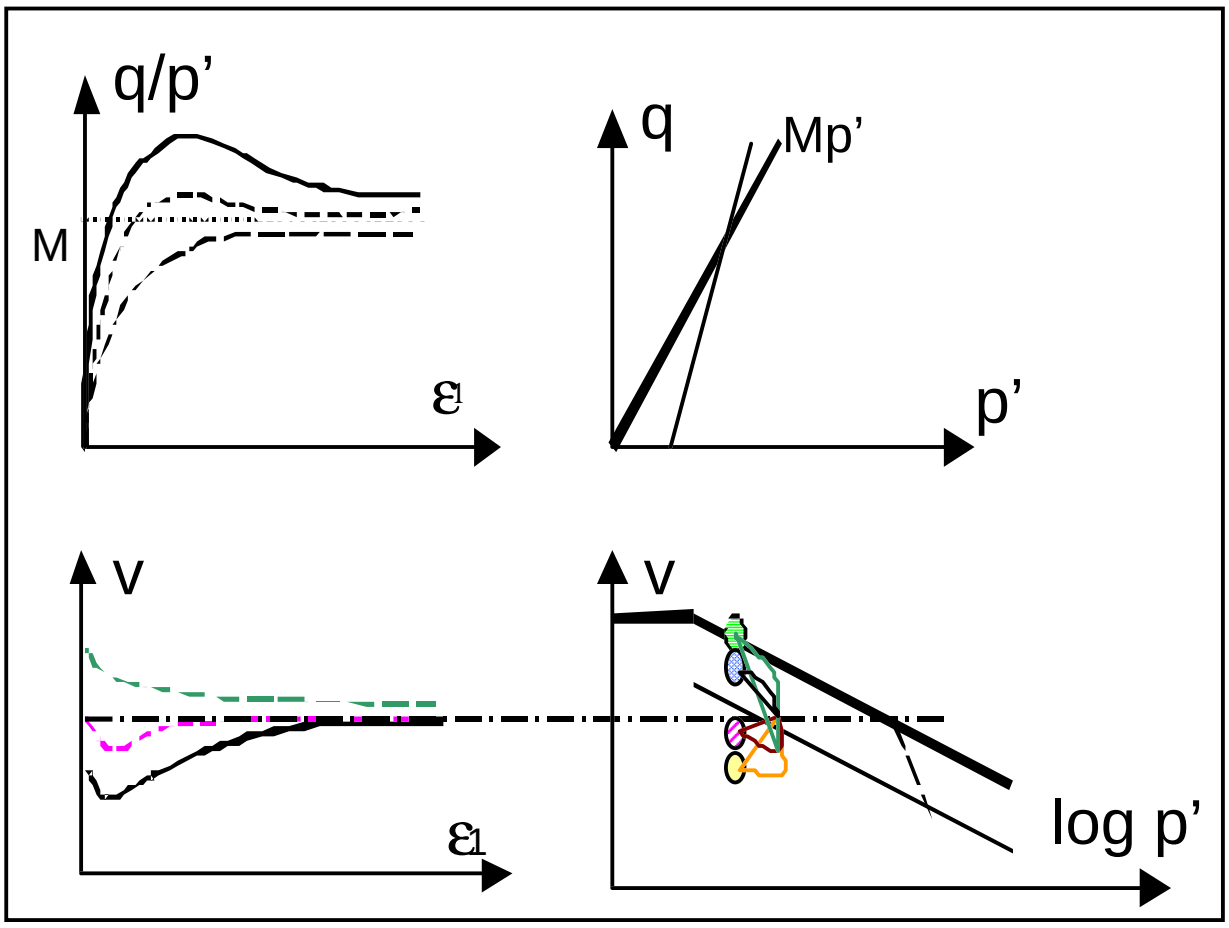

**Figure 5.2 :** *Résultats typiques d'un essai triaxial à $\sigma'_2=\sigma'_3=c^{ste}$ pour différentes valeurs de la densité initiale. La courbe en haut à droite est la trajectoire dans le plan* (q,p')*, celle en bas à droite la trajectoire dans le plan* (ln(p'), v)*. Ces trajectoires arrivent au même point critique en sens opposés soit par le haut, soit par le bas suivant la densité initiale (par le haut : états denses, plan* (q,p') *ou états lâches, plan* (ln(p'), v)*) ; par le bas : états denses, plan* (ln(p'), v ) *ou état lâche, plan* (q,p'), .

♣ *Chemins à volume constant (essais non drainés) :*

Comme dans ces tests le volume est maintenu constant, la projection de la trajectoire sur les plans (q,v) et (p',v) sont des droites verticales. Il suffit donc d'étudier la forme des trajectoires dans le plan (q,p'). De plus, lorsque l'état initial est normalement consolidé (très fortement surconsolidé), la trajectoire arrive à l'état critique par des valeurs de q et de p' décroissantes (croissantes), c'est-à-dire en sens inverse l'une de l'autre.

La Fig. 5.3 est un résultat typique. Elle montre que les trajectoires se terminent en longeant la droite q=M'p', en direction des q croissants (décroissants) pour les états surconsolidés (normalement consolidés). On peut trouver bien d'autres exemples, en particulier dans la thèse de Saïm pour du sable lâche (Figs. 3.25-à-3.27), pour du sable dense (Figs. 6.5-à-6.7, 6.10-à-6.24), ou pour des argiles Figs.3.21-à-3.23, 6.1, 6.2) .

On doit noter que la trajectoire dans le plan (q,p') présente souvent un changement de direction réellement discontinu à l'arrivée sur la droite q=M'p' ; il semble donc que ce point soit un point de bifurcation ; nous verrons dans le chap. 7 que ceci est conforme à la prédiction de notre modèle incrémentale simple que nous développerons dans ce chapitre 7. Il ne faut pas confondre ce point de bifurcation et celui de l'arrivée au point critique ; c'est malheureusement ce qui est couramment fait dans la littérature de la mécanique des sols.

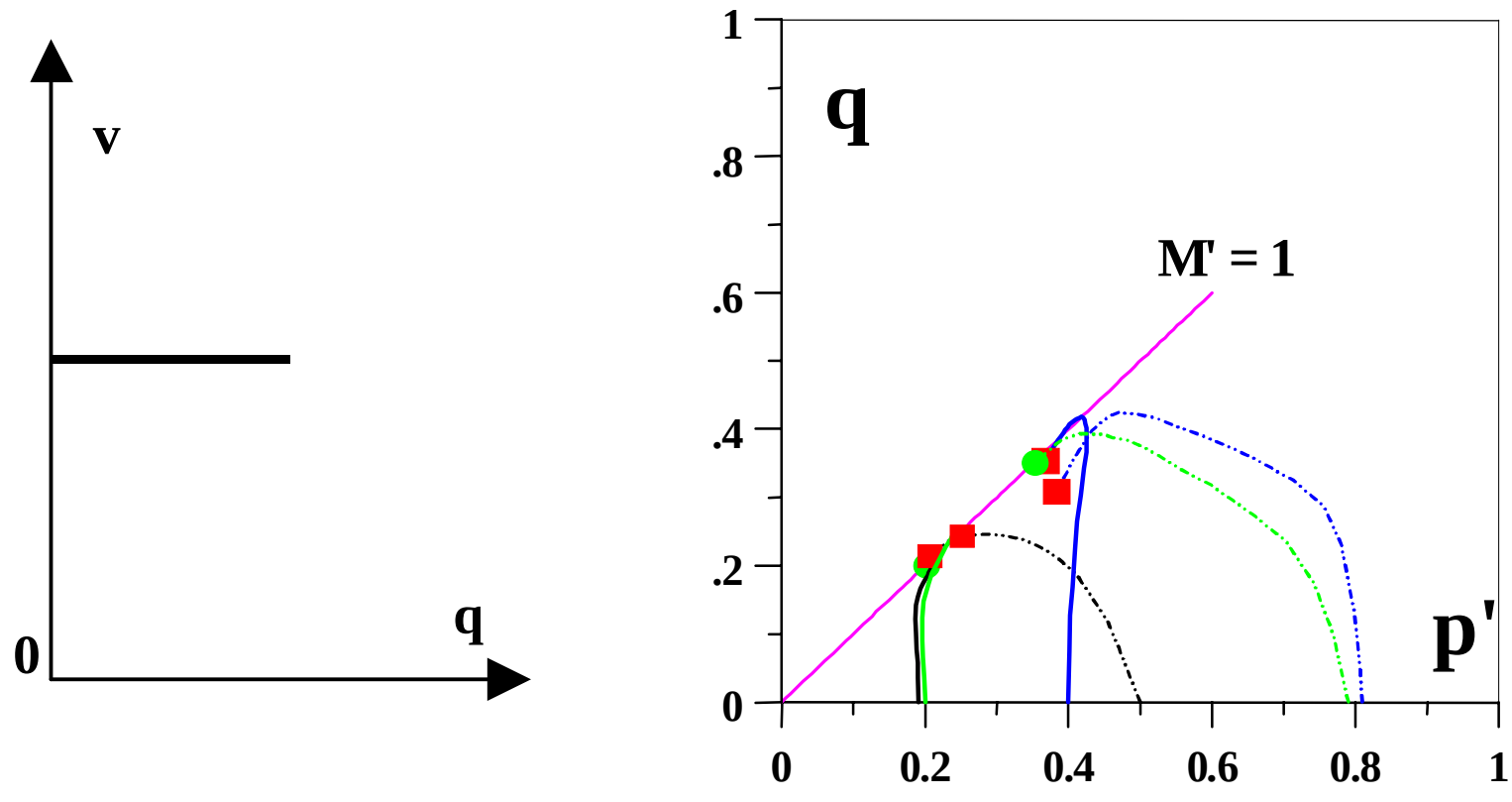

***Figure 5.3 :*** *Essais tiaxiaux à volume constant sur du sable d'Hostun pour différentes valeurs de la contrainte* p' *initiales : les états denses et lâches arrivent bien en sens opposés au point critique.*

Certains cas d'argiles très peu consolidées semblent plus complexes, car la trajectoire semble arriver perpendiculairement à la droite q=M'p', pour s'y arrêter; elle ne repart en longeant la droite q=M'p' vers la gauche que lorsque la déformation est devenue suffisamment importante. Ce redémarrage de la trajectoire est souvent interprétée comme étant lié au développement des hétérogénéités ; elle est associée aussi souvent à la localisation de la déformation. Si cette interprétation était correcte, il faudrait considérer que la surface constituant l'ensemble des trajectoires arrivant aux états critiques n'est pas régulière dans le cas des argiles. Dans le cas contraire, i.e. si le redémarrage de la trajectoire a un sens physique réel, il peut être interprété à l'aide de la théorie des bifurcations; nous verrons au chapitre 7 que cette bifurcation est du type trans, *cf.* Chap. 7 et appendice.

Dans le cas des sables lâches, l'arrivée à l'état critique se fait effectivement parallèlement à la droite q=M'p', dans le plan (q,p') ; l'incurvation de la trajectoire a lieu bien avant l'arrivée à la droite q=M'p' des états caractéristiques de telle sorte que la bifurcation n'est pas aussi nette que dans le cas des sables denses ou des argiles denses.
Ainsi les couples de trajectoires provenant des états lâches et denses arrivent bien en sens inverses l'un de l'autre pour les deux types d'essais triaxiaux que nous avons étudiés. C'est ce que nous cherchions à prouver. On peut donc conclure dans ce cas que les surfaces de Hvorslev et de Roscoe forment une seule et même surface et que cette surface est régulière ; cette surface n'a pas la forme d'un cône de sommet l'état critique, car les trajectoires arrivent en ce point toujours en sens inverses.

### *5.3.4. Quelques remarques*

L'identité entre ces deux surfaces de Roscoe et de Hvorslev n'a jamais été postulée avant Evesque (1999) à notre connaissance; ceci veut dire qu'elle n'a jamais été utilisée comme un test de la qualité des essais triaxiaux ; observer une telle identité grâce aux expériences contribue à la garantie de leur qualité et de leur cohérence. Elle montre donc la pertinence des données expérimentales sur lesquelles sont basées la mécanique des sols.
Cette remarque est importante, car certains résultats expérimentaux sont établis dans des conditions où l'évolution du système est peu stable, voir instable ; ils peuvent donc être l'objet de critiques. De telles critiques, même si elles sont sérieuses, ne forment pas un obstacle définitif, car ce n'est pas le seul domaine de la physique où l'on sait étudier des systèmes potentiellement instables: on sait par exemple préparer et étudier de l'eau surfondue (T<0°c), on sait que le flambage d'un tube soumis à une pression peut être retardé,….
Il est remarquable (et surprenant) que l'espace tri-dimensionnel {q, p', v} soit suffisamment grand pour décrire les propriétés mécaniques d'un milieu granulaire et leur évolution. En effet le système développe une anisotropie induite au fur et à mesure de la compression, or celle-ci n'est pas (et ne peut pas être) prise en compte dans cet espace.
L'adéquation de cet espace de dimension faible à la description de la mécanique des milieux granulaires est peut être liée aux raisons suivantes :
i) si le matériau développe de l'anisotropie au cours de sa déformation, l'évolution de cette anisotropie progresse de façon conjointe à celle de la déformation dans les essais triaxiaux classiques,
ii) que le nombre de types d'essais triaxiaux qui sont réellement utilisés est faible
iii) que l'anisotropie de l'état critique est bien défini, ainsi probablement que celle de ces états voisins
iv) que le mode de déformation lui-même dépend peu de l'anisotropie mais beaucoup de l'état de contrainte (via l'equation de Rowe), c'est pourquoi les états caractéristique et critique ont le même frottement et la même dilatance.

Cependant, il est clair qu'aucune étude expérimentale sérieuse n'a réellement cherché à établir si des trajectoires du même sable soumis au même essai et provenant de deux états initiaux différents pouvaient réellement se croiser ou non.

## Bibliographie :

# 6. Comportements sous chemins cycliques : contraction, dilatation et liquéfaction

## 6.1. Introduction : les différents cas :

Le comportement mécanique d'un milieu granulaire en réponse à des sollicitations cycliques dépend fortement des conditions choisies c'est-à-dire drainé ou non drainé... De plus nous verrons qu'il dépend fortement de l'existence d'une rotation des contraintes principales, ou d'une permutation des axes principaux. C'est pourquoi, il est essentiel de différencier le cas où le chargement imposé varie de manière cyclique autour d'un point de fonctionnement caractérisé par un déviateur non nul de celui où le chargement cyclique évolue autour d'une contrainte isotrope. Pour bien marquer cette différence, on parlera de chargement cyclique répété dans le premier cas et de chargement alterné dans le second cas.

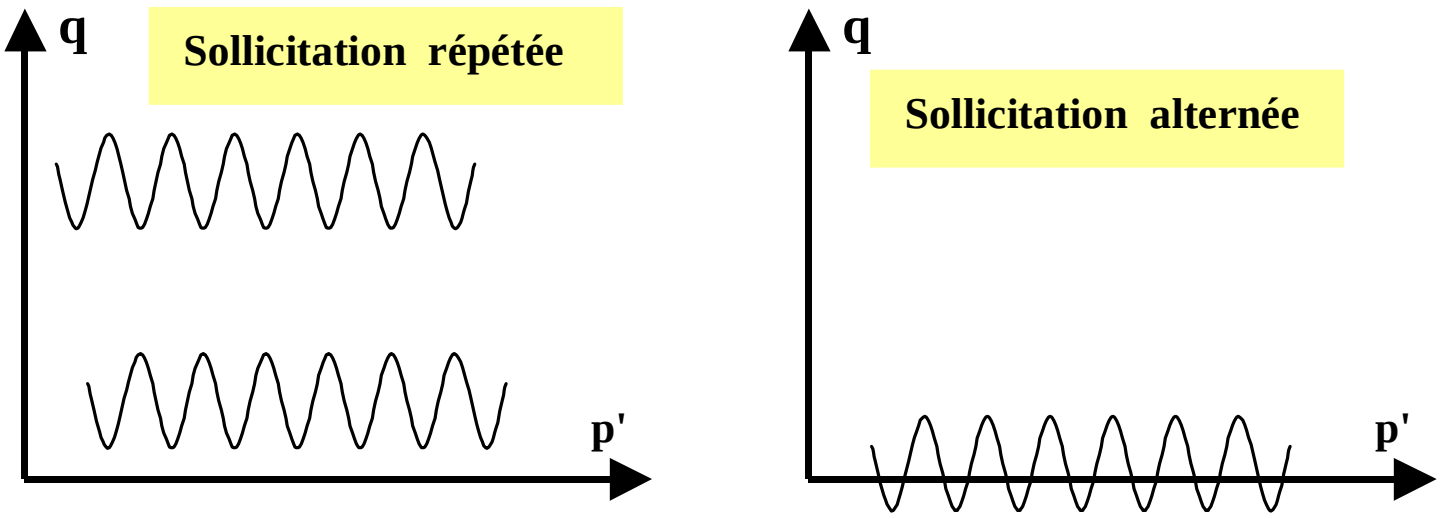

***Figure 6.1 :*** *Les deux types de sollicitations avec (6.1.b) et sans (6.1.a) changement de direction de la contrainte principale majeure.*

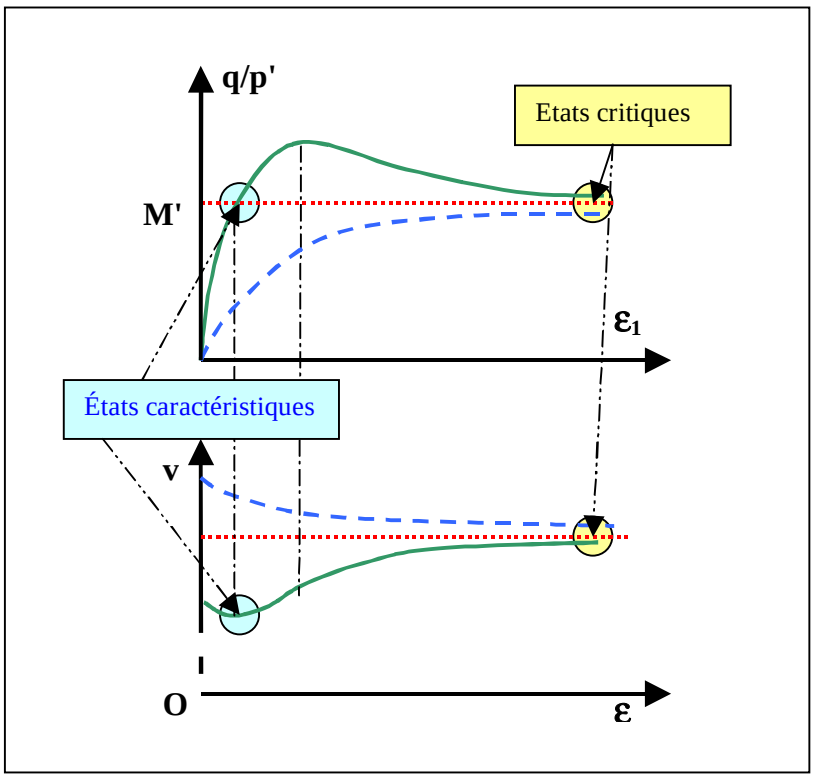

***Figure 6.2:*** *Rappel du comportement mécanique typique d'un milieu granulaire: notion d'état caractéristique et d'état critique. A l'état caractéristique, le matériau présente une variation de volume nulle et q/p'=M'. L'état critique est obtenu à très grande déformation. C'est l'état caractéristique le plus lâche pour une pression donnée* p' et q/p'=M'.

Pour comprendre la mécanique, il est important de rappeler la notion d'état caractéristique que nous avons développé dans le chapitre précédent et que nous rappelons dans la Fig. 6.2. L'état caractéristique est caractérisé par le même rapport q/p'=M' que celui de l'état de plasticité parfaite (état "critique"). Il a une dilatance nulle et il sépare la région dilatante caractérisée par q/p'>M' de la région contractante caractérisée par q/p'<M'). Ainsi, comme l'indique la Fig. 6.3, le plan (q,p') est donc divisé en deux zones par la droite M'p'. A gauche de cette droite le système est dilatant ; à droite, il est contractant.

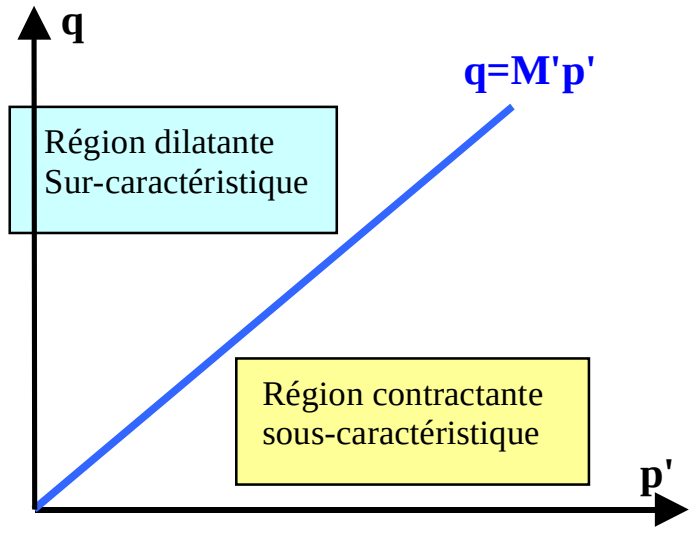

***Figure 6.3:** le plan* (p',q) *est divisé en deux zones, l'une dilatante, l'autre contractante par la droite des états caractéristiques* q=M'p'. *Le domaine dilatant est appelé aussi surcaractéristique, car il est situé au dessus de cette droite; le domaine contractant est appelé aussi sous-caractéristique.*

## 6.2. chemins drainés :

Si l'on fait subir des cycles de compression-décompression isotrope (à déviateur toujours nul) on obtient la compaction progressive de l'échantillon preuve que le comportement "élastique" que nous avions décrit pour la Fig. (4.1) n'est qu'approximatif.

### *6.2.1. comportement cyclique triaxial sur chemins répétés*

La déformation plastique est beaucoup plus importante lors du premier cycle que pour les suivants. La contrainte maximum atteinte $q_{max}$ joue le rôle de paramètre de mémoire : si l'on augmente l'amplitude du cycle, on voit une rupture de pente dans le diagramme contrainte-déformation (Fig. 6.4).

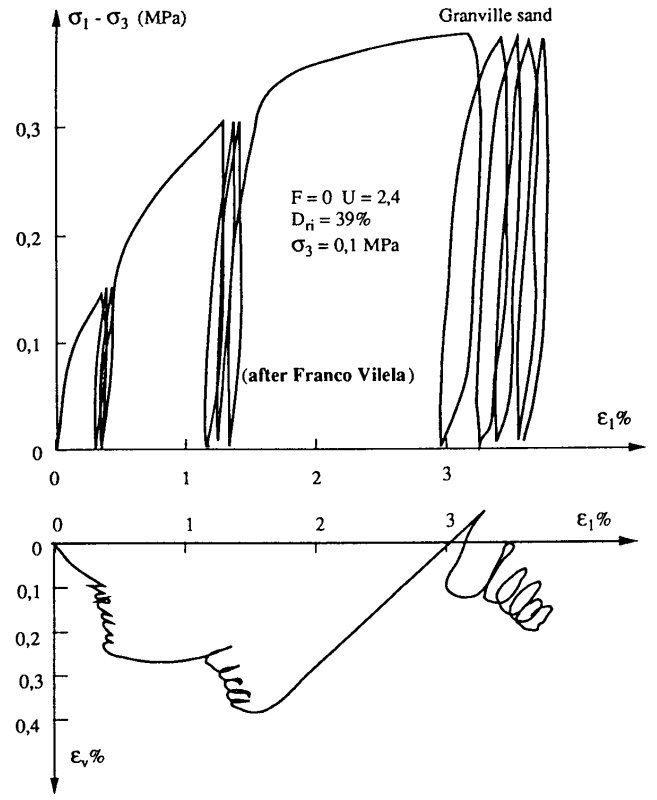

***Figure 6.4a:***
*Evolution de la déformation axiale* $\varepsilon_1$ *et de la déformation volumique* $\varepsilon_v$ *au cours* d'un *essai triaxial cyclique drainé et répété sur du sable de Granville. Plusieurs séries de cycles ont été imposés, à différents niveaux de* $q_{max}/p'$. *On remarque la densification successive, obtenue cycle après cycle, (d'après Biarez & Hicher 1994).*

Après ce premier cycle, les déformations sont plus régulières et plus faibles ; chaque cycle est contractant ou dilatant en moyenne suivant que la contrainte moyenne $q_{moyenne}$ est située en dessous ou au dessus de la droite q=M'p' de l'état caractéristique (Luong 1980). La partie du cycle au dessus de la droite caractéristique est dilatant (*cf.* Fig. 6.4).

♣ *Stabilisation de* v*:* Si q/p' est inférieur à M' la déformation axiale tend à se stabiliser et la compaction converge aussi, après un certain nombre de cycles. Il est intéressant de noter que cette compaction suit une loi temporelle en $\Delta v/v_o=-\alpha \mathrm{Log}(N)$ (Olivari 1973 & 1975) sur plusieurs décades ($N_{max}$=105). Elle semble donc être du même type que celle obtenue pour les expériences de densification au tap-tap (*cf.* Nowak *et al.* 1998) où l'on trouve une dépendance en fonction du nombre n de taps de la densité ρ:

$$(6.1) \qquad (\rho(t)-\rho_o)/(\rho_\infty - \rho(t)) = b \ln(1+n/n_o)$$

où $\rho_o$, $\rho_\infty$, b, $n_o$ sont des constantes à déterminer. Nous proposerons au chapitre 8 une explication théorique de cette loi ; celle-ci est basée sur des arguments microscopiques. Olivari (1973, 1975) trouve de même que la déformation axiale diverge pour les fortes valeurs du rapport q/p'.

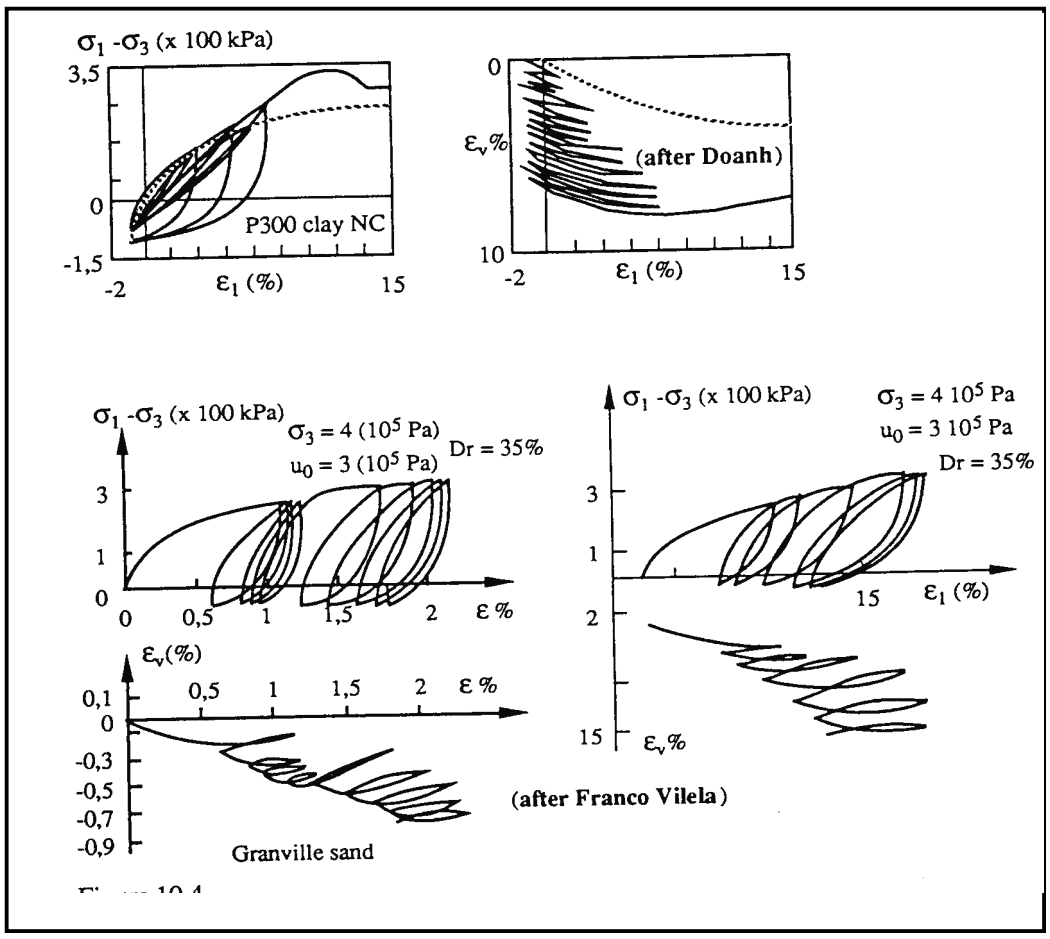

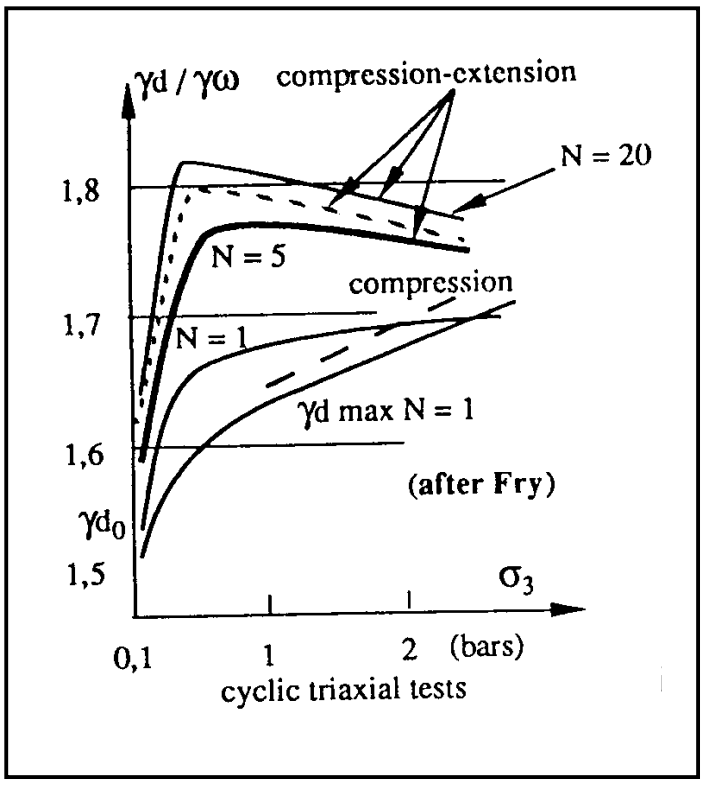

***Figure 6.4b et 6.4c:*** *Essais triaxiaux cycliques.(d'après Biarez & Hicher 1994)*

*Figure 6.4b: (Figure de gauche). Essais cycliques alternés sur de l'argile et du sable de Granville. Ces essais sont alternés, car* $\sigma_1$ *devient alternativement contrainte principale majeure puis mineure, quand* q *change de signe. On remarque l'ouvertire des cycles dans le plan* ($\varepsilon_1$, q)*, qui est beaucoup plus grande que dans le cas des cycles répétés (Fig. 6.4a). On voit l'efficacité des cycles alternés pour le compactage.*

***Figure 6.4c:*** *(Figure de droite). Variation de la densité d'un matériau après* N *cycles de cisaillements (*N=1 *ou* 5 *ou* 20) *en fonction de la pression* $\sigma_3$ *(en bars).*

### *6.2.2. comportements cycliques sur chemin avec variation des directions principales :*

♣ *chemins alternés*

Un changement de direction de contraintes principales (ou une permutation de celles-ci) induit une contraction de l'échantillon si l'on travaille près du point de contrainte isotrope ; en particulier, la contraction sera toujours plus importante pour des essais alternés que pour des essais répétés à amplitude de cycles égale (et petite) (*cf.* Fry 1979).

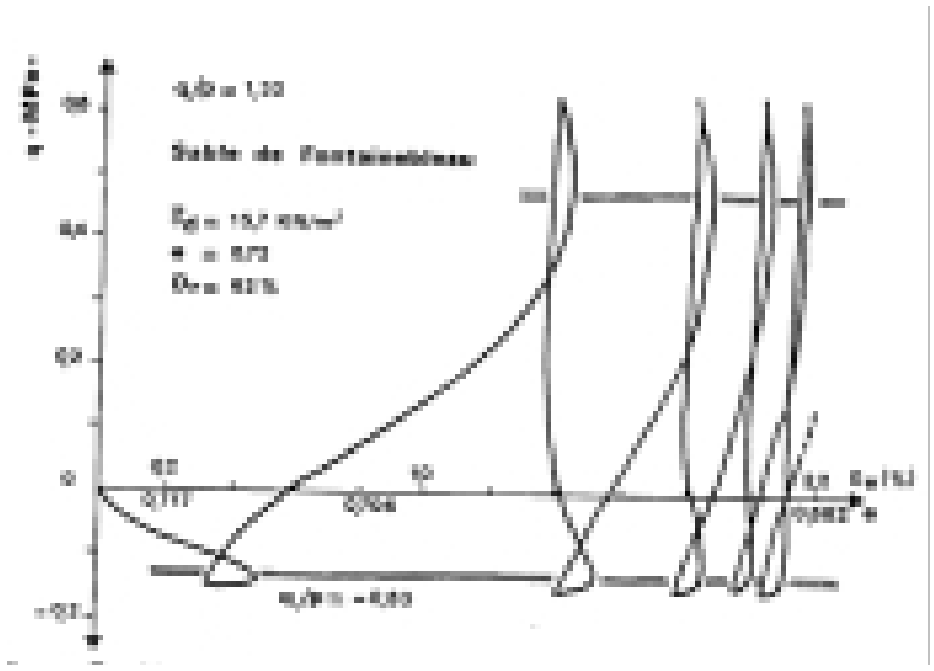

**Figure 6.5 :** *Essais cycliques drainés à amplitude de contrainte imposée; axe des absisses* $\varepsilon_v$ *, axe des ordonnées* q *en MPa. Matériau: sable. L'évolution du matériau se stabilise lentement.*

Si on travaille à amplitude de déformation $\varepsilon_1$ imposée, le cycle se déforme au bout d'un certain temps: il devient dilatant aux deux extrémités du cycle. En effet, le système se compacifie au fur et à mesure des cycles et il devient de plus en plus rigide de sorte que le rapport $q_{max}/p'$ croît au cours des cycles et finit par dépasser la valeur de la contrainte caractéristique q=M'p'. Des résultats similaires ont été obtenus par Luong (1980).

Enfin, un chargement cyclique prolongé dans le domaine sous-caractéristique, *i.e.* $q/p'<M'$, produit la contraction progessive du sable jusqu'à la stabilisation de la déformation axiale ; le même chargement cyclique dans le domaine sur-caractéristique produit une augmentation de volume jusqu'à la rupture.

♣ *chemins avec rotation des axes principaux*

Différentes expériences ont été réalisées soit à l'aide d'une cellule plane de cisaillement (parallélogramme) soit avec une cellule cylindrique creuse (Wong & Arthur 1986, Ishihara et Tohwata 1983). Elles démontrent l'existence de variations de volumes et une évolution des cycles contraintes déformations.

## 6.3. chemins non drainés

### *6.3.1. comportement cycliques sur chemins répétés :*

Dans ce cas, la différence principale de comportement lors des essais à chemins drainé et non drainé provient du fait qu'on autorise ou qu'on n'autorise pas la variation du volume de l'échantillon. Ainsi lorsqu'un cycle est dilatant en condition drainée, i.e. $(q/p')_{moyen}>M'$, le même cycle en condition non drainée cherchera à pomper de l'eau pour lui permettre d'augmenter le volume de ces pores; il s'en suit que la pression locale $u_w$ aura tendance à diminuer en condition non drainée; ceci fera augmenter la pression intergranulaire p' , de manière à garder $p'+u_w=c^{ste}$ ; au fur et à mesure des cycles, la pression p' augmentera et la pression $u_w$ diminuera, jusqu'au moment où $(q/p')_{moyen}=M'$; dans ce cas les cycles "perdent leur caractère dilatant" et la variation de pression moyenne p' et de volume v se stabilisent. La droite des états caractéristiques q=M'p' joue donc un rôle essentiel.

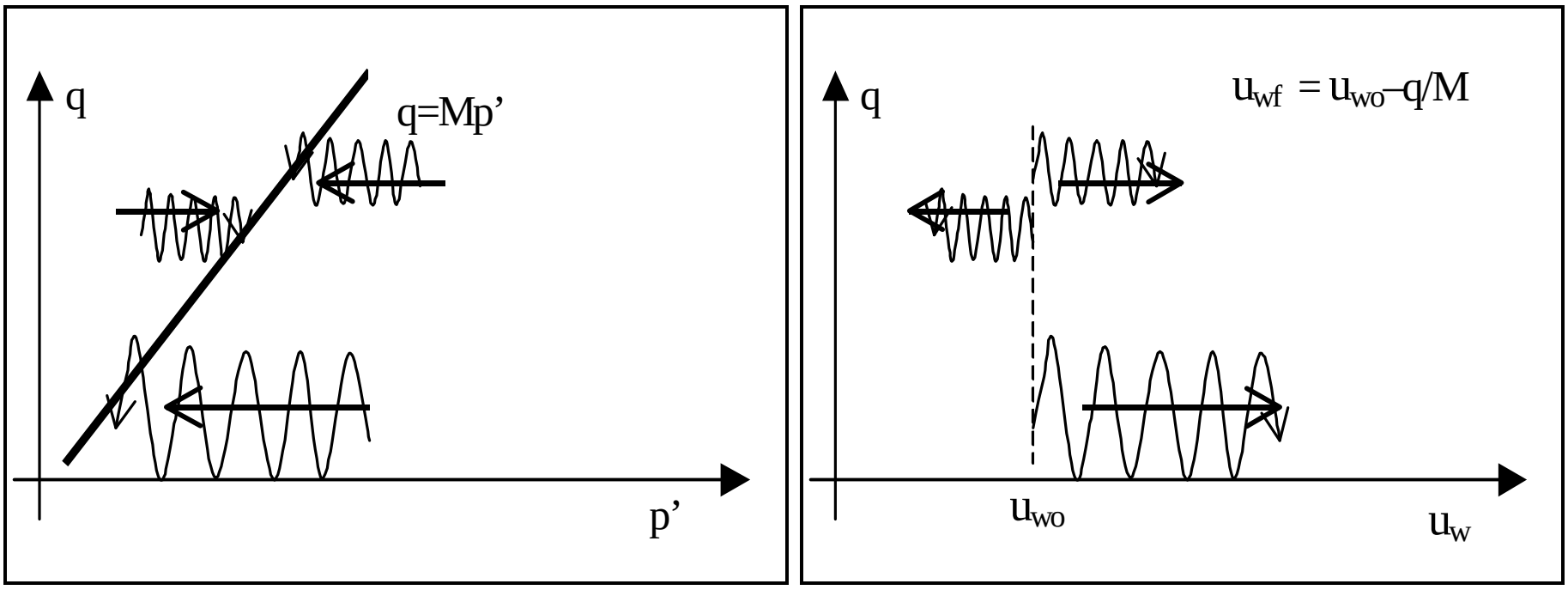

***Figure 6.6 :*** *Comportement cyclique sous condition non drainée : Cette condition s'obtient simplement en saturant les pores d'eau et en interdisant à cette eau de sortir de l'échantillon. Si on part à droite de l'état caractéristique, soit* q/p'<M' *, les cycles ont tendance à faire diminuer la pression inter-granulaire* p' *en augmentant la pression de l'eau* $u_w$ *; l'échantillon était relativement stable au départ, mais au bout d'un certain nombre de cycles, on obtient ainsi une mobilité cyclique car le rapport* $q_{max}/p'$ *croît. L'essai fini à* q/p'=M'. *Si on part à gauche de l'état caractéristique, soit* q/p'>M' *, l'échantillon est potentiellement instable, mais la pression intergranulaire augmente au cours des cycles, ce qui stabilise l'échantillon.*

De la même façon, lorsque le cycle est contractant en condition drainée, le même cycle en condition non drainée cherche à expulser de l'eau ; ainsi, en condition non drainée, la pression locale $u_w$ aura tendance à augmenter si $(q/p')_{moyen}<M'$; ceci fait diminuer la pression intergranulaire p' ,de manière à garder $p'+u_w=c^{ste}$; au fur et à mesure des cycles, la pression p' diminue et la pression $u_w$ augmente; $(q/p')_{moyen}$ évolue et croît jusqu'au moment où l'évolution perd le caractère contracatant, c'est-à-dire lorsque $(q/p')_{moyen}$ atteint la droite des états caractéristiques définie par q=M'p'.

**Ainsi, le point q=M'p' est un attracteur stable pour ce type d'essai**. On peut ainsi tracer le diagramme suivant de la Fig. 6.6.

Il est intéressant de remarquer à ce stade que la présence d'eau n'est pas obligatoire pour obtenir ce comportement, car il aurait suffit d'imposer des cycles de contraintes où q et p' varient exactement de la même façon que ce que nous avons décrits plus haut ; dans ce cas la variation de volume aurait été nulle. La saturation des pores par l'eau et les conditions non drainées sont une façon expérimentale simple de réaliser les conditions expérimentales recherchées.

### *6.3.2. comportements cycliques sur chemin avec variation de direction principales : liquéfaction :*

Lorsque les directions des contraintes principales varient (ou sont permutées) au cours du cycle, le matériau est sujet à un effet de mémoire beaucoup moins intense. C'est là l'effet essentiel. Il s'ensuit que les phénomènes décrits dans les sections précédentes sont amplifiés et/ou accélérés. C'est dans ce cas, en particulier, que la liquéfaction des sables est obtenue; pour cela il faut que le système ne soit pas drainé. Le phénomène de liquéfaction est important du point de vue technique compte tenu des dégats qu'il peut occasionner lors des séismes, ou simplement lors de la circulation d'engins sur du sable saturé d'eau (lors de débarquements militaires par exemple).

♣ *chemins alternés non drainés*

Lorsque les chemins de contrainte sont alternés, la pression intergranulaire peut diminuer fortement jusqu'à s'annuler périodiquement. Il y a alors absence de portance périodique. On comprend ainsi l'effet catastrophique qui en résulte sur les fondations des bâtiments par exemple. On constate aussi l'augmentation très forte de la pression de l'eau interstitielle, ce qui peut donner lieu à un courant d'eau vers le haut; c'est ce phénomène qui est responsable de l'apparition de geisers lors des séismes.

Le processus de liquéfaction peut être décomposé en trois grandes étapes.

♣ Durant la première phase (notée A, Fig. 6.7), la pression moyenne $u_w$ de l'eau s'accroît constamment et les déformations restent petites et cycliques.

♣ Dans la deuxième phase (notée B, Fig. 6.7), qui est une phase de transition, la pression $u_w$ devient telle que le système de contrainte (p',q) atteint par instant la droite caractéristique q=M'p'; le cycle ($\varepsilon_1$ , q) voit son amplitude croître énormément, le cycle (p',q) forme un 8 qui se déforme pour suivre sur une partie de la droite caractéristique; le point central du 8 est à peu près sur l'axe p', mais il se déplace vers l'origine (p'=0) de telle sorte que le 8 se déforme de plus en plus.

♣ Dans la troisième et dernière phase (notée C, Fig. 6.8), les cycles se stabilisent : la pression $u_w$ continue à varier périodiquement ; le cycle ($\varepsilon_1$ , q) prend la forme d'une marche d'escalier où le rôle de la hauteur serait joué par l'axe $\varepsilon_1$ ; l'hystérésis apparaît essentiellement aux valeurs maximum de $|\varepsilon_1|$; le cycle (p',q) passe deux fois par cycle par l'origine (p'=0, q=0) ; en fait le temps passé sur ce point est long car c'est pendant ces deux périodes que la majeure partie de la déformation $\varepsilon_1$ a lieu (c'est pourquoi il y a vraiment liquéfaction, *i.e.* perte de résistance au cisaillement); le 8 suit complètement la droite caractéristique. La déformation $\varepsilon_1$ croît.

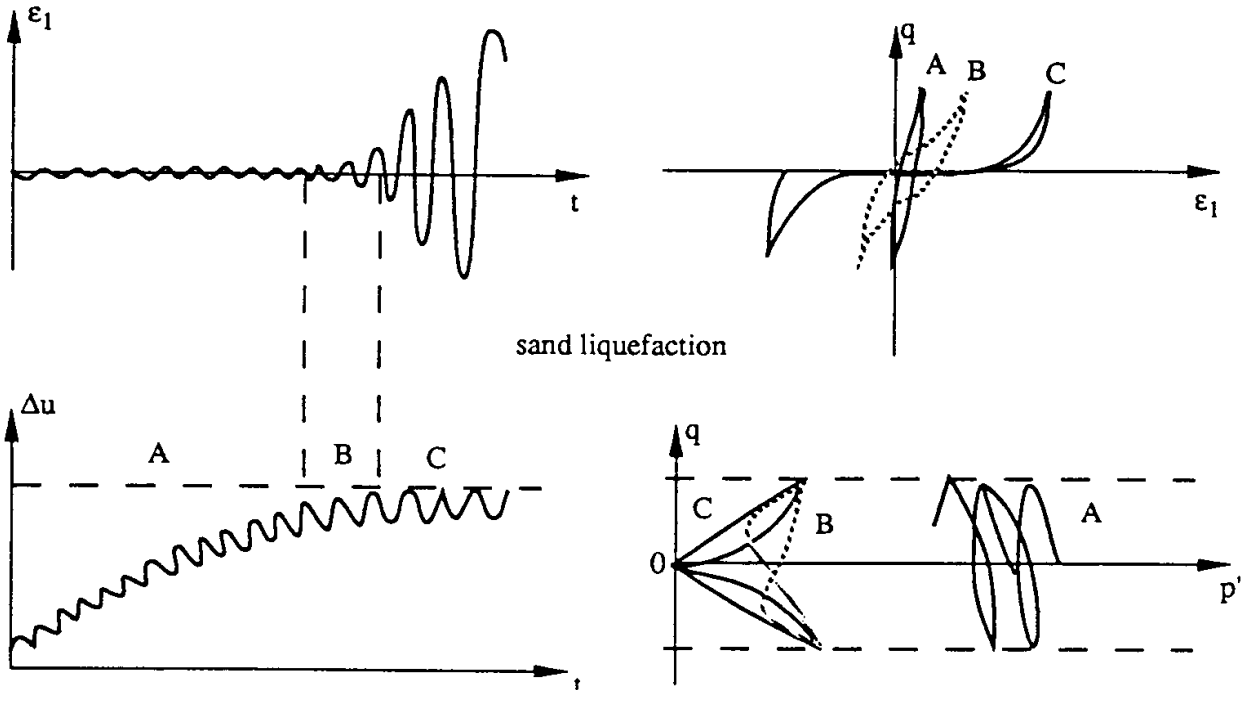

***Figure 6.7 :*** *Comportement typique d'un matériau granulaire en train de se "liquéfier ". La liquéfaction arrive parce que le système se déforme à volume constant. Pour comprendre et caractériser ce comportement, il est nécessaire de représenter les évolution de la déformation axiale* $\varepsilon_1$ *et de la pression de l'eau* $u_w$ *en fonction du temps* t, *ainsi que les variations du déviateur de contrainte* $q=\sigma'_1-\sigma'_3 = \sigma_1-\sigma_3$ *en fonction de* $\varepsilon_1$ *et de la contrainte moyenne inter-granulaire* p'.

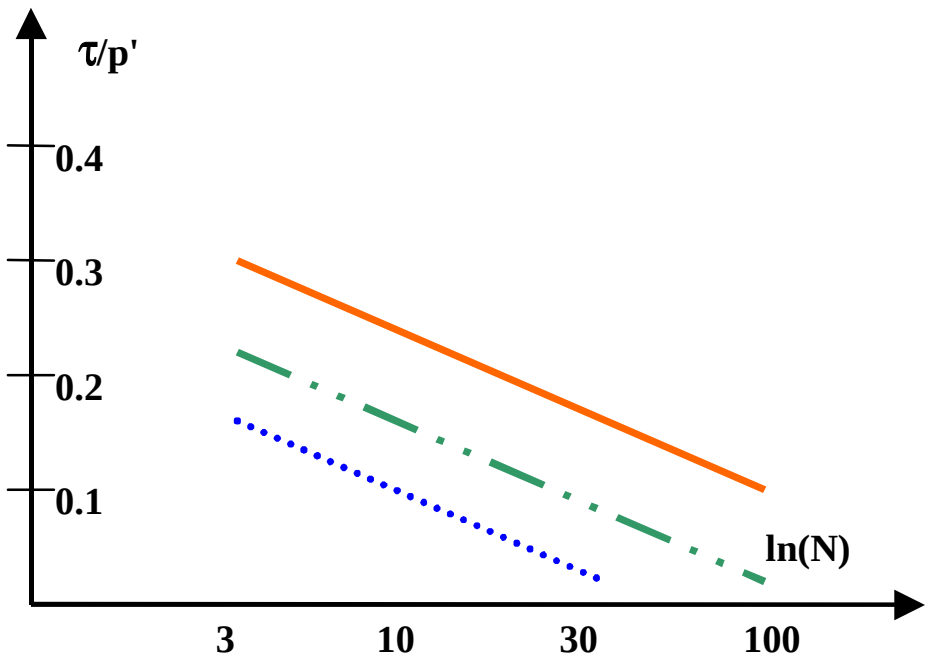

***Figure 6.8 :*** *Nombre de cycles* N *pour obtenir la liquéfaction (en échelle logarithmique) en fonction du taux de cisaillement horizontal* $\tau=\sigma_{xz}$ *pour différentes valeurs du rapport des contraintes horizontales* $\sigma_x$ *et verticales* $\sigma_z$ *de départ* $K_o=\sigma_x/\sigma_z$. ($K_o=1$ *trait plein* ———; $K_o=0,4$ *trait pointillé* …… ; $K_o=0,75$ *traits alternés —…—…—). On constate le même type de loi en* log(N) *ici que pour la densification en condition drainée (cf. § 6.2.1.). En fait, d'après Pecker (Figs.10,12,15, p.≈100), ces lois ne semblent pas être strictement linéaires, mais présentent une courbure légèrement positive. Elles dépendent aussi de la structure du sol. Plusieurs modèles plus ou moins phénoménologiques ont été proposés pour rendre compte de ce comportement ; on notera ceux à contrainte effective (Ishihara, Habib-Luong 1978), et celui de Martin-Finn-Seed (1975) qui conduit à une équation relativement complexe.*

On donne dans la Fig. 6.8, la variation du nombre de cycle nécessaire pour liquéfier un sable à une densité donnée en fonction de l'amplitude relative τ/p' du cisaillement imposé. La dépendance est logarithmique. Nous verrons au chapitre 8 qu'un argument microscopique similaire à celui utilisé pour expliquer l'Eq. (6.1) peut justifier la dépendance logarithmique. Le même type d'argument microscopique peut d'ailleurs être aussi utilisé pour expliquer la dépendance logarithmique du volume spécifique des états normalement consolidés et des états critiques en fonction de la pression moyenne p', *cf.* chapitre 8.

♣ *chemins sous cisaillement variable ou avec rotation des contraintes*
Dans le cas drainé, on avait trouvé que la rotation des directions des contraintes principales accroissait la vitesse de densification du matériau (§ 6.2.2.). De façon analogue, la rotation des contraintes facilitera la liquéfaction d'un sable en conditions non drainées (Biarez & Hicher 1994).

## 6.4. Tests expérimentaux :

Les essais de mécanique des sols sont réalisés pour caractériser soit la nature du matériau (distribution granulométrique, cohésion, nature chimique), soit sa tenue mécanique tant aux petites déformations qu'aux grandes. Mais caractériser les conditions mécaniques du passage de la statique aux grandes déformations, c'est caractériser la capacité à couler du matériau. Il est donc normal qu'un certain nombre des grandeurs que nous avons définies dans les chapitres précédents puissent servir aussi à définir un indice de coulabilité et soient donc relier à certains tests industriels.
Nous voulons aborder ce problème dans cette partie et montrer le lien qui existe entre ces tests industriels et les comportements caractéristiques mécaniques que nous venons d'exposer.

### *6.4.1. densification au tap-tap comme indicateur de coulabilité :*

Cette technique est utilisée en pharmacie pour mesurer expérimentalement la facilité avec laquelle un matériau se densifie. La technique consiste à faire subir à un matériau granulaire contenu dans une éprouvette un certain nombre de chocs verticaux et à mesurer le tassement. Pour cela, un mécanisme à came permet de laisser tomber l'éprouvette d'une certaine hauteur (5mm à 2cm typiquement) de façon répétitive. C'est une manière d'estimer l'écart entre la densité minimum du matériau et sa densité sous une certaine contrainte. Cette expérience caractérise donc la variation $v_{max}$-$v_{min}$ et la pente approximative de la droite (v *vs.* ln(p')). En d'autres termes, elle caractérise les états "critiques ".
Par ailleurs, on sait que pour qu'un échantillon puisse s'écouler, il faut qu'il puisse rentrer facilement dans son état de plasticité parfaite (état "critique "); ceci est d'autant plus facile que le matériau est peu dilatant ; en effet la dilatation requiert un certain travail d'autant plus important que la dilatance de l'échantillon est grande : $\delta W = p'\delta\varepsilon_v + q\,\delta\varepsilon_1$. Plus le système est "compactable", plus la différence de densité entre son état dense et son état lâche est grande, et plus l'énergie à lui fournir pour qu'il s'écoule est grande ; son état statique est ainsi très stable stable et sa coulabilité est faible.
Ainsi, cette expérience simple mesure donc approximativement le travail qu'il est nécessaire de fournir à l'échantillon pour rentrer dans son état de plasticité parfaite et est donc un indicateur de la coulabilité du matériau.

### *6.4.2. essai triaxial comme indicateur de coulabilité:*

Il est clair que le même raisonnement s'applique à l'essai triaxial, qui peut donc être utilisé comme test de coulabilité.

De plus certains matériaux granulaires peuvent avoir des comportements atypiques, provoqué par exemple par la forme allongée des grains… Il est alors possible que l'état caractéristique ne soit plus une droite dans le plan (q,p') , que le volume spécifique ne varie plus linéairement en fonction de ln(p') et que le milieu soit très dense à des pressions relativement faibles. On s'attend alors à un changement de comportement qu'on pourra détecter au triaxial de façon plus précise et que l'on doit prendre en compte pour déterminer le seuil d'écoulement (Luong 1989).

### *6.4.3. limites d'Atterberg* :

Cette méthode est utilisée pour caractériser les limites de liquidité et de plasticité des argiles. Elle ne peut pas être utilisé dans le cas de matériaux à grains trop gros.

♣ *limite de liquidité :* on assèche un échantillon ; puis on humidifie le matériau à étudier; on le place alors dans une coupelle mobile ; on fait une rainure dans le matériau et on laisse tomber la coupelle de quelques millimètres vingt cinq fois de suite sur une plaque dure. La limite de liquidité est la teneur en eau juste suffisante pour que la rainure disparaisse.

♣ *Limite de plasticité :* on assèche un échantillon ; puis on humidifie le matériau et on cherche à le rouler pour en faire un cylindre de diamètre 3mm ; si le matériau n'est pas assez humide, il se casse. La limite de plasticité est la teneur en eau pour laquelle la longueur minimum du cylindre est 1cm.

### *6.4.4. Densités maximum et minimum du sable :*

- *Pour déterminer la densité minimum*, on bâtit l'échantillon par la méthode de pluviation en utilisant une faible hauteur de chute et un débit faible.

*Pour la densité maximum*, on mesure la densité d'un échantillon après l'avoir fortement vibré (sous plusieurs g et sous un poids donné pendant un temps donné).

- *Remarque :* Comme nous l'avons vu, les pharmaciens quant à eux utilisent la méthode du tap-tap pour mesurer l'aptitude du milieu à se compacter. Cette méthode permet de faire subir à un récipient cylindrique vertical contenant le matériau granulaire une série de chocs calibrés correspondant à une hauteur de chute constante (de l'ordre de 1 à 2cm). Ils ont ainsi une estimation de la densité minimum et de la densité maximum dans la même expérience.

### *6.4.5. essai Proctor* :

On assèche un échantillon, puis on lui ajoute une quantité $m_e$ d'eau, et on le dame avec une certaine masse M, pendant un certain temps en laissant tomber la masse M d'une certaine hauteur. On mesure la densité sèche du matériau $\rho_s(m_e)$ à ce stade ; on répète pour d'autres quantité $m_e$ d'eau et on trace $\rho_s(m_e)$ . Ceci permet de déterminer la densité maximum et d'optimiser la quantité d'eau pour lubrifier les contacts lors de la compaction.

## Bibliographie

# 7. Modélisation du comportement mécanique d'un milieu granulaire

Nous avons vu dans les chapitres précédents que le nombre de paramètres nécessaires pour décrire l'état d'un milieu granulaire à l'équilibre mécanique est 3 : son volume spécifique v, le déviateur de contrainte axiale q et la pression granulaire p'. Ainsi, l'espace des phases d'un échantillon homogène de ce milieu est l'espace {v, q, p'} ou mieux {v, ln(q/M'), ln(p')}. Bien sur, la dimension de cet espace peut s'avérer insuffisant pour décrire certains effets, car il ne prend pas réellement en compte l'anisotropie; il faudrait alors lui adjoindre un certain nombre de paramètres supplémentaires. Cependant une description plus complète est très rarement nécessaire dans les cas étudiés au laboratoire pour lesquels les échantillons de matériau ont subi des histoires relativement simples. C'est pourquoi les phénomènes que nous décrirons dans ce chapitre doivent pouvoir se décrire essentiellement dans l'espace {v,q,p'} (ou {v,ln(q),ln(p')}.

Considérons une transformation du matériau à déformation axiale continûment croissante, les caractéristiques de l'échantillon au cours de cette déformation se modifient continûment ; l'évolution de l'échantillon apparaît donc comme une courbe à une dimension que nous appellerons trajectoire. A chaque point de cette trajectoire correspond une valeur de la déformation axiale. Si l'on veut faire l'analogie avec l'évolution des systèmes dynamiques, la déformation axiale est l'équivalent du temps ; une différence subsiste cependant : dans le cas présent, l'évolution du système est contrôlé par la déformation ; c'est-à-dire que si on s'arrête de déformer l'échantillon, son évolution stoppe et chaque état intermédiaire est donc un état intrinsèquement stable; dans le cas des systèmes dynamiques, cette liberté de bloquer l'écoulement du temps n'existe pas en général.

Pour pouvoir prédire la trajectoire, il faut connaître l'évolution du système lorsqu'on lui applique un incrément de contrainte ou de déformation ; il faut donc connaître la loi qui permet de déterminer $\delta\sigma'$ en fonction de $\delta\varepsilon$, ou l'inverse.

La recherche de cette loi sera réalisée dans la première partie de ce chapitre. Nous utiliserons pour cela le formalisme de la plasticité, dans sa formulation incrémentale. Nous utiliserons aussi la loi de Rowe, que nous avons décrite précédemment; celle-ci stipule que la dilatance est une fonction des contraintes $\sigma'_1$ et $\sigma'_2=\sigma'_3$. Cette association peut surprendre certains mécaniciens avertis, qui considèrent que l'existence de cette cette loi est la preuve tangible qu'il faille utiliser une approche de plasticité parfaite. C'est pourquoi une bonne partie de cette première partie a pour tâche de leur démontrer le contraire. Du point de vue de la nomenclature, nous pensons que l'approche utilisée peut s'appeler tout aussi bien "théorie incrémentale non linéaire", "théorie incrémentale linéaire par zones", "théorie hypoplastique" ou "théorie hypoélastique" ne sachant pas très bien les différencier.

Nous appliquerons cette théorie dans les parties suivantes et poserons les problèmes du comportement des échantillons homogènes dans le cadre de la théorie de l'évolution des systèmes dynamiques. Pour cela nous utiliserons le fait que la déformation axiale joue un rôle analogue au temps. Nous chercherons ainsi à résoudre le problème de l'évolution d'un échantillon soumis aux différents types de sollicitation monotone suivants : i) à une compression à contrainte latérale constante, qui nous servira de référence, ii) à une compression oedométrique, iii) compression à volume constant à, iv) une compression à contrainte p' constante, v) à une compression cyclique à volume constant, vi) ) à une compression cyclique à volume variable. Nous trouverons les points fixes stables de toutes ces sollicitations lorsque le matériau est forcé de rester homogène. Nous étudierons ce qu'implique cette modélisation à propos de l'arrivée des trajectoires au point critique; nous montrerons en particulier que le pseudo module d'Young E doit tendre vers 0 et le pseudo coefficient de Poisson $\nu$ vers 1/2, et que le rapport $E/(\nu-1/2)$ doit tendre vers une constante si les surfaces de Roscoe et de Hvorslev font partie de la même surface régulière dans l'espace des phases $\{v,\ln(q),\ln(p')\}$.
Nous montrerons ensuite que la modélisation incrémentale que nous avons utilisé permet de décrire, au moins qualitativement, la compaction, ou la décompaction, sous contraintes cycliques et les phénomènes de liquéfaction. Une approche microscopique de ce problème sera présentée dans le chapitre suivant. On verra alors que les lois de densification du milieu granulaire, *i.e.* $\delta e=-k \ln(\text{Nombre de cycles})$, découlent de cette approche.

## 7.1. Equation d'évolution :

### *7.1.1. Approche hypo-élastique ou incrémentale*

Cette approche est basée sur l'existence d'un déterminisme ; celui-ci postule que l'échantillon doit répondre d'une manière prévisible et déterministe à un incrément de contrainte connaissant les conditions aux limites. Par ailleurs, considérons un échantillon sous un état de contrainte $\boldsymbol{\sigma'}$ donné ; a priori, on doit pouvoir lui faire subir n'importe quel incrément infinitésimal $\boldsymbol{\delta\varepsilon}$ de déformation choisi à l'avance en lui appliquant un incrément de contrainte $\boldsymbol{\delta\sigma'}$ adapté. En d'autres termes, cela veut dire que l'évolution de l'échantillon est gouverné par une loi incrémentale g qui relie $\boldsymbol{\delta\sigma'}$, $\boldsymbol{\sigma'}$ et $\boldsymbol{\delta\varepsilon}$. Dans cette formulation, on néglige tous les effets retardés tels que les effets visqueux produits par des liquides interstitiels ou tels que les effets inertiels. On considère donc ici exclusivement le régime quasi-statique. Nous avons vu brièvement aux chapitres 1 et 2 comment introduire des termes inertiels et l'influence de l'eau dans cette formulation.
On peut donc écrire que l'évolution est pilotée par une équation du type :

$$g(\boldsymbol{\delta\sigma'},\boldsymbol{\sigma'}, \boldsymbol{\delta\varepsilon})=0 \qquad (7.1)$$

L'évolution réelle d'un sable et d'un milieu granulaire dépend en fait de l'histoire que le matériau a subi. En toute rigueur, on doit donc considérer que g dépend de cette histoire. Cependant, tenir compte réellement de cette histoire (de cet « écrouissage ») nécessiterait de rajouter des paramètres et donc d'augmenter le nombre de dimensions de l'espace des phases. Ceci rendrait très compliqué le formalisme et n'est peut-être pas nécessaire dans beaucoup de cas, d'autant que g contient déjà des comportements non linéaires puisqu'il dépend de $\boldsymbol{\sigma}'$. Nous négligerons donc ces effets d'histoire, au moins dans un premier temps, pour simplifier la modélisation .
Ainsi l'Eq. (7.1) peut se réécrire :

(7.2) $$\boldsymbol{\delta\varepsilon}=f(\boldsymbol{\delta\sigma}',\boldsymbol{\sigma}')$$

De plus, on sait que la réponse du matériau est indépendante de la vitesse de sollicitation, tant que cette dernière reste suffisamment lente ; c'est l'hypothèse du régime quasi-statique. Dans ce cas, posons $\boldsymbol{\delta\varepsilon}= \delta\lambda\ \partial\boldsymbol{\varepsilon}/\partial t$ et $\boldsymbol{\delta\sigma}'= \delta\lambda\ \partial\boldsymbol{\sigma}'/\partial t$ ; on doit pouvoir écrire :

(7.3.a) $$\delta\lambda\ \partial\boldsymbol{\varepsilon}/\partial t = f(\delta\lambda\ \partial\boldsymbol{\sigma}'/\partial t,\boldsymbol{\sigma}' ) = \delta\lambda\ f(\partial\boldsymbol{\sigma}'/\partial t,\boldsymbol{\sigma}' )$$

Soit :

(7.3.b) $$\partial\boldsymbol{\varepsilon}/\partial t= f(\partial\boldsymbol{\sigma}'/\partial t,\boldsymbol{\sigma}' )$$

Ainsi, la fonction f doit être homogène d'ordre 1. Ceci implique que l'amplitude de la réponse à un incrément $\boldsymbol{\delta\sigma}'$ dans une direction donnée constante doit être linéaire et proportionnelle à la norme de $\boldsymbol{\delta\sigma}'$. Cependant cette loi ne peut pas être totalement linéaire, car on sait qu'un matériau granulaire présente des phénomènes d'hystérésis et que la linéarité totale engendrerait une réversibilité totale. Pour pallier cet inconvénient, on doit utiliser pour f des lois hypoélastiques, hypoplastiques ou incrémentales (Darve, Tejchman 1997).
Dans le cas précis des matériaux granulaires, on sait qu'un phénomène d'hystérésis se développe spontanément dès que l'on applique des cycles alternativement $+\boldsymbol{\delta\sigma}'$ et $-\boldsymbol{\delta\sigma}'$, qui engendrent des réponses $\boldsymbol{\delta\varepsilon}$ et $\boldsymbol{\delta\varepsilon}' \neq -\boldsymbol{\delta\varepsilon}$ . Il faut donc tenir compte de ce point ; mais il est tentant de se limiter à cette non linéarité seulement et d'utiliser une loi incrémentale linéaire par zone avec un nombre de zones très faible ; nous nous limiterons par la suite à deux ou trois zones, mais n'en utiliserons qu'une réellement.
Dans cette approche, on découpe donc l'espace E de variation de $\boldsymbol{\delta\sigma}'$ en un petit nombre n de sous espaces $E_k$ disjoints et formant un espace complet ($E=\cup E_k$) dans lesquels il existe une loi rhéologique linéaire $f_k$ particulière ($f_k \neq f_{k'}$). De plus, les lois rhéologiques doivent être telles que $f_k(\boldsymbol{\delta\sigma}',\boldsymbol{\sigma}')=f_{k'}(\boldsymbol{\delta\sigma}',\boldsymbol{\sigma}')$ pour les $\boldsymbol{\delta\sigma}'$ appartenant à la frontière entre les sous espaces $E_k$ et $E_{k'}$, ceci pour assurer la continuité et l'unicité de la réponse.
On remarque aussi que la continuité de la réponse est assurée entre les réponses $+\boldsymbol{\delta\sigma}'$ et $-\boldsymbol{\delta\sigma}'$ , car la frontière a lieu pour $\boldsymbol{\delta\sigma}'=0$ .

### *7.1.2. Compatibilité entre approche incrémentalee et approche "plasticité parfaite":*

Darve a montré que l'approche incrémentale est capable de décrire la rhéologie de systèmes obéissant à une loi de plasticité parfaite à un mécanisme: l'opérateur $f_k$ est alors un opérateur de projection qui définit la normale à la surface de charge. Loret (1985a & b) quant à lui a démontré que l'approche incrémentale pouvait décrire les systèmes obéissant à la théorie de la plasticité parfaite et/ou à celle de l'élasto-plasticité à un ou plusieurs mécanismes ; dans le cas d'un système élasto-plastique à un mécanisme, la direction de la déformation plastique est contrôlée par la normale à la surface de charge et l'amplitude de déformation par la loi d'écrouissage de telle sorte qu'elle dépend linéairement de $||\delta\sigma'||$. Parce que les opérateurs de projection sont des opérateurs linéaires et parce qu'ils peuvent s'additionner, un échantillon qui obéit à une loi élasto-plastique à plusieurs mécanismes peut être toujours décrit dans le formalisme incrémental linéaire avec plusieurs zones.
Enfin, un milieu purement élastique peut être décrit dans le formalisme de l'hypo-élasticité , il suffit de réduire le nombre de zone à l'unité.

### *7.1.3. Principe de la modélisation :*

Dans le reste de cette section, nous décrirons le comportement du milieu granulaire dans ce formalisme, *i.e.* réponse incémentale linéaire par zone. De plus, pour limiter la complexité des comportements, et le nombre de paramètres à déterminer, nous choisissons un nombre n de zones faible : n=2 ou 3, et nous considérerons que tous les trajets étudiés appartiennent à la même zone, soit par exemple $d\varepsilon_1>0$.
Enfin, pour rendre compatible la description avec la nature tridimensionnelle de l'espace des phases, les réponses incrémentales seront choisies isotropes. Bien sur, cette approximation est probablement très sévère et doit produire quelques différences notables entre prédictions théoriques et résultats expérimentaux. Cependant, nous verrons que ces différences sont limitées dans beaucoup de cas et ne conduisent pas à des situations catastrophiques.
Dans ces conditions la loi incrémentale que nous devons considérer dans la zone d'intérêt est du type :

$$\begin{pmatrix}\delta\varepsilon_1\\ \delta\varepsilon_2\\ \delta\varepsilon_3\end{pmatrix} = -C_o \begin{pmatrix}1 & -\nu & -\nu\\ -\nu & 1 & -\nu\\ -\nu & -\nu & 1\end{pmatrix}\begin{pmatrix}\delta\sigma_1\\ \delta\sigma_2\\ \delta\sigma_3\end{pmatrix} \qquad (7.4)$$

Où $C_o$ a la dimension de l'inverse d'un module d'Young. Nous appellerons donc souvent $E_{plast}=1/C_o$ le pseudo module d'Young du matériau et $\nu$ son pseudo-coefficient de Poisson.
Ce pseudo-module d'Young et ce pseudo-coefficient de Poisson $\nu$ se déduisent des courbes expérimentales à contraintes latérales constantes q/p' en fonction de $\varepsilon_1$ , par exemple des Figs. 3.6, 3.11, 4.2,… :

$$E_{plast}=1/C_o=\delta\sigma'_1/\delta\varepsilon_1$$

Ces courbes montrent que $C_o$ dépend de q/p' , de p' et du volume spécifique v.

$C_o$ ne doit pas être confondu avec le module d'Young élastique qui varie en $p'^{1/2}$ , ni avec le module sécant .
Dans un premier temps nous ne nous occuperons pas de l'évolution de $C_o$ car il n'en existe pas d'expression mathématique simple. Ceci est lié au fait que la variation de E doit pouvoir tenir compte d'effets de mémoire ; E devra donc être la solution d'une équation d'évolution (intégrale). Nous aborderons ce problème dans le § 7.6.
Nous allons voir, par contre, que le pseudo coefficient de Poisson $\nu$ n'a pas un comportement aussi compliqué, c'est-à-dire qu'il ne présente pas d'effets de mémoire. Ceci est une conséquence de la loi de Rowe qui lie $\nu$ au rapport q/p'. De plus, nous allons voir aussi que $\nu$ gère presque complètement à lui tout seul l'évolution du matériau dans deux essais classiques : l'essai oedométrique et l'essai à volume constant. Nous allons donc proposer la modélisation de $\nu$ dans le § 7.1.4 ; puis nous développerons les prédictions de notre modèle dans le cas des essais oedométriques (§ 7.2) puis des essais à volume constant (§ 7.3). Nous généraliserons cette approche à d'autres essais, dits à dilatance imposée (§ 7.4). Nous rediscuterons de l'existence des surfaces de Roscoe et de Hvorslev (§ 7.5) ; nous montrerons que notre modélisation incrémentale les prédits et fixe une certaine relation entre l'évolution de $\nu$ et de $E_{plast}$ dans le voisinage des états critiques. Enfin, Nous proposerons une méthode pour modéliser les variations de $E_{plast}$ (§ 7.6). Elle utilise une formulation proposée par un autre modèle, appelé modèle de Hujeux.

### *7.1.4. Conséquence de la loi de Rowe :*

La loi de Rowe relie la variation de volume $\delta\varepsilon_v/\delta\varepsilon_1= (\delta\varepsilon_1+\delta\varepsilon_2+\delta\varepsilon_3)/\delta\varepsilon_1$ au rapport des contraintes q/p' dans un essai à pression $\sigma'_2=\sigma'_3$ constantes (ou $\delta\sigma'_2=\delta\sigma'_3=0$). Appliquons ces conditions expérimentales à la modélisation incrémentale (Eq. 7.4), on trouve:

$$(7.5) \qquad (\delta\varepsilon_v/\delta\varepsilon_1)_{\sigma'_2=\sigma'_3=cste} =1-2\nu$$

Ceci veut dire que la loi de Rowe se traduit dans notre modélisation par une variation du pseudo-coefficient de Poisson $\nu$ en fonction du rapport des contraintes q/p'. Compte tenu de l'Eq. (4.8) et du fait que la dilatance $K=- (\delta\varepsilon_v/\delta\varepsilon_1)_{\sigma'_2=\sigma'_3=c^{ste}} =2\nu-1$, on trouve que le pseudo-coefficient de Poisson doit obéir à :

$$(7.6) \qquad 2\nu = (\sigma_1'/\sigma'_3) \tan^2(\pi/4-\varphi/2)$$

On peut tester cette hypothèse expérimentalement en la confrontant aux résultats d'expériences à pression p' constante. La prévision théorique liée à l'Eq. (7.4) prévoit qu'une telle expérience a lieu à volume constant. Ceci est vérifié expérimentalement , au moins tant que le déviateur de contrainte reste suffisamment faible. De même, cette approche est confirmée par un autre résultat bien établi de la mécanique des sols expérimentale concernant les essais non drainés : dans tous ces essais, la pression reste constante lorsqu'on augmente le déviateur de contrainte tant que la valeur de q n'atteint pas q=p'/2 . Au dessus de cette valeur, les courbes dans le plan {q,p'} peuvent s'infléchir vers les p' plus petits lorsque le matériau est peu dense. Ce dernier point montre que le milieu développe une anisotropie.

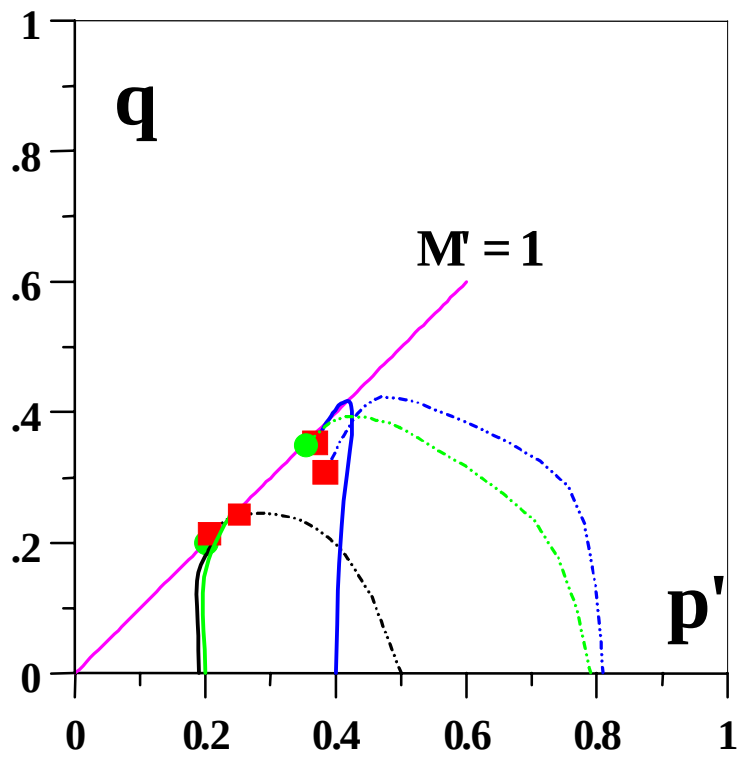

***Figure 7.1*** *: Essais tiaxiaux à volume constant sur du sable d'Hostun pour différentes valeurs de la contrainte* p' *initiales. Ils prouvent que le pseudo-coefficient de Poisson* ν *est approximativement constant pour les faibles contraintes p' et/ou les densités élevées, mais que le matériau lâche devient anisotrope pour* q/p'>1/2*.*

♣ *Remarque 1:* la valeur de q où l'infléchissement apparaît est d'autant plus grande que le milieu est dense et la pression p' est faible, c'est-à-dire d'autant plus grande que les déformations axiales sont faibles et que l'anisotropie des contacts reste faible.
Ceci donne donc une idée du domaine de validité de l'Eq. (7.4).

♣ *Remarque 2:* Les argiles se comportent de façon très similaire à des sables lâches de ce point de vue. Elles développent donc sûrement une anisotropie induite plus facilement.

♣ *Remarque 3:* Les courbes de la Fig. 7.1 représentent des essais à volume constant sur des échantillons initialement isotropes; ces courbes partent toutes verticalement, c'est-à-dire en gardant $p'=c^{ste}$ ; cette partie initiale de la trajectoire est incompatible avec une interprétation de la relation de Rowe basée sur une théorie plastique à un mécanisme, car celle-ci impliquerai que la réponse à q=0 ne devrait pas dépendre de la direction de l'incrément de contrainte ; elle prévoirait donc la même variation de volume pour un essai à $p'=c^{ste}$ et un essai à $\sigma'_3=c^{ste}$. Pour que la réponse à un incrément de q à dp'=0 corresponde à une variation de volume nulle, il faut que la réponse soit isotrope ; ceci est somme toute normal, pour un matériau isotrope soumis à une contrainte isotrope ; cependant, pour réaliser cette modélisation isotrope à partir d'une modélisation plastique, il faut introduire une série de mécanismes plastiques, ce qui se traduirait par une réponse qui obéirait à l'Eq. (7.4) automatiquement.

## 7.2. Application de la modélisation incrémentale isotrope à un essai oedométrique. Détermination de la constante $K_o$ des terres au repos (*i.e.* constante de Jaky)

### *7.2.1. Équation d'évolution :*

Nous allons maintenant déterminer le comportement d'un milieu granulaire dense, initialement soumis à une pression isotrope p' petite, et qui subit ensuite une compression oedométrique $\delta\varepsilon_2=\delta\varepsilon_3=0$. Par hypothèse, cette compression est réalisée en accroissant q (et $\sigma'_1$). La condition expérimentale $\delta\varepsilon_2=\delta\varepsilon_3=0$ combinée à la condition de symétrie axiale ($\delta\sigma'_2=\delta\sigma'_3$) et à la loi rhéologique définie par l'Eq. (7.4) impose que

$$0=-\nu\delta\sigma'_1+(1-\nu)\delta\sigma'_2$$

Ceci impose la relation entre les incréments de contraintes :

$$\delta\sigma'_2/\delta\sigma'_1=\delta\sigma'_3/\delta\sigma'_1=\nu/(1-\nu) \qquad (7.7)$$

L'Eq. (7.7) fixe donc l'évolution des contraintes si l'on connaît $\nu$. Le pseudo coefficient de Poisson $\nu$ dépend du rapport des contraintes (Eq. 7.6) à cause de la loi de Rowe. En remplaçant $\nu$ par sa valeur donnée par l'Eq. (7.6), l'Eq. (7.7) ne dépend plus que des contraintes et s'écrit:

$$\delta\sigma'_3/\delta\sigma'_1= (\sigma'_1/\sigma_3')\tan^2(\pi/4-\varphi/2) /(2- (\sigma'_1/\sigma'_3) \tan^2(\pi/4-\varphi/2)) \qquad (7.8)$$

Cette équation peut se réécrire en fonction des variables q,p' ; dans ce cas elle permet de déterminer complètement la projection dans le plan q,p' de la trajectoire représentant l'évolution du système. Pour déterminer totalement cette trajectoire, il faudrait déterminer l'évolution du volume en fonction de la contrainte q. Ceci peut être fait en intégrant les Eq. (7.4) si l'on connaît l'évolution de $C_o$ avec la contrainte. Il faut pour cela reprendre les résultats expérimentaux de la Fig. 4.2 ; nous ne le ferons pas .

En fait, dans le cas expérimental de l'oedomètre sous compression croissante que nous étudions ici, il existe une bijection entre la contrainte $\sigma'_1$ et la déformation axiale $\varepsilon_1$; par ailleurs, l'évolution des contraintes ne fait pas intervenir explicitement la variation de volume de l'échantillon comme le montre l'Eq. (7.7) ; on peut donc se contenter dans un premier temps de caractériser l'évolution du système dans le sous–espace {q,p'}. Nous allons montrer dans le paragraphe suivant que le rapport des contraintes tend vers une constante fixe, qui ne dépend que du frottement solide lorsque le système a subi une augmentation de charge suffisante.

### *7.2.2. l'évolution du rapport* **q/p'** *ou* **($\sigma'_3/\sigma'_1$)** *admet un point fixe stable qui est donc un attracteur*

L'Eq. (7.8) fixe l'évolution du rapport $\sigma'_3/\sigma'_1$. ***Question :*** ce rapport tend-il vers une valeur asymptotique lorsqu'on fait croître $\sigma'_1$ indéfiniment?

En fait, pour répondre à cette question, il faut trouver la valeur vers laquelle se rapport devrait tendre, puis démontrer que cette valeur est une valeur stable.
Ce rapport n'évoluera plus lorsque les contraintes et les incréments de contraintes seront telles que $\sigma'_3/\sigma'_1= \delta\sigma'_3/\delta\sigma'_1$ . Ceci revient donc à chercher les points fixes de l'Eq. (7.8). Ils sont donnés par les solutions de l'équation :

$$\delta\sigma'_3/\delta\sigma'_1=\sigma'_3/\sigma'_1$$

soit

(7.9a) $(\sigma'_1/\sigma'_3)\tan^2(\pi/4-\varphi/2) /(2- (\sigma'_1/\sigma'_3) \tan^2(\pi/4-\varphi/2))=(\sigma '_3/\sigma'_1)$

ce qui peut se réécrire en fonction de M' :

(7.9b) $2(1+M')(\sigma'_3/\sigma'_1)^2-(\sigma'_3/\sigma'_1) =1$

L'Eq. (7.9) n'a qu'une solution positive (donc physiquement acceptable) ; c'est:

(7.10a) $(\sigma'_3/\sigma'_1)_{oed} = [1+(9+8M)^{1/2}]/[4(1+M)]$

(7.10b) $(\sigma'_3/\sigma'_1)_{oed} = (1/4)[(1-\sin\varphi)/(1+\sin\varphi)][1+\{[9+7\sin\varphi]/[1-\sin\varphi]\}^{1/2}]$

où $M= (q/\sigma'_3)_{\text{état critique}}=2\sin\varphi/(1-\sin\varphi)$ et $M'=q_c/p'$. On notera $K_o$ la valeur $(\sigma'_3/\sigma'_1)_{oed}$ .
On va maintenant montrer que ce point fixe est stable, c'est-à-dire qu'il attire l'évolution du rapport des contraintes . Pour cela il faut montrer que le rapport $(\partial\sigma'_3/\partial\sigma'_1)/(\sigma'_3/\sigma'_1)$ est inférieur à 1 si le rapport $(\sigma'_3/\sigma'_1)/(\sigma'_3/\sigma'_1)_{oed}$ est plus grand que 1 , et que le rapport $(\partial\sigma'_3/\partial\sigma'_1)/(\sigma'_3/\sigma_1)$ est plus grand que 1 si $(\sigma'_3/\sigma'_1)/ (\sigma'_3/\sigma'_1)_{oed}$ est inférieur à 1. Ceci se vérifie facilement quelque soit $\sigma'_3/\sigma'_1$.
Cependant, comme l'évolution est continue, et qu'il n'y a qu'un seul point fixe, on peut même se contenter de vérifier que l'Eq. (7.8) ramène bien le rapport de contrainte vers le point fixe pour des cas particuliers seulement.

♣ *évolution du point q=0* : dans ce cas, on a $\sigma'_3/\sigma'_1 =1$ ; l'Eq. (7.8) impose donc que $(\delta\sigma'_3/\delta\sigma'_1)= \tan^2(\pi/4-\varphi/2)/(2- \tan^2(\pi/4-\varphi/2))$; ce rapport est bien inférieur à 1, et le rapport des contraintes $(\sigma'_3/\sigma'_1)$ décroît donc, se rapprochant de $K_o$.

♣ *évolution du point* q=M'p' *de dilatance nulle*: dans ce cas, on a $(\sigma'_3/\sigma'_1)<K_o=(\sigma'_3/\sigma'_1)_{oed}$, $\nu= ½$ et $\delta\sigma'_3/\delta\sigma'_1 =1$ ; donc le rapport des contraintes augmente et se rapproche de la valeur $K_o$.
On peut alors tracer qualitativement l'évolution du système dans l'espace des phases. Partons d'un état {q=0, p', v} et augmentons $\sigma'_1$ ; dans ce cas, q augmente car la pression moyenne p' évolue moins vite que $\sigma'_1$ et le rapport des contraintes $\sigma'_3/\sigma'_1$ se rapproche du rapport $K_o$ caractéristique du chargement oedométrique. La projection de la trajectoire dans le plan {q,p'} rejoint donc la projection dans le plan {q,p'} de la droite des états normalement consolidés sous compression oedométrique. Pendant ce temps, le volume se contracte relativement lentement de telle sorte que la trajectoire réelle ne rejoint la droite réelle des états normalement consolidés sous compression oedométrique que beaucoup plus tard, c'est-à-dire nettement après que les deux projections (de la trajectoire et de la droite de compression oedométrique) dans le plan {q,p'}se soient rejointes.

### *7.2.3. Confrontation prédiction théorique et expérience :*

On peut confronter cette prédiction théorique aux résultats expérimentaux. On sait en effet depuis Jaky que le rapport des contraintes $\sigma'_3/\sigma'_1$ expérimrental tend vers une valeur asymptotique bien définie notée $K_{Jaky}$ ou $K_o$ aux fortes contraintes. Cette valeur dépend de l'angle de frottement φ :

(7.11) $$K_o=K_{Jaky}= 1-\sin\varphi$$

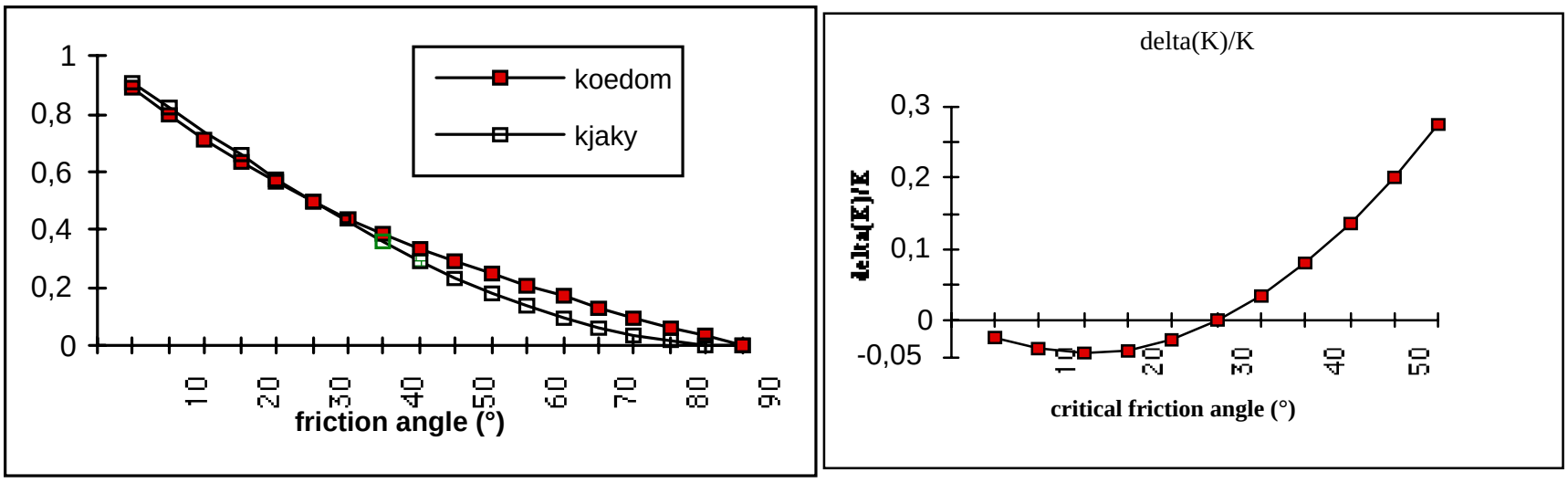

***Figure 7.2 :*** *Variation du rapport des contraintes $\sigma'_3/\sigma'_1$ dans un essai oedométrique : figure de gauche : comparaison entre les valeurs théoriques (Eq. 7.10) et expérimentales (formule de Jaky (Eq. 7.11) ; figure de droite : différence entre l'Eq. (7.10) et l'Eq. (7.11).*

Cette loi est obtenue par calage sur les résultats expérimentaux. Nous représentons dans la Fig. 7.3 la comparaison des Eqs. (7.10) et (7.11). Leur accord valide notre modèle.

### *7.2.4. effet de l'anisotropie :*

La déformation uniaxiale du matériau doit se traduire par la génération d'une anisotropie des contacts ; cette dernière est d'autant plus grande que les déformations sont grandes. Or les déformations sont d'autant plus grandes que le matériau est lâche, à q/p' donné. L'état correspondant à un essai oedométrique (rapport de contrainte $K_o$ ) est tel que q/p'<M'. Il se situe donc aux petites déformations, *i.e.* avant le minimum de volume des courbes de la Fig. 4.2. Ce minimum se situe donc très souvent à des déformations axiales inférieures à 2% dès que le sable est légèrement densifié au départ. Ceci veut dire que la modélisation proposée devrait être valable dans la quasi totalité des cas.

On peut conforter cette conclusion aux expériences: en effet, d'après nos résultats, le rapport q/p' est de l'ordre de 3/4 dans l'état oedométrique asymptotique puisque $\sigma'_3/\sigma'_1 \approx$ ½ dans cet état. Or, nous avons vu que l'isotropie de la réponse incrémentale doit se traduire par la constance de p' quand q croit dans un essai triaxial à volume constant, (c'est-à-dire dans le cas représenté dans la Fig. 7.1). L'examen des courbes de la Fig. 7.1 montre que la réponse reste isotrope au delà de q/p'≈½ , ce qui correspond approximativement au fonctionnement oedométrique (q/p'≈ 3/4 ). Ceci confirme donc que l'approche proposée est valide, même pour les échantillons relativement lâche.

Cependant, si l'anisotropie devenait grande, il faudrait utiliser une équation incrémentale de forme plus générale . Nous donnons la forme qui correspond à la symétrie du problème dans l'Eq. (7.12). Dans cette équation, le coefficient $\nu$ obéirait à la loi de Rowe, mais les variations de $\nu'$ , $\nu''$ et $\alpha$ restent à déterminer expérimentalement. En fait, on peut montrer que $\nu'=\nu$, car la matrice qui opère dans la même zone incrémentale doit être telle que le travail ne doit pas dépendre du chemin infinitésimal suivi dans une même zone; ceci impose que $\delta W= \boldsymbol{\delta\sigma}' \boldsymbol{\delta\varepsilon} = \Sigma C_{ij}\, \delta\sigma'_i\, \delta\sigma'_j$ ne doit dépendre que du $\boldsymbol{\delta\sigma}'$ réel entre état initial et état final, soit $\nu'=\nu$. Enfin, le calcul montre que l'Eq. (7.12) conduit au même rapport oedométrique que l'Eq. (7.10) si $1-\nu=\alpha-\nu''$ .

(7.12.a) $$\begin{pmatrix} \delta\varepsilon_1 \\ \delta\varepsilon_2 \\ \delta\varepsilon_3 \end{pmatrix} = -C_o \begin{pmatrix} 1 & -\nu' & -\nu' \\ -\nu & \alpha & -\nu'' \\ -\nu & -\nu'' & \alpha \end{pmatrix} \begin{pmatrix} \delta\sigma_1 \\ \delta\sigma_2 \\ \delta\sigma_3 \end{pmatrix}$$

(7.12.b) $$\nu'=\nu$$

Lorsqu'on ne dispose que d'essais axisymétriques, on ne peut différentier les actions de $\sigma'_2$ et $\sigma'_3$ , car $\sigma'_2 = \sigma'_3$ ; il en résulte qu'on peut choisir $\alpha$ arbitrairement et poser $\alpha=1$.

Enfin, lorsqu'on néglige l'anisotropie, on peut intégrer numériquement l'équation d'évolution du rapport des contraintes (Eq. 7.8). Le résultat obtenu pour un frottement $\varphi=30°$ et pour deux tenseurs de contraintes initiales différents ($\sigma'_2/\sigma'_1=1$) et ($\sigma'_2/\sigma'_1=0,3$) est reporté dans la Fig. 7.3. On constate que l'évolution du rapport $\sigma'_2/\sigma'_1$ converge bien continûment vers la valeur $K_o=½$ prévue, par valeur décroissante (croissante) quand le rapport initial $(\sigma'_2/\sigma'_1)_i$ est plus grand (petit) que $K_o$.

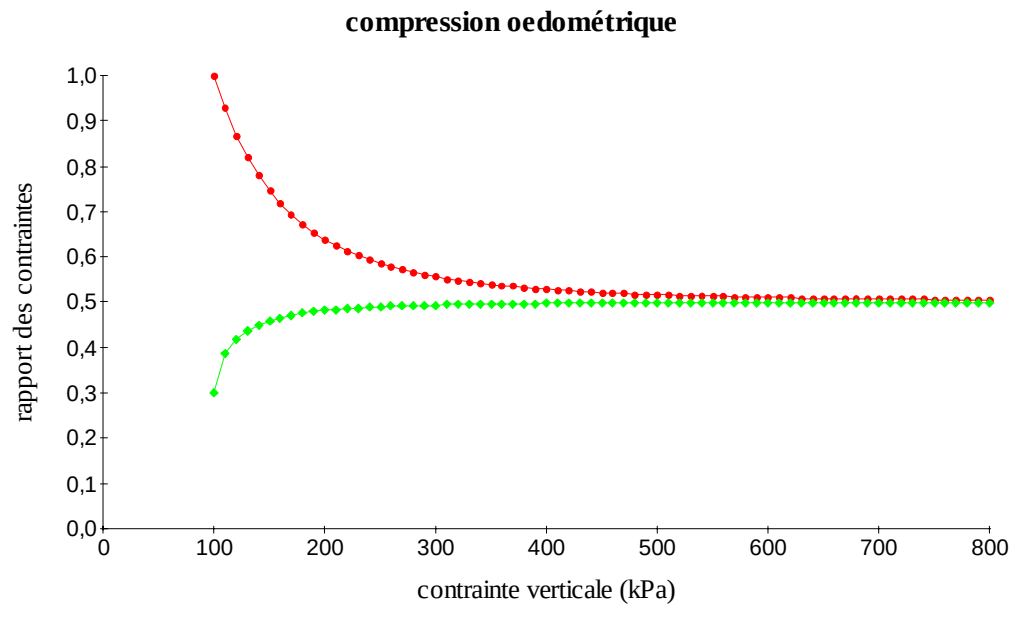

***Figure 7.3 :*** *évolution du rapport des contraintes d'un essai oedométrique obtenues par intégration numérique de l'Eq. (7.8) pour deux valeurs différentes du rapport des contraintes.*

## 7.3. Essai de compression uniaxiale à volume constant (Essai non drainé)

D'après l'Eq. (7.4), qui suppose que la réponse incémentale de l'échantillon est isotrope, la variation de volume s'écrit:

(7.13) $$\delta\varepsilon_v=\delta v/v=3(1-2\nu)\delta p'$$

L'Eq. (7.13) permet ainsi de conclure que pour qu'un échantillon dont la réponse incrémentale est isotope subisse une déformation à volume constant, il faut que la transformation vérifie :

(7.14) $$(1-2\nu)\delta p'=0$$

Ceci implique soit que la déformation se fait à pression moyenne constante, soit que le pseudo coefficient de Poisson ν est égal à ½ .
Dans le cas d'un sable, la relation de Rowe permet de montrer que pour que ν soit égal à ½, il faut que le rapport des contraintes q/p' soit : q/p'=M' , c'est-à-dire que l'échantillon soit dans un de ces états caractéristiques.
Au début de la compression à volume constant, q≈0 et ν est donc différent de ½ ; l'Eq. (7.14) impose donc que la compression se fait à pression p' constante. La trajectoire dans le plan (q,p') doit donc être verticale. C'est effectivement ce qu'on observe dans la Fig.7.1. Cette condition ($p'=c^{ste}$) reste valable tant que le système ne développe pas une réponse anisotrope ; pour cela il faut que les déformations restent faibles, c'est-à-dire que le milieu soit suffisamment dense au départ et soumis à une pression p' suffisamment faible (*cf.* discussion de la section précédente).
Supposons que cette condition d'isotropie de la réponse incrémentale soit satisfaite tout au long de l'essai. Le milieu évolue et son point de fonctionnement décrit la droite $p'=c^{ste}$ ; cette trajectoire croise la droite q=p'M' après un certain temps. Arrivé à ce point, le système est libre d'évoluer dans n'importe quelle direction puisque la condition (7.14) est toujours satisfaite (on rappelle que ν=½ pour q/p'=M').
Considérons alors que la trajectoire prenne une direction quelconque définie par {δq,δp'}. Dès que l'incrément (δq,δp') devient suffisamment grand et que δq/δp'≠M', le pseudo coefficient de Poisson ν devient différent de ½ et la condition δv=0 réimpose δp'=0 . Au contraire si δq/δp'=M' la condition (7.14) reste toujours valable et le système peut poursuivre son évolution dans cette direction. A l'intersection, il y a donc deux trajectoires possibles : celle à δp'=0 et celle à δq/δp'=M'; La trajectoire qui sera effectivement choisie est celle qui demande le moins d'énergie.

### *7.3.1. Les états caractéristiques sont des points de bifurcation des trajectoires à volume constant :*

On peut donc conclure que lorsque la trajectoire atteint la droite q=M'p', elle atteint un point de bifurcation ; à cet endroit elle peut soit poursuivre dans la direction $p'=c^{ste}$, soit partir dans la direction q=M'p'. Le choix doit se faire du point de vue énergétique : la trajectoire effectivement choisie doit être celle qui dissipe le moins d'énergie lors de l'accroissement de contrainte. L'énergie dissipée s'écrit :

(7.15) $\delta W = \delta\sigma'_1\delta\varepsilon_1 + \delta\sigma'_2\delta\varepsilon_2 + \delta\sigma'_3\delta\varepsilon_3 = \delta\sigma'_1\delta\varepsilon_1 + 2\delta\sigma'_3\delta\varepsilon_3$

Ceci se réécrit compte tenu de l'Eq. (7.4)

(7.16) $\delta W = (\delta\sigma'_1)^2 + 2(1-\nu)\,(\delta\sigma'_2)^2 - 4\nu\,\delta\sigma'_1\delta\sigma'_2$

La déformation à pression constante impose $\delta\sigma'_1+\delta\sigma'_2+\delta\sigma'_3=0$ , soit:

(7.17a) $\delta\sigma'_1=-2\delta\sigma'_2$

celle gardant le système dans l'état caractéristique impose $\delta q=M'\,\delta p'=M\delta\sigma'_3$:

(7.17b) $\delta\sigma'_1=(1+M)\,\delta\sigma'_2$ avec $M\approx 2$

A $\delta\sigma'_1$ donné, les deux incréments $\delta\sigma'_2$ ont donc approximativement les mêmes valeurs absolues, mais des signes contraires. L'Eq. (7.16) permet donc de conclure que la déformation à volume constant dissipe plus que la déformation gardant l'échantillon dans un état caractéristique, *i.e.* q=M'p'. C'est donc cette dernière qui devrait être spontanément sélectionnée.

Prenant M=2 et ν=½ , on trouve que la différence de dissipation est :

(7.18) $\delta W_{\,p=cste} - dW_{\text{état caractéristique}} \cong (5/18)(1+13\nu)(\delta\sigma'_1)^2 \cong 2(\delta\sigma'_1)^2$

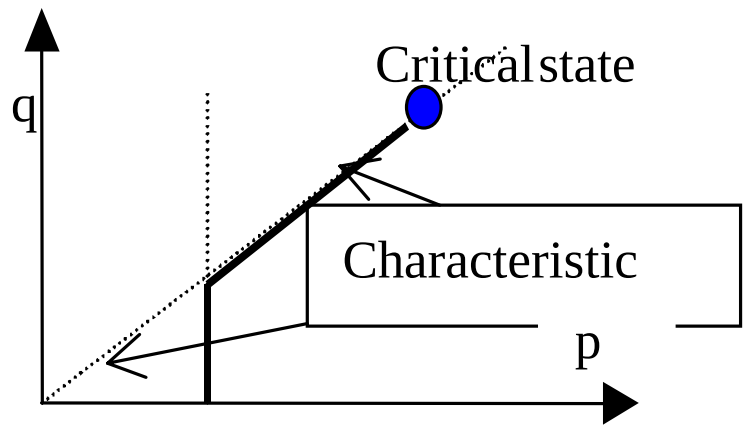

***Figure 7.4.a :*** *prévision du comportement oedométrique à partir d'une loi incrémentale isotrope*

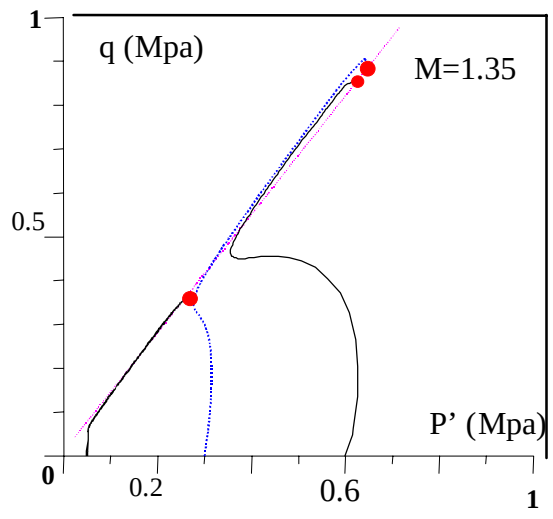

***Figure 7.4.b :*** *Essais expérimentaux à volume constant*

### *7.3.2. cette bifurcation est transcritique*

La bifurcation s'opérant à q=M'p' dans un essai à volume constant (*i.e.* non drainé) est donc de type transcritique compte tenu de ses caractéristiques; en effet il y a interversion des trajectoires stables. On prévoit donc qu'un essai à volume constant typique sur un sable dense se comporte comme dans la Fig. 7.4.a.

### *7.3.3. Évolution d'un échantillon dense:*

Si l'on suppose que la réponse de l'échantillon reste isotrope et qu'on néglige les effets liés à l'apparition de l'anisotropie, cette approche prévoit donc que le volume reste constant et que la trajectoire est contenu dans un plan $v=c^{ste}$ ; dans ce plan $\{q,p', v=c^{ste}\}$ la trajectoire commence par monter verticalement ($p'=c^{ste}$) puis tourne à droite et suit une ligne d'états caractéristiques. Cette trajectoire se termine au point critique, car il n'y a plus d'état stable après celui-là. On prévoit donc qu'un essai à volume constant typique se comporte comme dans la Fig. 7.4.a. La Fig. 7.4.b reporte trois exemples de comportements expérimentaux; seuls les cas les moins denses (pour une pression donnée) obliquent vers la gauche montrant qu'une anisotropie s'est établie.

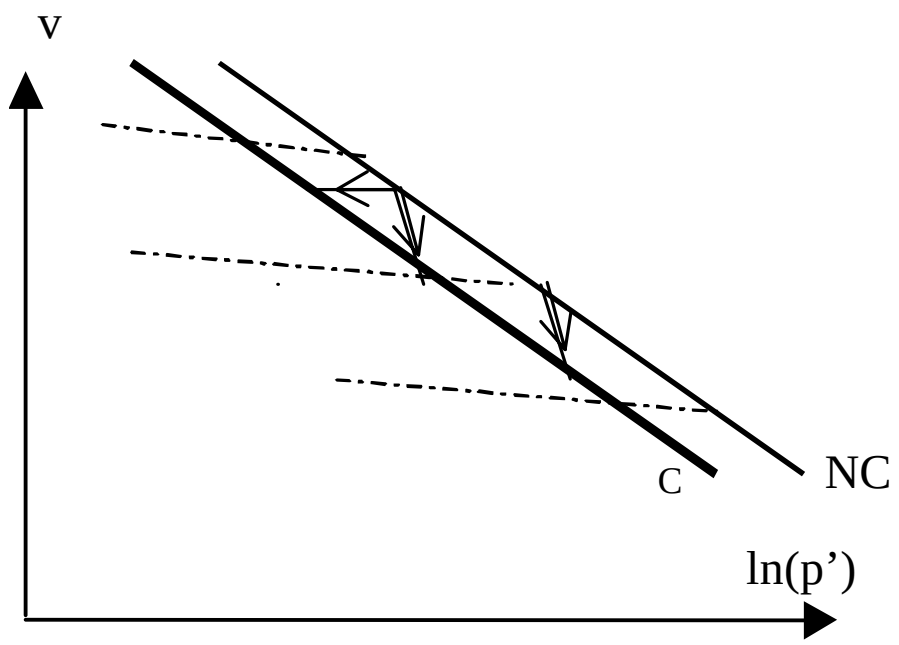

***Figure 7.5:*** *on a reporté sur cette figure le volume spécifique* v *des états normalements consolidés sous contrainte isotrope* p' *(i.e.* q=0*) et sous contrainte critique (i.e.* p', q=Mp', *états critiques). Ceci démontre que les états normalement consolidés à une pression* p' *sont toujours moins denses que les états critiques, cf. Fig. 4.1.b.*

### *7.3.4. Evolution d'un échantillon lâche : La trajectoire peut-elle tourner sur la gauche quand elle atteint la droite des états caractéristiques?*

La Fig. 7.5 reporte la variation du volume spécifique d'un échantillon normalement consolidé sous pression isotrope (p',q=0) et sous pression critique (p',q=M'p'). Ainsi, la densité critique est donc toujours supérieure à la densité normalement consolidé, à pression p' donné. Par ailleurs, la Fig. 7.5 indique aussi que la densité de ces états (normalement consolidés ou critiques) croît lorsque la pression augmente. Compte tenu de ces faits expérimentaux, la pression p' doit diminuer dans un essai à volume constant (*i.e*. non drainé) sur des milieux très lâches (*i.e*. normalement consolidés).

Appliqué au modèle incrémental isotrope, l'ensemble de ces faits impose le schéma suivant (*cf.* Fig. 7.6.a) : la trajectoire doit bifurquer soit à droite soit à gauche lorsqu'elle arrive à la droite des états caractéristiques, suivant que le tas est respectivement dense ou lâche. La trajectoire tourne à droite si le volume spécifique initial est inférieur au volume critique $v_c$(p') final; elle tourne à gauche dans le cas contraire.

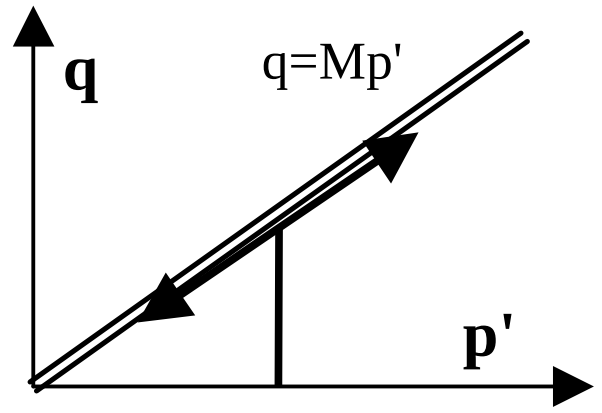

***Figure 7.6.a:*** *Trajectoire prévue dans le plan* (q,p') *par la modélisation incrémentale isotrope pour des essais non drainés. La trajectoire bifurque à droite si* $v_{initial}<v_c(p'_{initiale})$ *; elle bifurque à gauche dans le cas contraire.*

Cependant, lorsque la trajectoire doit bifurquer à gauche, l'évolution du matériau devient instable. En effet, l'Eq. (4.6) de stabilité $\delta^2W= \boldsymbol{\sigma}' \boldsymbol{\varepsilon}$ devient négatif dans le cas présent d'une réponse isotrope axi-symétrique ($\varepsilon_2=\varepsilon_3$ , $\sigma'_2=\sigma'_3$) et d'un essai non drainé ($\varepsilon_v= \varepsilon_1+ \varepsilon_2 + \varepsilon_3=0$) :

$$\delta^2W=2\ \delta q\ (\varepsilon_1-\varepsilon_2)/3 + \delta p'\ \varepsilon_v = (4/3)\ \varepsilon_1\ \delta q \qquad (7.19)$$

Le travail du second ordre $\delta^2W$ devient donc négatif dès que la trajectoire oblique à gauche. Cette partie de trajectoire, qui ne pourrait être observée que pour des matériaux très lâches est en fait instable. Avant d'arriver au point de bifurcation, le matériau peut évoluer spontanément, par exemple en devenant inhomogène, *i.e.* localisation de la déformation. C'est pourquoi les trajectoires des matériaux lâches obliques vers la gauche bien avant d'arriver au point de bifurcation théorique.
Pour exprimer ce fait dans le cadre de la théorie des sytèmes dynamiques on peut dire que le point d'intersection entre la droite q=M'p' et la droite p'=c$^{ste}$ est un point répulseur dans la direction de la droite q=M'p' si le matériau est lâche. Dans ces conditions, il est naturel que la trajectoire oblique à gauche avant ce point.
En fait il existe une autre explication à ce virement vers la gauche que nous discuterons dans le § 7.3.6 suivant ; c'est le développement d'une anisotropie induite par la déformation.

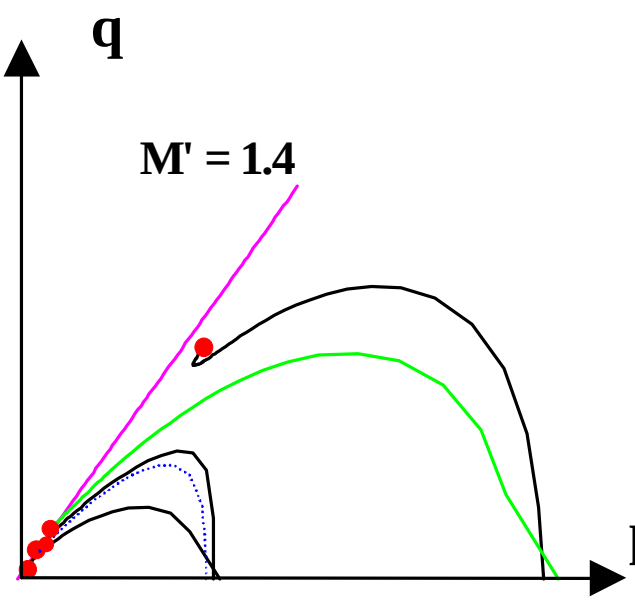

***Figure 7.6.b:*** *Essais expérimentaux non drainés sur du sable d'Hostun lâche, (d'aprèsFlavigny et Megachou)*

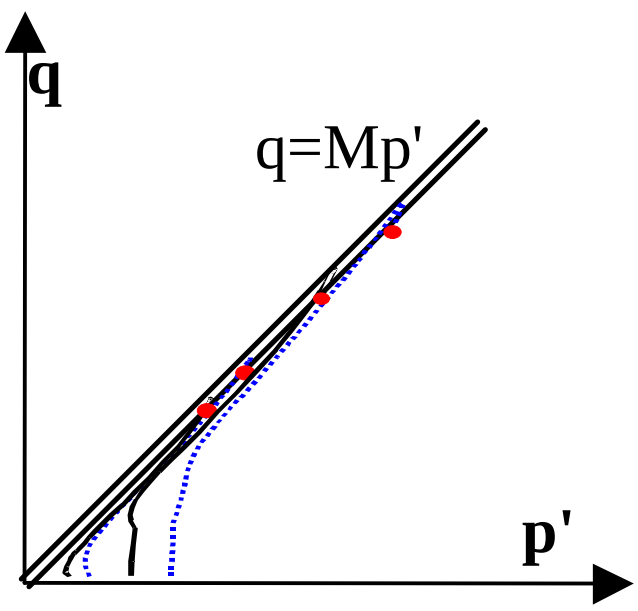

***Figure 7.6.c:*** *Essais expérimentaux non drainés sur du sable d'Hostun lâche, (d'après Mokham)*

Le modèle isotrope prévoit donc un infléchissement vers la gauche dans le plan (q,p') des trajectoires des essais à volume constant pour les matériaux lâches, puis une arrivée au point critique. Dans le cas des matériaux denses, la trajectoire doit passer par le point de bifurcation puis emprunter la droite caractéristique vers les déviateurs croissants. Nous reportons dans la Fig. 7.6.b des résultats expérimentaux qui montrent que ces phénomènes sont bien observés.
Cette corrélation est d'ailleurs acceptée en mécanique des sols (*cf.* Fig. 7.6.d); elle est assez bien décrite dans le livre de Biarez & Hicher (1994). L'interprétation qu'on donne ici de ces faits expérimentaux a donc l'avantage d'être simple est en concordance avec les faits expérimentaux. Il existe bien sûr des cas expérimentaux pathologiques, et le comportement réel peut être plus compliqué ; mais avant de décrire ces effets dans le § 7.3.6, nous discuterons brièvement des problèmes liés au développement de l'anisotropie.

### *7.3.5. Rôle de l'anisotropie :*

Considérons la classe des échantillons très lâches à une pression p' donnée. On est d'amblé face à un dilemme si l'on considère que le matériau et sa réponse reste isotrope : en effet, dans ces conditions la variation des volume de l'échantillon doit être nulle pour une transformation où l'on maintient la pression p' constante. Dans ce cas, cela voudrait dire que les états critiques et les états normalement consolidés qui correspondant à la même pression p' doivent avoir le même volume spécifique. La Fig. 7.5 démontre que ceci est faux. Cela implique donc que le matériau ne reste pas isotrope près de l'état critique et qu'il doit développer de l'anisotropie induite.
Par contre, on a vu que lorsque la matériau est suffisamment dense, et qu'il est soumis à un essai non drainé, il peut atteindre la droite q=M'p' des états caractéristiques sans s'être trop déformé et en restant donc isotrope ; il développera donc ensuite son anisotropie pour arriver à l'état critique. Un point remarquable est que la droite caractéristique q=M'p' reste indépendante de l'anisotropie induite, puisque l'échantillon poursuivra sa trajectoire sur cette droite jusqu'à ce qu'il atteigne l'état critique.

***Question : comment intégrer ces faits ?***
L'Eq. (7.12) avec $\nu'=\nu$ permet de décrire les essais développant de l'anisotropie ; la condition $\delta\varepsilon_v=0$ devient :

(7.20) $$0= \delta\varepsilon_v=2\ \delta q\ (1-\alpha-\nu+\nu'')/3 + dp'\ (1+2\alpha-4\nu-2\nu'')$$

On a alors les différentes réponses possibles suivantes :

♣ Lorsque la réponse matériau est isotrope, le coefficient devant $\delta q$ est nul, car $\alpha=1$ et $\nu=\nu''$ ; ceci impose donc que la trajectoire est telle que $\delta p'=0$ , soit: $p'=c^{ste}$.

♣ Cette trajectoire reste aussi la solution si la réponse du matériau devient anisotrope, tant que la matrice est telle que $(1-\alpha-\nu+\nu'')=0$ .

♣ En revanche, la trajectoire passe par une tangente horizontale, parallèle à l'axe p', dans le plan (p',q) si la condition $(1+2\alpha-4\nu-2\nu'')=0$ est réalisée.

♣ Enfin, dans le cas général, la trajectoire est donnée par :

(7.21.a) **trajectoire :** $$q/p'= \delta q/\delta p'=3(1+2\alpha-4\nu-2\nu'')/ [2(1-\alpha-\nu+\nu'')]$$

♣ ***Etats caractéristiques :*** Aux grandes déformations, le rapport q/p' pourrait tendre vers une limite donnée par la résolution de l'Eq. (7.20) et telle que

(7.21.b) **trajectoire :** $$q/p'= \delta q/\delta p'=3(1+2\alpha-4\nu-2\nu'')/ [2(1-\alpha-\nu+\nu'')]$$

Cependant, ceci ***ne peut pas expliquer l'existence de la bifurcation*** de trajectoire observée lors des essais à $v=c^{ste}$ des échantillons très denses. A cause de ceci, on doit conclure que la fin de la trajectoire est caractérisée par q/p'=M', et que ce rapport impose une variation nulle du volume de l'échantillon. Ceci impose donc que l'échantillon dans l'état q/p'=M' est dans un état carctéristique. Cet état est caractérisé par $(1-\alpha-\nu+\nu'')=0$ et par $(1+2\alpha-4\nu-2\nu'')=0$. Ces deux conditions imposent $\nu=½$ et $\alpha-\nu''=½$

*Ceci est donc la généralisation de la notion d'état caractéristique.*

(7.22) **Etats caractéristiques q/p'=M' :** $\nu=½$ et $\alpha-\nu''=½$

Cette définition est compatible avec la loi de Rowe.

On voit ainsi que l'on peut et que l'on doit généraliser la notion d' «état caractéristique ». Cette généralisation permet de décrire le comportement non drainé des états denses, même anisotropes; elle reste aussi valable pour les matériaux peu denses et permet de décrire les essais non drainés aussi dans ce cas.

♣ ***Remarque :*** On devrait réviser notre position, si le comportement non drainé longeait une droite de pente différente de q/p'=M' et/ou si cette droite ne passait par l'origine, comme certains auteurs semblent le croire, *cf.* § 3.4.2. et Fig. 3.8.

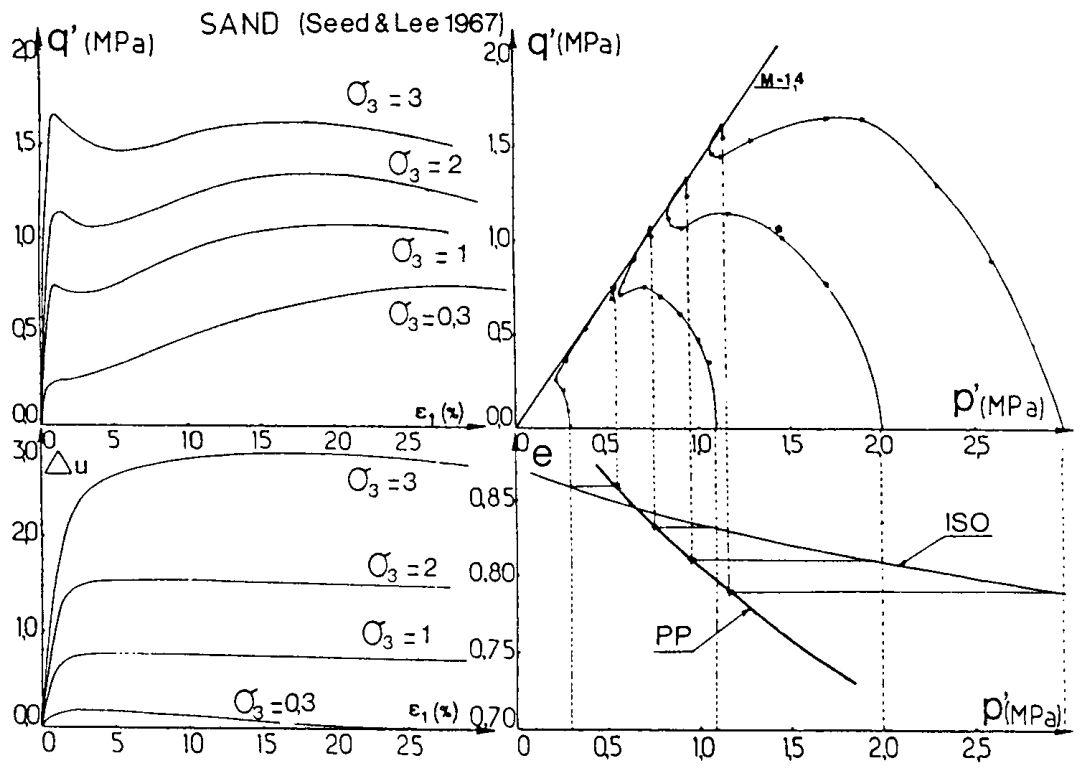

**Figure 7.6.d:** *Résumé des comportements des milieux granulaires sous essais triaxiaux drainés; d'après Biarez & Hicher, Fig. 5.34*

## *7.3.6. Cas pathologiques :*

Les différents types de trajectoire que nous venons de décrire sont relativement souvent observés expérimentalement (*cf.* Figures de Saim (1997), de Biarez & Hicher (1994), Fig. 7.4,…). Cependant, on observe de temps en temps des trajectoires qui dévient par rapport à cette « norme ». Certaines, qui correspondent à des échantillons de densité intermédiaire, coupent la droite q=M'p' puis s'infléchissent vers la droite pour suivre enfin la droite des états caractéristiques. Ceci peut s'expliquer de deux manières différentes: (i) par le fait que l'expérimentateur a peut-être bridé son expérience pour l'obliger à suivre la trajectoire $p'=c^{ste}$, (ii) par l'existence d'une distribution hétérogène, cette hétérogénéité pouvant être présente dès le départ ou pouvant s'être développée spontanément au cours de l'expérience.

Dans d'autres cas, pour des échantillons relativement lâches et surtout dans le cas des argiles, la trajectoire monte verticalement comme il est normal, puis s'incurve vers la gauche ($\delta p'<0$) avant d'arriver à la droite des états caractéristiques. A ce stade, on observe alors très souvent la bifurcation transcritique précédemment décrite, mais on peut observer aussi la pathologie décrite au paragraphe précédent. Dans ce cas, l'incurvation à gauche de la trajectoire résulte probablement de la création d'une réponse anisotrope induite par la compression, ou de la génération d'une inhomogénéité. Des études complémentaires sont nécessaires.

## 7.4 Généralisation à des essais à dilatance imposée

Le modèle précédent se généralise à tout essai pour lequel la variation de volume en fonction de $\varepsilon_1$ est imposé. En effet, l'Eq. (7.4) combinée à la valeur de la dilatance K permet de calculer l'incrément de contrainte pour un pseudo-coefficient de Poisson. Ce dernier varie avec l'état de contrainte imposé au système. Un calcul similaire à celui développé dans la section 7.3 précédente permet donc de calculer l'évolution des contraintes et le rapport asymptotique des contraintes obtenues après forte compression.

Il faut cependant noter que toutes les variations de volume $K=-(\partial\varepsilon_v/\partial\varepsilon_1)$ ne sont pas admissibles de telle sorte que le problème ne doit admettre de solution que pour une gamme raisonnable de dilatance K; pour les autres valeurs de K, le système doit développer des instabilités mécaniques.

Lorsque le système admet une solution, son évolution sera similaire à celle trouvée dans la section 7.2 pour l'essai oedométrique ($K_{oed}=-1$): i) ainsi, il existera un rapport de contrainte asymptotique $(\sigma_3/\sigma_1)_K$ obtenu aux fortes contraintes ; ii) cet état semblera être une valeur attractrice vers laquelle tout système évolue spontanément. On n'observera pas en général de bifurcation, comme celle que nous venons de décrire pour le cas des essais non drainés. En effet, cette évolution est pathologique.

En effet, si K est fixé, on a $\delta\varepsilon_2=(K+1)/2\ \delta\varepsilon_1=X\ \delta\varepsilon_1$ avec $X=-(K+1)/2$. En utilisant l'Eq. (7.4), la condition $\delta\varepsilon_2=X\ \delta\varepsilon_1$ impose que l'une des deux relations suivantes soient vérifiées: soit $C_o=0$ ,soit $(1-\nu+2X\nu)\delta\sigma_2=(X+\nu)\delta\sigma_1$ . La première solution, *i.e.* $C_o=0$, n'est obtenue que lorsque les courbes q/p' vs. $\varepsilon_1$ passe par leur maximum, ou lorsque le milieu est dans son état critique; ce ne peut donc être qu'une solution transitoire. La deuxième solution, *i.e.* $(1-\nu+2X\nu)\delta\sigma_2=(X+\nu)\delta\sigma_1$ règle donc l'évolution des contraintes, comme dans le cas oedométrique. Comme $\nu$ dépend de $\sigma'_1/\sigma'_2$ d'après l'Eq. (7.6), l'équation différentielle $(1-\nu+2X\nu)\delta\sigma_2=(X+\nu)\delta\sigma_1$ peut être intégrée, au moins numériquement, et l'évolution des contraintes est totalement déterminée. Cette évolution doit tendre vers une limite; celle-ci s'obtient en imposant: $\delta\sigma_2/\delta\sigma_1=\sigma_2/\sigma_1$ .

Cependant, il existe un cas particulier, lorsque les deux coefficients $(1-\nu+2X\nu)$ et $(X+\nu)$ sont nuls en même temps. Dans ce cas, l'incrément de contraintes $\delta\sigma_1$ et $\delta\sigma_2$ peuvent être n'importe quoi. Le système peut alors évoluer en gardant ces deux coefficients nuls. Cette possibilité correspond en fait à une évolution pathologique rare; en effet, ceci ne peut arriver que si X=ν=½, c'est à dire pour le cas non drainé.

## 7.5. Surface(s) de Hvorslev et de Roscoe

### *7.5.1. Rappel de la définition :*

Les états critiques sont définis par les Eqs. (4.1) pour lesquelles η=1, soit {q=M'p' , $v=v_c=v_{co}-\lambda\ln(p'/p'_o)$}, les états normalement consolidés sous pression isotrope p' sont définis par les Eqs. (4.1) pour lesquelles η=0, soit {q=M'p' , $v=v_c=v_{co}-\lambda\ln(p'/p'_o)$}. Considérons des états très fortement consolidés ou au contraire normalement ou faiblement consolidés. On a vu qu'ils n'ont pas le même comportement au triaxial : les uns doivent se dilater pour arriver au point critique, les autres non. C'est pourquoi on considère qu'ils forment deux classes différentes, avec des comportements différents qu'on caractérise grâce à leur trajectoire dans l'espace des phases (q,p',v). Nous avons vu aux chapitres 3 et 5 que c'est ce qu'ont proposé Roscoe et Hvorslev les premiers.
Si on considère un état critique donné et l'ensemble des essais triaxiaux de compression uniforme, l'ensemble des trajectoires qui arrivent à cet état critique et qui partent des états normalement consolidés forment une surface de l'espace (q, p', v) appelée surface de Roscoe .
De la même façon, Hvorslev a montré que l'ensemble des trajectoires qui partent de tous les points surconsolidés et qui arrivent à un même point critique convergent vers ce point en décrivant une surface de l'espace (q,p',v). En d'autres termes, si l'on considère deux états surconsolidés différents soumis au même test triaxial, la trajectoire de l'espace (q,p',v) correspondant à chacun de ces deux tests seront différentes dans un premier temps, puis se rejoindront pour converger vers l'état critique. On considère que la partie commune de l'évolution commence juste après le maximum de q/p'.
Ces deux surfaces de Hvorslev et de Roscoe sont représentées en général comme étant distinctes, c'est-à-dire comme faisant un angle entre elles (*cf.* Figs. 11.13-11.16 & 13.4 de Atkinson & Bransby (1977)) . Nous avons vu cependant qu'un argument de la théorie des systèmes dynamiques permettait de prévoir que ces deux surfaces appartenaient à la même surface (*cf.* 5.2.2). Nous voulons confirmer ce point de vue, et montrer que le modèle incrémental proposé dans ce chapitre 7 est compatible avec les analyses précédentes.

### *7.5.2. Application du modèle incrémental aux surfaces de Hvorslev et de Roscoe :*

La modélisation que nous avons choisie est basée sur une réponse incrémentale, linéaire par zone. Elle fait correspondre à un incrément des paramètres de contrôle (par exemple δq,δp') une réponse incrémentale ($\delta\varepsilon_1$, $\delta\varepsilon_v$). Le cas qui nous intéresse est axi-symétrique ; dans ce cas, cette modélisation impose que l'espace des paramètres de contrôle sur lesquels on agit est bidimensionnel; ceci implique que la réponse qu'elle produit est elle aussi bidimensionnelle ; cependant, les deux incréments de contrôle peuvent être n'importe quelle paire d'incréments prise parmi les 4 incréments possibles ($\delta q,\delta p',\delta\varepsilon_1,\delta\varepsilon_v$) (soit par exemple ($\delta q,\delta p'$), ($\delta q,\delta\varepsilon_v$), …) et la réponse correspondant aux 2 incréments restants (soit par exemple ($\delta\varepsilon_1,\delta\varepsilon_v$), ($\delta p',\delta\varepsilon_1$),… respectivement). Il y a donc $C_4^2$=6 paires d'incréments possibles.
Pour montrer la compatibilité entre l'approche des systèmes dynamiques et celle de l'hypo-élasticité, nous considérons un point de l'espace (v,q,p') ; ce point caractérise l'état du système granulaire, dont l'évolution est fixée par la matrice hypo-élastique. Compte tenu des propriétés de cette matrice, qui fait correspondre les incréments de déformation aux incréments de contrainte et réciproquement, l'ensemble des points qui peuvent être atteints à partir d'un état forment donc un espace bidimensionnel ; cet ensemble est défini par les coordonnées des incréments de l'une des 6 paires possibles. C'est donc une surface.
On déduit ainsi que l'ensemble des trajectoires qui partent ou arrivent d'un état donné doivent appartenir à (et définir) une surface. Cependant, compte tenu des règles de l'hypo-élasticité par zones, cette surface est régulière par zones et doit présenter des discontinuités d'inclinaison aux jonctions entre zones . Par contre, si l'on se limite à l'étude de la topologie des trajectoires restant dans la même zone, la surface constituée par l'ensemble de ces trajectoires doit être régulière. C'est pourquoi la surface de Roscoe est régulière et celle de Hvorslev l'est aussi.

♣ *Autre surface : celle qui part d'un état normalement consolidé :* A partir de la modélisation incrémentale, on déduit donc que l'ensemble des trajectoires partant d'un état normalement consolidé et obéissant à une loi de chargement simple forme une surface 2d. Cette surface contient la courbe des états normalement consolidés puisque ceux-ci sont obtenus par une compression isotrope à partir d'un état initial isotrope et très lâche. La trajectoire partant de ces états normalement consolidés aboutissent finalement à un état critique dont la densité dépend du mode de chargement choisi ($\sigma'_3=c^{ste}$, $p'=c^{ste}$,…). La droite des états critiques appartient donc à cette surface. Ainsi, l'existence de la surface de Roscoe est une conséquence de la modélisation incrémentale et de l'existence des états critiques et caractéristiques.
A ce stade il est utile de faire quelques remarques :
• Remarque 1 : la modélisation incrémentale prévoirait que le volume d'un échantillon normalement consolidé isotrope soumis à un incrément de contrainte isotrope ( $\delta q=0$ et $\delta p'\neq 0$) devrait varier linéairement avec $\delta p'$. En fait on constate qu'il suit une loi logarithmique. Ceci tend à indiquer que le pseudo module d' Young varie linéairement avec p' ; en effet les états normalement consolidés obéissent à $v=v_{co}-\lambda\ln(p')$, ce qui donne en différentiant $dv= -\lambda dp'/p' = -v/E\ dp'$ ; soit $E\approx v_{co}\ p'/\lambda$. Le module d'Young ne devrait donc pas varier comme $p'^{1/2}$ comme les autres résultats expérimentaux semblent l'indiquer (*cf.* Eq. 4.3). Ce point correspond à une inconsistance du modèle; cette inconsistance existe aussi pour les modèles cam-clay,….
• Remarque 2 : si l'on soumet un échantillon normalement consolidé isotrope à un incrément ($\delta q\neq 0$ , $\delta p'=0$), il ne doit pas varier de volume. C'est ce qu'on observe expérimentalement . Par contre, si on lui applique un incrément ($\delta q\neq 0$ , $\delta\sigma'_2= \delta\sigma'_3=0$), il doit varier de volume; c'est ce qu'on observe ; mais la projection dans le plan (v,p') de la trajectoire doit suivre celle de l'état normalement consolidé sous compression isotrope puisque $\delta q$ ne doit pas provoquer un incrément de volume. Ceci est effectivement observé expérimentalement, tant que le rapport q/p' reste faible (*cf.* Figs. 4.2, 4.5). Voir aussi § 7.3.5. pour une discussion plus complète.

• Remarque 3 : Si l'on cherche à préserver la forme de la dépendance du volume spécifique avec la pression, qui est logarithmique, ($v_{nc/c}(\eta)= v_{nc/c}(p_o)- \lambda \ln(p'/p'_o)$) et qu'on cherche à généraliser cette dépendance à des valeurs de q différents de 0, tout en gardant la forme logarithmique précédente, on tombe sur l'Eq. (4.4) car le développement limité autour de $q/p'= \eta=0$ de cette fonction ne doit faire intervenir que des puissance paires de $\eta$.

$$v_{\text{normalement consolidé anisotrope}}(\eta)= v_c(p_o)- \lambda \ln(p'/p'_o) - \lambda_d \ln(1+\eta^2/M'^2) \qquad \text{avec } \eta=q/p'$$

• Remarque 4 : On ne peut donc pas densifier un échantillon en le soumettant à un essai triaxial monotone à pression constante, sauf en lui imposant de grands déviateurs de contrainte. On sait par contre le densifier à l'aide de cycles de contrainte.

♣ *Surface de Hvorslev :* notre modèle incrémental, linéaire par zones, prévoit une évolution irréversible du matériau; cependant une de ces conséquences est de rendre possible la détermination de l'ensemble des états de départ qui peuvent conduire à un état donné, si l'état considéré est caractérisé par une seule loi incrémentale quel que soit l'état initial précédent; cet ensemble est alors obtenu en inversant le sens des trajectoires par la pensée . Tel semble être le cas d'un état critique, puisque son pseudo coefficient de Poisson est ½ . Ainsi, on peut conclure que l'ensemble local des états voisins d'un état critique qui peuvent aboutir à cet état critique est une surface de dimension 2.

♣ *Convergence des trajectoires vers une surface passant par l'état critique* : En fait la situation est plus compliquée qu'il n'y paraît : les expériences montrent que l'arrivée au point critique se fait dans un espace à 2 dimensions. Dans la modélisation hypo-élastique, cet espace est caractérisé par un pseudo coefficient de Poisson $\nu=1/2$ et un pseudo module d'Young E égal à 0 ; ces deux coefficients dépendent de (q/p'-M'). Si l'un ou l'autre de ces coefficients convergeait vers leur valeur asymptotique plus vite que l'autre, l'arrivée au point critique devrait se faire par une seule trajectoire ; cette trajectoire serait horizontale ($v=c^{ste}$) si $\nu$ tend vers ½ plus vite que E tend vers 0 ; elle serait verticale dans l'autre cas.

Pour démontrer ce point simplement, supposons que la loi incrémentale est isotrope ; dans ce cas elle a la forme de l'Eq. (7.4). On obtient :

$$\delta\varepsilon_v=(1-2\nu)\delta p'/E \propto (q/p'-M')^{a-b}\ \delta p \qquad (7.23.a)$$

où l'on a posé que E et $1-2\nu$ variaient respectivement comme:

(7.23.b) $(1\text{-}2\nu) \propto (q/p'\text{-}M')^{a}$

(7.23.c) $E \propto (q/p'\text{-}M')^{b}$

et où le signe $\propto$ signifie « proportionnel à ».
L'équation de Rowe implique que a=1 ; le fait que l'espace d'arrivée au point critique est bidimensionnel implique donc que a-b=0 et donc que b=1 aussi.

• *Remarque 1:* Pour se convaincre que l'espace d'arrivée au point critique est bidimensionnel, il est utile de rappeler qu'il existe bon nombre d'essais triaxiaux à pression constante, à volume constant, à $\sigma'_3=\sigma'_2=c^{ste}$,…, qui convergent systématiquement vers l'état critique et de manière différentes, *cf.* essais à $v=c^{ste}$ et essais à $\sigma'_3=c^{ste}$.
• *Remarque 2:* La relation de Rowe montre que la même loi (7.23.b), avec le même coefficient de proportionnalité, est valable en approchant de l'état critique à partir des états normalement consolidés et des états surconsolidés. Ainsi, le fait que le même type d'essai arrive toujours avec la même direction mais en sens opposé lorsqu'on part des états normalement consolidés ou des états surconsolidés montrent que la même loi (7.23.c), avec le même coefficient de proportionnalité, est valable en approchant de l'état critique à partir des états normalement consolidés et des états surconsolidés.

***En conclusion*** : Cette modélisation incrémentale, linéaire par zone, prévoit que l'ensemble des trajectoires aboutissant à un état critique est au plus une surface de dimension 2, si le fonctionnement mécanique des états caractéristiques est bien défini et si le rapport $(1\text{-}2\nu)/E$ du pseudo coefficient de Poisson $\nu$ et du pseudo module d'Young E tend vers une limite ; sinon , *i.e.* si le rapport $(1\text{-}2\nu)/E$ tend soit vers 0 soit vers l'infini, cet ensemble se réduirait à une courbe unique; ceci n'est pas observé. En conséquence, la modélisation que nous avons proposée est totalement compatible avec l'existence des surfaces de Hvorslev et de Roscoe et les exposants a et b ci-dessus sont a=b=1.
Enfin, on sait d'un point de vue expérimental que le fonctionnement mécanique de l'état critique est indépendant des conditions initiales (et en particulier de la densité initiale). Ceci impose donc que la matrice pseudo élastique est la même en ce point quelle que soit la densité initiale ; ceci impose donc que l'arrivée des trajectoires correspondant à un même test triaxial ($p'=c^{ste}$ ou $v=c^{ste}$ ou $\sigma'_3=c^{ste}$,…) a la même direction quelle que soit la densité initiale du tas, qu'elle soit lâche ou dense. L'arrivée a lieu par contre en sens inverse l'un de l'autre dans ces deux cas. Il en résulte donc que la transition entre les surfaces de Hvorslev et de Roscoe est régulière et que ces deux demi surfaces forment une seule et même surface.
Dans cette modélisation, la topologie des trajectoires est relativement simple ; elle n'est bien définie que pour des essais de compression uniforme, car il faut que le système évolue dans la même zone incrémentale; on suppose que le pseudo coefficient de Poisson du matériau ne dépend que des contraintes q et p'.
Dans la réalité, la modélisation complète doit peut-être être plus compliquée. En effet, nous avons négligé souvent les comportements anisotropes et nous nous sommes limités à des lois de chargement simples, de telle sorte que la variable déformation $\varepsilon_1$ contrôle aussi la variable d'anisotropie induite. Dans les cas pratiques, ceci n'est pas toujours le cas ; et les trajectoires d'un matériau peuvent être différentes si ce dernier a ou n'a pas été soumis à des cycles, même s'il semble que le pseudo coefficient de Poisson $\nu$ du milieu ne dépende effectivement que du rapport des contraintes q/p', car nous allons voir que le pseudo module d'Young est une variable dont le comportement est plus complexe et dépend du passé de l'échantillon.

Pour ces raisons, il est donc probable que l'espace (q,v,p') n'est pas suffisant pour rendre compte complètement des caractéristiques mécaniques du matériau ; il semble par contre bien adapté à des lois de chargement simple et monotone.

## 7.6. Évaluation du pseudo module d'Young :

### *7.6.1. Le modèle Hujeux:*

Pour modéliser le comportement du pseudo-module d'Young, il suffit de partir d'un modèle existant. Nous choisissons de partir du modèle Hujeux (Hujeux 1985, Aubry et al. 1985), qui est un modèle élasto-plastique isotrope à deux mécanismes, car ce modèle décrit assez bien la plupart des comportements mécaniques des milieux granulaires. En particulier, il est capable de décrire l'hystérésis et l'évolution plastique; il prend en compte le caractère soit contractant, soit dilatant du matériau et il respecte l'évolution vers les états critiques:

(7.24) $$v_c=v_{co} - \lambda \ln(p'/p'_o)$$ aux grandes déformations

Les notations sont identiques à celles que nous avons déjà utilisées. Le modèle Hujeux contient deux mécanismes d'écrouissage, l'un est une compression isotrope, l'autre est un mécanisme purement déviatoire. La fonction de charge F est donnée par:

(7.25) $$F = q/p' - M' [1 - b \ln(p'/p'_{co})-(b/\lambda)\varepsilon_{v,p}] \varepsilon_{d,p}/(a+\varepsilon_{d,p}) = 0$$

avec $\sin\varphi=3M'/(6+M')$. Les notations sont les suivantes: $\varepsilon_{v,p}$ ($\varepsilon_{v,e}$) et $\varepsilon_{d,p}$ ($\varepsilon_{d,e}$) sont les parties isotrope and deviatoire de la déformation plastique (élastique). La déformation totale est la somme de ces deux déformations. Dans le modèle Hujeux, les déformation $\varepsilon$ sont considérées comme positives lorsque le volume de l'échantillon croît et sa taille augmente. Dans ce cas, la constante a est négative. Les relations qui permettent de passer des composantes tensorielles déviatoires et isotropes aux composantes exprimées dans les axes principaux sont rappelées dans la table 7.1.

Les valeurs typiques des paramètres a, b, $\lambda$, M', $\varphi$ entrant dans la loi Hujeux sont donnés dans la Table 7.2 pour différents sols et milieu granulaire.

Le paramètre a contrôle l'amplitude de la déformation déviatoire (ou $\varepsilon_1$) à partir de laquelle le matériau passe du comportement contractant au comportement dilatant. $\lambda$ contrôle l'évolution du volume spécifique en fonction de la pression moyenne p'; $p'_{co}$ caractérise donc le volume spécifique initial de l'échantillon; c'est la pression de surconsolidation, lorsque le sol est de l'argile. Plus $p'_{co}$ est grand, plus le matériau est dense et plus il est dilatant. b contrôle l'amplitude du pic $q_{max}/p'$ du déviateur de contrainte.

| | |
|---|---|
| $\varepsilon_d = \varepsilon_{d,tot} = \varepsilon_{d,e} + \varepsilon_{d,p}$ | $\varepsilon_v = \varepsilon_{v,tot} = \varepsilon_{v,e} + \delta\varepsilon_{v,p}$ |

| | | | |
|---|---|---|---|
| $\varepsilon_1 = [2\varepsilon_d+\varepsilon_v]/3$ | $\varepsilon_2=\varepsilon_3= [\varepsilon_v - \varepsilon_d]/3$ | $\varepsilon_d = \varepsilon_1 - \varepsilon_2$ | $\varepsilon_v = \varepsilon_1 + 2\varepsilon_3$ |
| $\sigma'_1 = 2q/3+p$ | $\sigma'_2=\sigma'_3=p' -q/3$ | $q= \sigma'_1 - \sigma'_2$ | $p'= [\sigma'_1+2 \sigma'_2]/3$ |

***Table 7.1:*** *Relations entre les composantes des tenseurs de déformations totale, élastique et plastique (ligne du haut). Relations entre les composantes des tenseurs de déformation et de contrainte exprimées soit par leur termes deviatoires et isotropes soit dans les axes principaux (deuxième ligne: déformation; troisième ligne: contrtainte).*

| | sables | argiles |
|---|---|---|
| a | -0.03=-3% | -.03 to |
| b | 0.12 - 0.2 | 1 |
| $M'=q_c/p'$ | 1,2 | 1.2 |
| $\varphi$ | 30°-40° | 30° |
| $\lambda$ | 0.06 | 0.1 |

***Table 7.2:*** *Valeurs typiques des paramètres entrant dans le modèle Hujeux.*

En outre, le modèle Hujeux est un modèle élasto-plastique, régi par la théorie de la plasticité. Il requiert donc d'introduire (i) une loi d'écoulement plastique, qui gère la déformation plastique, (ii) un module d'Young $E_e$ et (iii) un coefficient de Poisson $\nu_e$ élastiques; ces deux coefficients gèrent la déformation élastique . $E_e$ et $\nu_e$ ne doivent pas être confondus avec le pseudo module d'Young $1/C_o$ et le pseudo coefficient de Poisson $\nu$ que nous avons déjà introduits; c'est $C_o$ pour lequel nous recherchons une loi de variatio. Les coefficients $E_e$ et $\nu_e$ sont supposés constants. Ceci conduit aux trois équations suivantes:

(7.26) $q = \varepsilon_{d,e}[E_e /(1+\nu_e)] \quad \Leftrightarrow \quad \varepsilon_{d,e} = [(1+\nu_e)/E_e]\, q$

(7.27) $p' = \varepsilon_{v,e}[E_e /\{3(1-2\nu_e)\}] \quad \Leftrightarrow \quad \varepsilon_{v,e} = [3(1-2\nu_e)/E_e]\, p'$

(7.28) $\delta\varepsilon_{v,p} = \delta\varepsilon_{d,p}\, (M'-q/p')$

On reconnaît dans l'Eq. (7.28) la loi d'écoulement introduite par Roscoe pour le modèle Granta Gravel (Schofield & Wroth 1968). Par la suite, nous allons supposer que la partie élastique de la déformation est négligeable, soit $E_e \rightarrow\infty$ et $\nu_e$ reste fini.

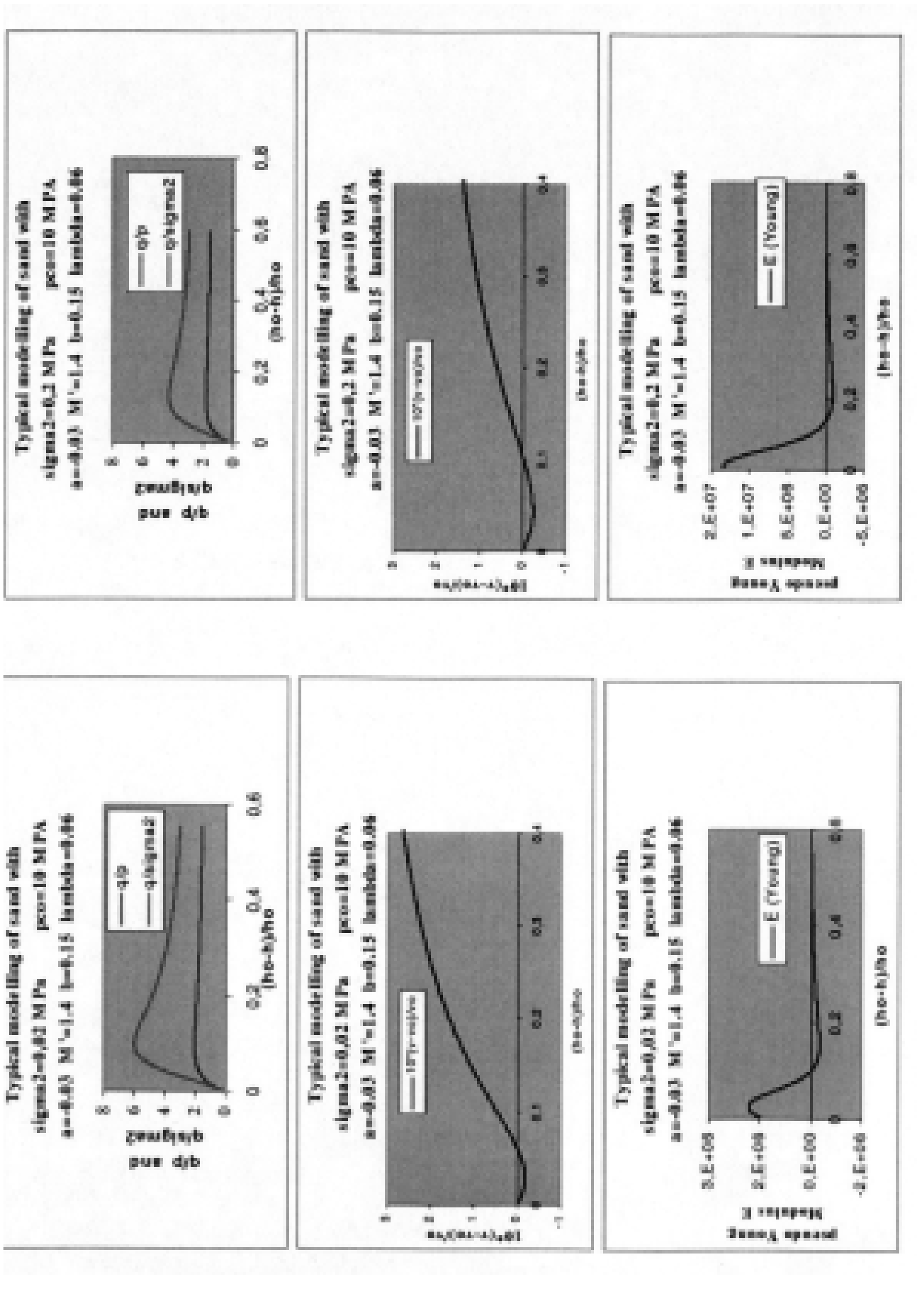

***Figure 7.7.a:*** *Comportements typiques prévus par le modèle Hujeux pour du sable tassé à la même densité initiale* ($p'_{co}$ =100 M Pa) *dans un essai triaxial à contrainte radiale* $\sigma'_2$ *constante;* $\sigma'_2$ = 0.02 M Pa; $\sigma'_2$ = 0.2 M Pa.

*Valeurs des paramètres:* a=-0.03 ; M'=1.4 ; b=0.15; $\lambda$=0.06; $p'_{co}$=10 M Pa

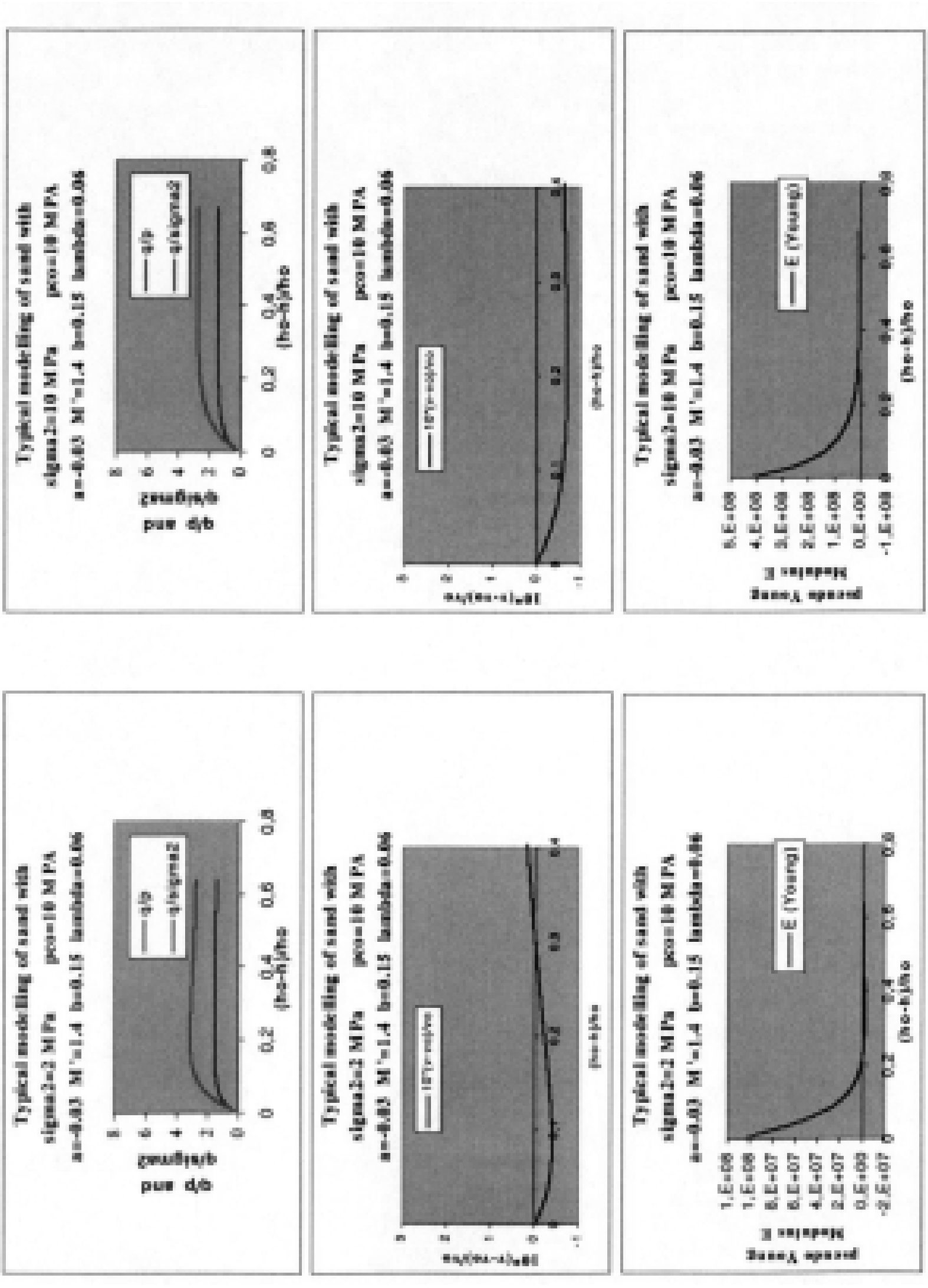

***Figure 7.7.b:** Comportements typiques prévus par le modèle Hujeux pour du sable tassé à la même densité initiale* ($p'_{co}$ =100 M Pa) *dans un essai triaxial à contrainte radiale* $\sigma'_2$ *constante;* $\sigma'_2$ =2 M Pa ; $\sigma'_2$ =10 M Pa.
*Valeurs des paramètres:* a=-0.03 ; M'=1.4 ; b=0.15; $\lambda$=0.06; $p'_{co}$=10 M Pa

### *7.6.2. Déterminer la loi de variation de $C_o$:*

Considérons un échantillon soumis à une contrainte ($\sigma'_1$,$\sigma'_2$=$\sigma'_3$) ou (q,p') qui a été déformé ($\varepsilon_1$,$\varepsilon_2$=$\varepsilon_3$) ou ($\varepsilon_d$,$\varepsilon_v$). Le problème est d'identifier les valeurs des coefficients de l'Eq. (7.4) qui le décrivent, ainsi que leurs évolutions avec la contrainte. Nous avons déjà trouvé la variation de $\nu$ . Il reste à trouver celle de $C_o$. Pour cela, on peut procéder à un essai de compression à $\sigma'_2=\sigma'_3=c^{ste}$ et identifier terme à terme.
On peut aussi supposer que les modélisations existantes sont correctes et en choisir une (Hujeux), puis identifier la variation de $C_o$ à celle que l'on calculera en utilisant le modèle choisi. C'est ce que nous allons faire avec le modèle Hujeux.
Comme nous avons supposé que la partie élastique de la déformation est négligeable, on peut identifier la déformation plastique à la déformation totale; soit: $\delta\varepsilon_{d,p}=\delta\varepsilon_d$ et $\delta\varepsilon_{v,p}=\delta\varepsilon_v$ . La déformation étant plastique, la fonction de charge doit être vérifiée pour un $\delta$q et un $\delta$p' considéré. De plus comme nous avons choisi le chemin $d\sigma'_2=d\sigma'_3=0$, cela impose dp'=dq/3 , *cf.* Tableau (7.1), et $\delta\varepsilon_v=\delta\varepsilon_d$ (M-q/p') à cause de l'Eq. (7.28). Ainsi, la condition dF=0 permet d'écrire:

(7.29.a) $$-C_o(1+\nu) = \Big[1 - (1/3) \Big\{\{M(1-b) - Mb \ln(p'/p'_{co}) - (Mb/\lambda)\, \varepsilon_v \} *[\varepsilon_{d,p} /(a+\varepsilon_d)] \Big\}\Big] \Big/ \Big\{(M'p'-q)(M'b/\lambda) [ \varepsilon_d /(a+\varepsilon_d)]+ [a/(a+\varepsilon_d)^2] \{M'p' – M'bp' \ln(p'/p'_{co}) - (M'b/\lambda)\, p'\, \varepsilon_{v,p}\}\Big\}$$

avec

(7.29.b) $$\nu= \sigma'_1/[2\sigma'_3(1+M')]= (2q+3p') /[(6p' -2q)(1+M')]$$

Ce qui permet de trouver $C_o$.

***Remarques:*** On peut noter que d'après le modèle Hujeux, $C_o$ est approximativement inversement proportionnel à l'amplitude du champ de contrainte $\sigma'_2=\sigma'_3$. On remarque aussi que la réponse incrémentale dépend de la contrainte, mais aussi de la déformation, car la déformation axiale gère l'amplitude de la variation de volume, et que la variation de volume intervient dans F. Ainsi, l'évolution de $C_o$ dépend du chemin suivi. Ceci complique la prédiction, mais c'est cohérent avec ce que l'on sait des lois rhéologiques et du comportement des matériaux granulaires.

### *7.6.3. Comportements typiques obtenus avec ce modèle sur des essais à $\sigma_2=^{cste}$:*

Comme on peut calculer $C_o$ par cette méthode (Eq. 7.29), on peut maintenant simuler tout type de réponse en intégrant l'évolution donnée par l'Eq. (7.4). En particulier, on peut simuler des essais de compression à $\sigma'_2=c^{ste}$. On reporte sur la Fig. 7.7 trois exemples types, où l'évolution de q/p', de $v/v_o$ et de $1/C_o$ sont tracés en fonction de la variation relative de la hauteur de l'échantillon $(h-h_o)/h_o=-\varepsilon_1$. Ces trois exemples correspondent à 3 valeurs différentes de $\sigma'_2$ , pour la même pression de surconsolidation $p'_{co}$ =10 MPa. Comme on pouvait s'y attendre, les résultats dépendent du rapport $\sigma'_2/p'_{co}$. car ils dépendent de la différence entre le volume spécifique initial $v_o$ (ou $p'_{co}$) et celui de l'état critique $v_c$ (ou $p'_c$). Plus l'échantillon est dense (ou $p'_{co}$ est grand), et plus $\sigma'_2$ est faible, plus la dilatance est grande et plus la variation totale de volume l'est aussi. Pour des valeurs de$\sigma'_2$ grandes, i.e. $p'_{co}=\sigma'_2$, l'échantillon est toujours contractant tout le long de la compression et le rapport q/p' reste toujours inférieur à M'; il atteint M' par valeur inférieure aux grandes déformations. Il faut noter que des valeurs de $p'_{co}$ plus petites que $\sigma'_2$ ne sont pas réalisables en pratique, mais peuvent être simulées (*cf.* Fig. 7.8); il faut donc faire attention de ne pas traiter des cas non physiques.

Les simulations confirment que $1/C_o=E$ est proportionnel à la contrainte $\sigma'_2$, approximativement. L'évolution de q/p' a toujours lieu dans la même gamme de déformation $\varepsilon_1$ (1-15%), quoique l'évolution précise dépende aussi du rapport $p'_{co}/\sigma'_2$.

On remarque aussi que lorsque le rapport $p'_{co}/\sigma'_2$ est très grand, le pseudo module d'Young $E=1/C_o$ croît aux petites déformations. C'est lié à la contraction du volume dans la première phase de déformation; cette contraction produit un raidissement de la réponse de l'échantillon. L'augmentation de E est relativement plus faible quand $\sigma'_2$ est plus grand (à $p'_{co}$ donné) car E est lui-même proportionnel à $\sigma'_2$ , ce qui conduit à un dE/E plus petit. Cet effet ne semble pas avoir été noté par les auteurs du modèle Hujeux (Hujeux 1985, Aubry et al. 1985); c'est qu'il est très souvent petit et qu'il peut être masqué lorsque le modèle tourne avec un module élastique en plus dans la simulation.

On peut aussi chercher à comprendre le fonctionnement du modèle Hujeux (Fig. 7.7): aux grandes déformations, i.e. $||\varepsilon_d||>>a$; dans ce cas, le terme $\varepsilon_d/(a+\varepsilon_d)$ est à peu près constant; ainsi le système obéit à la loi de Granta gravel (Schofield & Wroth 1968), et la réponse dépend du volume v spécifique par rapport à $v_c$. Ainsi, le modèle d'Hujeux est équivalent à Granta gravel quand $||\varepsilon_d||>>a$. Par contre, quand $||\varepsilon_d||$ devient petit, *i.e.* $||\varepsilon_d||<a$, le terme $\varepsilon_d/(a+\varepsilon_d)$ domine dF. Ceci modifie la réponse et l'écarte de celle du modèle de Granta gravel: le modèle Hujeux impose un processus de déformation pratiquement indépendant de v dans ces conditions.

Compte tenu de tout ceci, il est facile d'imaginer des modifications possibles à apporter au modèle Hujeux pour qu'il donne une variation non linéaire de q/p' *vs.* $\varepsilon_1$ dans le régime des $\varepsilon_1$ petits, *i.e.* $||\varepsilon_d||<<a$ : on a juste à remplacer le terme $\varepsilon_d/(a+\varepsilon_d)$ de F par un terme $\varepsilon_d^n/(a+\varepsilon_d)^n$ . Dans ce cas, cependant, $C_o$ ne restera pas proportionnel au champ de contrainte (q,p') , dans le domaine de déformation intéressant, *i.e.* $||\varepsilon_d||<<a$; de plus le domaine de variation intéressant de $\varepsilon_1$ dépendra du niveau de la contrainte, c'est à dire qu'il variera avec p'.

### *7.6.4. Domaine de validité des paramètres* **a, b, $\lambda$ et $p'_{co}$***:*

Pour finir, il est bon de remarquer que le modèle Hujeux est sensible à la valeur des paramètres introduits (sic!). Il peut donc donner des comportements non réalistes si le domaine de variations de $\varepsilon_1$ n'est pas approprié ($\varepsilon_1$ doit rester dans le domaine [0,1]) et si les valeurs des paramètres sont mal choisis. Nous donnons de tels exemples dans la Fig. 7.8.

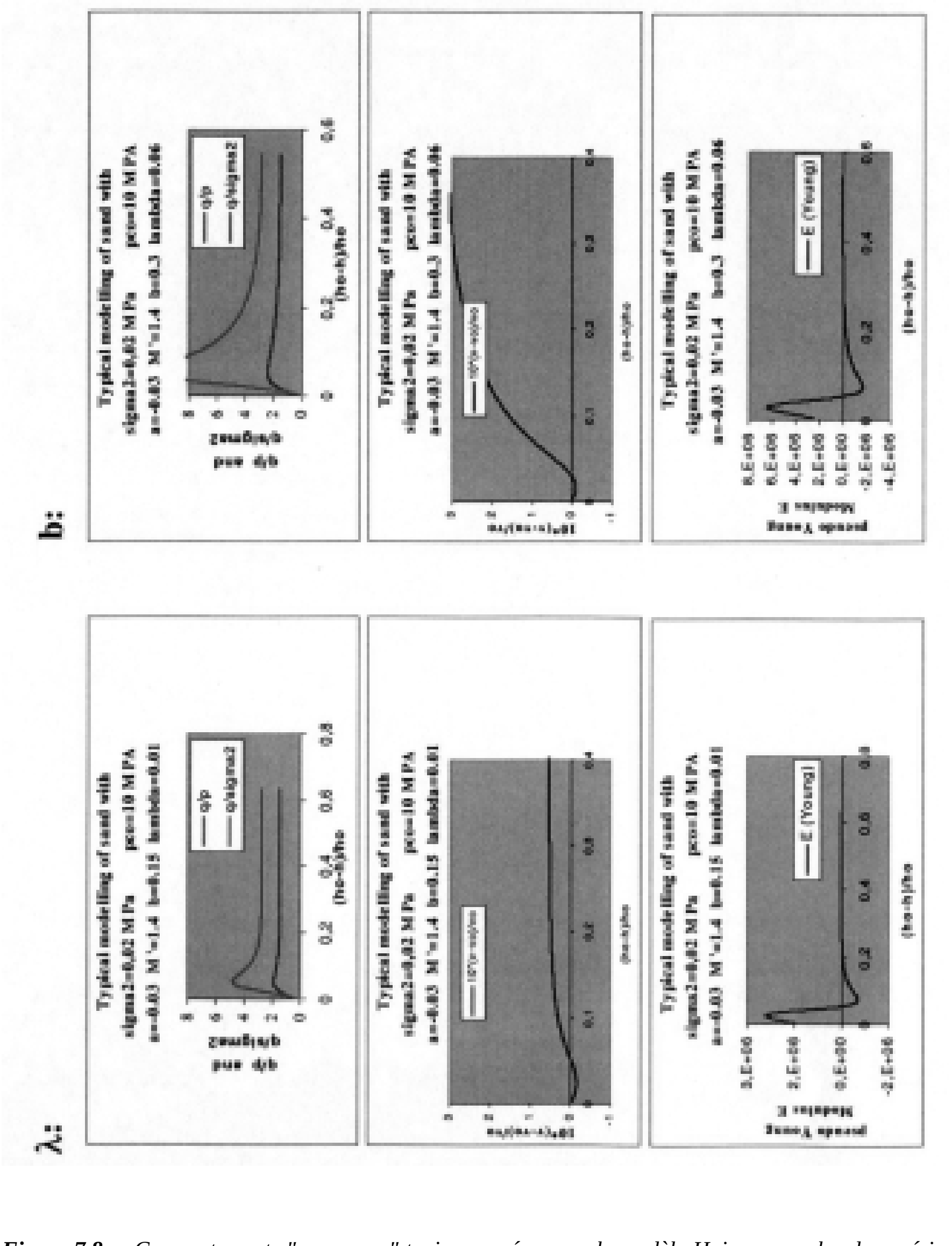

***Figure 7.8.a:*** *Comportements "anormaux" typiques prévus par le modèle Hujeux pour les deux séries de paramètres mal adaptés suivants:*

a=-0.03 ; M'=1.4 ; b=0.15 ; $\lambda$=0.01; $p'_{co}$=10 MPa ; $\sigma'_2$= 0.02 MPa

a=-0.03 ; M'=1.4 ; b=0.3 ; $\lambda$=0.06; $p'_{co}$=10 MPa ; $\sigma'_2$= 0.2 MPa .

## 7.7. Comportement sous chargements cycliques :

### 7.7.1. Comment décrire les comportements sous chargements cycliques:

On sait que le comportement sous chargement cycliques est contrôlé principalement par:

(i) la valeur moyenne <q>/p' du rapport q/p' à laquelle les cycles sont imposés,

(ii) par l'amplitude $\Delta q$ des cycles.

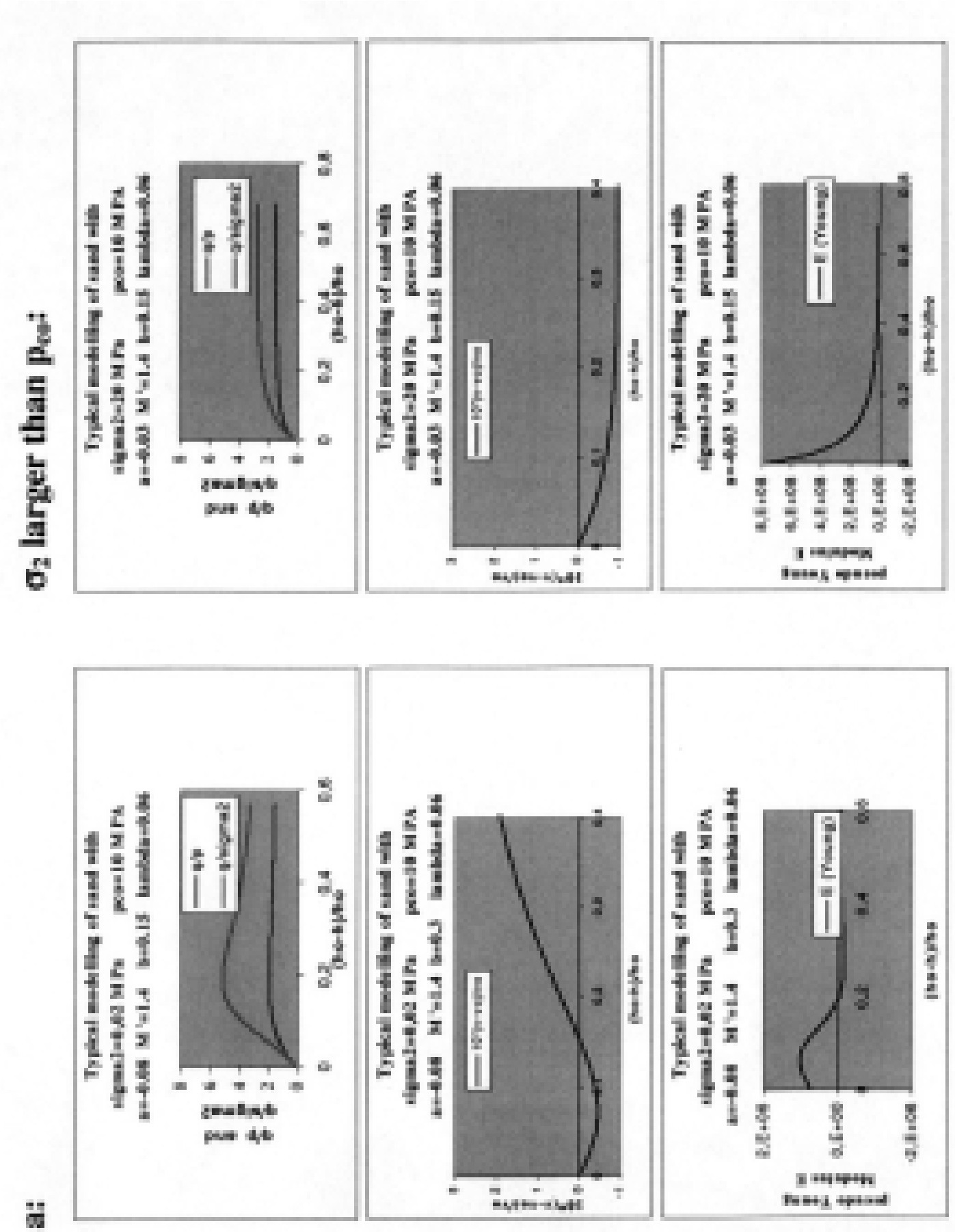

***Figure 7.8.b:*** *Comportements "anormaux" typiques prévus par le modèle Hujeux pour les deux séries de paramètres mal adaptés suivants:*

a=-0.08 ; M'=1.4 ; b=0.15 ; $\lambda$=0.01; $p'_{co}$=10 MPa ; $\sigma'_2$= 0.02 MPa

a=-0.03 ; M'=1.4 ; b=0.15 ; $\lambda$=0.06; $p'_{co}$=10 MPa ; $\sigma'_2$= 20 MPa .

En particulier, la nature contractante ou dilatante de l'effet moyen de chaque cycle dépend de la valeur de <q>/p' , comparé à M'. Si M'-<q>/p' <0, les cycles ont lieu dans le domaine sous-caractéristique et ils sont contractants; dans le cas contraire, domaine sur caractéristique, les cycles ont tendance à faire dilater le matériau. On sait aussi que l'effet moyen varie comme le logarithme du nombre N de cycles effectués:

(7.30) $$\Delta_N v = \Delta_1 v / \ln(N)$$

Enfin, plus l'amplitude relative du cycle Δq/p' est grand plus l'effet (contraction ou dilatation) est grand; la valeur de Δq-<q> est aussi importante car un changement de la direction des contraintes principales (soit changement de direction, soit permutation des axes principaux majeurs et mineurs) amplifie énormément les effets d'un cycle.
On sait que le modèle Hujeux est capable de décrire ces phénomènes de manière relativement réaliste. On peut aussi rendre compte de ces phénomènes dans le modèle incrémental (Eq.7.4); pour cela il faut modifier la valeur du paramètre d'écrouissage entrant dans F, *i.e.* Eq. (7.25), à chaque cycle, *cf.* § 7.7.2.

Montrons tout d'abord que le modèle incrémental est compatible avec les faits observés, c'est-à-dire avec le fait que les cycles sont soit contractants, soit dilatants, suivant la valeur du rapport <q>/p'. Pour cela on se limite à des essais de compression cyclique uniaxiale réalisés à $\sigma'_2=\sigma'_3=c^{ste}$ sur un échantillon drainé. Durant la 1ère demi-période, l'échantillon se déforme plastiquement; pendant la seconde, il se déforme élastiquement; cependant, comme $E_e\rightarrow\infty$, la déformation élastique est quasi nulle durant cette deuxième demi-période, de telle sorte que le comportement global résulte du comportement dans chaque première demi-période. En se limitant au cas de cycles de petites amplitudes, *i.e.* <q><<p', le comportement de ces demi-périodes est donc soit contractant soit dilatant suivant que le rapport <q>/p' est plus petit ou plus grand que M', car l'on peut écrire d'après l'Eq. (7.4):

(7.31) $$\Delta v=-\int 3C_o(1-2\nu)dp'=-\int C_o(1-2\nu)dq \approx C_o(1-2\nu)\Delta q$$

avec dp'=dq/3. Ceci donne:

(7.32.a) $$\Delta v >0 \quad \text{if} \quad \nu>1/2 \text{ or } \langle q\rangle/p'>M'$$

(7.32.b) $$\Delta v <0 \quad \text{if} \quad \nu <1/2 \text{ or } \langle q\rangle/p'<M'$$

ce qui est effectivement observé.
De plus, si l'on part de l'Eq. (7.25) et que l'on suppose (i) que les variations Δq et $\Delta\varepsilon_d$ de q et $\varepsilon_d$ sont petites, (ii) que $\varepsilon_d$<<a, (iii) <q><<M'p', on obtient que l'amplitude $\Delta\varepsilon_d$ des cycles sont données approximativement par:

(7.33) $$\Delta\varepsilon_d \approx [\Delta q/(M'p')]/\{1+(b/\lambda)(v_c-v)\}=C_o\Delta q$$

***essais cycliques sur échantillon non drainé:*** Dans le cas des essais non drainés cycliques, la condition expérimentale impose δv=0. Si on considère un incrément de contrainte $\delta\sigma'_1$, l'application de l'Eq. (7.4) avec dv=0 impose une variation $\delta\varepsilon_1$ de déformation axiale et une variation $\delta\sigma'_2$ de $\sigma'_2$, puisque $\delta\varepsilon_v=0$; on trouve ainsi que $\delta\sigma'_2 >0$ si le coefficient de Poisson $\nu>½$ et le contraire, *i.e.* $\delta\sigma'_2 <0$, si le coefficient de Poisson est $\nu<½$. Le premier cas arrive lorsque $<q>/p' > M'$, le second lorsque $<q>/p' < M'$. Ainsi, si l'on maintient la pression totale $\sigma_2=c^{ste}= \sigma'_2 +u_w$ constante , on prévoit bien:

(7.34.a) $\Delta u_w <0$ if $\nu>1/2$ or $<q>/p'>M'$

(7.34.b) $\Delta u_w >0$ if $\nu <1/2$ or $<q>/p'<M'$

Ce qui est effectivement observé expérimentalement.
Ce résultat est en accord avec les principes de la liquéfaction: en effet celle-ci ne s'obtient que lorsque $<q>\approx 0$; dans ce cas les cycles sont tels que $<q>/p'<M'$ de telle sorte que la pression de l'eau interstitielle croît à chaque cycle (Eq. 7.34.b) et la pression intergranulaire $\sigma'_2$ diminue d'autant. Ce processus s'arrête lorsque p' =0 dans une grande partie du cycle. Dans cet état le sable ne peut plus mobiliser du frottement solide, c'est donc devenu un liquide.

### *7.7.2. comment modéliser l'effet des cycles successifs:*

Ainsi, la plupart des caractéristiques du comportement des milieux granulaires sous chargements cycliques sont correctement décrits par cette modélisation incrémentale (Eq. 7.4), dès lors que l'on suppose que le pseudo coefficient de Poisson $\nu$ du matériau varie avec le rapport des contraintes q/p' . Il faut aussi considérer que l'échantillon se déforme plastiquement pendant 1/2 période (charge) puis élastiquement pendant l'autre (décharge).
L'hypothèse d'une réponse incrémentale isotrope semble très souvent suffisante; ceci est d'ailleurs en accord avec l'espace des phases (q,v,p') utilisé pratiquement pour représenter l'état du milieu granulaire.
Pour modéliser l'écrouissage en fonction du nombre de cycles, il faut que la fonction de charge F contienne un processus d'écrouissage; ceci est fait dans le modèle Hujeux en modifiant l'Eq. (7.25) en introduisant une variable d'écrouissage évoluant en fonction du nombre de cycles. La variable d'écrouissage que nous pourrions retenir pour notre modélisation pourrait être soit la valeur de $\varepsilon_d$ qui peut évolué après chaque cycle (par exemple modifié $\varepsilon_d$ en $\varepsilon_d$-$\Delta\varepsilon_d$ dans l'Eq. (7.25)), soit en modifiant $\varepsilon_v$ par $\varepsilon_{vo}$ lnN , soit $\varepsilon_d$ par $\varepsilon_{do}$ lnN, soit encore plus simplement, en divisant $C_o$ par lnN.
Enfin quand q passe par 0, la valeur de $\varepsilon_d$ peut-être remise à 0 pour amplifier les effets liés à des variations de direction des contraintes principales, et pour annuler les effets de mémoire.
Lorsque les cycles ne passent par q=0, , il faut choisir de toutes les façons la variation de la variable d'écrouissage de telle sorte que les effets d'amplitudes de cycles et du nombre N de cycles soit bien décrits.

## 7.8. Conclusion :

Le modèle proposé semble donc très efficace et simple pour représenter les comportements rhéologiques des milieux granulaires; il est basé sur une modélisation incrémentale à plusieurs zones, dont l'une contient tous les essais de compression, $\varepsilon_1<0$.

Dans la zone de compression, le modèle suppose un comportement isotrope défini par un pseudo coefficient de Poisson $\nu$ et un pseudo module d'Young $E=1/C_o$. Ces coefficients sont choisis pour décrire les essais triaxiaux à $\sigma'_2=\sigma'_3=c^{ste}$; à cause de cela le coefficient de Poisson $\nu$ ne doit dépendre que du rapport des contraintes $\sigma'_1/\sigma'_3$ (loi de Rowe). Nous avons trouvé la loi de variation de $C_o$ en utilisant le modèle Hujeux; il n'est pas exclu que des expressions de $C_o$ plus simples, mais tout aussi réalistes, puissent être trouvées à partir d'autres modèles. Nous avons discuté la forme des variations de $C_o$ en fonction des paramètres de la loi Hujeux.

Du fait de l'acceptation des résultats expérimentaux triaxiaux à $\sigma'_2=\sigma'_3=c^{ste}$ comme guide de notre raisonnement, notre modèle reconnaît implicitement l'existence des états critiques, puisque ces états sont les états limites des essais à $\sigma'_2=\sigma'_3=c^{ste}$. Le modèle reconnaît donc implicitement aussi les notions de frottement solide et de dilatance; il ne reconnaît pas cependant un rôle essentiel à la théorie de la plasticité parfaite.

Grâce à ce modèle, on peut expliquer très simplement la plupart des essais de compression simple.

- ***essai oedométrique:*** on retrouve que le rapport des contraintes tend vers $\sigma'_3/\sigma'_1=f(\varphi)\approx 1-\sin\varphi$. Ceci est lié au fait que le système choisi le coefficient de Poisson $\nu$ tel qu'il impose $\delta\sigma'_3/\delta\sigma'_1= \sigma'_3/\sigma'_1$ , sachant que $\sigma'_3/\sigma'_1=\nu/(1-\nu)$. Ceci suffit à fixer le rapport des contraintes.
- ***essai non drainé:*** on trouve que la trajectoire dans le plan (q,p') doit commencer à $p'=c^{ste}$, puis subir une bifurcation et suivre la droite des états caractéristiques $q=M'p'$. On a généralisé la notion d'état caractéristique aux cas d'échantillons ayant une réponse anisotrope. Plus fondamental, on a aussi associé la notion d'état caractéristique (i) à l'existence d'un pseudo coefficient de Poisson $\nu$ et (ii) au fait que $\nu=½$ pour ces états. Ceci s'avère capital pour comprendre l'intérêt de la droite caractéristique pour comprendre le processus de liquéfaction.
- ***surfaces de Hvorslev et de Roscoe:*** on a montré que l'arrivée au point critique dans l'espace (q,v,p') devait se faire soit sur une courbe, soit sur une surface 2d.Compte tenu de l'expérience, la surface 2d a été sélectionnée ce qui montre que $\nu-½$ et $C_o$ tendent tous deux vers 0 de la même manière, c'est-à-dire que $(\nu-½)/C_o \to c^{ste}$, avec $c^{ste} \neq 0$ et $\neq\infty$. De plus, on a conclu que la manière d'arriver à l'état critique devait être indépendante du sens de l'arrivée, *i.e.* côté dense, côté lâche. On en a conclu que les surfaces de Hvorslev et de Roscoe devaientt être identiques et ne former qu'une seule et même surface. Ceci semble vérifier expérimentalement..

- ***Compatibilité avec l'espace des phases (vq,p'):*** ce modèle est compatible avec l'espace des phases utilisé pour décrire l'état d'un milieu granulaire en mécanique des sols.
- ***dans les essais cycliques,*** on décrit la charge comme appartenant à la zone de compression plastique, la décharge comme appartenant à la zone de décompression élastique. Dans ces conditions les effets moyens d'un cycle sont bien décrits: contractance des essais cycliques drainés si q/p'-M'<0, dilatance dans le cas contraire; compatibilité avec le principe de la liquéfaction des sols dans des essais cycliques non drainés.

En conclusion notre modèle est original par sa simplicité; il est probablement assez proche de celui de Duncan, mais il choisit un coefficient de Poisson variable qui le rend beaucoup plus descriptif. De même, il est au moins aussi simple que le modèle de Granta gravel, mais décrit beaucoup mieux les petites déformations et aussi bien les grandes. Enfin, il est assez proche d'autres modèles (Hujeux,…), mais il n'en a pas la complexité.

### *Bibliographie :*

# 8. Quelques réflexions sur le passage « micro-macro »

Les matériaux granulaires sont des matériaux à structure discontinue. Dans ces conditions, on peut se demander ce que vaut une approche "mécanique des milieux continus" pour décrire leur comportement.

En fait, cette question n'a probablement pas grand sens , puisque la plupart des matériaux sont discontinus à l'échelle microscopique: gaz, liquide, solides ont tous une dimension critique au dessous de laquelle la structure n'est pas continue. Or dans la plupart de ces cas on sait qu'une représentation continue peut représenter correctement leur comportement et prévoir ceux-ci à l'aide d'équations déterministes.

La question essentielle est en fait de savoir si l'on peut réduire le comportement d'un grand nombre d'objets (grains dans le cas des milieux granulaires, atomes ou molécules dans le cas d'un gaz ou d'un liquide, micro-cristaux dans le cas d'un poly-cristal) à celui d'un certain nombre de grandeurs macroscopiques représentatives et de définir ces grandeurs. Si l'on arrive à faire cela, on a réussi à réduire la dimension de l'espace des phases. Dans le cas d'un gaz parfait monoatomique à l'équilibre, on sait que cette procédure existe et que l'on peut réduire les 6N degrés de libertés (ou coordonnées) correspondant à la description de l'état mécanique des N atomes à la donnée des 3 grandeurs suivantes: pression p, volume spécifique v et température T.

Lorsque le nombre d'éléments d'un système est grand, une telle procédure de réduction doit exister sous peine d'être incapable d'appréhender le système lui-même, en tant que système. En d'autres termes, on doit a priori pouvoir définir ce système par un certain nombre de grandeurs moyennes qui ont un sens physique. Du nombre de paramètres minimum nécessaires à la description du système vont dépendre (i) la difficulté de sa représentation macroscopique et (ii) la complexité de sa loi d'évolution.

Lorsque l'évolution d'un système semble relativement simple et stable dans une certaine gamme de conditions, c'est que les lois qui le représentent doivent être relativement simples et stables et que le nombre de paramètres représentatifs nécessaires est petit. Ceci est d'autant plus vrai que l'on sait que des systèmes non linéaires qui ne sont gérés que par trois degrés de liberté seulement peuvent déjà engendrer du chaos déterministe, avec des attracteurs étranges, c'est-à-dire qu'ils peuvent déjà avoir des comportement "étranges", qui semblent imprévisibles... avec des bifurcations, (*cf.* Appendice 1),…

Inversement, si l'on arrive à représenter un système contenant un grand nombre N d'éléments, chacun ayant x degrés de liberté, par un petit nombre de paramètres macroscopiques, c'est que les xN degrés de liberté internes obéissent à des lois statistiques. La connaissance de ces lois statistiques permet alors de relier ces xN degrés microscopiques aux grandeurs macroscopiques. L'objet de la physique statistique est alors de définir ou de déterminer les lois statistiques qui régissent les degrés de libertés internes, et de les relier aux grandeurs macroscopiques.

Lorsque ceci est réalisé, on a réussi à réduire la représentation complète du système aux seules grandeurs macroscopiques; c'est-à-dire qu'on a non seulement réduit le système à la connaissance d'un petit nombre de paramètres, qui le caractérise à l'échelle macroscopique, mais on a aussi su remonter aux variables microscopiques .
On voit ainsi le rôle capital joué par l'ensemble des grandeurs macroscopiques pertinentes: elles servent à prédire le comportement macroscopique et à brosser la physique statistique d'un système complexe à grand nombre de degré de liberté. Il est donc essentiel de les définir et les connaître correctement.

Dans le cas du sable ou des milieux granulaires, les lois de comportements que nous avons décrites dans les chapitres précédents (i) montrent que ces milieux ont des comportements relativement simples à appréhender, (ii) que ceux-ci s'analysent dans l'espace (q,v,p'=σ'), *i.e.* (déviateur de contrainte q, volume spécifique v, contrainte moyenne intergranulaire p'=σ'=trace(**σ**')/3) et (iii) que tout ceci nous permet de définir précisément les grandeurs macroscopiques pertinentes. On peut donc se lancer maintenant dans la recherche d'une description plus microscopique, et chercher à relier les grandeurs microscopiques précédentes aux grandeurs macroscopiques précitées.
On utilisera pour cela des concepts classiques de physique statistique; par exemple, nous supposerons que les grandeurs physiques que nous considérerons obéissent bien aux lois physiques normales (équation d'équilibre, loi de l'action et de la réaction,…) mais aussi que ces grandeurs microscopiques obéissent à la loi des grands nombres et à un principe de désordre optimal.
Dans la première section, nous rappelleront les principes de base de la physique statistique (loi des grands nombres, désordre maximum,…) en traitant un exemple particulier, celui du gaz parfait; cela permettra aussi de rappeler une méthode classique de physique statistique, la méthode des multiplicateurs de Lagrange, et de rappeler l'importance de la ***distribution de Poisson***. En fait, la ***distribution exponentielle*** que l'on obtiendra dans cet exemple, et que l'on obtient très souvent dans beaucoup d'autres exemples, est lié au fait qu'on s'intéresse à un problème discret dont chaque élément peut avoir une taille variable, mais dont la valeur moyenne est fixée.
Nous appliquerons cette procédure au cas d'un milieu granulaire. Ainsi, dans une seconde partie, nous chercherons à déterminer la distribution des forces de contacts dans un tel milieu; nous verrons alors que la contrainte moyenne joue le rôle d'une température $T_\sigma$. La Section 3 cherchera à modéliser les fluctuations de la taille des pores dans un milieu granulaire; là aussi le volume spécifique moyen sera relié à une nouvelle "température" $T_\varpi$.
Dans ces deux dernières sections (Sections 2 & 3), les différentes statistiques seront supposées indépendantes. Bien entendu ceci n'est pas réaliste d'un point de vue mécanique: les lois macroscopiques montrent par exemple que le volume spécifique d'un matériau normalement consolidé varie  avec la pression moyenne p'=σ' qui lui est appliquée. La quatrième partie cherchera donc à liée l'évolution des deux statistiques; pour cela nous généraliserons une approche proposée initialement par Boutreux & de Gennes (1997) pour modéliser la densification d'un milieu granulaire contenu dans une éprouvette verticale soumise à des chocs verticaux répétitifs, *i.e.* méthode expérimentale de caractérisation des poudres utilisée en pharmacie, appelée tap-tap.
Avant de commencer, il nous paraît nécessaire de rappeler les domaines d'application de la distribution exponentielle et de la distribution gaussienne. Nous verrons que la distribution exponentielle est la distribution obtenue lorsqu'on s'intéresse à la probabilité d'occurrence d'un seul événement simple, dans le cas où l'on s'intéresse à un problème discret dont chaque élément peut avoir une taille variable, mais dont la valeur moyenne est fixée. Au contraire, la distribution gaussienne est obtenue par application du théorème central limite. Dans ce dernier cas, on s'intéresse à déterminer la probabilité d'occurrence d'un événement complexe, c'est-à-dire d'un événement formé de la conjonction de $N_1$ événements aléatoires; on montre, *i.e.* théorème central limite, que lorsque les événements sont indépendants et que la loi d'occurrence des événements simples est relativement régulière, l'on obtient une ***distribution gaussienne*** dont la largeur $\delta L$ est proportionnelle $\delta L \sim (N_1)^{1/2}$ et la valeur moyenne <L> proportionnelle à $<L> \sim N_1$, de telle sorte que la largeur relative $\delta L/<L>$ est proportionnelle à $\delta L/<L> \sim (N_1)^{-1/2}$.
La distribution exponentielle n'a donc rien d'extraordinaire; elle indique simplement qu'on a affaire à un ensemble d'événements simples et discrets qui sont soumis à un certain nombre de conditions statistiques, chacune agissant comme une contrainte, et dont l'une porte sur la valeur moyenne; elle est donc liée à la discrétisation d'un phénomène. C'est elle, a priori que l'on doit obtenir lorsqu'on remplace un champ de contrainte par un réseau de forces de contact, pour peu que ces contacts soient réels.

## 8.1. Rappels : Physique statistique d'un gaz parfait; l'approche de Boltzmann

Considérons un gaz parfait, formé de N molécules identiques. Les molécules sont animées d'une vitesse $v_\alpha$ qui varie (i) dans le temps et (ii) de molécule en molécule; la vitesse $v_\alpha$ correspond à une énergie cinétiques $\varepsilon_\alpha=1/2mv_\alpha^2$; on doit déterminer la distribution $p(v_\alpha)$. Pour cela considérons le cas N très grand et discrétisons le problème; les énergies prennent des valeurs discrètes *i.e.* $\varepsilon_i=\varepsilon_o+i\delta\varepsilon$, où i varie de 1 à l'infini; on suppose de plus que $\delta\varepsilon$ est suffisamment grand pour que chaque case $\varepsilon_i$ contienne un grand nombre $N_i$ de particules. On doit donc distribuer les N particules dans les différentes cases d'énergie. Sachant que les particules sont discernables, le nombre W de complexions possibles correspondant à une même distribution $\{\ldots,N_i,\ldots\}$ de particules dans les cases $\varepsilon_i$ est:

(8.1.a) $$W = N!/ (N_1!\ldots N_i!\ldots)$$

En physique statistique, on suppose toujours que l'état observé correspond à un des états les plus probables, car N est toujours très grand; ceci impose que l'état observé est un de ceux pour lequel W est maximum, ou, ce qui est pareil, pour lequel ln(W) maximum. Cependant cet état le plus probable doit aussi être tel qu'il est constitué de N particules exactement et qu'il a une énergie totale fixée $U=\Sigma_i \ \varepsilon_i N_i$. Ces deux dernières conditions s'écrivent:

(8.1.b) $N= \Sigma_i N_i$

(8.1.c) $U = \Sigma_i \ \varepsilon_i N_i$

On peut réécrire ces 3 conditions, en utilisant la formule de Stirling : $\ln N! \approx N\ln N-N$:

(8.2.a) $d(\ln W)=0= \Sigma_i \ln(N_i)\, dN_i$

(8.2.b) $0= \Sigma_i \ dN_i$

(8.2.c) $0= \Sigma_i \ \varepsilon_i \, dN_i$

Une manière de condenser ces trois conditions est d'utiliser la méthode des multiplicateurs de Lagrange (*cf.* Bruhat 1968), c'est-à-dire de multiplier les deux dernières équations par deux coefficients $\alpha$ et $\beta$ et de les additionner à la première; la condition obtenue doit être vraie quels que soient $\alpha$ et $\beta$. Les conditions (8.2) deviennent:

(8.3) $0= -\Sigma_i \{\ln(N_i) - \alpha - \beta \varepsilon_i \} dN_i$

Cependant, le nombre total de variables indépendantes est égal au nombre de cases. Il est donc égal au nombre de variables $N_i$ . On peut donc choisir comme variables indépendantes tous les $N_i$, ce qui impose d'exprimer $\alpha$ et $\beta$ en fonction des $N_i$. En posant $A=\exp(-\alpha)$, on trouve que l'Eq. (8.3) impose que A et $\beta$ sont données par:

(8.4.a) $N_i = A \exp\{-\beta\varepsilon_i \}$

et qu'ils doivent vérifier $N=\Sigma_i N_i$ et $U=NkT= \Sigma_i N_i \varepsilon_i$ ; ceci fixe la valeur de A (ou $\alpha$) et de $\beta$ :

(8.4.b) $A=N/(kT)$ et $\beta =1/(kT)$.

T est ici l'énergie cinétique moyenne portée par une molécule de gaz.

On se propose maintenant d'utiliser la même approche pour trouver la distribution des forces de contacts dans un milieu granulaire. On part du principe que l'on connaît la contrainte moyenne $\sigma'=\mathrm{Trace}(\sigma')/3$ et que le milieu est homogène et isotrope à l'échelle macroscopique.

## 8.2. Distribution des forces de contact dans un matériau granulaire isotrope homogène sous contrainte isotrope (Evesque 199a):

Dans cette section, on va chercher à déterminer la distribution des modules des forces de contacts dans un milieu granulaire. Pour simplifier la modélisation et obtenir des calculs exacts, on considérera exclusivement un milieu granulaire homogène et isotrope à l'échelle macroscopique soumis à une contrainte granulaire statique $\sigma'$=Trace($\sigma'$)/3 isotrope. De plus ce milieu sera constitué de grains identiques de forme sphérique, rayon R=D/2.

D'un point de vue expérimental, on sait depuis longtemps que la force appliquée à chaque contact est une variable aléatoire qui varie grandement (Dantu 1957 & 1968, Drescher & de Josselin de Jong 1972, Rothenberg & Barthurst 1992, Barthurst & Rothenberg 1988) . Du point de vue théorique, on sait aussi (Landau & Lifschitz 1990, Christoffersen *et al.* 1981, Kantani 1981) qu'il existe une relation générale qui permet de déterminer la contrainte moyenne intergranulaire $\sigma'_{ij}$ appliquée au volume V, connaissant la distribution des forces de contacts $\mathbf{f}^c$ appliquée au contact c de position $\mathbf{r}^c$ :

$$(8.5) \qquad \sigma'_{ij}=(1/V)\Sigma_{\text{contacts c dans volume V}}\, f^c_i\, r^c_j$$

Dans cette équation $f^c_i$ ($r^c_j$) est la composante i (j) de la force $\mathbf{f}^c$ de contact (de la position $\mathbf{r}^c$ du contact c).

Si l'on prend maintenant en compte (i) le principe d'action et de réaction, (ii) que tous les grains ont même diamètre D et (iii) que chaque grain est en équilibre, *i.e.* $\Sigma_{\text{force appliquée sur un grain}}\mathbf{f}^c=0$, $\Sigma_{\text{appliquée sur un grain}}\mathbf{f}^c \wedge \mathbf{r}^c=0$ , l'Eq. (8.5) se réécrit:

$$(8.6) \qquad \sigma'_{ij} = (D/V)\, \Sigma_{\text{tous les contacts c du volume V}}\, f^c_i\, n^c_j$$

où $n^c_j$ est la composante j du vecteur unitaire $\mathbf{n}^c$ normal à la surface de contact au point de contact c. Comme l'échantillon est supposé homogène et isotrope et qu'on lui applique une contrainte moyenne isotrope ($\sigma'_{ij}=\delta_{ij}$) , on peut écrire $\sigma'_{ij}$ sous la forme:

$$(8.7) \qquad \sigma'_{ij} = (D/V)\, \Sigma_{|f|}\, |\mathbf{f}|\{\Sigma_{n^c_j \text{ qui sont soumis à } |f|}\; \Sigma_{u^c_i \text{ subissant un } |f| \text{ bien défini}}\, u^c_i\, n^c_j\}$$

où $\mathbf{u}^c$ est le vecteur unitaire parallèle à la force $\mathbf{f}^c$. A cause de l'hypothèse d'isotropie, la dernière somme de l'Eq. (8.7) est nulle si $i\neq j$. Supposons de plus que l'orientation moyenne relative de la force de contact par rapport à la direction de contact soit indépendante de l'amplitude de la force $\mathbf{f}^c$; dans ce cas, on a $<u^c_i\, n^c_j> =<u^c_i\, n^c_i>\delta_{ij}= \delta_{ij}<\mathbf{u}^c \cdot \mathbf{n}^c>/3=\delta_{ij} <\cos\theta>/3$ où $\theta$ est l'angle entre $\mathbf{f}^c$ et la normale au contact c. Ainsi, l'Eq. (8.7) devient:

$$(8.8) \qquad \sigma'_{ij} = (D/V)\, \delta_{ij}\, \Sigma_{|f|}\, |\mathbf{f}| <\cos\theta>/3$$

Cette équation se réécrit:

$$(8.9.a) \qquad \sigma'_{ij} = D<\cos\theta>/(3V)\, \delta_{ij}\, \Sigma_{|f|}\, |\mathbf{f}|\, N_{|\mathbf{f}|}$$

où $N_{|\mathbf{f}|}$ est le nombre de contacts soumis à une force de module compris entre $|\mathbf{f}|$ and $|\mathbf{f}|+|\delta\mathbf{f}|$.

Pour trouver la distribution des forces, on peut maintenant procéder comme dans le cas du gaz parfait: on suppose que le désordre est parfait, *i.e.* optimum; on discrétise le problème, c'est-à-dire qu'on considère que les valeurs des modules $|\mathbf{f}|$ ne prennent que des valeurs discrètes $|\mathbf{f}|_k = k\ \delta f$; on doit donc répartir N modules de force dans des cases différentes et trouver le nombre W de complexion optimum, sachant que la contrainte moyenne est fixée par l'Eq. (8.9.a) et que le nombre de contacts est imposé. On a donc aussi:

(8.9.b) $\ln(W) = \ln(N!) - \Sigma_{|f|} \ln(N_{|\mathbf{f}|}!)$

(8.9.c) $N = \Sigma_{|f|} N_{|\mathbf{f}|}$

On dérive les Eq. (8.9) et on applique la méthode des multiplicateurs de Lagrange; on trouve donc:

(8.10) $N_{|\mathbf{f}|} = A_\sigma \exp\{-\beta_\sigma |\mathbf{f}|\}$

Où $A_\sigma$ et $\beta_\sigma$ sont à déterminer grâce aux contraintes (8.9.a & c). Posant $\sigma'_{11}+\sigma'_{22}+\sigma'_{33}=\mathrm{Tr}(\boldsymbol{\sigma}')$ et la densité n de contacts, *i.e.* n=N/V, on obtient la distribution:

(8.11) $p(|\mathbf{f}|) = N_{|\mathbf{f}|}/N = [nD\langle\cos\theta\rangle/\mathrm{Tr}(\boldsymbol{\sigma}')] \exp\{-n\,|\mathbf{f}|\, D\langle\cos\theta\rangle/\mathrm{Tr}(\boldsymbol{\sigma}')\}$

***Discussion:***

♣ La forme de la distribution obtenue, *i.e.* distribution exponentielle, est en accord avec les résultats expérimentaux (Gherbi *et al.* 1993, Bagi 1993) et les simulations numériques (Tkachenko & Witten 2000, Radjai *et al.* 1997 & 1998). Ce calcul donne la même distribution que le modèle scalaire de Liu et al. (1995), mais il est beaucoup plus simple.

♣ L'Eq. (8.11) resterait valable si tous les contacts étaient à la limite de glissement; dans ce cas, on devrait remplacer $\langle\cos\theta\rangle$ par $\cos\varphi$, où $\varphi$ est l'angle de frottement solide.

♣ Bien sur, l'approche que nous venons de développer est simpliste, car elle ne prend pas en compte un certain nombre d'effets: elle ne tient pas compte de l'histoire du matériau qui peut le rendre inhomogène,…, qui peut faire dépendre $\langle\cos\theta\rangle$ de $|\mathbf{f}|$ .

♣ Plus ennuyeux est le fait que l'on ne tient pas compte de la dimension de l'espace. En effet, dans le cas d'un gaz, la distribution exponentielle doit être modifiée pour prendre en compte la dégénérescence des états possibles ayant un même module de vitesse v. C'est pourquoi, la distribution p(v)dv des vitesses dans un gaz varie en fait comme $dv\ v^2 \exp(-mv^2/kT)$. De la même façon ici, la densité des forces doit dépendre aussi de la dégénérescence d'orientation de **f**; cependant cette dégénérescence est liée en partie à l'angle que fait **f** avec la normale **n** au contact; sa variation peut donc être plus complexe que dans le cas d'un gaz. Cependant, on s'attend que le terme exponentiel domine la variation de $p(|\mathbf{f}|)$ aux grandes valeurs de $|\mathbf{f}|$. C'est pourquoi la loi exponentielle ne gère bien que la queue de la distribution, la distribution aux faibles valeurs de $|\mathbf{f}|$ pouvant être modifiée par un terme $|\mathbf{f}|^{d-1}$, où d est la dimension de l'espace , *i.e.* d=2 ou 3.

♣ De même on ne prend pas en compte l'existence de corrélations locales entre les forces; mais ce dernier point ne devrait pas être important dans notre approche comme nous allons le montrer: en effet, la sommation sur les contacts est prise dans un un grand volume et elle ne fait intervenir que des forces simples et non des produits de plusieurs forces. Les termes sommés ne sont donc que des termes à "un corps" et non à "plusieurs corps"; dans ces conditions, la sommation ne fait pas intervenir de corrélations entre forces appliquées à deux points différents; ainsi dès que le volume test est plus grand que la longueur de corrélation la moyenne ne doit pas dépendre de corrélations locales. On peut donc conclure que le modèle proposé doit être valable dès que la longueur de corrélation entre les forces n'est pas infinie. Dans le cas contraire, où cette longueur devient infini, on devrait aussi pouvoir observer une sorte d'ordre global.

♣ De toutes les façons, comme nous le rappelions dans le début de ce chapitre, il semble qu'une queue de distribution exponentielle est un résultat robuste, qui provient de l'existence d'une discrétisation associée (i) à un processus générant un désordre maximum et (ii) à l'existence d'un certain nombre de conditions, *i.e.* contraintes, dont l'une gère la valeur moyenne des événements.

♣ Nous avons vu qu'une distribution réelle différente de la loi exponentielle pour des valeurs de $|\mathbf{f}|$ de inférieures à la valeur moyenne peut s'expliquer par une densité anormal d'états dégénérés. Cependant, il faut citer une autre explication potentielle proposée récemment par O'Hern *et al.* (2000): elle peut être aussi lieu par une transition de blocage (ou d'embouteillage, *i.e.* jamming en anglais); cette transition serait analogue à la transition vitreuse qui bloque les molécules d'un liquide.

***Remarque:***

Enfin, il est important de faire la remarque suivante pour démontrer la pertinence de la notion de température granulaire, et d'associer cette température granulaire à la contrainte moyenne. Pour cela, il est bon de rappeler que la pression dans un gaz est donné par la quantité de mouvement $\Delta(mv)/\delta t$ transportée par les molécules qui traverse une surface $\delta S$ par unité de temps $\delta t$. Comme le nombre de molécules traversant la surface dS pendant dt est $\Sigma_v\ (n(\mathbf{v}))(\boldsymbol{\delta S}\bullet\mathbf{v}\delta t)$, où $n(\mathbf{v})$ est la densité de molécules ayant la vitesse $\mathbf{v}$, la quantité de mouvement recherchée est (Bruhat 1968):

$$p = \langle mv^2/2 \rangle = (N/V)kT \qquad (8.12)$$

On trouve ainsi l'équation du gaz parfait. Ainsi, dans le gaz parfait, il y a équivalence entre pression est température; dans le cas du milieu granulaire, la pression nous sert à définir la température $T_\sigma$ des forces granulaires.

## 8. 3. Distribution des vides dans un milieu granulaire:

On peut utiliser la même procédure, avec des multiplicateurs de Lagrange, pour déterminer la distribution des vides dans un milieu granulaire sec. C'est ce que nous allons faire maintenant. On rappelle tout d'abord les différentes définitions : l'indice des vides $e=V_{vides}/V_{solide}$ est le rapport entre le volume occupé par les pores $V_{vides}$ et le volume occupé par le solide ; la porosité $\phi = V_{vides}/V_{tot} = V_{vides}/(V_{vides}+V_{solide})=e/(1+e)$; le volume spécifique $v=1/\rho$ est l'inverse de la densité; c'est le volume total divisé par le produit de la densidé du solide $\rho_s$ par le volume de solide: $v=(1+e)/\rho_s= (V_{vides}+V_{solide})/(\rho_s V_{solide})$.
Considérons un matériau granulaire avec des grains rigides, en l'absence de toute cohésion. Dans ce cas il contient des trous et la densité de son empilement peut se caractériser par l'indice des vides e. De plus, pour cet ensemble de grains, il existe une densité maximum qui se caractérise par $e_{min}$, et une densité minimum caractérisée par $e_{max}$,; $e_{max}$ doit exister puisqu'on a supposé qu"il n'y a pas de cohésion. $e_{min}$ et $e_{max}$ dépendent certainement de la distribution granulométrique et de la forme des grains. De plus, il est possible que la configuration permettant la fabrication de $e_{min}$ ne soit pas unique; dans ce cas il faudrait en tenir compte dans l'approche suivante et développer une approche avec un nombre de complexion variable pour $e_{min}$. Nous n'envisagerons pas cette hypothèse ici.
Toutes les densités intermédiaires sont a priori possible $e_{min}<e<e_{max}$ ; à une densité intermédiaire e, $e_{min}<e < e_{max}$,le matériau présente un désordre topologique lié à l'introduction de trous plus ou moins grands dans l'empilement le plus dense. Considérons maintenant ce réseau de trous nouveaux comme des défauts locaux de taille variable; appelons ces défauts des "trous" et caractérisons les par leur volume $\varpi$. Cherchons à évaluer la distribution $N_\varpi$. du nombre de trous dans le volume V. Si nous nous plaçons dans l'ensemble microcanonique, le nombre total de trous est $N=\Sigma N_\varpi$ et le volume spécifique du matériau est fixé; on a donc aussi: $(e-e_{min})V_{sol}=\Sigma\, \varpi N_\varpi$ qui est fixé.
Ainsi, en appliquant un raisonnement similaire à ceux que nous venons de développer dans les sections 8.1 et 8.2, on trouve que l'optimisation du nombre de complexion, subordonnée à la satisfaction des deux contraintes v fixé et N fixé, impose la distribution exponentielle :

(8.13) $$p(\varpi) = N_\varpi/N = A\exp\{-\beta_\varpi\varpi)$$

où $\alpha$ (tel que $A=\exp[\alpha]$) et $\beta_\varpi$ sont les deux multiplicateurs de Lagrange. $1/\beta_\varpi=\varpi_v$ définit la taille moyenne des trous $\varpi_v$. Jusqu'à présent nous n'avons pas introduit de taille typique microscopique; cette taille doit être fixée pour définir la taille typique des défauts. On peut donc introduire le volume $v_g$ d'un grain comme taille typique, car la taille d'un pore est très souvent plus petite que la taille d'un grain. On détermine A et $\beta_\varpi$ en imposant à l'Eq. (8.11) de vérifier les deux contraintes $N=\Sigma\, N_\varpi$ et $(e-e_{min}) V_{solide}=\Sigma\, \varpi N_\varpi$ . On trouve:

(8.14.a) $$p(\varpi) = [(e-e_{min})v_g)]^{-1} \exp\{-\varpi/[(e-e_{min})v_g)]\}$$

avec

(8.14.b) $$\varpi_v=(e-e_{min})v_g$$

La distribution des trous suit donc une loi exponentielle; c'est ce qu'ont supposé Boutreux & de Gennes (1997) dans leur calcul pour modéliser la densification des milieux granulaires par le tap-tap. Ce modèle sera décrit un peu plus loin. Pour l'instant nous allons nous inspirer de leur approche pour chercher à expliquer la variation de l'indice des vides des états normalement consolidés en fonction de la pression $\sigma'=p'=\mathrm{Trace}(\boldsymbol{\sigma}')/3$.

***Remarque:***
Si le nombre de complexion de l'état $e_{min}$ était dégénéré, on pourrait probablement rendre compte de cette dégénérescence par l'introduction d'un nombre supplémentaire de défauts ou "trous"; le matériau étant désordonné, c'est trous devraient avoir a priori des tailles différentes, distribuées d'une certaine manière, de manière à vérifier les même contraintes que précédemment. Il est donc possible que la configuration $e_{min}$ puisse elle aussi être caractérisée par une température finie $T_{v_{min}}$ . Dans un tel cas $\varpi_v$ représenterait un écart de température $\Delta T_v$ et non une température absolue.

## 8.4. Equations des états normalement consolidés et des états critiques (Evesque 1999b)

Dans cette partie, et dans un premier temps, on travaillera par analogie et on utilisera l'approche proposée par Boutreux et de Gennes (1997) pour la modélisation de la densification au tap-tap ; cela permettra de proposer un modèle microscopique donnant l'équation des états normalement consolidés et des états critiques.
Cependant, on montrera que ce modèle ne peut s'étendre facilement au cas des argiles lâches, pour lesquelles les variations de volume sont très grandes. Ainsi, puisque les argiles et les sables obéissent aux mêmes lois de variations du volume spécifique en fonction de la pression $p'=\sigma'$ (Eq. 3.9 ou 4.1), on sera amené à chercher une autre explication. Celle-ci sera basée sur l'analyse dimensionnelle. Les paramètres de la nouvelle solution seront alors identifiés à ceux de la loi précédente de manière à interpréter leur provenance physique.

### *8.4.1. Approche "mécanique statistique"*

On constate expérimentalement (Eqs. 3.9 ou 4.1) que lorsqu'on impose un accroissement de pression $\delta\sigma'$ à un état normalement consolidé, celui-ci se densifie spontanément : $\delta v<0$ ou $\delta e<0$. Ceci peut s'exprimer en d'autres termes: lorsqu'on augmente la pression de $\delta\sigma'$, il y a (i) disparition de quelques trous et (ii) réarrangement de la distribution des trous.
On va maintenant proposer une modélisation simple de ce phénomène à partir de considérations entropiques et statistiques à l'échelle microscopique. On suivra pour cela un raisonnement analogue à celui proposé par Boutreux et de Gennes (1997) pour modéliser la densification dans une expérience de tap-tap.
Pour cela, on utilise les résultats de la section précédente décrivant la statistique des trous dans un milieu granulaire. De ce fait, on peut considérer que les fluctuations locales de densité sont associées à une distribution $p(\varpi)$ de trous de différentes taille $\varpi$, dont la taille moyenne est $\varpi_v$ ; $\varpi_v$ est relié à la densité moyenne 1/v de l'échantillon et à la taille moyenne des grains (Equation 8.14):

$$(8.14) \qquad p(\varpi)= (1/\varpi_v) \exp-(\varpi/\varpi_v)$$

où $\varpi_v =(e-e_{min})v_g$ est le paramètre qui caractérise la loi de Poisson.
A l'instant initial, la densité initiale est $\rho_o=\rho_s v_g/v_o$, et la taille moyenne des trous est donc $\varpi_{vo}$. On sait que le système se contracte lorsqu'on impose un accroissement de pression $\delta p'=\delta\sigma'$; c'est une contraction irréversible qui provient du fait que certains grains perdent leur stabilité et viennent détruire des trous lorsqu'on déforme le matériau. Les trous détruits de manière irréversible dans ce processus sont essentiellement des trous de grande taille, car les trous de petite taille, s'ils peuvent être détruits lors de la déformation, peuvent aussi être créés lors de la destruction d'un grand trou. L'évolution de la distribution des trous est donc un processus complexe; cependant, elle doit obéir à une loi simple, qui privilégie un désordre statistique; pour tenir compte de ce fait, nous considérerons donc que la distribution statistique des trous reste une distribution exponentielle au cours de l'évolution du matériau. En d'autres termes, nous supposerons par la suite i) que la partie irréversible de la déformation est liée à la disparition irréversible des grands trous ; cependant, nous supposerons aussi que ii) la forme de la loi statistique reste la même lors de ce processus de densification et qu'elle reste une loi de Poisson, *i.e.* $p(\varpi)= (1/\varpi_v) \exp(-\varpi/\varpi_v)$ . Ainsi on suppose que l'on peut écrire :

$$(8.15.a) \qquad \delta\varpi_v= -\alpha' p(\varpi>\varpi_1)\delta \sigma'$$

avec

$$(8.15.b) \qquad p(\varpi>\varpi_1)= \int_{\varpi_1}^{\infty} (1/\varpi_v)\exp(-\varpi/\varpi_v)dv_t =\exp(-\varpi_1/\varpi_v)$$

Par hypothèse, on a aussi $\varpi_1\approx v_g$ , puisque la taille des trous considérés doit être plus grand que le volume d'un grain. Dans ces conditions la loi d'évolution s'écrit donc :

$$(8.16) \qquad \exp(v_g/\varpi_v) \delta\varpi_v= -\alpha' \delta\sigma'$$

$\varpi_v$ représente la taille moyenne des trous sous la contrainte $\sigma'$; il est relié à l'indice des vides moyen e par l'Eq. (8.14.b)

$$\varpi_v=(e-e_{min})v_g$$

On peut donc écrire

$$(8.17) \qquad \exp[1/(e-e_{min})] de=-(\alpha'/v_g) d\sigma'$$

Le terme exponentiel doit être grand et doit varier vite quand e varie. Dans le cas d'un sable ce n'est pas étonnant car l'indice des vides varie peu et la probabilité de trouver un trou de la taille de $v_g=d^3$ est petite. Dans ces conditions, on peut supposer que la variation de e est faible et écrire $e=e_o+\delta e$ et $de=d(\delta e)$. Ceci donne:

de $\exp[1/(e-e_{min})]=-(\alpha'/v_g)d\sigma' \approx d(\delta e) \exp[1/(e_o-e_{min})] \exp(\delta e/(e_o-e_{min})^2]$

Si on pose $\beta'=(\alpha'/v_g)\exp[-1/(e_o-e_{min})]$, on tombe ainsi sur l'équation :

(8.18) $-(\beta')d\sigma'= d(\delta e) \exp[\delta e/(e_o-e_{min})^2]$

qui donne après intégration:

(8.19) $-(\beta')(\sigma'-\sigma'_o)= (e_o-e_{min})^2 \{\exp[\delta e/(e_o-e_{min})^2] -1\}$

Soit, en négligeant $\sigma'_o$ devant $\sigma'$ et 1 devant l'exponentielle:

(8.20.a) $\delta e \propto - \lambda \ln (\sigma')$

avec

(8.20.b) $\lambda=(e_o-e_{min})^2$

On obtient donc la dépendance recherchée. L'Eq. (8.20.b) décrit raisonnablement bien la variation de l'indice des vides (et du volume spécifique) des états normalement consolidés. Un raisonnement analogue conduit à l'équation de la ligne des états critiques, Eq. (4.9), ou aux lignes des états normalement consolidés sous déviateur de contrainte $\eta=q/\sigma'$ constant. Cependant, ces calculs ne permettent pas de placer ces lignes les unes par rapport aux autres; ils permettent par contre de pressentir que la pente $\lambda_\eta$ de toutes ces droites (dans le plan (v,ln($\sigma'$)) est la même, indépendante de $\eta$. Puisqu'elle est liée à l'amplitude de variation de e: $\lambda_\eta=(e_o-e_{min})^2$ .

♣ ***Test expérimental de $\lambda$:***
Si on prend un ensemble de billes identiques, on sait que la porosité $\phi$ peut varier de 0.45 à 0.26; ceci indique que e varie de 0.82 à 0.35, puisque $e=\phi/(1-\phi)$. L'application de l'Eq. (8.20.b) prévoit donc que $0.22<\lambda_{\text{billes identiques}}<0.8$, puisque $e_{min}$ est sûrement plus petit que 0.35 et sûrement plus grand que 0.
Si l'on essaie d'appliquer cette formule aux cas expérimentaux, on trouve souvent que la pente $\lambda$ est de l'ordre de 0.06; ceci correspond à $\lambda=(e_{max}-e_{min})^2=0.06$; soit à $e_{max}-e_{min}=0.25$ . Ceci correspond bien à ce que l'on observe expérimentalement, *cf.* Biarez & Hicher (1994).
Il est possible de trouver une loi un peu plus général pour l; en effet, la distribution de taille des trous $\varpi$ nécessite d'introduire un volume typique microscopique; nous avons choisi la taille d'un grain $v_g$, de telle sorte que $\varpi_v=(e-e_{min})v_g$ . De même, nous avons considéré que l'évolution irréversible de la distribution des trous était lié à la destruction de trous plus grand que $\varpi_1$, et nous avons posé $\varpi_1=v_g$. Or il est possible que ces deux tailles soient en fait légèrement différentes. Dans ce cas, *i.e.* $\varpi_1\neq v_g$, on trouve:

$$\lambda= (e-e_{min})^2 v_g/\varpi_1 .$$

♣ ***Remarques :*** Les approximations que l'on a faites dans le calcul précédent ne sont probablement pas correctes dans toute la gamme d'indice des vides susceptible d'être atteint. Il est donc intéressant de proposer une approche alternative plus générale. L'approche que nous allons proposer plus loin permettra de définir une équation différentielle pour la variation de l'indice des vides. On pourra choisir l'équation différentielle qui sera compatible avec le modèle micromécanique ci-dessus; ceci permettra d'élargir le domaine de variation de l'approche précédente.
De même, en théorie, l'approche qui vient d'être développée devrait pouvoir s'appliquer aux argiles, puisque leurs états normalement consolidés sont régis par la même loi de variation avec la contrainte moyenne $\sigma'$. En fait, le développement limité utilisé pour passer de l'Eq. (8.17) à l'Eq. (8.18) n'est pas acceptable dans ce cas, car les variations de e deviennent beaucoup trop importante. Pour que le modèle soit admissible, il faudrait considérer que la taille $v_g$ d'un grain typique varie aussi avec la pression $\sigma'$ dans le cas d'une argile. Cependant, cela nécessiterait de connaître la loi de variation de $v_g$ avec $\sigma'$, ce que nous ne connaissons pas.
Nous allons donc proposer maintenant une approche théorique macroscopique du phénomène de compaction, ce qui nous permettra aussi de consolider le modèle micro-mécanique.

### *8.4.2. Approche macroscopique: Apport de l'analyse dimensionnelle*

On peut chercher à résoudre la contradiction existant sur le comportement des argiles en proposant une autre base au problème précédent. Pour cela il faut raisonner à l'aide de l'analyse dimensionnelle. On constate tout d'abord qu'un sol normalement consolidé doit a priori se contracter sous l'influence d'un accroissement $\delta\sigma'$ de pression $\sigma'$. Ceci veut dire qu'il faut chercher une relation entre ces deux grandeurs.

Commençons d'abord par analyser le cas d'un sable: Il est clair que la déformation locale d'un grain a peu à voir avec les variations de volume observées; en d'autres termes, ce qui compte du point de vue physique, c'est la variation de densité, mais pas du tout la déformation locale, liée aux forces de contact. Ainsi, le module d'Young ne doit pas être considéré comme un paramètre important, ni tout autre grandeur physique caractérisant une propriété mécanique du matériau formant le grain. Pour caractériser les variations de volume de l'échantillon, il semble donc naturel de se tourner soit vers la variation de sa densité, soit vers celle de son indice des vides; en fait, la densité fait intervenir une propriété du matériau et il est donc préférable de sélectionner l'indice des vides e. De toutes les façons ces deux grandeurs sont sans dimension. Comme la pression est, elle, une grandeur avec une dimension, la relation qui relie la variation d'indice des vides aux variations de pression doit être de la forme :

$$(8.21) \qquad \delta e= k_o\, \delta\sigma'/\sigma' + \Sigma_{ij}\, k_{ij}\, (\delta\sigma'/\sigma')^i\, e^j + \ldots$$

Le premier terme du membre de droite correspond au développement à l'ordre le plus bas, les termes sous le signe $\Sigma$ sont des termes d'ordre plus élevé, qui font intervenir la non linéarité de la réponse avec la pression {termes $(\delta\sigma'/\sigma')^i$} et des interactions à plusieurs corps {termes $e^j$}. On voit mal quelle interprétation physique donner à ces termes successifs: tout d'abord les termes $(\delta\sigma'/\sigma')^i$ où i>1 correspondent à des variations du $2^{ème}$, $3^{ème}$ ordre et sont donc négligeables; par ailleurs, si les $k_{1j}$ (i=1,j ≠0) étaient non nuls, cela démontrerait l'existence de processus qui font intervenir plusieurs trous (processus a) ou plusieurs grains (processus b). Dans ce cas on s'attend à ce que le nombre de contacts par grains joue un rôle; celui-ci est de l'ordre de 3 à 5; ainsi cela laisse prévoir soit que $k_{1j}\neq 0$ pour j=0 ou j=1 (pour un mécanisme de compaction indépendant du nombre de contact) soit 3<j<5 (pour un mécanisme de compaction lié au nombre des contacts sur un grain).
La première solution $k_{ij}\neq 0$ pour j=0, i=0 est la plus simple; elle a l'avantage de ne tenir compte ni des forces de contacts, ni du nombre de voisin des grains. C'est celle-ci que nous privilégierons.
Si j=1, la solution prévoit une décroissance de e proportionnelle à $\sigma'^{\,a}$. Ce type de comportement correspond plutôt à l'écrasement d'une structure fractale; ce serait le cas des argiles étudiées par Allain et al. (1995 & 1996). Ces variations sont trop importantes pour concerner un empilement granulaire dense.
Dans certains cas, les grains peuvent se casser, et les phénomènes deviennent plus complexes. Ce ne sera pas le cas envisagé ici, mais cela est observé effectivement à très forte pression ($\sigma'$>100 MPa) sur les sables; le problème spécifique des argiles naturelles sera traité un peu plus loin. Celui des argiles très pures obtenues par sédimentation a déjà été mentionné plus haut; ces argiles présentent une structure fractale à courte échelle (*i.e.* à l'échelle d'un grain) et la taille même du grain varie avec la pression, car des grains fractals de dimension trop grande sont trop ténus pour supporter des pressions excessives. On trouve alors que la taille du grain varie comme une loi de puissance avec la pression ( *cf.* Allain *et al.* 1995 & 1996) . Nous ne traiterons pas ce cas , mais le lecteur intéressé pourra le traiter lui-même en modifiant en conséquence les équations d'évolution de l'indice des vides que nous établirons plus loin pour les argiles naturelles, *i.e.* Eqs. 8.23 à 8.24.

♣ ***Cas d'un sable:*** Pour traiter le cas d'un sable, on part donc de la forme la plus simple de l'Eq. (8.21), c'est-à-dire avec les conditions $k_{ij}=0$ si $i\neq 0$ et $j\neq 0$, $k_o\neq 0$, ($k_o<0$). L'Eq. (8.21) devient:

(8.22.a) $$\delta e= k_o\, \delta\sigma'/\sigma'$$

Elle a pour solution:

(8.22.b) $$e= e_o + k_o \ln(\sigma'/\sigma'_o)$$

En identifiant les Eqs. (8.22.b) et les Eqs. 3.9 ou 4.4 ou 4.9 terme à terme, on trouve $k_o=-\lambda$. On retrouve donc bien la forme générale de l'Eq. (8.20.a). On constate que l'Eq. (8.22.a) ne fait pas intervenir la taille du grain; elle est donc indépendante de cette taille.

♣ ***Cas des argiles naturelles:*** A priori l'Eq. (8.22) précédente ne tient pas compte d'une diminution possible du volume des grains. On constate en contre partie que l'Eq. (8.22) reste aussi valable pour une argile naturelle; on est donc amené à se demander si un modèle plus général, intégrant une loi d'évolution de la taille des grains, ne pourrait pas être construit et permettre de décrire des variations rapides de l'indice des vides.
Un tel modèle doit prendre en compte l'évolution réelle de la porosité du matériau. Il faut donc discerner entre la porosité des pores, $\phi_s$, c'est-à-dire la porosité structurelle des vides entre les grains, et celle $\phi_g$ des grains d'argile eux-mêmes. L'évolution de la porosité $\phi_s$ d'une argile est liée à l'indice des vides; elle doit être la même que celle d'un matériau granulaire; mais la porosité $\phi_g$ pourrait varier totalement différemment.
Si l'on décompose e en $e_s+e_g$, on aura donc:

(8.23.a) $$\delta e_s = k_s\, \delta\sigma'/\sigma'$$

si l'on suppose en outre que l'on peut écrire:

(8.23.b) $$\delta e_g = k_g\, \delta\sigma'/\sigma'$$

l'équation d'évolution de e s'écrit:

(8.23.a) $$\delta e = k_o\, \delta\sigma'/\sigma'$$

avec

(8.23.b) $$k_o = k_s + k_g$$

et l'évolution de e suit bien la loi:

(8.23.c) $$e = e_o + k_o \ln(\sigma'/\sigma'_o)$$

♣ ***Remarque:*** On peut utiliser le modèle micromécanique du § 8.4.1 pour justifier la forme de l'équation différentielle (Eqs 8.22 & 8.21) en faisant un développement limité pour calculer la variation $d\delta$ dans le modèle micromécanique précité. Mais ce D.L. n'est valable que dans une toute petite gamme $\Delta e$ de l'indice des vides. On se sert alors de l'approche macroscopique pour augmenter le domaine de validité de l'équation d'évolution de e.

♣ ***Estimation de $\lambda=-k_o$:***
On peut utiliser les résultats du § 8.4.1 pour trouver la valeur de $k_s$. On trouve ainsi:

(8.24) $$-k_s = \lambda = (e_o - e_{min})^2$$

### *8.4.3 Approche macroscopique: Effet d'une contrainte anisotrope:*

Les calculs que nous avons exposés n'ont pris en compte que la pression isotrope $\sigma'$ et aucun déviateur q.
Considérons maintenant une expérience à déviateur $q/\sigma'=\eta=c^{ste}$, pendant laquelle q et $\sigma'$ varient proportionnellement. Une telle expérience peut se caractériser par la variation de $\sigma'$, puisqu'on connaît $q=\eta\sigma'$ connaissant $\sigma'$. Ainsi, le raisonnement utilisé pour écrire l'Eq. (8.21) reste valable quand $\eta\neq 0$, ainsi que celui qui permet de passer de l'Eq. ( 8.21) à l'Eq. (8.22); la valeur de $k_o$ peut être différente et dépendre de $\eta$.

Supposons que la valeur de $k_o$ dépende de $\eta$, cela imposerait que la projection sur le plan {v,ln($\sigma$')} de ces deux variations se coupent en un point donné. L'espace des phases {v,q, ln($\sigma$')} étant 3d, ceci est donc possible. Cependant, cela voudrait dire que les états normalement consolidés pourraient être soit contractant, soit dilatant lorsqu'on augmente q. En effet, en appelant $e_1$ l'indice des vides pour lequel le croisement a lieu, les matériaux dont l'indice des vides e serait plus grand que $e_1$, $e>e_1$, devrait se contracter pour arriver à l'état critique, tandis que ceux tels que $e<e_1$ devrait se dilater, ou vice versa suivant l'évolution de $k_o$ en fonction de $\eta$ . Or nous avons vu que la dilatation d'un matériau lorsqu'on augmente q, indique un matériau instable. Il est donc très peu probable que cette situation puisse exister.
On en déduit donc que $k_o$ doit être indépendant de $\eta$.

Revenons maintenant sur la nature des trajectoires conduisant des états normalement consolidés vers les états critiques par augmentation de $\eta$. On sait depuis le paragraphe § 7.5 que du fait de l'isotropie du matériau initial le volume spécifique $v_{nc}$ doit varier dans le voisinage de $\eta$=0 comme:

(8.25.a) $$v_{nc}(\eta)\approx v_{nc}(0) - \lambda_d\,\eta^2/M'^2$$

Par ailleurs, les hypothèses du § 8.4.2 imposent que la loi de variation de $v_{nc}(\eta)$ doit être indépendante de la valeur de $\sigma$', et ne doit dépendre que de $\eta$. Ceci implique que l'on peut écrire:

(8.25.b) $$v_{nc}(\eta)=v_{nc}(0) + f(\eta)$$

Ces deux formulations sont donc compatibles, l'Eq. (8.25.a) étant la plus restrictive. On remarque enfin que l'Eq. (8.25.a) est bien compatible avec la loi observée expérimentalement (Eq. 4.1.a), et l'explique en partie:

(8.25.c) $$v_{nc}(\eta)=v_o - \lambda \ln(p') - \lambda_d \ln(1+\eta^2/M'^2)$$

## 8.5. Comportement sous chargement cyclique: modélisation de la densification spontanée

### *8.5.1. Le modèle de Boutreux & de Gennes pour la densification au tap-tap :*

Comme nous l'avons déjà dit, Boutreux & de Gennes (1997) ont proposé un modèle pour la densification d'un milieu granulaire par le tap-tap. Cet appareil est constitué par une éprouvette verticale contenant le milieu granulaire à caractériser; ce milieu est au départ à la densité $1/v_o=\rho_o$; l'expérience consiste à faire subir au milieu un certain nombre N de chocs verticaux en laissant tomber l'éprouvette d'une hauteur constante. On note la variation de volume spécifique v ou d'indice des vides e de l'échantillon à la fin de l'expérience.
Si l'on postule comme nous l'avons déjà fait dans § 8.4.1 (i) que le milieu granulaire se caractérise par un ensemble de défauts appelés trous, (ii) que la statistique $p(\varpi)$ de la taille $\varpi$ de ces trous est donnée par une loi exponentielle de taille moyenne $\varpi_v=(e-e_{min})v_g$ , (iii) que la densification est un processus irréversible contrôlé par la disparition des grands trous, (trous $\varpi > \varpi_1=v_g$ plus grands que la taille d'un grain $v_g$), (iv) que la disparition de ces grands trous modifie la taille moyenne $\varpi_v$ des trous, (v) tout en en gardant la forme exponentielle de la loi de distribution, on peut écrire en prenant le nombre N de tap comme une variable continu e(*cf.* § 8.4.1):

(8.26.a) $$d\varpi_v=-\alpha'' \, dN \int p(\varpi>\varpi_1)d\varpi = -\alpha'' \, dN \exp-(v_g/\varpi_v)$$

avec

(8.26.b) $$\varpi_v=(e-e_{min})v_g$$

Comme précédemment ceci se réécrit d'une façon analogue à l'Eq. (8.17):

(8.27) $$\exp[1/(e-e_{min})]de = - (\alpha''/v_g) \, dN$$

soit, en faisant le changement de variable $u=1/(e-e_{min})$

(8.28.a) $$u^{-2} \exp(u)du=(\alpha''/v_g) \, dN$$

et en intégrant par rapport à u et par rapport àN, en considérant que le terme $u^{-2}$ varie lentement par rapport au terme exponentiel:

(8.28.b) $$u_o^{-2} [\exp(u)-\exp(u_o)]=(\alpha''/v_g) \, N$$

soit en négligeant le terme $\exp(u_o)$ qui est petit devant exp(u):

(8.28.c) $$\exp(u)\approx(\alpha''/v_g)u_o^2 \, N = \exp[1/(e-e_{min})]$$

ou finalement:

(8.29.a) $$1/(e-e_{min}) \approx \ln(N)+\ln(N_o)$$

où l'on a posé:

(8.29.b) $$N_o=(\alpha''/v_g)u_o^2=\alpha''/[v_g(e_o-e_{min})^2]$$

On a donc trouvé grâce à cette méthode la loi de compaction d'un matériau granulaire par le tap-tap; elle donne la variation de l'indice des vides en fonction du nombre N de taps.

### *8.5.2. Variation du volume de l'échantillon en fonction du nombre de cycles dans un essai triaxial:*

Étudions maintenant la loi de compaction d'un milieu granulaire soumis à un cisaillement cyclique dans un appareil triaxial. Du point de vue expérimental, on sait que la densité de l'échantillon soumis à des sollicitations cycliques varie comme le logarithme du nombre n de cycles (Olivari 1973, 1975), *cf.* aussi Figs. 6.4.c & 6.7 du chapitre 6. Pour expliquer ce phénomène on peut tenter des approches similaires à celles proposées dans la section précédente (§ 8.5.1). Nous suivrons tout d'abord une approche "mécanique statistique" similaire à celle du § 8.5.1 , puis nous tenterons une analyse dimensionnelle, similaire à l'approche proposée au § 8.4.2.

♣ *Approche mécanique statistique:* Comme précédemment, on part de l'hypothèse que le milieu granulaire est désordonné, avec des trous et des fluctuations locales de densité. On caractérise ces fluctuations locales de densité par une distribution $p(\varpi)$ de trous de différentes taille $\varpi$, dont la taille moyenne, notée $\varpi_v$ , est reliée à la densité moyenne 1/v de l'échantillon et à la taille moyenne des grains:

(8.30.a) $$p(\varpi)= (1/\varpi_v) \exp-(\varpi/\varpi_v)$$

avec

(8.30.b) $$\varpi_v=(e-e_{min})v_g$$

Un traitement similaire au traitement de la section § 8.5.1 donne la loi de variation de e:

(8.31.a) $$\exp(v_g/\varpi_v)\, d\varpi_v = -\alpha\, dn$$

où $v_g = \varpi_1$ est la taille d'un grain, $\alpha$ un coefficient et n le nombre de cycles.

Si $v_g$ est grand devant $\varpi_v$, c'est à dire si les événements qui permettent la compaction sont peu probables, le terme exponentiel varie très rapidement ; il varie donc beaucoup plus rapidement que $d\varpi_v$. Notant $\varpi_{vo}$ la taille moyenne des trous de la distribution initiale, on peut alors écrire $\varpi_v = \varpi_{vo}+\delta\varpi_v$ . Un développement limité autour de $\varpi_{vo}$ donne:

$$\exp(\varpi_g/\varpi_v)= \exp\{ v_g/(\varpi_{vo}[1+\delta\varpi_v/\varpi_{vo}])\}=\exp\{(v_g/\varpi_{vo})[1- \delta\varpi_v/\varpi_{vo}+(\delta\varpi_v/\varpi_{vo})^2 -...]\}$$
$$=\exp(v_g/\varpi_{vo}) \exp(-v_g\delta\varpi_v/\varpi_{vo}^2)$$

la variation de $\varpi_v$ dans le terme exponentiel sera importante si $v_g\delta\varpi_v/\varpi_{vo}^2>1$ .
L'équation d'évolution de $\varpi_v$ devient : $\exp(v_g/\varpi_{vo}) \exp(-v_g\delta\varpi_v/\varpi_{vo}^2)\, d\varpi_v = -\alpha\, dn$ qu'on peut réécrire:

(8.32.a) $$\exp\{-\delta e/(e_o-e_{min})^2\}\, de = -\beta\, dn$$

avec

(8.32.a) $$\beta= (\alpha/v_g) \exp\{-(e_o-e_{min})^{-1}\}$$

compte tenu de l'Eq. (8.30). On va chercher une solution du type:

(8.33) $$\delta e=-\gamma \ln(n/n_o) \quad \text{soit } de = -(\gamma/n)\, dn$$

qui est compatible avec l'observation expérimentale puisque $\delta e$ est proportionnel à $\delta v$. On obtient ainsi en posant $\Delta=e_o-e_{min}$ et en combinant les Eqs. (8.32 & 8.33) :

(8.34) $$-\beta dn = de \exp\{(\gamma /\Delta^2)\ln(n/n_o)\}$$

(8.34) $-\beta dn = -(\gamma/n)\, dn \quad \exp\{\ln[(n/n_o)^{\delta}]$

avec

(8.35) $\delta = \gamma/\Delta^2 = \gamma/(e_o - e_{min})^2$

On trouve ainsi:

(8.36) $\beta\, \delta n = (\gamma/n)\,(n/n_o)^{\delta}\, dn$

qui impose en identifiant terme à terme:

(8.37) $\delta=1$ & $\beta = \gamma/n_o$

soit, d'après l'Eq. (8.35) et l'Eq. (8.32):

(8.38.a) $\gamma = \Delta^2 = (e_o - e_{min})^2$

(8.38.b) $n_o = \gamma/\beta = (e_o - e_{min})^2 \{ \exp\{(e_o - e_{min})^{-1}\}\} (v_g/\alpha)$

(8.38.c) $\delta e = -\gamma \ln(n/n_o)$

où e est l'indice des vides.

♣ *Remarques:*

- Ce modèle est un modèle qui postule une évolution irréversible du paramètre volumique mais qui impose une entropie maximale (le système est donc à désordre optimal ou à l'équilibre « thermodynamique »). Il évolue constamment vers une densité plus grande. Cependant, comme la variation de volume est logarithmique en $\ln(n_o)$, l'évolution ralentit après un certain nombre de cycles; a priori, plus l'amplitude du cycle est grande, plus les déformations qu'ils engendrent le sont aussi de telle sorte qu'il vaut mieux utiliser des amplitudes de cycles plus grandes pour arriver plus vite à de fortes densités .

- Cependant, on sait du point de vue expérimentale, que (i) l'amplitude du cycle de contrainte, (ii) la nature alternée ou répétée de ce cycle jouent des rôles importants et assez complexes dans le processus de compactage, *cf.* Figs. 6.1, 6.4, 6.5, 6.6, 6.7 du chapitre 6.

- Il est donc possible que la densité optimum dépende de l'amplitude des cycles (ou des vibrations) et varie aussi avec la nature du cycle : $v_\infty$=f(amplitude), *cf.* Fig. 6.4.c.

- Ce phénomène est probablement aussi observé dans le cas du tap-tap, où on atteint une courbe limite $\rho(\Gamma)$ réversible pour les accélérations $\Gamma$ très grandes ; de plus on observe que $\rho(\Gamma)$ décroît quand $\Gamma$ croît. Cependant, pour calculer $\rho(\Gamma)$ il faudrait faire un bilan détaillé (i) des trous détruits lors d'un cycle par l'effondrement des grains dans des trous trop grands et (ii) des petits trous créés de façon entropique par les mouvements cycliques, (iii) de la création d'une population de trous hors d'équilibre ; ceci ne peut être pris en compte dans l'état actuel de la théorie.

♣ *Approche macroscopique: apport de l'analyse dimensionnelle*

On peut utiliser un raisonnement analogue à celui explicité dans le § 8.4.2. On arrive ainsi à une équation reliant l'indice des vides e et le temps t du type:

(8.39) $$de=k'dt/t =k''dn/n$$

où n est le nombre de cycle. On obtient ainsi directement l'équation recherchée:

(8.40) $$e-e_o = v-v_o = k'' \ln(n)$$

Ce comportement est observé expérimentalement, *cf.* Fig. 6.8. On peut lui donner une interprétation microscopique en utilisant l'approche "mécanique statistique" proposée précédemment.

### 8.6. Les cycles sont-ils tous contractant?

Le modèles que nous venons de développer font intervenir une contraction du matériau à chaque cycle ou incrément de contrainte. En fait, on sait grâce aux études des lois macroscopiques que cela n'est pas vrai. Par exemple un cycle n'est contractant que s'il se trouve dans le domaine sous caractéristique, c'est à dire tel que $q/\sigma'=\eta<M'$. Dans le cas contraire il est dilatant, *cf.* Fig. 8.1.

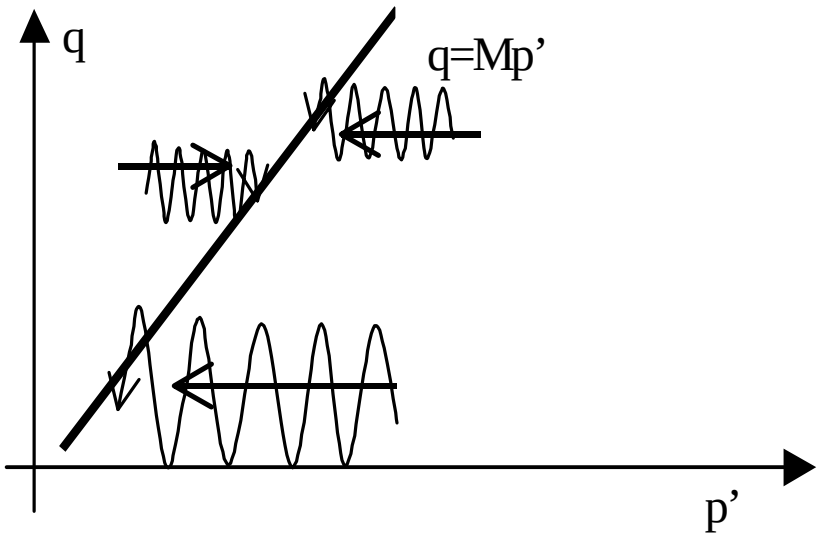

***Figure 8.1:*** *Comportement cyclique sous condition non drainée : le cycle de contrainte limite tend au bout d'un certain nombre de cycles à se centrer sur la droite caractéristique* $q=M'p'= M'\sigma'$. *Comme dans le cas de la Fig.* 6.6*, les cycles se décomposent en une compression et une extension uniaxiales.*

***Essai alterné :*** *Si on part à droite de l'état caractéristique, les cycles ont tendance à faire diminuer la pression inter-granulaire et à faire augmenter la pression* $u_w$ *de l'eau ; l'échantillon était relativement stable au départ, mais au bout d'un certain nombre de cycles, on obtient ainsi une mobilité cyclique. Si on part à gauche de l'état caractéristique, l'échantillon est potentiellement instable, mais la pression intergranulaire augmente au cours des cycles, ce qui stabilise l'échantillon.*

***Essai répété :*** *le cycle limite tend à se centrer sur le point* $q=p'=\sigma'=0$ *; les contraintes s'annulent périodiquement , la friction solide granulaire perd de sa force et le sable est liquéfié.*

Ainsi, les modèles que nous avons proposés dans les sections précédentes doivent être modifier quand on rentre dans le domaine surcaractéristique. De même, l'amplitude de la contraction dépendra du rapport $q/\sigma'=\eta$. Ceci veut dire que les coefficients $\alpha$, $\alpha'$ et $\alpha''$ , *cf.* Eqs. (8.16), (8.27) & (8.31), doivent dépendre de $\eta$-M', et ce linéairement pour qu'ils changent de signe pour $\eta$=M'.  Dans toutes ces équations, il faudrait remplacer $\alpha$, $\alpha'$, $\alpha''$, par $\alpha\eta$/M', $\alpha'\eta$/M', $\alpha''\eta$/M' respectivement.
Ce changement de signe peut expliquer en particulier le phénomène de dilatance observé dans les essais cycliques dans le domaine surcaractéristique ou dans les expériences de tap-tap, pour lesquelles on observe une augmentation du volume spécifique lorsque l'amplitude des chocs devient très importante.

## *Bibliographie :*

# 9. Conclusion

En guise de conclusion nous allons proposer une relecture de ce livre dans un ordre différent. Cela permettra de proposer de donner un éclairage nouveau à la mécanique des milieux granulaires et de montrer qu'elle est gérée par un petit nombre de principes et de concepts relativement simples. Nous espérons que cela donnera lieu au développement d'une nouvelle méthode, plus pédagogique et rationnelle, pour enseigner la mécanique des sols et des milieux granulaires quasi statiques.
Pour cela, nous limiterons l'exposé à des situations axi-symétriques.

## 9.1. Quelques remarques liminaires

***Le volume spécifique est une caractéristique mécanique du milieu granulaire:*** De toute évidence, le même milieu granulaire peut être compacté à différente densité ; et son état le plus lâche ne peut pas avoir une densité nulle. Pour un matériau donné, il existe donc une densité minimum et une densité maximum. L'état de densité minimum correspond à l'état normalement consolidé le plus lâche.
On peut soumettre un milieu granulaire à une contrainte. Sous l'action de cette contrainte, il se déforme.
***Contrainte effective :*** Lorsque le volume des pores est saturé par un fluide à la pression $u_w$, et que l'on reste en régime quasi statique, la contrainte totale exercée sur le milieu (grains + fluide) se décompose en deux, pression du fluide + contrainte intergranulaire, et seule la contrainte intergranulaire contrôle la mécanique du milieu granulaire.
***Les états normalement consolidés :*** Nous avons vu au chapitre 8 grâce à des arguments microscopiques, que la densité minimum devait dépendre de la pression et que la dépendance devait être logarithmique. Ainsi, pour une pression donnée, il existe un état le plus lâche qui a une densité minimum. Cet état correspond donc à l'état normalement consolidé sous pression isotrope.

## 9.2. La modélisation

***Espace des phases :*** Nous postulons maintenant que l'espace des phases caractérisant un matériau granulaire est l'espace à 3 dimensions (v,q,p.), où v est le volume spécifique, q le déviateur de contrainte et p' la pression intergranulaire moyenne. Compte tenu du paragraphe précédent, il vaut mieux choisir l'espace {v,ln(q), ln(p')}. Les états normalement consolidés sont donc une droite dans cet espace.
***Modélisation incrémentale :*** De toute évidence, un matériau granulaire se déforme lorsqu'il est soumis à une variation de contrainte. On postule que cette loi de variation est déterministe ; elle peut être non linéaire. On rend compte de cette loi par une formulation incrémentale : $\delta\varepsilon=g(\delta\sigma)$

L'hypothèse de réponse quasi statique impose que la fonction g est homogène de degré 1, car la réponse doit être indépendante de la vitesse de déformation.
On doit maintenant tenir compte de différentes propriétés :
♣ **états critiques :** Aux grandes déformations, la mécanique est contrôlée par un frottement solide d'angle φ; il fixe le rapport des contraintes q/p'=M' aux grandes déformations : M'=. Le volume spécifique $v_c$ du milieu n'évolue plus ; mais il dépend de p', de manière logarithmique ; c'est l'état critique. L'état critique a une densité inférieure à l'état normalement consolidé, mais sa dépendance en fonction de p' est la même que ce dernier. Ce point s'explique probablement par le modèle microscopique du chapitre 8.
♣ **hystérésis :** la réponse à un incrément de chargement et de déchargement ne sont pas identiques ; ceci implique que la fonction g n'est pas complètement linéaire ; elle peut être linéaire par zone, avec un minimum de deux zones. C'est ce que nous postulerons pour simplifier.
♣ **Existence de E et ν :** Dans la zone de charge, on postulera que la réponse est isotrope ; on la quantifiera par un pseudo module d'Young E et un pseudo coefficient de Poisson ν.
♣ **Loi de Rowe : ν=f(q/p') :** Expérimentalement, on trouve une relation entre la variation de volume spécifique $\partial v/\partial \varepsilon_1$ à contrainte $\sigma'_2=\sigma'_3=c^{ste}$ et le rapport des contraintes q/p' ; cette loi expérimentale s'appelle la loi de Rowe. On en déduit que le pseudo coefficient de Poisson dépend de q/p'.
♣ **E(q,p', v) :**
♣ **loi d'écrouissage 1 :** Effet de mémoire sur E car au déchargement E garde la même valeur qu'à la fin du chargement (réponse quasi élastique), quelque soit le Δq, tant qu'il n'y a pas de rotation des axes principaux, ni interversion des axes; le comportement cyclique, et pas d'effet sur n.
♣ **loi d'écrouissage 2 :** pas d'effet sur ν. Preuve : ν ne dépend que de q/p', même dans les cas de chargements cycliques.

### 9.3. Les prévisions :

On trouvera les prévisions du modèle détaillée dans le chapitre 7, ainsi que des comparaisons entre ces prévisions et les résultats typiques de mécanique des sols.

Ce modèle explique :
- ♥ la constante $K_o=\sigma'_3/\sigma'_1 \approx 1-\sin\varphi$ des terres au repos. Le fait que le rapport tend vers deux valeurs sans rapport l'une avec l'autre dans un essai oedométrique en charge ou en décharge.
- ♥ la trajectoire dans le plan (p',q) d'un essai à volume constant. L'existence de la trajectoire $p'=c^{ste}$ aux faibles q, puis de la bifurcation sur la droite des états caractéristiques.

Ce modèle permet de généraliser la notion d'état caractéristique.

Ce modèle reproduit les essais triaxiaux classiques (à $\sigma'_3=c^{ste}$).

Ce modèle prévoit la complémentarité des surfaces de Hvorslev et de Roscoe : ce sont deux parties de la même surface. Ceci est vérifiée expérimentalement.

Le modèle micro-mécanique prévoit

- ♣ la dépendance linéaire du volume spécifique de l'état critique $v_c$ et de l'état normalement consolidé $v_{nc}$ en ln(p').
- ♣ la dépendance linéaire du tassement en fonction du nombre de taps pour l'expérience du tap-tap.
- ♣ la dépendance linéaire de la densification pour des sollicitations cycliques drainés en fonction du nombre de cycles N

Ce modèle est compatible :

- ♦ avec les résultats expérimentaux sur les comportements cycliques drainés, non drainés, sur- et sous- caractéristiques.
- ♦ avec les principes de la liquéfaction

## 9.4. Les questions :

Pourquoi le pseudo coefficient de Poisson $\nu$ ne dépend-il que du rapport des contraintes q/p' ?

Pourquoi la réponse du matériau reste-t-elle si longtemps isotrope ?

Comment calculer le frottement solide mesuré, à partir des caractéristiques mécaniques des grains ?

Comment calculer le pseudo module d'Young E.

Le régime quasi statique est-il réellement quasi-statique, c'est-à-dire ne présentant pas de rupture d'équilibre, ou est-il un régime dynamique intermittant ?

Ainsi, la lecture des chapitres 7 et 8 devraient pouvoir se faire directement, et livrer au lecteur les caractéristiques principales de la mécanique des milieux granulaires.

# Appendice A1:
# Théorie des systèmes dynamiques :

## A1.1. Espace des phases :

Un système physique se caractérise par un ensemble de grandeurs physiques représentatives ; il est aussi contraint par un ensemble de paramètres que nous appellerons *paramètres de contrôle*. L'évolution de ce système sera donc caractérisée par l'évolution de cet ensemble de grandeurs physiques $X_i$, compte tenu des paramètres de contrôle. L'ensemble minimum de ces grandeurs forment un espace $(\ldots,X_i,\ldots)$ qu'on appelle espace des phases. Un point dans cet espace caractérise donc l'état du système au moment considéré ainsi que son évolution instantanée. L'évolution de ce système pourra se représenter par une trajectoire dans cet espace des phases, et cette trajectoire caractérisera complètement l'évolution.
D'un point de vue pratique, si l'espace des phases est à n dimension et que le système est déterministe, l'évolution du système est définie par une série de n équations différentielles du premier ordre couplées :

(A1.1a) $$dX_i/dt=f_i(\ldots,X_j,\ldots)$$

Les $(\ldots,X_i,\ldots)$ peuvent être écrits sous forme vectorielle. On notera **X** le vecteur de composantes $(\ldots,X_i,\ldots)$ ; l'Eq. (A1.1a) s'écrit ainsi sous la forme condensée suivante:

(A1.1b) $$d\mathbf{X}/dt=\mathbf{f}(\mathbf{X})$$

On distingue les systèmes *discrets* des systèmes *répartis*. Les premiers sont définis par un ensemble fini de scalaires $X_i$ . Dans les systèmes *répartis,* les $X_i$ représentent des champs variant dans l'espace et les fonctions f couplent ces champs entre eux et leurs variations spatiales.
D'après les équations (A1.1), l'évolution du système est totalement déterminée par la connaissance des $X_i$ et des conditions initiales; il résulte de ceci que si les fonctions f ne dépendent pas explicitement du temps, deux trajectoires dans l'espace ne peuvent en général pas se couper. C'est pourquoi les systèmes sont appelés déterministes. Dans le cas contraire, i.e. si **f** dépend du temps, on doit rajouter explicitement le temps à l'espace des phases pour conserver cette propriété des trajectoires ; ceci augmente la dimension de l'espace des phases d'une unité.
Le système est dit *autonome,* si **f** ne dépend pas du temps.

## A1.2. Quelques exemples :

*Particule dans un champ de force :* Un des exemples les plus simples est le problème unidimensionnel d'une particule libre de masse m dans un champ de force stationnaire donné (masse fixée à un ressort par exemple). Lorsque le champ de force ne dépend que de la position et de la vitesse de la particule, l'évolution de ce système se caractérise à tout instant par la donnée de la vitesse v et de la position x de la masse, ou mieux par son impulsion p=mv et sa position x. Si on ne connaît pas le champ de force, l'étude de l'évolution du système permettra de le déterminer ; en revanche, si ce champ de force est totalement connu (comme dans le cas d'une masse liée à un ressort), la trajectoire de la particule pourra se calculer connaissant les coordonnées du point initial dans l'espace des phases.
Les équations du mouvement sont décrits dans le formalisme de Hamilton par le hamiltonien $\boldsymbol{H}=p^2/2m+kx^2/2$ où k est la raideur du ressort, auquel il faut adjoindre les équations d'évolution de Hamilton:

(A1.2) $dx/dt=\partial \boldsymbol{H}/\partial p$ & $dp/dt=-\partial \boldsymbol{H}/\partial x$

Un calcul simple montre que les trajectoires dans l'espace des phases (x,p) sont des cercles.
Si la raideur du ressort est non linéaire, la dynamique peut se compliquer; en particulier elle peut engendrer l'existence de deux puits si le hamiltonien s'écrit $\boldsymbol{H}=p^2/(2m)+k(x^2/2-\alpha x^4/4)$ avec $\alpha$ positif.
Un autre exemple est celui d'une masse au sommet d'une colline, pour laquelle la dynamique est décrite par $\boldsymbol{H}=p^2/2m-kx^2/2$. Dans ce cas les trajectoires sont des hyperboles.

*Réseaux électriques :* D'autres exemples peuvent être pris dans d'autres domaines de la physique . L'analogie entre les problèmes mécaniques vibratoires et les réseaux électriques est bien connue; la dynamique des réseaux électriques est une source d'illustration sans fin de nos propos. Là aussi, l'évolution du potentiel électrique $V_i$ en chaque nœud i et du courant $I_{ij}$ en chaque bras peut se mettre sous la forme d'une série d'équations différentielles du premier ordre ; le nombre maximale des équations différentielles indépendantes qui gèrent l'évolution d'un réseau est une caractéristique de ce réseau ; c'est la dimension de l'espace des phases de ce réseau.

*Gaz de particules :* La dimensionalité de l'espace des phases d'un gaz de N particules indépendantes est 6N : 3N qui correspondent aux 3N coordonnées des particules et 3N correspondent à leur vitesse. Cependant ce nombre de paramètres indépendants peut être réduit dans certains cas : si l'on suppose que ce gaz est homogène et en équilibre thermodynamique, et que le nombre de particules N est grand, la physique de cet ensemble est gouvernée par un ensemble de deux paramètres qui peuvent être la pression et la température par exemple , la densité moyenne 1/v se déduisant de ces deux grandeurs par l'équation d'état pv=kT.
Si le gaz reste homogène, mais que son état est perturbé, son évolution sera caractérisé par l'évolution de deux grandeurs seulement, p etT par exemple; il faudra connaître les équations qui gouvernent $\partial p/\partial t$ et $\partial T/\partial t$. S'il ne reste pas homogène, il faudra introduire des champs p(r) et T(r).

## A1.3. Points fixes :

Les *points fixes* sont des points de l'espace des phases qui n'évoluent pas; ils correspondent donc à $\partial X_i/\partial t=0$ quelque soit i. Ils sont obtenus en résolvant l'Eq. (A1.2) où le membre de gauche de l'égalité est 0, soit:

$$0=f_i(\ldots,X_j^{(\text{point fixe})},\ldots) \qquad \text{(A1.3a)}$$

Comme les fonctions f sont en général non linéaires, cette série d'équations présentent plusieurs solutions que nous noterons $\mathbf{X}^{(k)} = (..., X_i^{(k)} ,...)$. Il existe deux types de points fixes, les points fixes stables et ceux qui sont instables. Les premiers sont tels que si l'on décale la position du système d'un incrément infinitésimal $\mathbf{X}-\mathbf{X}^{(k)}=\boldsymbol{\delta}\mathbf{X}= (...,\delta X_i^{(k)},...)$, la dynamique du système cherchera à rapprocher ce dernier du point fixe, c'est à dire à faire diminuer la norme $|\boldsymbol{\delta}\mathbf{X}|$. Les points instables sont tels que $|\boldsymbol{\delta}\mathbf{X}|$ croît spontanément. En fait, la dynamique est en général moins simple que cela, car chacune des directions $X_i$ peut avoir deux comportements différents comme nous allons le montrer.

En effet, pour étudier la dynamique au voisinage d'un point quelconque, et en particulier au voisinage de ces points fixes, on se ramène en général à linéariser le problème; on écrit ainsi les Eq. (A1.2) sous la forme:

$$d(\delta X_i)/dt=f_i(...,X_j+\delta X_j,...) = f_i(...,X^{(k)}{}_j,...) + \sum_j \delta X_j \partial f_i/\partial X_j + \ldots = \sum_j \delta X_j \partial f_i/\partial X_j$$

soit:

$$d(\underline{\boldsymbol{\delta}\mathbf{X}})/dt = \underline{\underline{\mathbf{grad}}}_x(\underline{\mathbf{f}}) \bullet \underline{\boldsymbol{\delta}\mathbf{X}} \qquad \text{(A1.3b)}$$

On tombe donc sur une équation linéaire si on se limite au premier ordre de perturbation. On peut montrer que la solution de A1.3b s'écrit formellement $\boldsymbol{\delta}\mathbf{X}(t)=U_t\,\boldsymbol{\delta}\mathbf{X}(0)$ ou l'opérateur d'évolution $U_t$ est donné par $U_t=\exp[\text{grad}_x(f)t]$ . Par exemple, on peut souvent diagonaliser $\underline{\underline{\mathbf{grad}}}_x(\underline{\mathbf{f}})$ ; on trouve ainsi les directions propres et les valeurs propres de la dynamique; il résulte ainsi qu'à chacune des valeurs propres est associée une direction propre; si cette valeur propre est positive l'écart dans cette direction croît exponentiellement; dans le cas contraire, il décroît. Le point fixe est donc instable dans les directions propres présentant des valeurs propres positives, et stable dans les directions caractérisées par une valeur propre négative. Il peut apparaître aussi des valeurs propres complexes, ce qui montre que certaines directions ont des dynamiques couplées, de type spirale…. Nous donnons un exemple au § A1.11.5.

En fait l'approximation linéaire (A1.3b) reste valable pour les modes instables tant que l'amplitude de $\boldsymbol{\delta}\mathbf{X}$ reste petite. Lorsqu'ils deviennent trop grands les développements aux deuxième et troisième ordre deviennent nécessaires, ce qui couple les modes entre eux ainsi que leur évolution.

Un exemple de point fixe totalement stable est celui d'une masse au fond d'une vallée avec dissipation d'énergie; un exemple de point fixe totalement instable est celui d'une masse au sommet d'une colline; un exemple de point partiellement stable et partiellement instable est celui d'un point col.

### A1.4. Paramètres de contrôle :

Dans certains cas expérimentaux, certains degrés de liberté du système sont gelés, c'est-à-dire qu'ils sont maintenus constants pendant l'expérience ; c'est par exemple le cas pour l'énergie totale d'une particule isolée. C'est aussi le cas si on étudie comportement isotherme d'un gaz parfait ainsi que ses transformations isothermes : dans ce cas la température $T_o$ du gaz est fixée et l'évolution du système est un problème à 1 dimension qui se caractérise par l'évolution de la pression p(t) au cours de temps. Un degré de liberté gelé est appelé aussi un paramètre de contrôle. Ainsi les conditions de fonctionnement réel du système seront spécifiées par un ensemble de paramètres de contrôle.

### A1.5. Systèmes hamiltoniens - systèmes dissipatifs :

Un système est dit hamiltonien ou conservatif lorsqu'il conserve l'énergie totale ; il est dit dissipatif dans le cas contraire. Lorsqu'il est conservatif la fonction de Hamilton $\boldsymbol{H}(\ldots,p_i,q_i,\ldots)$ existe et permet de décrire l'évolution de la manière suivante :

$$\text{(A1.2)} \qquad dq_i/dt=\partial\boldsymbol{H}/\partial p_i \quad \& \quad dp_i/dt=-\partial\boldsymbol{H}/\partial q_i$$

Considérons un élément $\Omega(t=t_o)$ de l'espace des phases à l'instant initial ; il représente les conditions initiales d'un certain nombre de points ; chaque point va suivre une trajectoire spécifique au cours du temps ; un instant plus tard cet élément de volume $\Omega(t)$ aura évolué et se sera déformé; mais son volume v restera constant ; ceci est le **théorème de Liouville** qui résulte de la nullité de la divergence de d**X**/dt :

$$\text{(A1.4)} \quad dv/dt = \int_{\Omega(t)} \mathrm{div}(d\mathbf{X}/dt)dv=\int_{\Omega(t)} \Sigma_i(\partial/\partial q_i\bullet\partial\boldsymbol{H}/\partial p_i - \partial/\partial p_i\bullet\partial\boldsymbol{H}/\partial q_i)dv =0$$

La constance de l'énergie impose une relation entre les coordonnées de l'espace des phases relatives à l'évolution d'un système donné. La trajectoire d'un système non dissipatif appartient donc à une surface de dimension n-1 ; celle-ci se réduit à une courbe lorsque l'espace des phases est à deux dimensions (cas d'une particule dans un champ par exemple).

L'évolution d'un système peut obéir à plusieurs invariants à la fois (i.e. plusieurs relations de conservation); chacun d'eux introduit des relations supplémentaires entre les coordonnées réduisant d'autant la dimension de la surface contenant la trajectoire : s'il y a k relations de conservations distinctes, la surface contenant la trajectoire est de dimension n-k. Comme exemple, on peut citer le cas d'un ensemble isolé de particules en interaction , pour lequel il y a conservation de l'impulsion totale et du moment cinétique total (l'énergie n'est conservée que si les collisions sont élastiques). Il se peut que chacun de ces invariants agissent sur des groupes distinctes de variables; on peut alors séparer les équations de la dynamique, ce qui facilite l'analyse. On dit que le système est séparable.

Si les conditions initiales d'un système conservatif sont connues avec une certaine précision, on peut appliquer le théorème de Liouville pour conclure que le volume total de l'incertitude reste constant. Ceci dit, ce fait cache en général une réalité plus complexe lorsque l'évolution du système est non linéaire, car l'incertitude élémentaire peut croître dans une direction et décroître dans une autre; (nous avons déjà vu un phénomène similaire quand nous avons établi les Eq. (A1.3)); ainsi, si le processus de déformation se répète dans le temps, l'incertitude croît exponentiellement dans certaines directions, et décroît dans d'autres ; le coefficient caractérisant cette croissance (ou décroissance) est appelé **coefficient de Liapounov**.
Si on laisse se poursuivre l'évolution , le phénomène se complique encore en général: car l'évolution d'un système non linéaire peut être telle que les trajectoires se replient sur elles-mêmes. Après un grand nombre de repliement, on a l'équivalent d'un processus de pâte feuilletée . Cette analogie très imagée est plus que formelle, car le volume d'un élément de matière de la pâte reste constant au cours au temps : considérons par exemple une pâte de volume total V et considérons les points de cette pâte, localisés à t=0 dans un volume $\delta v$ autour d'un point x ; lorsqu'on commence à pétrir la pâte, celle-ci s'aplatit dans une direction ; la taille de l'élément considéré décroît dans une direction et croît dans les autres à cause de la conservation du volume ; puis vient le processus de repliement ; ce processus laisse le volume $\delta v$ invariant, mais crée des strates distinctes de telle sorte qu'il deviendra presque impossible de préciser la position finale d'un point connaissant sa position initiale au fur et à mesure des repliements.

La conservation de l'élément de volume de l'espace des phases au cours de l'évolution est une conséquence de la conservation de l'énergie. Elle n'est donc plus réalisée dans le cas des systèmes dissipatifs ; le volume élémentaire représentatif de départ peut alors soit diminuer, soit augmenter au cours du temps. Par exemple, l'évolution d'un oscillateur mécanique linéaire s'écrit :

(A1.5a) $dq/dt=p/m$

(A1.5b) $dp/dt=-kq$

dans le cas où il y a conservation de l'énergie. Si l'oscillateur est soumis à des pertes visqueuses, ces équations deviennent :

(A1.5c) $dq/dt=p/m$

(A1.5d) $dp/dt=-kq-\gamma p$

Les trajectoires sont des cercles s'il n'y a pas de pertes ; elles sont des spirales qui convergent vers 0 si les pertes sont positives ($\gamma>0$) et des spirales divergeant à l'infini si $\gamma$ est négatif. On vérifie sur cet exemple que le point fixe est bien le point définit par q=0, p=0.
Si l'oscillateur est couplé à une force extérieure périodique f(t), il faut rajouter f dans le membre de droite de l'Eq. (A1.5d) et on obtient un cycle limite.

### A1.6 Remarque à propos des systèmes conservatifs :

Comme nous l'avons vu, un système isolé de particules a une dynamique réversible par rapport au temps qui se décrit à l'aide du formalisme de Lagrange ou par celui de Hamilton. Ce dernier est basé sur l'existence de la fonction de Hamilton ***H*** qui représente l'énergie du système. ***H*** peut dépendre explicitement du temps ou non.
Dans le formalisme de Lagrange, les dérivées par rapport au temps sont considérées explicitement et la dépendance temporelle des mouvements est obtenue directement . Ceci n'est plus le cas dans le formalisme de Hamilton où les dérivées temporelles sont remplacées par les variables $p_i$. ; il apparaît en contrepartie une série d'équations différentielles qui couplent les $p_i$ et les $q_i$. de telle sorte que la moitié seulement des coordonnées de l'espace des phases peuvent être choisies librement ; ainsi, du fait de ces conditions restrictives, seules les transformations appartenant à une certaine classe de transformation, appelées *transformations canoniques*, peuvent être appliquées aux coordonnées de l'espace des phases en respectant le formalisme de Hamilton. Les séries de coordonnées $\{\ldots,(p_i,q_i),\ldots\}$ qui obéissent aux équations de Hamilton sont appelées coordonnées canoniques.
Pour que le volume de l'espace des phases soit conservé dans un système Hamiltonien (ou conservatif), il faut qu'il soit exprimé dans un système de coordonnées canoniques.

### A1.7 Eléments de dynamique qualitative :

Nous allons introduire quelques notions et définitions de la théorie qualitative des systèmes dissipatifs. Nous avons déjà introduit la notion de **points fixes** . On les appelle aussi **points singuliers**, **points critiques** ou **points d'équilibre**. Pour étudier la dynamique du système, nous avons linéarisé les équations d'évolution ; ce faisant nous nous sommes placés dans l'**espace tangent** et avons utilisé l'**opérateur linéaire tangent**. La dynamique dans cet espace se caractérise par un ensemble de vecteurs propres associés à des valeurs propres ; ces valeurs propres décrivent la dépendance temporelle de l'évolution de chacune des directions. Cet espace se décompose en général en deux variétés, l'une de ces variétés est caractérisée par des valeurs propres négatives ; elle est appelée **variété stable** (ou **sous-espace stable)**, l'autre par des valeurs propres positives ; elle est appelée **variété instable** ou **sous-espace instable**.

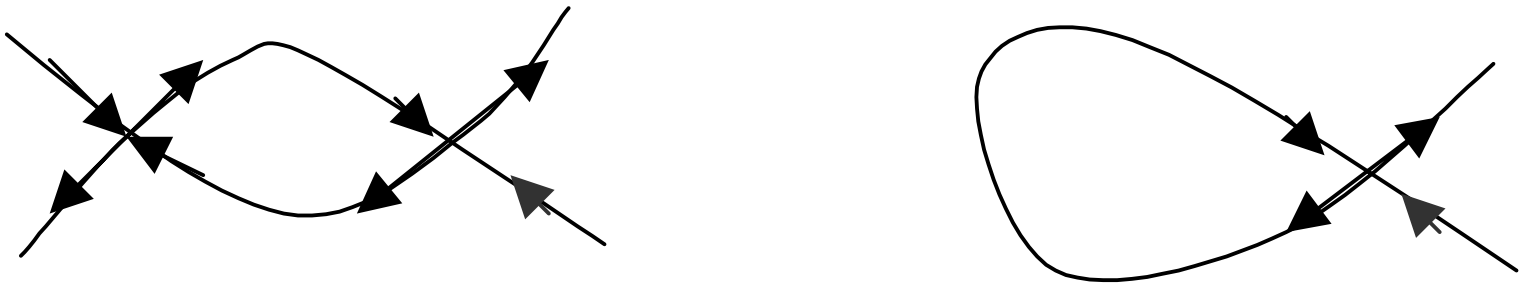

***Figure A1.1 :*** *trajectoires hétéroclines et trajectoires homoclines*

Lorsque la variété instable de l'espace tangent relatif à un point fixe de l'espace des phases est l'ensemble vide, le point est appelé **puits** (**sink** en anglais) ; un puits est donc tel que toutes les directions de l'espace sont stables en ce point. Lorsque la variété stable d'un point fixe est l'ensemble vide, le point est appelé **source**. Si certaines directions sont stables, d'autres instables, le point fixe est appelé **point co*l*** ou **point selle** (saddle en anglais).
Il existe aussi quelques règles topologiques caractérisant les variétés :
Les variétés stables des différents points fixes n'ont pas le droit de se couper, à cause de l'unicité des trajectoires passant par un point (déterminisme) ; en effet, si elles se coupaient, les points auraient deux futurs. De même, et pour des raisons similaires, deux variétés instables n'ont pas le droit de se couper. Par contre les variétés stables peuvent couper les variétés instables. Lorsque ces trajectoires relient deux points fixes différents, les trajectoires sont dites **hétéroclines** ; lorsqu'elles partent et arrivent au même point fixe, elles sont dites **homoclines**.

**Zones absorbantes, et ensembles limites :** Pour un système conservatif, le théorème de Liouville garantit la conservation du volume ; mais ce n'est plus le cas d'un système dissipatif. On appelle zone absorbante une zone où l'énergie décroît en moyenne au cours du temps ; ceci revient à considérer que le volume de l'espace des phases diminue au cours du temps dans ces zones ; les trajectoires se confinent dans un volume qui diminue continûment et tend finalement vers 0. Il atteint alors un **ensemble limite.** Cet ensemble limite peut être composé de plusieurs points fixes et de trajectoire les reliants. Dans ce cas la dynamique ultime convergera vers l'un de ces points appelé **attracteu***r*. Un attracteur est donc un sousensemble indécomposable en plusieurs parties ; il a un bassin d'attraction spécifique. Une zone absorbante est donc composée du bassin d'attraction de un ou plusieurs attracteurs ; quand il y en a plusieurs, il y a des points cols qui permettent la connexion entre eux.

**réduction de la dimension de l'espace des phases :** nous allons montrer que près d'un point fixe la dynamique est piloté par un petit nombre de modes; ainsi, on peut réduire la dimension de l'espace des phases aux alentours. Ceci provient du fait que l'espace tangent se décompose en *sous-espaces stable/central/instable* associés à des valeurs propres $s_n=\sigma_n+i\omega_n$ caractérisées par une *partie réelle* $\sigma_n$ *négative/nulle/positive.* Dans ces conditions, la dynamique linéaire imposée par l'opérateur linéaire tangent dans les directions stables convergent vite et la dynamique réelle de ces modes stables devient très vite contrôlée par les termes non linéaires couplant ces modes stables aux modes centraux et instables; l'évolution de ces modes n'a plus à être pris en compte et le système devient piloter par ses modes instables et centraux. Ce faisant on a donc bien réduit l'espace des phases.
On cherche ensuite à extrapoler cette approche au domaine non linéaire. On trouve alors la forme normale des équations qui contient des non linéarités.

## A1.8. Attracteur

Nous allons ici préciser la notion d'attracteur que nous avons introduite précédemment ; pour cela, il est utile de discuter quelques cas particuliers. Commençons par celui d'un pendule simple couplé avec dissipation, ou ce qui est équivalent, celui d'une particule dans un puits de potentiel parabolique à une dimension; si ce pendule ne dissipait pas et qu'il n'était pas excité, ses trajectoires dans l'espace des phases (q,p) ne dépendraient que des conditions initiales; ce seraient des cercles et le comportement serait périodique. Mais ce pendule dissipe; ainsi, lorsqu'il n'est pas excité, il perd constamment de l'énergie cinétique au cours du temps et sa trajectoire devient une spirale qui se terminera au point d'équilibre ($q_{équilibre}$,p=0); on peut donc considérer que le point ($q_{équilibre}$,p=0) est un point attracteur de la dynamique du pendule lorsque ce pendule n'est pas couplé à l'extérieur. On peut généraliser cette notion et dire que tout point fixe stable d'un système dissipatif isolé est un attracteur de sa dynamique; il peut cependant exister plusieurs attracteurs différents s'il existe plusieurs minima distincts, chacun ayant un bassin d'attraction spécifique.

Si maintenant ce pendule simple n'est plus isolé, mais qu'on le soumet à une excitation périodique, il se met en mouvement et atteint au bout d'un certain temps un régime stationnaire en général périodique; dans l'espace des phases, sa trajectoire prend donc finalement la forme d'un cercle, après un régime transitoire; ce cercle est indépendant des conditions initiales choisies; il est donc un attracteur de la dynamique.

On peut généraliser ces concepts à des cas plus compliqués, comme par exemple celui d'une particule dans un potentiel présentant deux minimums et un col. Dans le cas où le système n'est pas excité et n'est pas dissipatif, les trajectoires de basse énergie tournent autour de l'un ou de l'autre des minimums, les trajectoires de plus grandes énergies tournent autour des deux puits. Lorsque le système dissipe et qu'il n'est pas excité périodiquement, toutes les trajectoires se terminent au fond de l'un ou de l'autre des puits; s'il est excité périodiquement, sa trajectoire est périodique au bout d'un certain temps, centrée sur l'un ou l'autre puits si l'énergie est suffisamment faible, ou sur les deux dans le cas contraire.

Des cas plus compliqués apparaissent dès que le nombre de degrés de liberté est plus élevé. On peut montrer par exemple que les trajectoires représentant le comportement asymptotique des systèmes dissipatifs sont confinées dans des sous espaces de dimensionalité inférieure à l'espace des phases; ce peut être des formes simples (tores,...) dans des cas simples, en particulier lorsque les réponses sont linéaires. Mais ces lieux peuvent être aussi beaucoup plus compliqués lorsque le système a un comportement non linéaire; en particulier on observe souvent que ces lieux ont des **dimensionalités non entières**; de tels sous-espaces sont **fractals**; on les appelle alors des attracteurs étranges ; les comportements associés son dits « chaotiques ».

### A1.9. Remarque sur les attracteurs des systèmes non linéaires

On peut bien entendu engendrer une réponse cyclique dans un système qui répond linéairement en lui imposant une excitation cyclique; mais de nombreux systèmes physiques simples sont suffisamment non linéaires pour qu'ils puissent engendrer des réponses cycliques lorsqu'on les soumet à une excitation continue simple: le son produit par l'effet Larsen lorsqu'on couple un micro et un amplificateur, ou le courant alternatif produit par un générateur Basse Fréquence sont autant d'exemples pris en électronique; le son produit par le mouvement de l'archer sur le violon, celui produit par l'air qui excite la vibration de la hanche d'une clarinette, le phénomène du stick-slip, celui de la houle, la convection de Rayleigh-Bénard sont autant d'exemples issus de la mécanique et de la mécanique des fluides. En optique on pourrait citer le cas des lasers, des effets de saturations d'absorption.
De la même façon, la génération d'une dynamique complexe de type chaotique ne nécessite pas une excitation périodique; elle nécessite simplement que le système réponde d'une façon non linéaire de manière importante et que son nombre de degré de liberté (i.e. dimension de l'espace des phases) soit suffisant (i.e. plus grand que 3).

### A1.10. Section de Poincaré

Lorsqu'on soumet un système dissipatif à une excitation périodique de suffisamment faible intensité il répond en général de manière périodique après un certain temps d'adaptation. Ce n'est qu'en poussant l'intensité d'excitation qu'on obtient des réponses plus complexes. Dans certains cas cela se traduit par l'apparition de fréquence plus élevées (doublement de fréquence,...) et s'explique simplement par un terme non linéaire; dans d'autres au contraire, on voit apparaître des réponses à plus basses fréquences (doublement de période,....). Ce dernier phénomène est plus inquiétant car il permet d'engendrer des temps caractéristiques nettement plus longs que les périodes de base que l'on impose au départ; ce mécanisme risque donc de rendre le système imprévisible.
Par ailleurs, nous avons montré plus haut que l'espace des phases que nous devions considérer lorsqu'on excite un système périodiquement est son espace des phases normal auquel on doit rajouter une dimension supplémentaire : le temps. Ceci complique la représentation.
On peut chercher à s'affranchir de l'étude temporelle complète grâce à l'utilisation d'une méthode stroboscopique à la fréquence d'excitation ; dans ce cas, on ne relève que les états du système correspondant à la même phase $\phi$. Si cet état évolue à chaque période, le système n'est pas stationnaire; dans le cas contraire, il l'est. Le procédé revient à faire une coupe à $\phi=c^{ste}$ dans l'espace des phases; c'est pourquoi on appelle ce procédé "coupe de Poincaré", puisqu'il a été introduit par Poincaré.
Dans certains cas, il est préférable de procéder autrement. Par exemple dans le cas où l'on voudrait étudier les rebonds d'une balle sur une plaque vibrante, l'instant où les impacts ont lieux sont des instants privilégiés de la dynamique. Ils n'ont pas lieu périodiquement; si z est la hauteur de la balle à l'instant t et $z_1$ celle de la plaque, ces instants, qui correspondent à $z=z_1$, définissent une surface et on peut caractériser la trajectoire de la balle par ses instants de rebond. Ce faisant on a usé d'un procédé similaire au précédent et on parle aussi de coupe de Poincaré dans ce cas. De façon plus général, on peut définir la coupe de Poincaré comme le procédé qui consiste à caractériser la dynamique d'un système par l'ensemble des points d'intersection de sa trajectoire avec un plan ou une surface de l'espace des phases. Une coupe de Poincaré réduit l'information que contient la trajectoire complète. Elle peut donc ne pas être représentative de toute la dynamique. Il est nécessaire de faire varier la surface de coupe (ou la phase du stroboscope) pour avoir une représentation plus réaliste.

## A1-11. Bifurcations ponctuelles

### *A1.11.1. Notion de bifurcation :*

Prenons l'exemple de l'équilibre d'une masse m au sommet d'une tige rigide verticale de longueur l. Cette tige est fixée à un axe de rotation horizontale à son bout inférieur; elle est maintenue dans cette position à l'aide d'un ressort exerçant un couple de torsion c et présentant un fort frottement visqueux $\nu d\theta/dt$. Considérons le problème mécanique de ce système en négligeant les effets d'inertie, c'est-à-dire en ne considérant que le problème quasi-statique de l'équilibre mécanique. Si $\theta$ est l'angle que fait la tige avec la verticale, il s'écrit:

$$\nu d\theta/dt=-c\theta+mgl\sin\theta \qquad \text{(A1.6a)}$$

En développant $\sin\theta$ au troisième ordre et en posant $\mu=(mgl-c)/\nu$, on obtient l'équation:

$$d\theta/dt=\mu\theta-mgl/(6\nu)\theta^3 \qquad \text{(A1.6b)}$$

Les points fixes de cette équation sont $\theta_0=0$ si $\mu<0$ et $\theta_0=0$ et $\theta_1=\pm(6\nu\mu/[mgl])^{1/2}$ si $\mu>0$. L'étude de la dynamique aux alentours des points fixes montrent que la solution $\theta=0$ est stable pour $\mu<0$ et instable pour $\mu>0$; de la même façon, en utilisant les Eqs. (A1.3), on montre que les solutions $\theta_1=\pm(6\nu\mu/[mgl])^{1/2}$ pour $\mu>0$ sont stables. En $\mu=0$, il y a donc changement de la solution stable. On appelle ce phénomène une bifurcation.

### *A1.11.2. Bifurcation fourche supercritique :*

L'exemple précédent montre que le système présente deux évolutions distinctes de la grandeur $\theta$ en fonction du paramètre de contrôle $\mu$, l'un pour les $\mu$ négatifs, l'autre pour les $\mu$ positifs. Cependant cette variable $\theta$ ne présente pas d'évolution discontinue en fonction de $\mu$ ; dans ce cas, on dit que la bifurcation est supercritique ; de plus, cette bifurcation a la forme d'une fourche. On l'appelle donc *bifurcation fourche supercritique*. On l'a représenée en Fig. A1.2.a.

De plus, l'Eq. (A1.6b) montre que le temps caractéristique de la réponse est $1/\mu$ ; ainsi, il tend vers l'infini au point de bifurcation. Ce ralentissement, appelé en général **ralentissement critique**, est une des caractéristique fondamentale des bifurcation supercritiques ; physiquement, il provient du fait que le système doit choisir entre des états quasi-équivalents ; son choix peut-être modifié par les moindres fluctuations.
Pour illustrer ce propos, reprenons l'exemple de la tige chargée d'une masse m (Eq. A1.6) et chargeons la de plus en plus à son sommet, tant que sa charge ne dépasse pas c/gl, la position verticale est stable; mais dès que la masse dépasse cette valeur, la position se déplace vers la droite ou vers la gauche spontanément. Plus la charge est grande plus l'inclinaison de la barre est grande. A $\mu=0$, il y a donc un changement de comportement. Si l'on se place à $\mu=0$, la position d'équilibre est $\theta=0$, mais la force de rappel du ressort est nulle de telle sorte que si l'on écarte légèrement la masse de sa position d'équilibre, elle aura beaucoup de mal à revenir à son point d'équilibre.

*Pour généraliser ces résultats*, on constate qu'un système physique peut changer de comportement ou d'état en modifiant un de ses paramètres de contrôle. La dynamique de ce changement d'état est contrôlé par des équations d'évolution du type Eqs. (A1.3). Ces dernières peuvent représenter l'évolution de grandeurs locales ou de grandeurs globales ; elles peuvent donc décrire des systèmes évoluant dans l'espace et dans le temps ; dans ce cas elles donnent lieu à une dynamique spatio-temporelle plus ou moins complexe .

Mais nous nous sommes limités ici à des cas où ces équations représentent des grandeurs globales ou moyennes, ou des systèmes locaux ; c'est ce dernier cas que nous venons de décrire par une bifurcation fourche. Cette bifurcation, et les autres que nos allons décrire par la suite, sont caractéristiques des non linéarités des équations d'évolution (A1.3). Ainsi, la nature même de la bifurcation fourche est définie par l'équation différentielle qui la pilote, c'est-à-dire l'Eq. (A1.6) dans le cas présent. En particulier, à chaque type de non linéarité correspond un type de bifurcation.
De façon plus générale, la bifurcation supercritique de type fourche est contrôlée par l'équation différentielle de type :

$$dx/dt=\mu x-x^3 \qquad \text{(A1.7a)}$$

### *A1.11.3 Bifurcation sous-critique :*

D'autres cas physiques sont contrôlés par une équation différentielle du type :

$$dx/dt=\mu x+x^3-x^5 \qquad \text{(A1.7b)}$$

Les points fixes de cette équation sont :

| | |
|---|---|
| $x_1=0,$ | si $\mu<-1$ |
| $x_1=0,\ x_{2,3}=\pm[1+(1+\mu)^{1/2}]/2\ ,\ x_{4,5}=\pm[1-(1+\mu)^{1/2}]/2$ | si $-1<\mu<0$, |
| et $x_1=0,\ x_1=0,\ x_{2,3}=\pm[1+(1+\mu)^{1/2}]/2$ | si $0<\mu$ |

L'étude de la stabilité linéaire au voisinage de chacun des points fixes montre que le point $x_1$ est stable si $\mu<0$ mais instable si si $\mu>0$, que les points $x_{2,3}$ sont stables quand ils existent (i.e. si $\mu>-1$, et que les points $x_{4,5}$ sont toujours instables quand ils existent (i.e. $-1<\mu<0$). Elle est représentée en Fig. A1.2.b.

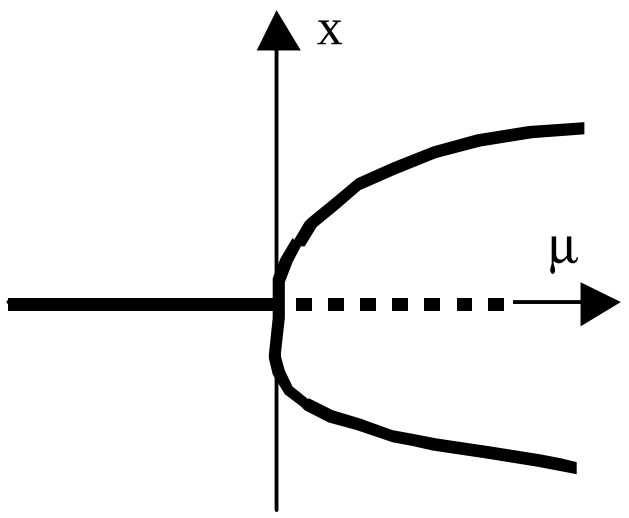

***Figure I-2.a :*** *Bifurcation fourche supercritique :* $dx/dt=\mu x-x^3$

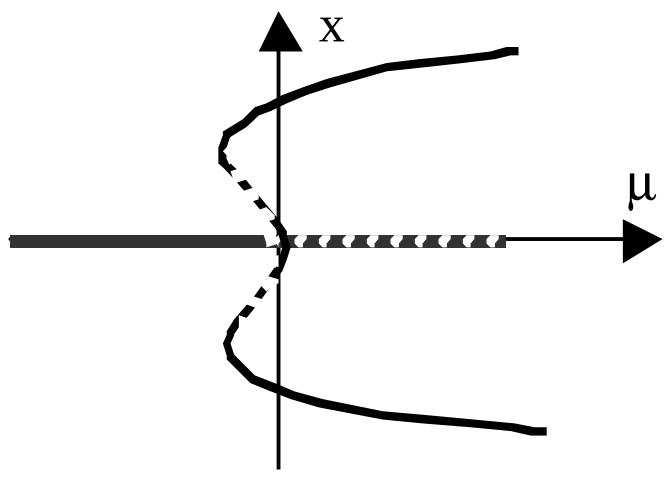

***Figure I-2.b:*** *Bifurcation souscitique:* $dx/dt=\mu x+x^3-x^5$

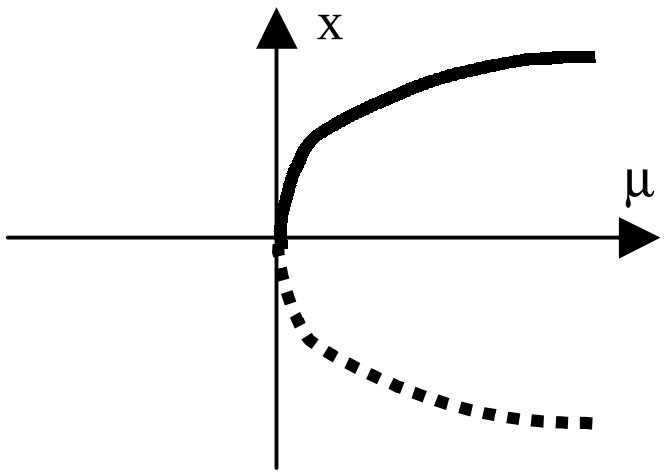

***Figure A1.2.c :*** *Bifurcation nœud-col :* $dx/dt=\mu -x^2$

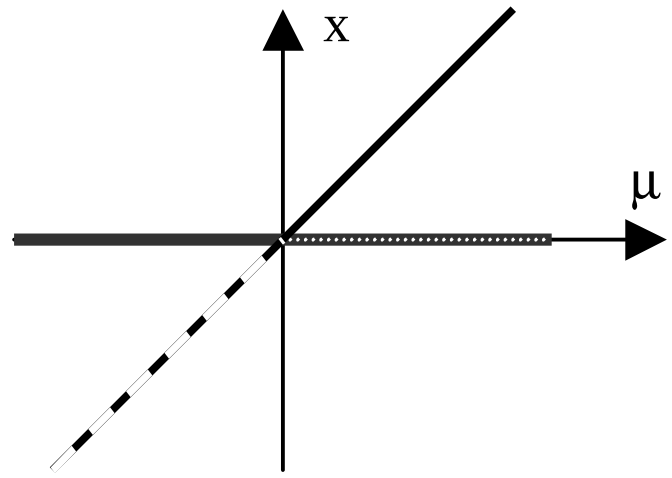

***Figure A1.2.d :*** *Bifurcation transcritique :* $dx/dt=\mu x -x^2$

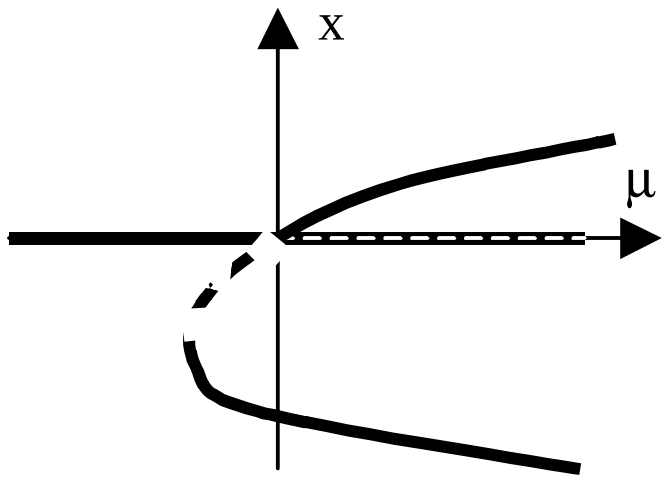

***Figure A1.2.e*** *: Bifurcation:* $dx/dt=\mu x -x^2-x^3$

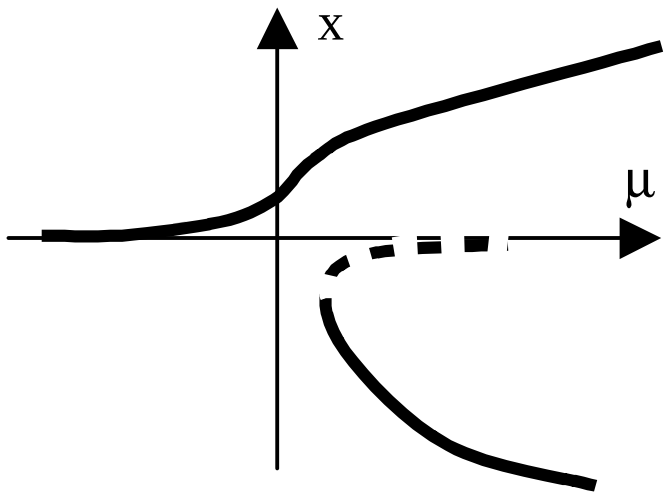

***Figure A1.2.f*** *: Bifurcation fourche imparfaite:* $dx/dt=\mu x -x^2$

L'évolution du système est alors la suivante : Si l'on part d'un μ fortement négatif, la seule solution stationnaire possible est $x_1=0$; lorsqu'on augmente μ, cette solution reste stable tant que μ est négatif et le système gardera ce comportement. La solution $x_1$ devient instable à $\mu>0$ ; le système saute alors brusquement et choisit l'une des deux solutions $x_{2,3}$ . Au fur et à mesure que $\mu$ croît, le système suit la solution $x_2$ ou $x_3$ ; il évolue donc lentement. Si on redescend la valeur du paramètre de contrôle, le système évolue continûment dans le sens inverse. Il évolue continûment lorsque $\mu$ passe la valeur $\mu=0$ puisque la solution $x_2$ (ou $x_3$) reste stable. Arrivé à $\mu<-1$, les solutions $x_2$ et $x_3$ n'existent plus; le système saute donc à la valeur $x_1=0$ qui redevient le seul comportement stable.

L'équation (A1.7a) décrit donc un système qui présente à la fois un **comportement hystérétique et une instabilité**. Un tel type de comportement est dit **souscritique**.

Les graphes de ces deux bifurcations sont représentés dans les Figures (A1.2a) et (A1.2b) ci-dessus en fonction du paramètre de contrôle $\mu$.

### *A1.11.4. Autres types de bifurcations faisant intervenir une seule grandeur*:

Il existe d'autres types de bifurcations correspondant à d'autres équations d'évolution. Nous les résumons sous forme de graphes dans les Figs. (A1.2.c-f) suivantes. Dans toutes ces figures, les points fixes stables sont indiqués par des traits pleins épais, les points fixes instables par des traits épais en pointillés. Nous donnons les formes normales des équations qui les gouvernent dans la légende de ces figures.

### *A1.11.5. Bifurcation faisant intervenir deux grandeurs couplées:*

D'autres bifurcations font intervenir deux coordonnées couplées. C'est le cas de la bifurcation de Hopf dont un exemple physique est l'oscillateur de Van der Pol. Dans ces cas, les états stationnaires attracteurs ne sont plus obligatoirement des points fixes, mais des cycles limites.

La forme normale de la bifurcation de Hopf est

$$dz/dt=\mu z -|z|^2 z \qquad \text{(A1.8a)}$$

où z est une variable du plan complexe faisant intervenir deux coordonnées séparées : $z=x+iy$. De la même façon $\mu$ est complexe : $\mu=\mu_r+i\mu_i$ . Si on pose $z=xe^{i\theta}$ , l'équation dynamique de z se transforme dans les deux équations :

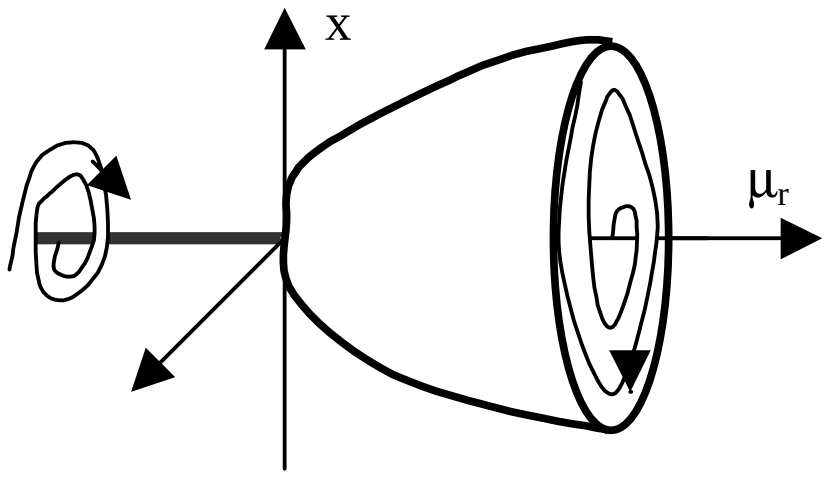

***Figure A1.3:*** *Bifurcation de Hopf :* $dz/dt=\mu z -|z^2|z$, avec $z=xe^{i\theta}$
*soit :* $dx/dt=\mu_r x - x^3$ et $d\theta/dt=\mu_\iota$

(A1.8b) $dx/dt=\mu_r\, x - x^3$

(A1.8c) $d\theta/dt=\mu_\iota$

La coordonnée θ croît ou décroît linéairement avec le temps, tandis que la coordonnée x décrit une bifurcation fourche supercritique. Les trajectoires sont donc des spirales convergeant vers le point x=0 si $\mu_r$ est négatif, ou convergeant vers le cycle x= $\mu_r^{1/2}$ si $\mu_r$ est positif. Le sens de la rotation est donné par $\mu_i$ .
On peut transformer cette bifurcation de Hopf en bifurcation souscritique en changeant le signe du terme $|z|^2z$ et en ajoutant un terme $-|z|^4z$ .

### *A1.11.6 Analogie avec les transitions de phase :*

Lorsqu'un échantillon subit une transition de phase, on constate qu'une de ses propriétés physiques change de type de comportement lorsqu'un paramètre de contrôle passe par une valeur critique. Pour définir une transition de phase, il faut donc un paramètre de contrôle, telle que la température, et un paramètre d'ordre qui caractérise la propriété physique ; celui-ci peut être le volume spécifique moyen pour la transition liquide-gaz et les transitions solide-solide, ou l'aimantation spontané m pour les transitions magnétiques. Par ailleurs, suivant les transitions, on observe soit une variation brutale, soit une discontinuité de pente du paramètre d'ordre en faisant varier le paramètre de contrôle. Lorsque la variation du paramètre d'ordre est brutale (c'est-à-dire discontinue), on parle de transition du premier ordre ; lorsque cette variation est continue, mais qu'elle présente une discontinuité de pente, on parle de transition de phase du deuxième ordre.

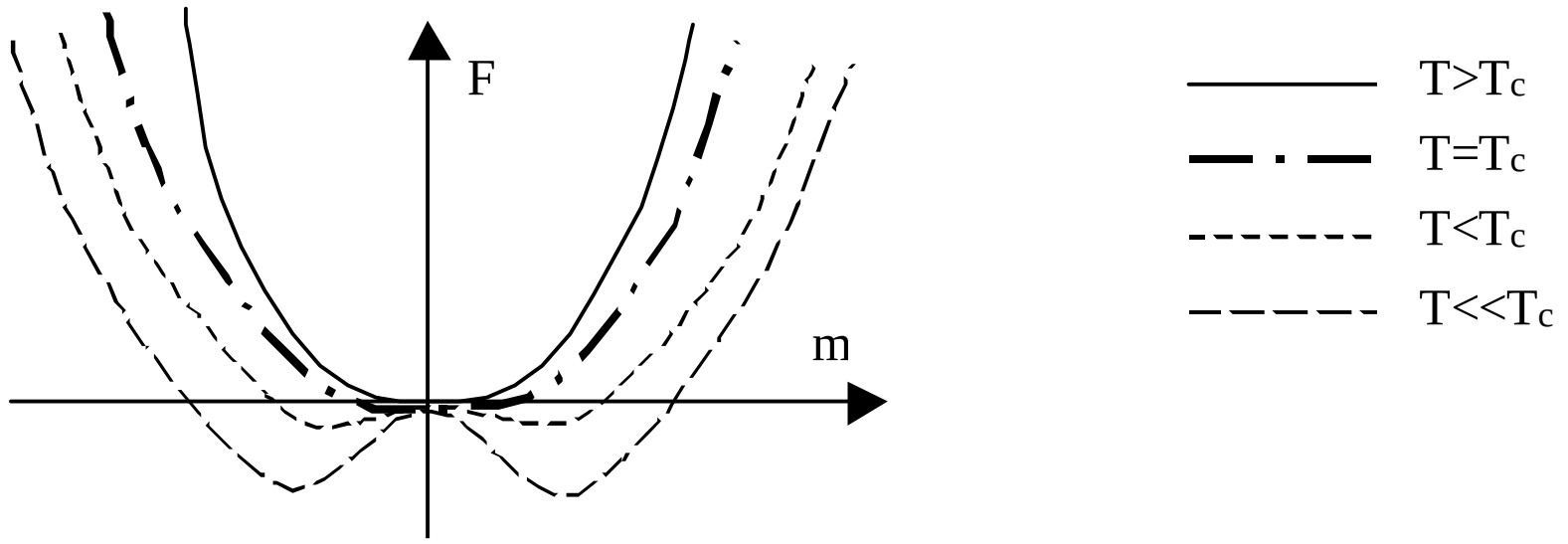

***Figure I-4*** *: Energie libre F d'un système présentant une transition ferromagnétique-paramagnétique en fonction de l'aimantation spontanée m, pour différentes température. En première approximation, F est décrit par l'équation :* $F=a(T)m^2+b(T)m^4$

Des exemples de transitions de phases du premier ordre sont les transitions liquides-gaz, liquide-solide, gaz-solide, solide-solide pour lesquels le paramètre d'ordre est le volume spécifique. Ce dernier varie bien de façon discontinu au point de transition.
Des exemples de transition de phase du deuxième ordre sont les transitions liquide-gaz au point critique, certaines transitions ordre-désordre, les transitions ferromagnétiques-paramagnétiques, la transition de percolation, les transitions conducteur-supraconducteur et fluide-superfluide,… Leurs paramètres d'ordre respectifs sont la densité du milieu, la fonction de corrélation (ou probabilité qu'un site A soit occupé par un atome de type A), l'aimantation spontanée, la conduction , « la concentration en états supraconducteurs » et la « concentration en liquide superfluide » (la définition de ces deux derniers termes mériteraient d'être précisée) …

On peut dresser une analogie entre les bifurcations supercritiques et les transitions de phase du deuxième ordre. Nous la ferons à partir de la transition ferromagnétique-paramagnétique d'un système d'Ising; dans ce cas, tous les moments magnétiques locaux ont un module fixe m et ne peuvent être orientés que dans une seule direction (Ox); on a ainsi $m_y=m_z=0$, $m_x=\pm M$; chaque moment magnétique interagit avec ses $N_1$ plus proches voisins et l'énergie d'interaction pour chaque couple est $\pm\mu M^2$ (négative si les deux moments ont le même sens, positive si les deux moments sont en sens inverse). On peut calculer l'énergie libre F(m) d'un tel système présentant une aimantation moyenne m; F est la somme d'un terme d'énergie qui pousse les moments à être orientés dans le même sens et d'un terme entropique qui les poussent à se mettre dans un sens différents; elle dépend de la température T. On trouve que F(m) présente un seul minimum à m=0 lorsque la température T est plus grande qu'une certaine valeur $T_c \approx N_1 \mu M^2/k_B$, où $k_B$ est la constante de Boltzman .
En revanche, quand $T<T_c$, on trouve que F(m) présente deux minimums $F(m_1)=F(m_2)$ placés en $m_1$ et $m_2$ , symétriquement l'un de l'autre par rapport à l'origine (soit $m_2=-m_1$) et tel que le point F(0) est devenu un maximum local; on trouve aussi que $m_1$ et $m_2$ tendent vers 0 ainsi que $F(m_1)-F(0)$ et $F(m_2)-F(0)$.

On peut donc dire que la position d'équilibre thermodynamique stable de ce système est soit $m_1$ soit $m_2$ lorsque $T<T_c$, et qu'elle est m=0 pour $T>T_c$. Supposons maintenant qu'on soumette le système à un champ magnétique H; ce dernier modifiera la fonction F et déplacera la position d'équilibre du système, créant une aimantation moyenne différente de celle de l'équilibre sans champ. Si on coupe brusquement le champ (i.e. plus rapidement que le temps de réponse des moments magnétiques), l'aimantation spontanée hors d'équilibre subsistera quelques instants, ce qui veut dire que l'énergie libre du système ne sera plus minimum; une dynamique se créera pour rejoindre l'équilibre thermodynamique; elle sera pilotée par la courbure de l'énergie libre. L'évolution de l'aimantation pourra donc se formaliser d'une manière très sensiblement équivalente à celle d'un particule dans un champ de force.
Dans cet exemple, on peut donc analyser la transition de phase à $T_c$ comme un processus de bifurcation supercritique qui apparaît lors du refroidissement de l'échantillon au dessous de $T_c$ ; ce sera une bifurcation de type fourche supercritique qui se caractérisera par un ralentissement critique. Pour cela il suffit de considérer que le paramètre de contrôle est la température et que la dynamique que l'on étudie est celle de l'aimantation spontanée. Cette analogie est donc plus que formelle.

Cependant, cette interprétation cache une difficulté, car la dynamique d'un ensemble de particules couplées (i.e. ici les moments magnétiques de chaque atome) est plus complexe que celle d'un oscillateur simple; en particulier elle donne lieu à des effets coopératifs qui n'existent pas dans le cas de l'oscillateur simple; de tels effets coopératifs renforcent la lenteur , caractérisée par le temps caractéristique $\tau(T-T_c)$, de la réponse; ils modifient aussi la loi reliant l'aimantation spontanée à l'équilibre $m(T-T_c)$ à la différence de température $T-T_c$ , ainsi que celle de la susceptibilité magnétique $\chi(T-T_c)$. Ces lois sont alors contrôlées par des exposants critiques différents des exposants obtenus pour des dynamiques à une particule. Il n'empêche que ces phénomènes de transition de phase peuvent se formaliser totalement dans le langage des systèmes dynamiques et avec le même type d'équations.

En d'autres mots, la différence essentielle entre ces deux types de problèmes provient du processus de réduction de la dynamique de l'espace des phases aux seules quantités pertinentes (variétés instables). En effet, cette réduction de l'espace aux seules grandeurs physiques mesurables et pertinentes concerne un grand nombre de variables équivalentes dans le cas des transitions de phases (les moments magnétiques locaux dans la transition ferro-paramagnétique) ; si la transition est du $2^{ème}$ ordre, ces quantités fluctuent de façon corrélées et la réduction se fait d'autant plus difficilement qu'on approche du point critique; on s'aperçoit alors que le temps de convergence pour calculer ou simuler l'évolution de ces grandeurs diverge anormalement lorsqu'on s'approche du point critique car cela requiert de tenir compte précisément des configurations d'un nombre de plus en plus grand de particules (tels les moments magnétiques de l'exemple ci-dessus); en d'autres termes encore, près de $T_c$ , les moyennes locales ne sont plus représentatives et la théorie de champ moyen plus adéquate pour calculer l'évolution réelle.
Au contraire, dans un système dynamique classique cette réduction se fait facilement car seul un petit nombre de paramètres impose leur dynamique au reste des variables.

## Bibliographie :

# Appendice A2:
# Quelques techniques expérimentales pour caractériser le milieu granulaire :

Un certain nombre de techniques expérimentales ont déjà été décrites à la fin du chapitre 6 , *i.e.* § 6.4. Ces techniques avaient pour but d'apprécier la « coulabilité » d'un matériau granulaire. Elles cherchaient donc toutes à apprécier les qualités mécaniques du matériau ; ainsi, on peut chercher à relier les grandeurs qu'elles mesurent aux caractéristiques mécaniques mesurer par un triaxial. Nous avons vu en particulier que ces techniques mesurent essentiellement les volumes spécifiques maximum et minimum du milieu.
Dans cet appendice, nous décrivons brièvement d'autres techniques expérimentales, couramment utilisées dans l'industrie, sur le terrain ou au laboratoire pour caractériser d'autres caractéristiques physiques des matériaux granulaires.

## A2.1. **distribution granulométrique** :

Elle se détermine par tamisage pour les grains supérieur à 40mm et par sédimentologie pour les tailles inférieures. On la caractérise par le rapport $d_{60}/d_{10}$ des diamètres des mailles du tamis qui laissent passer respectivement 60% et 10% en masse des grains.
On peut aussi utiliser un granulomètre laser qui mesure par diffraction la taille des grains.

## A2.2. nature chimique

La nature chimique des matériaux en présence influence nettement les propriétés mécaniques. Par exemple, on constate très souvent qu'un milieu granulaire est caractérisé par les mêmes valeur du frottement solide qu'il soit sec ou saturé d'eau. Ceci ne peut s'expliquer qu'en considérant que le milieu "sec" est la plupart du temps déjà recouvert d'une couche moléculaire d'eau qui a modifié les qualités de frottement de la surface.

## A2. 3. essais mécaniques :

Essai triaxial, essai oedométrique, boîte de cisaillement, boîte de cisaillement annulaire, boîte de cisaillement homogène, pénétromètre, pénétromètre dynamique, pressiomètre…

## A2.4. influence de l'eau :

Un milieu granulaire naturel contient toujours une certaine quantité d'eau condensée dans ses pores. C'est donc en général un milieu triphasique grain-eau-air.

Lorsque l'eau est en très faible quantité, elle ne recouvre pas la totalité de la surface. De plus la rugosité de cette dernière empêche la migration de l'eau et confine l'eau à certains endroits, empêchant cette dernière d'atteindre les contacts; le milieu semble donc sec avec des forces capillaires quasi nulles. Cependant, lorsque la proportion d'eau est augmentée, les “puits” rugueux de la surface des grains sont saturés en eau et l'eau excédentaire peut migrer venant mouiller les points de contacts. Des forces capillaires sont développées au sein du matériau.
Si l'on augmente encore la proportion d'eau, certains des pores du milieu granulaire sont saturés d'eau, d'autres partiellement de telle sorte que les milieux gazeux et liquide peuvent être tous les deux percolants et les pores le siège de courants. Pour décrire la mécanique d'un tel système, il sera nécessaire d'introduire trois phases différentes avec des mécaniques couplées. Outre le tenseur des contraintes granulaires, il faudra introduire la pression de l'eau $u_w$ et du gaz $u_g$, ainsi que les deux lois d'écoulement et les équations de conservation de la matière.

Toujours est-il qu'il s'avère souvent important de mesurer la teneur en eau d'un matériau. Pour cela, il existe trois méthodes principales :

### *A2.4.1 Pycnomètre :*

Un pycnomètre est petit récipient en verre de volume et de poids calibrés qu'on remplit de sable. On pèse le tout, puis on pompe et on le passe à l'étuve ; une dernière pesée permet de connaître la masse $\delta m$ d'eau évaporée . On tire ainsi des trois pesées la masse du sable, la masse d'eau, la teneur en eau,…. C'est une méthode simple, facile d'emploi et fiable, qui ne peut pas être automatisée.

### *A2.4.2 Mesure capacitive de la teneur en eau :*

Le principe est de mesurer la capacité d'un condensateur de géométrie donnée et contenant le sable à étudier. En effet, l'eau est un matériau de forte permittivité électrique, grandeur qui affecte la valeur C du condensateur. En théorie, C devrait être proportionnelle à la teneur en eau. Dans la pratique, il faut étalonner la mesure en fonction du sable et de sa densité ; il faut aussi travailler à 50kHz pour éviter les problèmes de polarisation du matériau.
La mesure de résistance électrique est moins précise à cause des problèmes de polarisation.

### *A2.4.3. psychromètre*

Un psychromètre est un double thermocouple, l'un travaille à température ambiante, dans le sable. L'autre est aussi placé dans le sable, mais il est d'abord refroidi par effet Pelletier pour qu'une fine pellicule d'eau se condense dessus (l'épaisseur de la couche est de quelques nm) . On mesure ensuite la température d'équilibre lors de l'évaporation de l'eau et on en déduit la pression de vapeur saturante.
Dans la pratique, il est nécessaire d'étalonner soigneusement l'appareil.
Si l'ambiance est très sèche, il existe des psychromètres plus précis qui fonctionnent avec des transistors.

On peut chercher à mesurer la porosité du matériau par la porosimétrie à mercure ou à He, ou l'absorption des rayons $\gamma$, à mesurer la surface spécifique « des grains » par des mesures d'isotherme d'adsorption. On peut aussi mesurer la perméabilité du matériau à l'eau en mesurant le rapport entre le débit traversant une section du matériau et le gradient de pression qui lui est appliqué. On peut aussi mesurer des vitesses de diffusion de molécules d'eau grâce à l'IRM.
Enfin, il existe des techniques de visualisation 3-d (scaner, IRM) ce qui permet de visualiser les localisations des déformations.

# Bibliographie Générale:

# Table des matières:

**Appendices:**

The electronic arXiv.org version of this paper has been settled during a stay at the Kavli Institute of Theoretical Physics of the University of California at Santa Barbara (KITP-UCSB), in june 2005, supported in part by the National Science Fundation under Grant n° PHY99-07949.

*Poudres & Grains* can be found at :
http://www.mssmat.ecp.fr/rubrique.php3?id_rubrique=402